\pdfoutput=1
\documentclass[12pt]{cernrep}

\usepackage{subfigure}
\usepackage{floatflt}
\usepackage{multirow}
\usepackage{amsmath}
\usepackage{amssymb}
\usepackage[english]{babel}
\usepackage{epsfig}
\usepackage{booktabs}
\usepackage{cite}
\usepackage{graphicx}
\usepackage{color}
\usepackage{slashed}

\input commands.tex

\begin{document}

\setcounter{tocdepth}{0}
\pagestyle{myheadings}
\thispagestyle{empty}

\title{
NEW PHYSICS AT THE LHC: A LES HOUCHES REPORT\\[4mm]
Physics at TeV Colliders 2007 -- New Physics Working Group\\ 
}

\author{
\textbf{G.~Brooijmans}$^{1}$, 
\textbf{A.~Delgado}$^{2}$,  
\textbf{B.A.~Dobrescu}$^{3}$, 
\textbf{C.~Grojean}$^{4,5}$, 
\textbf{M.~Narain}$^{6}$, 
J.~Alwall$^7$, 
G.~Azuelos$^{8,9}$, 
K.~Black$^{10}$, 
E.~Boos$^{11}$, 
T.~Bose$^{6}$,  
V.~Bunichev$^{11}$, 
R.S.~Chivukula$^{12}$, 
R.~Contino$^4$, 
A.~Djouadi$^{13}$, 
L.~Dudko$^{14}$,  
J.~Ferland$^{8}$, 
Y.~Gershtein$^{15}$, 
M.~Gigg$^{16}$, 
S.~Gonzalez de la Hoz$^{17}$, 
M.~Herquet$^{18}$, 
J.~Hirn$^{19}$, 
G.~Landsberg$^6$,
K.~Lane$^{20,21}$, 
E.~Maina$^{22}$, 
L.~March$^{17}$,
A.~Martin$^{19}$, 
X.~Miao$^{23}$, 
G.~Moreau$^{13}$,
M.M.~Nojiri$^{24}$, 
A.~Pukhov$^{25}$, 
P.~Ribeiro$^{26}$, 
P.~Richardson$^{4,14}$, 
E.~Ros$^{17}$,
R.~Rosenfeld$^{27}$, 
J.~Santiago$^{3,28}$, 
V.~Sanz$^{20}$, 
H.J.~Schreiber$^{29}$, 
G.~Servant$^{4,5}$, 
A.~Sherstnev$^{14,30}$, 
E.H.~Simmons$^{12}$,
R.K.~Singh$^{13,21}$,
P.~Skands$^{3,4}$,
S.~Su$^{23}$, 
T.M.P.~Tait$^{31, 32}$, 
M.~Takeuchi$^{33}$, 
M.~Vos$^{17}$, 
D.G.E.~Walker$^{34,35}$. 
}


\institute{\vspace{1cm}
\textbf{convenor} of {\em Non SUSY New Physics} working group\\
\vspace{.3cm}
$^1$ Physics Department, Columbia University, New York, NY 10027, USA\\ 
$^2$ Dpt. of Physics, University of Notre Dame, Notre Dame,  IN 46556, USA \\
$^3$ Fermilab, PO Box 500, Batavia, IL 60510, USA\\ 
$^4$ Physics Department, Theory Unit, CERN, CH-1211 Geneva 23, Switzerland\\
$^5$ IPhT, CEA-Saclay, Orme des Merisiers, F-91191 Gif-sur-Yvette Cedex, France\\  
$^6$ Department of Physics, Brown University, Providence, RI 02912 , USA\\
$^7$ SLAC, 2575 Sand Hill Road, Menlo Park, CA 94025-7090, USA\\
$^8$ Universit\'e de Montr\'eal, Montr\'eal, Canada\\
$^9$ TRIUMF, Vancouver, Canada\\
$^{10}$ Lab. for Particle Physics and Cosmology, Harvard University, Cambridge, MA 02138, USA\\
$^{11}$ Skobeltsyn Institute of Nuclear Physics, MSU, 119992 Moscow, Russia\\
$^{12}$ Dpt. of Physics and Astronomy, Michigan State University, East Lansing, MI 48824, USA\\
$^{13}$ LPT, CNRS and U. Paris--Sud, F-91405 Orsay Cedex, France\\
$^{14}$ IPPP, University of Durham, South Rd, Durham DH13LE, UK\\
$^{15}$ Department of Physics, Florida State University, Tallahassee, FL32306, USA\\
$^{16}$ IPPP, Durham University, South Rd, DH1 3LE, UK\\ 
$^{17}$ IFIC - centre mixte Univ. Val\`encia/CSIC, Valencia, Spain\\
$^{18}$ CP3, Universit\'e catholique de Louvain, B-1348 Louvain-la-Neuve, Belgium\\ 
$^{19}$ Department of Physics, Sloane Lab, Yale University, New Haven CT 06520, USA\\
$^{20}$ Department of Physics, Boston University, Boston, MA 02215, USA\\
$^{21}$ LAPTH, F-74941, Annecy-le-Vieux, France\\
$^{22}$ INFN and  Universit\`a di Torino, 10125 Torino, Italy\\
$^{23}$ Department of Physics, University of Arizona, Tucson, AZ 85721, USA\\
$^{24}$ Theory Group, KEK, Tsukuba, 305-0801, Japan\\
$^{25}$ Faculty of Physics 1, Moscow State University, Leninskiye Gory, Moscow, 119992, Russia\\
$^{26}$ LIP, Av. Elias Garcia 14, 1000-149 Lisboa, Portugal\\
$^{27}$ IFT, Universidade Estadual Paulista, S\~ao Paulo, Brazil\\
$^{28}$ ITP, ETH, CH-8093 Z\"urich, Switzerland\\
$^{29}$ DESY, Deutsches Elektronen-Synchrotron, D-15738  Zeuthen, Germany\\
$^{30}$ Cavendish Laboratory, Cambridge University, Madingley Road, Cambridge CB3 0HE, UK\\
$^{31}$ Department of Physics and Astronomy, Northwestern University, Evanston, IL 60208, USA\\
$^{32}$ Argonne National Laboratory, Argonne, IL 60439, USA\\
$^{33}$ Yukawa Institute for Theoretical Physics, Kyoto University, Kyoto 606-8502, Japan\\
$^{34}$ Department of Physics, University of California, Berkeley, CA 94720, USA\\
$^{35}$ Theoretical Physics Group, Lawrence Berkeley National Lab., Berkeley, CA 94720, USA
\vspace{1cm}
}

\maketitle

\vspace{1cm}

\begin{abstract}
We present a collection of signatures for physics beyond the 
standard model that need to be explored at the LHC. 
The signatures are organized according to the experimental 
objects that appear in the final state, and in particular 
the number of high $p_T$ leptons. 
Our report, which includes brief experimental and theoretical reviews 
as well as original results, summarizes the activities of the 
``New Physics'' working group for the ``Physics at TeV Colliders"
workshop (Les Houches, France, 11--29 June, 2007).
\end{abstract}

\vspace{3cm}

\section*{ACKNOWLEDGEMENTS}

We would like to heartily thank the funding bodies, the organisers 
(P.~Aurenche, G.~B\'elanger, F.~Boudjema, J.P.~Guillet, S.~Kraml, R.~Lafaye, M.~M\"uhlleitner, E.~Pilon, P.~Slavich and D.~Zerwas), the staff and the other
participants of the Les Houches workshop for providing a stimulating and
lively environment in which to work. 

\newpage

\tableofcontents

\newpage

\part{Introduction}

{\it G.~Brooijmans, A.~Delgado, B.A.~Dobrescu, C.~Grojean and M.~Narain}

The exploration of the energy frontier will soon enter a dramatic 
new phase. With the startup of the LHC, planned for later this year,
collisions at partonic center-of-mass energies above the TeV scale 
will for the first time be observed in large numbers.

The Standard Model currently provides an impressively accurate 
description of a wide range of experimental data. Nevertheless, 
the seven-fold increase in the center-of mass energy compared to the 
current highest-energy collider, the Tevatron, implies that the 
LHC will probe short distances where physics may be fundamentally 
different from the Standard Model.
As a result, there is great potential for paradigm-changing discoveries, but at the 
same time the lack of reliable predictions for physics at the TeV scale
makes it difficult to optimize the discovery potential of the 
LHC. 

The TeV scale has been known for more than 30 years to be the  
energy of collisions required for revealing the origin of electroweak 
symmetry breaking. The computation of the amplitude for longitudinal 
$WW$ scattering \cite{Lee:1977eg} shows that perturbative unitarity 
is violated unless certain new particles exist at the TeV scale. 
More precisely, either a Higgs boson or some spin-1 particles 
that couple to $WW$ (as in the case of Technicolor or Higgsless models)
are within the reach of the LHC, or else quantum field 
theory is no longer a good description of nature at that scale. 

Given that ATLAS and CMS are multi-purpose detectors, it is commonly believed that
they will provide such an in-depth exploration of the TeV scale that the nature 
of new physics will be revealed. Although this is likely to be true, one
should recognize that the backgrounds will be large, and an
effective search for the manifestations
of new physics would require a large number of analyses dedicated to particular 
final states.
Other than the unitarity of longitudinal $WW$ scattering,
there are no clear-cut indications of what the ATLAS and CMS 
experiments might observe. 
Furthermore, recent theoretical developments have shown that the range of 
possibilities for physics at the TeV scale is very broad.
Many well-motivated models predict various new particles which
may be tested at the LHC. Hence, it would be useful to analyze as many of them as 
possible in order to ensure that the triggers are well-chosen and that 
the physics analyses have sufficient coverage.

The purpose of this report is to provide the LHC experimentalists 
with a collection of signatures for physics beyond the Standard 
Model organized according to the experimental objects that appear 
in the final state. The next four sections are focused on final
states that include, in turn, three or more leptons, 
two leptons, a single lepton, and no leptons.
Section 6 then describes an interface for event generators used 
in searches for physics beyond the Standard Model.

Whatever the nature of TeV scale physics is, the LHC
will advance the understanding of the basic laws of physics.
We hope that this report will help the effort of the particle physics community 
of pinning down the correct description of physics at the TeV scale.


%






\AddToContent{G.~Brooijmans, A.~Delgado, B.A.~Dobrescu, C.~Grojean and M.~Narain}
\setcounter{figure}{0}
\setcounter{table}{0}
\setcounter{section}{0}
\setcounter{equation}{0}
\setcounter{footnote}{0}
\clearpage

\superpart{Multi-Lepton Final States}


\part{Four leptons + missing energy from one UED}

{\it M.~Gigg and P.~Ribeiro}

\begin{abstract}
Minimal Universal Extra Dimensions (MUED) models predict the presence of massive
Kaluza-Klein particles decaying to final states containing 
Standard Model leptons and jets.
The multi-lepton final states provide the cleanest signature. The
ability of the CMS detector to find MUED final state signals with
four electrons, four muons or two electrons and two muons
was studied. The prospect of distinguishing between MUED and 
the Minimal Supersymmetric Standard Model (MSSM)
is then discussed using simulations from the event generator Herwig++.
\end{abstract}

\section{DISCOVERY POTENTIAL FOR THE FOUR LEPTON FINAL STATE }

\subsection{Introduction}
\label{sec:ued4lepton_intro}

The Universal Extra Dimensions (UED) model~\cite{Appelquist:2000nn} is an
extension of the sub-millimeter extra dimensions model
(ADD)~\cite{Arkani-Hamed:1998rs,Arkani-Hamed:1998nn}
in which all Standard Model (SM) fields, fermions as well as bosons,
propagate in the bulk. In the minimal UED (MUED) scenario~\cite{Cheng:2002ab}    
 only one Extra Dimension (ED) compactified on an orbifold is needed to create 
an infinite number of excitation modes of Kaluza-Klein (KK) particles 
with the same spin and couplings as the corresponding SM particles. 
The mass spectrum of the KK particles is defined by three free parameters:
 $\rm R^{-1}$, the size of the ED, given in terms of the  compactification radius;
 $\Lambda$R, the number of excitation modes (KK levels) allowed in the effective theory; 
 and $\rm m_{H}$, the SM Higgs boson mass. KK partners are indicated with the
subscript related to the $n$-th mode of excitations (e.g. at the first
level they are $\rm g_1,Z_1,u_{L1},e_{R1},\gamma_1$).
A direct search for MUED in the multi-lepton channel at Tevatron energy of 1.8 TeV~\cite{Lin:2005ix}
set a lower bound on the size of ED of $\rm R^{-1}>280~\rm{GeV}$. Also,  constraints from dark matter  infer $600~< \rm{R^{-1}} < 1050~\rm{GeV}$ ~\cite{Servant:2002aq}. 

In this section a summary report on the discovery potential of the CMS experiment~\cite{CMS_TDR1} for MUED 
is presented. The complete analysis is described in~\cite{CMS_Note_2006-008}. 
The experimental signatures for production  of first level KK states at hadron colliders are isolated leptons and/or jets
radiated in the cascade decay process, in addition to the transverse
missing energy carried away by the lightest KK particle (LKP).
These characteristics were exploited to discriminate the signal from the background.
The four lepton final state constitutes the cleanest channel. The KK mass spectrum, however, is highly degenerate
since the masses of the KK particles with respect to the corresponding SM particles
at tree level are  $m_{n}^2={n}^2/R^2+m_{SM}^2$, where $n$ is the excitation mode.
Furthermore, radiative corrections do not introduce an additional large splitting and typically,
within the same excitation mode,
there is a difference of about 100 GeV between the heaviest and the lightest KK particle.  
Therefore, the average values of the lepton momentum and the missing transverse
energy are typically smaller than average values which characterise searches for supersymmetric events.

\subsection{Signal and background processes} \label{sec:ued4lepton_processes}

The MUED signal is produced in a $\rm{pp}$ collision as a pair of two
KK strongly interacting particles, gluons ($\rm g_1$) or quarks ($\rm q_1$).
Three significant subprocesses were considered: 
\begin{center}
$   pp  \rightarrow g_{1} g_{1},  
~~ pp  \rightarrow Q_{1}/q_{1} Q_{1}/q_{1}, 
~~ pp  \rightarrow g_{1} Q_{1}/q_{1},$
\end{center}
Singlet and doublet KK quarks of the first generation were taken into
account. Four points of the MUED parameter space have been chosen for the study: 
$\rm{m_H=120\,GeV}$, $\rm{\Lambda R=20}$ and 
$\rm{R^{-1}\in\{300, 500, 700, 900\}\,GeV}$.
The total cross section strongly depends on
the compactification radius being equal to 2190, 165, 26 and 5.86 pb 
for $\rm{R^{-1}}=300,500,700$ and $900$ GeV respectively. 
The four lepton final state signature can provide a discrimination against the 
SM background and is considered as in the following:
\begin{equation} 
\label{eq:ued4lepton_brs}
  g_1 \rightarrow  Q_1 Q\,,
~~ Q_1 \rightarrow  Z_1 Q\,,
~~ Z_1 \rightarrow  L_1 \ell^{\pm},
~~ L_1 \rightarrow  LKP(\gamma_1)\ell^{\mp}.
\end{equation} 
The KK gluon ($\rm g_1$) decays into a KK quark ($\rm q_1$) and a SM anti-quark;
then, the $\rm q_1$ decays into the KK boson (Z$_{1}$)
and a SM quark. Subsequently,  Z$_{1}$ decays into a pair
of leptons, one being a KK lepton ($\rm l_1\equiv$  singlet $\rm l_{R1}$ or, 
mainly, doublet $\rm l_{L1}$). 
Finally,  $\rm l_1$ can decay
only into the LKP photon ($\gamma_1$) and a SM lepton. 
Instead, if $\rm q_1$ is produced initially, then the decay cascade is shorter.
The B.R. of the four lepton final state is about $\ 10^{-4}-10^{-3}$. 
Within a decay branch the pair of SM leptons ($ \ell^{\pm} \ell^{\mp}
$)  has the same flavour and opposite sign.  Three possible combinations
of four leptons arise, namely 4$e$, 4$\mu$ and 2$e$2$\mu$, studied in
three separated channels. Signal events were generated with CompHEP
 with particle definitions and Feynman rules taken from 
\cite{Cheng:2002iz} at the LO approximation. 
The background to MUED signals results from SM processes with four leptons
in the final state. The dominant sources are from the continuum production of
$\rm{ (Z^{*}/\gamma^{*})( Z^{*}/\gamma^{*})}$ 
and real $\rm{ZZ}$ production, from processes involving pair production of heavy quark 
flavours such as $\rm{ t \bar{t}}$ and $ \rm{b\bar{b}b\bar{b}}$, and
the associated production of $\rm{Z b\bar{b}}$. Background events were generated with PYTHIA 
and ALPGEN.
Signal and background events were processed with full detector simulation
 using official CMS software (OSCAR version 3.6.5). Underlying events
from minimum bias interactions were superimposed to generated events,
assuming an average number of~5 inelastic collisions,
including diffractive interactions, at each beam crossing, simulating the
effect of pile-up at $\rm {\mathcal{L}}=2\cdot 10^{33}\rm{cm^{-2}s^{-1}}$
(LHC low luminosity scenario).  
The reconstruction of physics objects was based on the dedicated CMS 
software ORCA (version 8.7.3/4). 


\subsection{Event Selection} \label{sec:ued4lepton_eventSelection}

First, Level 1 (L1) and High Level trigger (HLT) requirements for 
$\rm {\mathcal{L}}=2\cdot 10^{33}\rm{cm^{-2}s^{-1}}$ are applied to
the simulated events.
We then require the presence of at least two pairs of OSSF leptons.
The leptons should be isolated (4 iso) and are required to be within
the following kinematical boundaries ($\varepsilon_{2}$):

\begin{tabular}{ll}
 $\bullet$&
 electrons with $p_{T}$ $>$7.0 GeV and $|\eta|<$ 2.5,\tabularnewline
 $\bullet$&
 muons ~~\, with $p_{T}$ $>$5.0 GeV and $|\eta|<$ 2.4. \tabularnewline
\end{tabular}
 
Because a substantial fraction of the background leptons results from b-quark
leptonic decays, we reject events where one or more b-jets are identified (Bveto).
Due to the soft KK mass spectrum, leptons from the MUED cascade (eq.~\ref{eq:ued4lepton_brs})
have on average lower transverse momentum than some of the
background channels, like for example the background from top quark decays. 
For this reason we apply upper bound cuts on the lepton transverse momentum (lept $\rm{p_{T}}$) of 
70, 60, 40, 30 GeV for the $1^{st}, 2^{nd}, 3^{rd}, 4^{th}$ lepton sorted in $p_{T}$, respectively.
A missing transverse energy cut of ${\rm E_T\hspace{-0.4cm}/\hspace{0.2cm}}>60$
GeV proves to be important especially
for high $\rm{R^{-1}}$ values where the ${\rm E_T\hspace{-0.4cm}/\hspace{0.2cm}}$ is higher due to the massive LKPs, as
the background is significantly rejected with respect to a  small reduction
of signal events. Finally, we apply a selection on the invariant
mass of the lepton pairs, according to which an event is rejected 
if it has one or more OSSF lepton pair with 
$\rm M_{inv} < 5$ GeV or $\rm M_{inv} > 80$ GeV,
aimed at rejecting the $\rm{ZZ}$ background. The selection cuts were chosen so
 that the signal efficiency is maximum for ($\rm{R^{-1}=900}$ GeV),
where the signal cross section is lowest.
 
\begin{figure}[ht]
\begin{center}
\begin{tabular}{cc}
\includegraphics[width=0.45\textwidth]{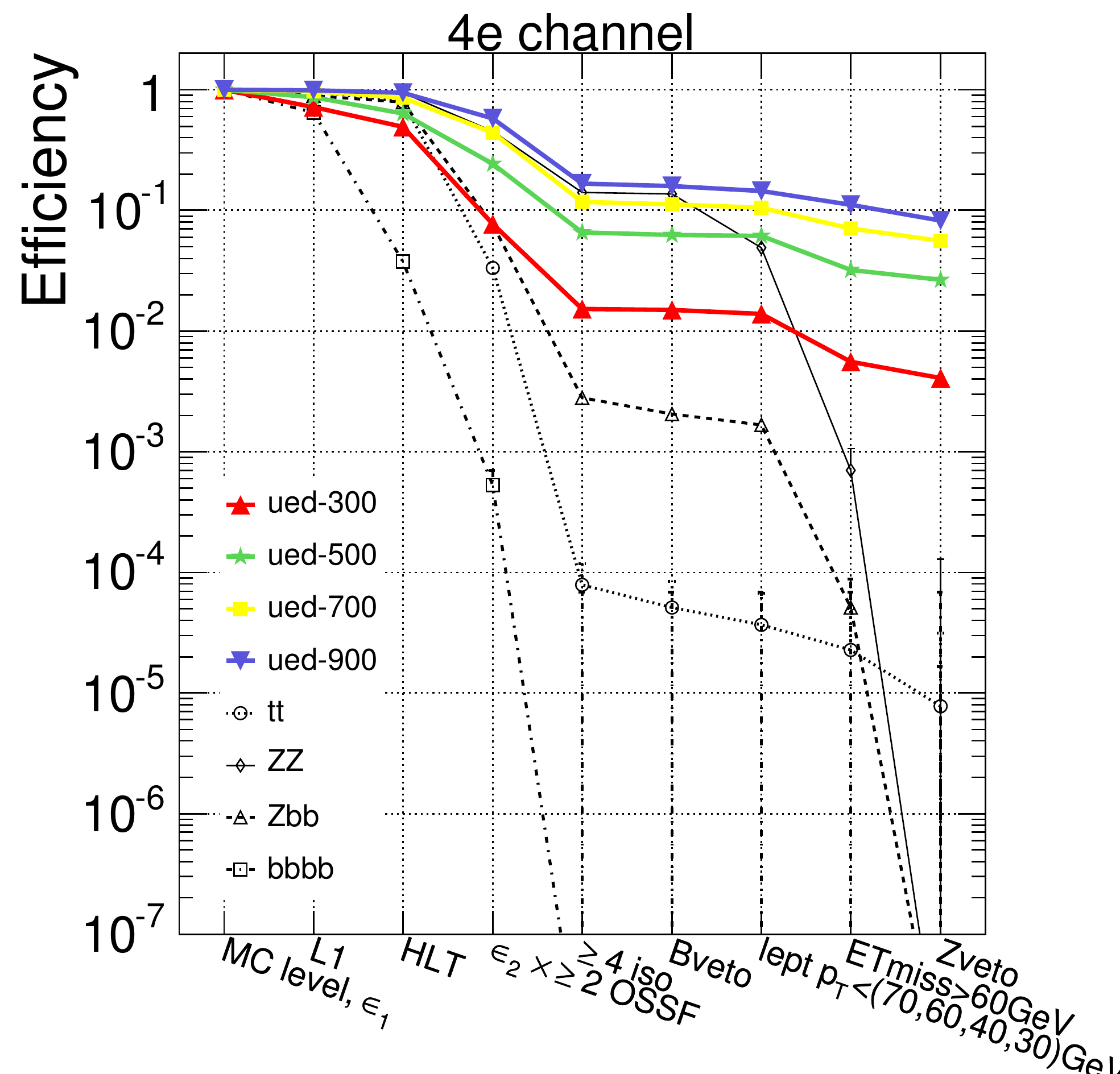}&
\includegraphics[width=0.45\textwidth]{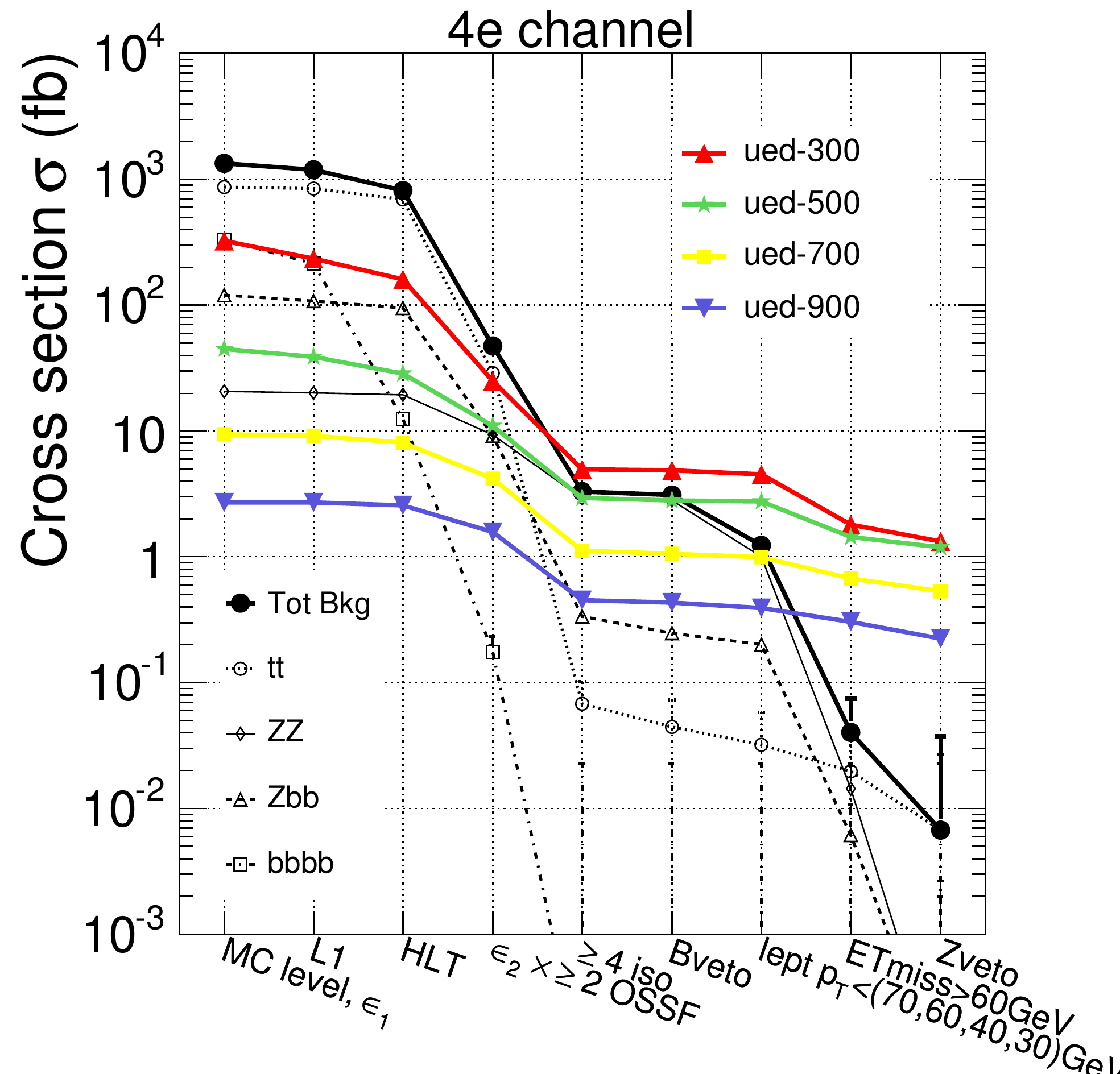} 
\tabularnewline
\includegraphics[width=0.45\textwidth]{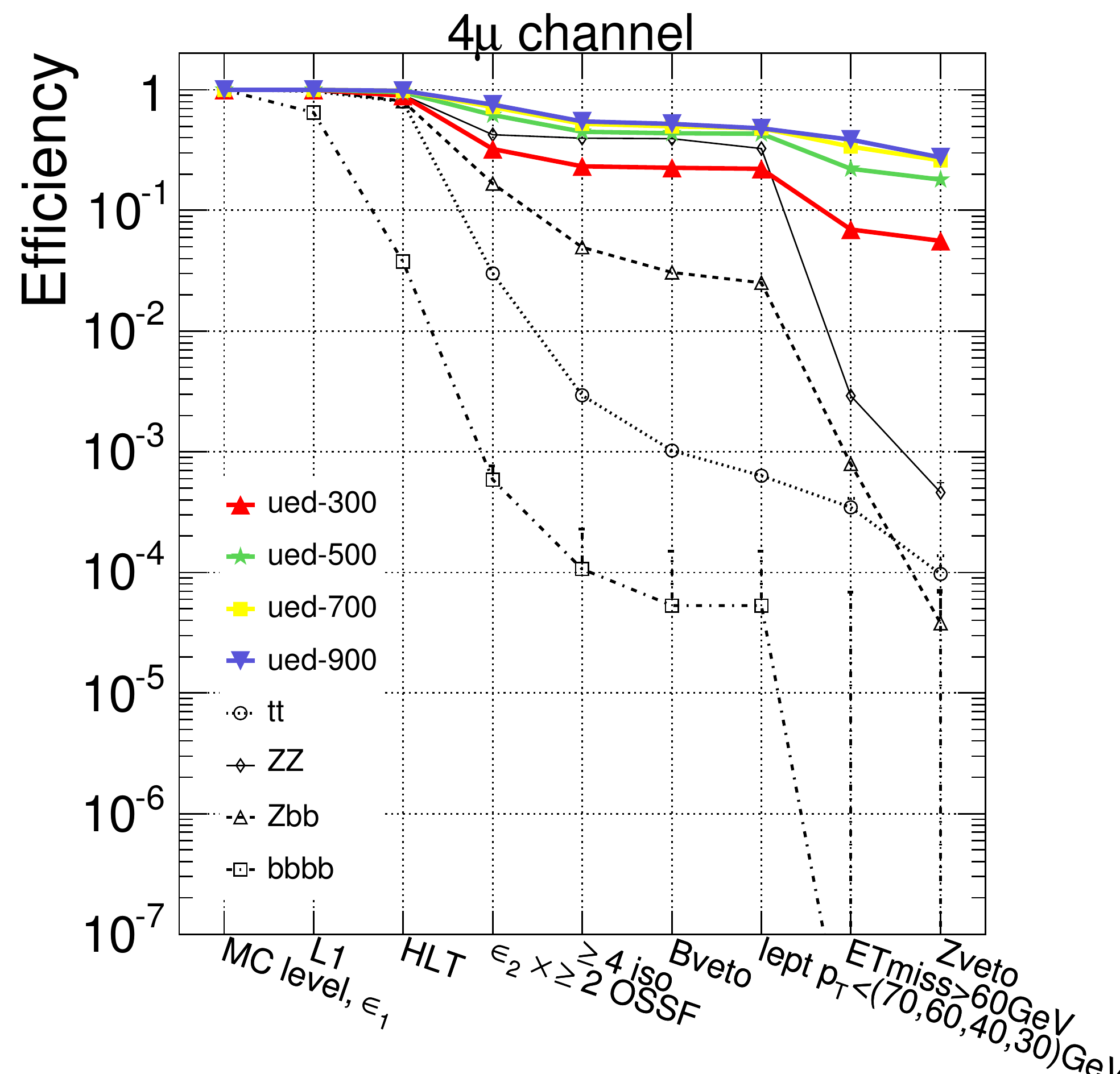}&
\includegraphics[width=0.45\textwidth]{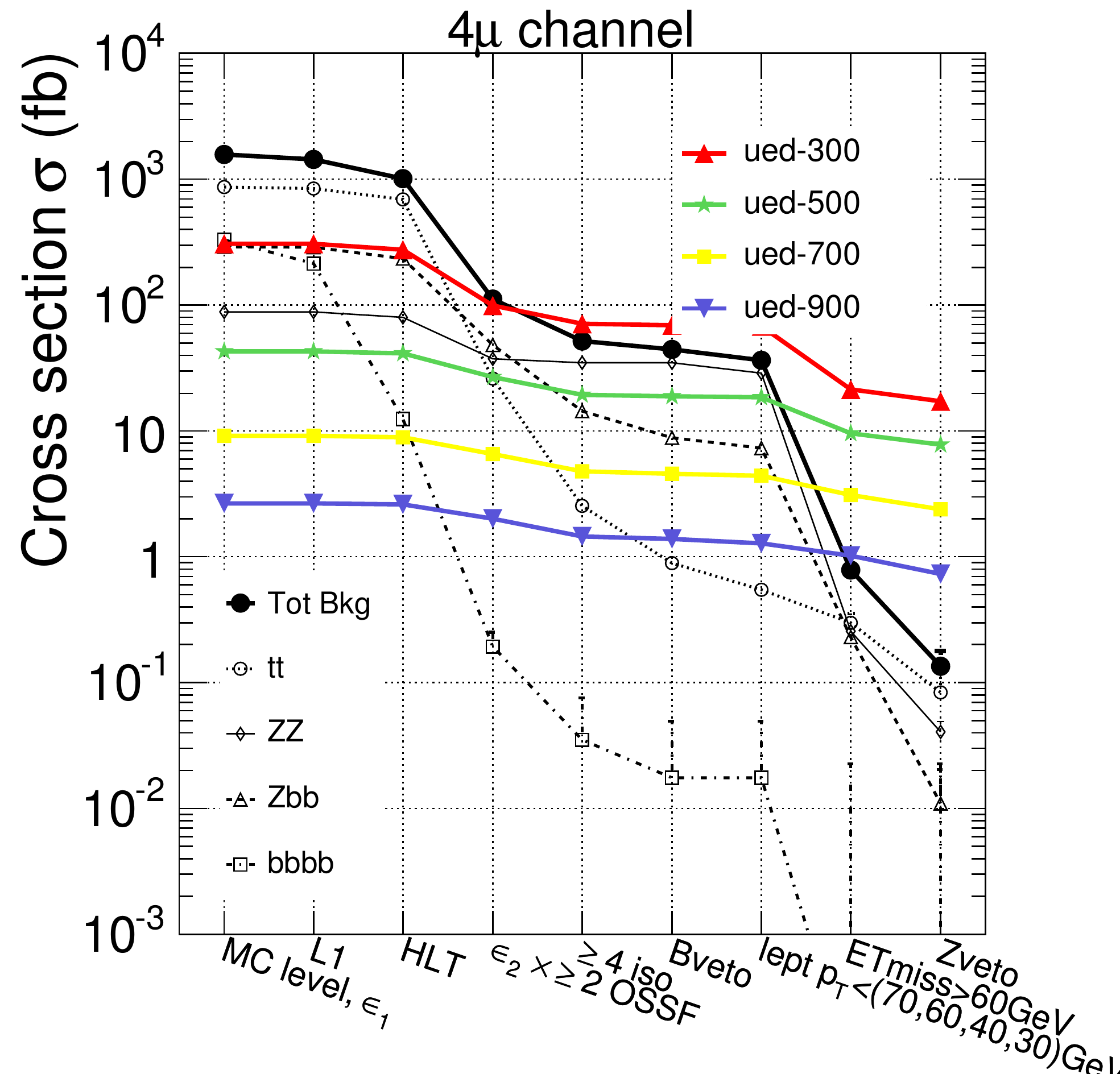} 
\tabularnewline
\includegraphics[width=0.45\textwidth]{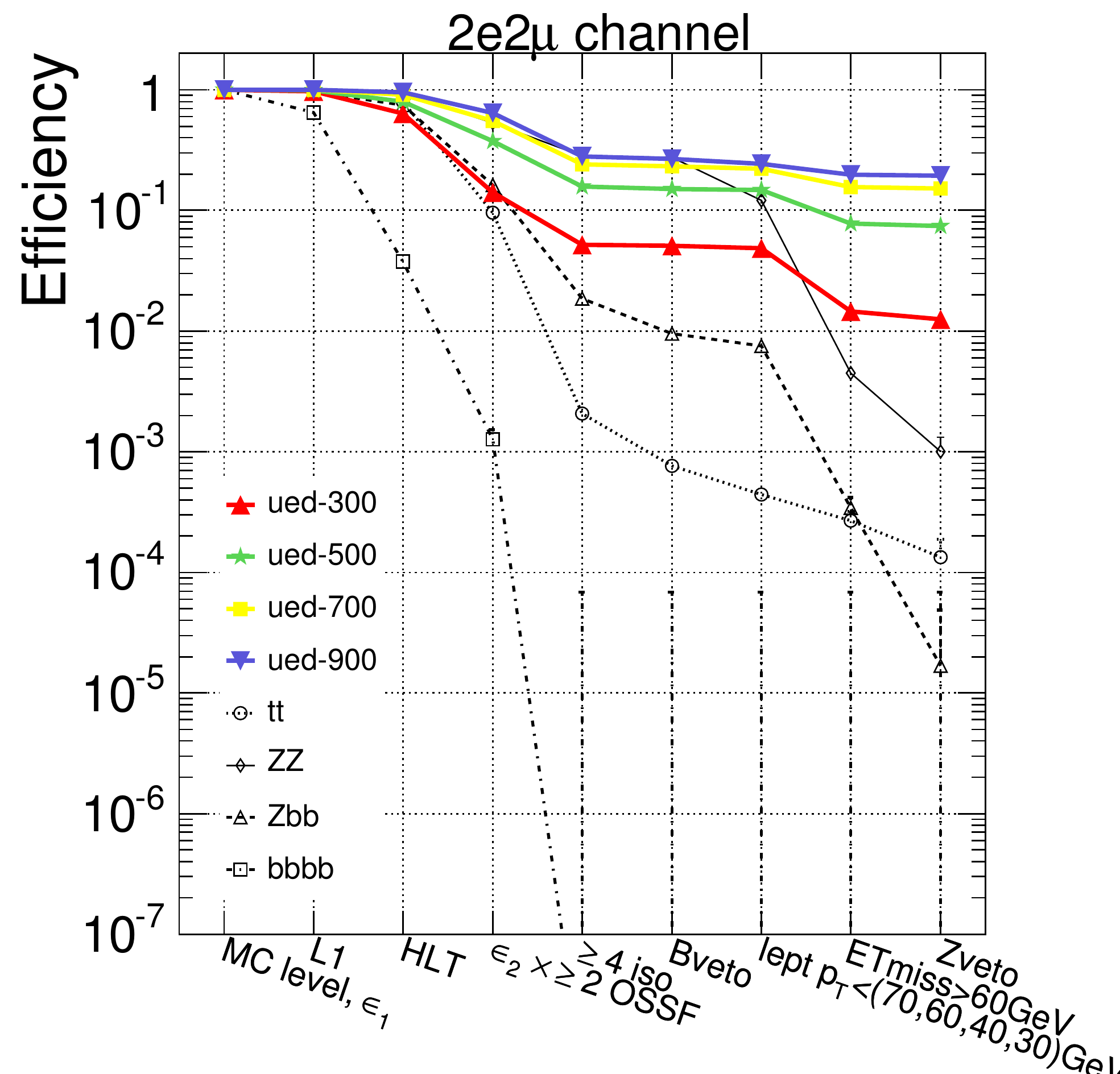}&
\includegraphics[width=0.45\textwidth]{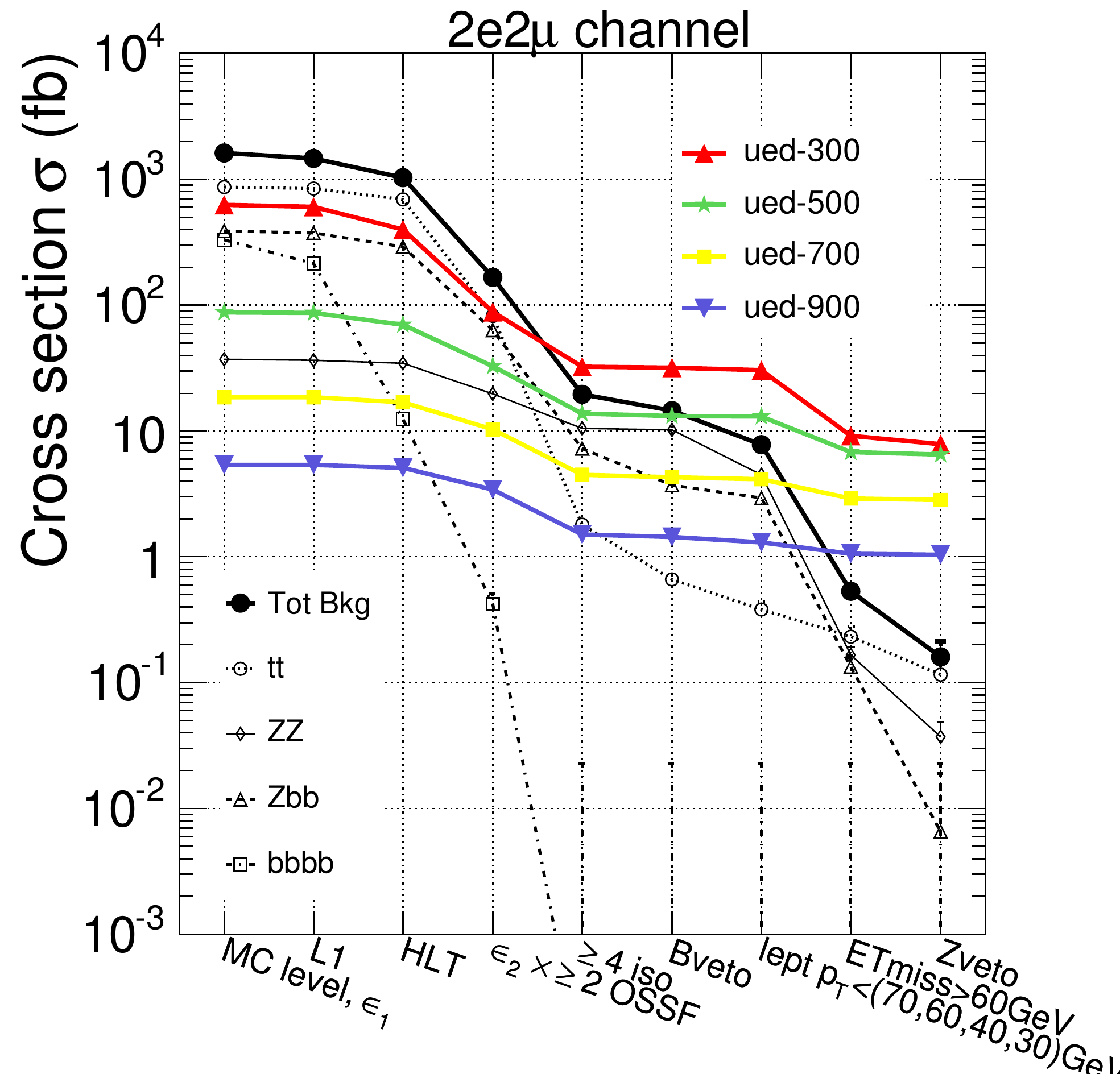} 
\tabularnewline
\end{tabular}
\end{center}
\caption{Cumulative efficiency and  cross section after
each selection cut for the MUED signal 
($\mathrm{R^{-1}}\in\{300, 500, 700, 900\}$ GeV, $\rm m_{H}=120$ GeV, $\Lambda\mathrm{R}=20$)
and the background for all channels. Only the upper statistical uncertainties are shown.
}
\label{fig:ued4lepton_cuts-effsig}
\end{figure}

The summary of all selection cuts is presented in figure~\ref{fig:ued4lepton_cuts-effsig} 
in two ways: as an efficiency of each cut after
the previous one (left), and as a cross section after each cut (right). After all selection
cuts, a S/B greater than five is achieved for all studied points of the parameter space. 

\subsection{Results}\label{sec:ued4l_results}
\begin{figure}[ht]
\begin{center}
\includegraphics[width=0.70\textwidth]{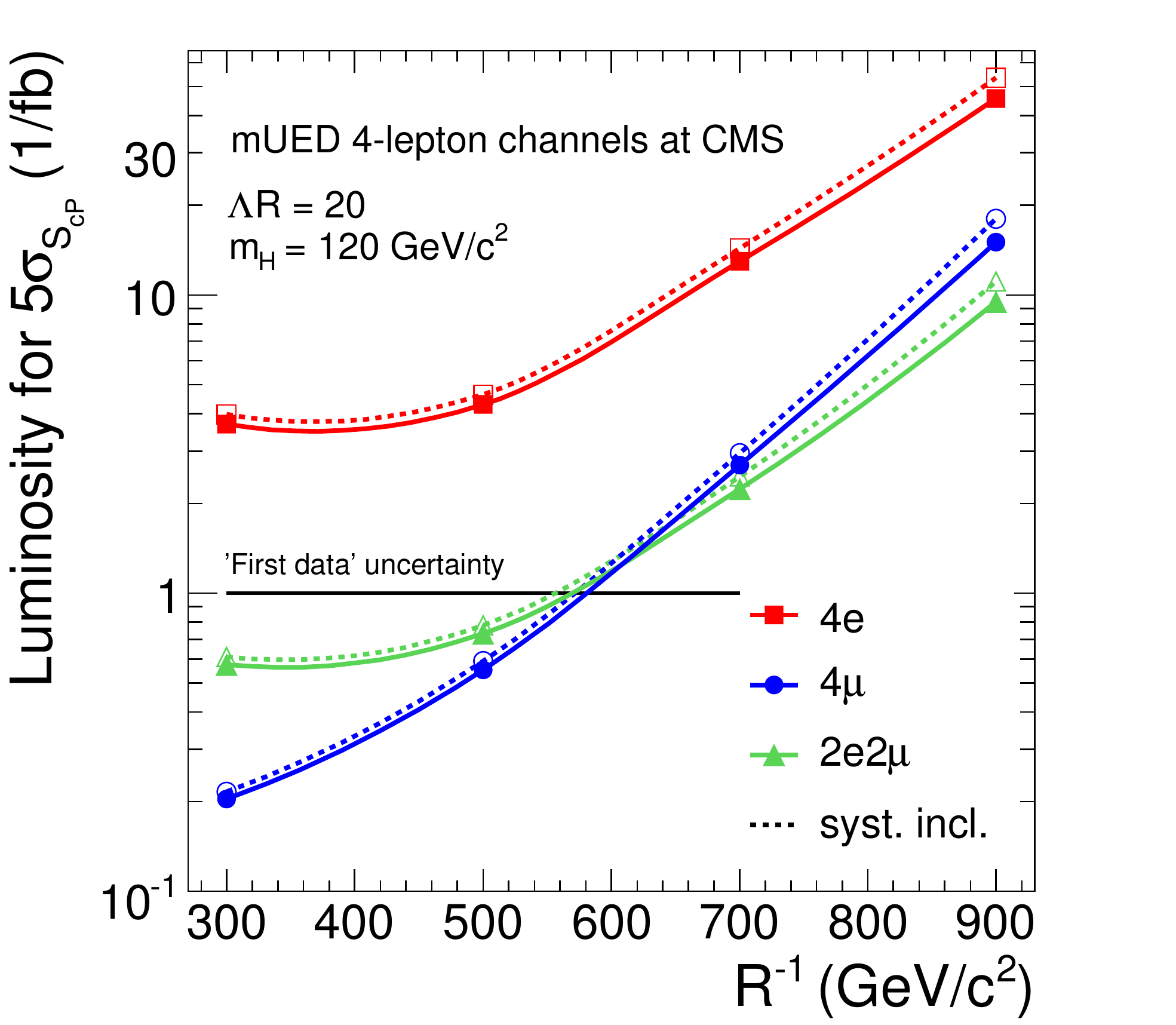}
\hspace{1cm}
\caption{The discovery potential of MUED signals 
($\mathrm{R^{-1}}\in\{300, 500, 700, 900\}$ GeV,  $\rm m_{H}=120$ GeV, $\Lambda\mathrm{R}=20$) 
in the four-lepton channels is defined as the integrated luminosity needed to measure 
a signal with a significance ($\rm S_{cP}$) of five standard deviations.
The dashed (solid) lines show results including (not including) systematical
uncertainties.
The uncertainties due to the limited understanding of the detector performance 
and characteristic of the early phase of the LHC data taking are not considered and
may limit the sensitivity below one $fb^{-1}$(horizontal 'First data uncertainty' line).
}
\label{fig:ued4lepton_lum}
\end{center}
\end{figure}

The CMS discovery potential of MUED in the four lepton channel, defined as the integrated luminosity 
needed to measure a signal with a significance ($\rm S_{cP}$) of five standard deviations
is shown in figure~\ref{fig:ued4lepton_lum}. 
The significance  estimator S$\rm _{cP}$
gives the probability to observe a number of events equal or greater than 
$\rm N_{obs}=N_{Signal}+N_{Bkg}$, assuming a background-only hypothesis, 
converted to the equivalent number of standard deviations of a Gaussian distribution.
The dashed (solid) lines show results including (not including) systematical
uncertainties. The systematic uncertainties include a 20\% uncertainty on the background cross section,
the effect of jet energy scale on the missing energy distribution (3-10\%, $p_{T}$ dependent)
and a 5\% uncertainty in the b-tagging algorithm efficiency.
For the three four-lepton channels ( 4$e$, 4$\mu$ and 2$e$2$\mu$) 
and for the integrated luminosity of 30 $fb^{-1}$ ,
the signal significance  is above the background by a few standard deviations and therefore
the MUED signal could be detected at the CMS experiment during the first few years of data taking.
In the  4$\mu$ and the 2$e$2$\mu$ channels alone, a significance of five
standard deviations for $R^{-1}=500$ GeV could be reached with less than one $fb^{-1}$ of data.

\section{MODEL DISCRIMINATION}\label{sec:ued4lepton_disc}
The four lepton channel described above can also produced in supersymmetry 
(SUSY). The focus of this section will be in comparison
of different signatures from SUSY and UED using the monte carlo event 
generator \textsf{Herwig++}.

As of version 2.1 of \textsf{Herwig++}~\cite{Bahr:2007ni} BSM physics was included for
the first time with both the minimal supersymmetric standard model (MSSM) and 
minimal universal extra dimensions models implemented including spin 
correlations in production and decay~\cite{Gigg:2007cr}.\footnote{All plots are made with version 2.1.1 of \textsf{Herwig++}\ which included some minor bugfixes.}
This allows the comparative study of both
models within the same general purpose event generator. To compare the two models
in the most sensible manner the mass spectra should be the same. The simplest
way to achieve this is to chose a scale for the MUED model and then
adjust the parameters in the MSSM so that the relevant masses are matched.
Two scales were chosen for this work, $\mathrm{R^{-1}=500\,GeV}$ and
$\mathrm{R^{-1}=900\,GeV}$, both with $\Lambda R=20$ which produced the
mass spectra shown in tables~\ref{tab:ued4lepton_500} and~\ref{tab:ued4lepton_900}.
The MSSM spectrum file and decay tables were produced using SDECAY version 
1.3~\cite{Muhlleitner:2003vg}.

\begin{table}[!htb]
\begin{center}
\begin{tabular}{|c|c|c|c|c|c|}
\hline
$g_1$ & $q_{L1}$ & $u_{R1}$ & $d_{R1}$ & $l_{L1}$ 
& $\gamma_1$ \\
\hline
626.31 & 588.27 & 576.31 & 574.90 & 514.78  & 500.98 \\
\hline
\end{tabular}
\end{center}
\vspace{-5mm}
\caption{The UED mass spectrum for $\mathrm{R^{-1}=500\,GeV}$ and $\Lambda R=20$.
The SUSY counterparts are matched to this. All values are in GeV.}
\label{tab:ued4lepton_500}
\end{table}

\begin{table}[!htb]
\begin{center}
\begin{tabular}{|c|c|c|c|c|c|}
\hline
$g_1$ & $q_{L1}$ & $u_{R1}$ & $d_{R1}$ & $l_{L1}$ 
& $\gamma_1$ \\
\hline
 1114.25 & 1050.50  & 1028.84 & 1025.28 & 926.79  & 900.00 \\
\hline
\end{tabular}
\end{center}
\vspace{-5mm}
\caption{The UED mass spectrum for $\mathrm{R^{-1}=900\,GeV}$ and $\Lambda R=20$.
The SUSY counterparts are matched to this. All values are in GeV.}
\label{tab:ued4lepton_900}
\end{table}

\begin{figure}[!htb]
\begin{center}
\includegraphics[angle=90,width=0.412\textwidth]{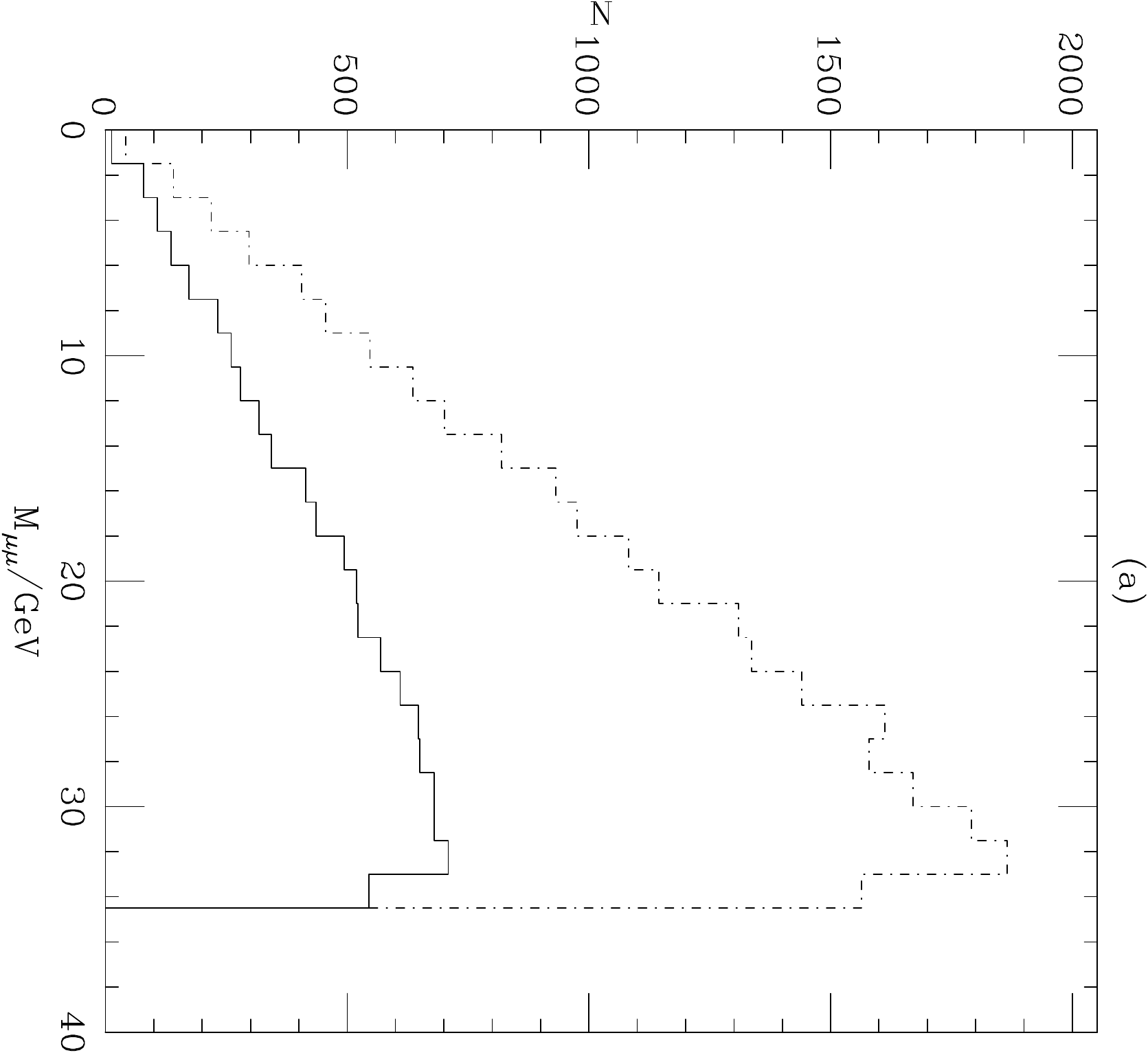} 
\hspace{1.5mm}
\includegraphics[angle=90,width=0.4\textwidth]{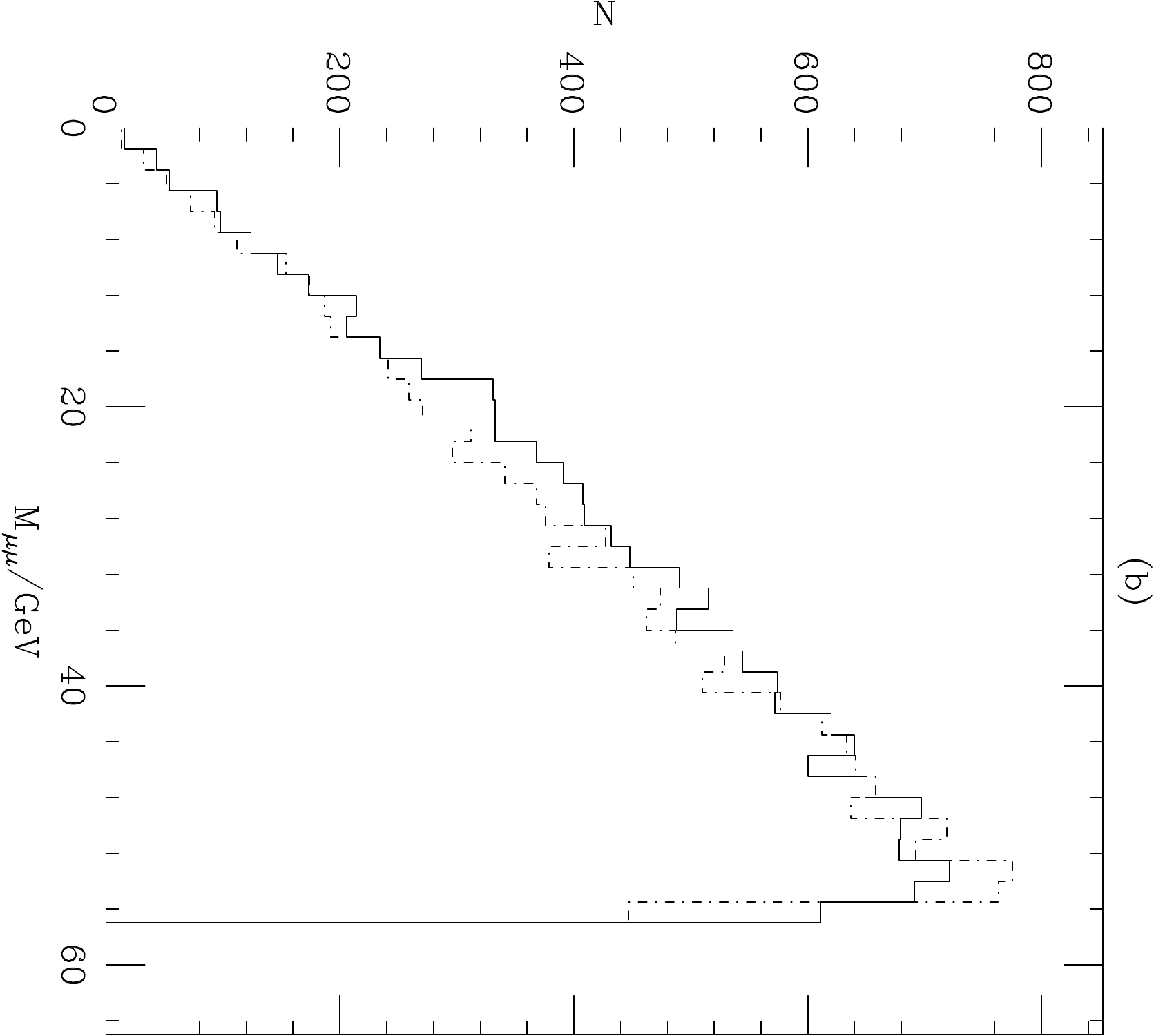}
\end{center}
\vspace{-7mm}
\caption{The invariant mass of the di-muon pair coming from each chain
in a four muon event for (a) $\mathrm{R^{-1}=500\,GeV}$ and (b) 
$\mathrm{R^{-1}=900\,GeV}$. Solid: UED, dot-dash: SUSY.}
\label{fig:ued4lepton_dimuon}
\end{figure}

The previous section tells us that the four lepton channel could give
a sizeable signal compared to the standard model background. Plotting the invariant mass
distribution of the di-muon pairs in a four muon final state event 
for MUED and SUSY gives the results shown in 
figure~\ref{fig:ued4lepton_dimuon}. It is apparent that larger values of the
masses within the spectrum make it more difficult to distinguish between the two 
models. Moreover, in the $\mathrm{R^{-1}=500\,GeV}$ case
the shapes of the distributions are similar it is just the overall
number of events in the SUSY case that is larger due to the size of the relative 
branching ratios. Given the possibility that the distributions could
be so similar it will be necessary to make use of other combinations
of invariant mass plots. The most logical is the invariant mass of a quark
plus one of the lepton or antileptons.\footnote{The theoretical distributions
for the case where one distinguishes between quark and antiquark
are given in ~\cite{Smillie:2005ar} and will not be reproduced here.}
Since it is possible to distinguish between leptons and antileptons in a 
detector one can make separate distributions for the 
jet\footnote{We are working at the parton 
level so we define a jet as simply a quark or an antiquark.} plus lepton
and the jet + antilepton cases. Since these now take into account the
helicity of the quark these distributions will be more sensitive to
spin effects.

\begin{figure}[!htb]
\begin{center}
\includegraphics[angle=90,width=0.412\textwidth]{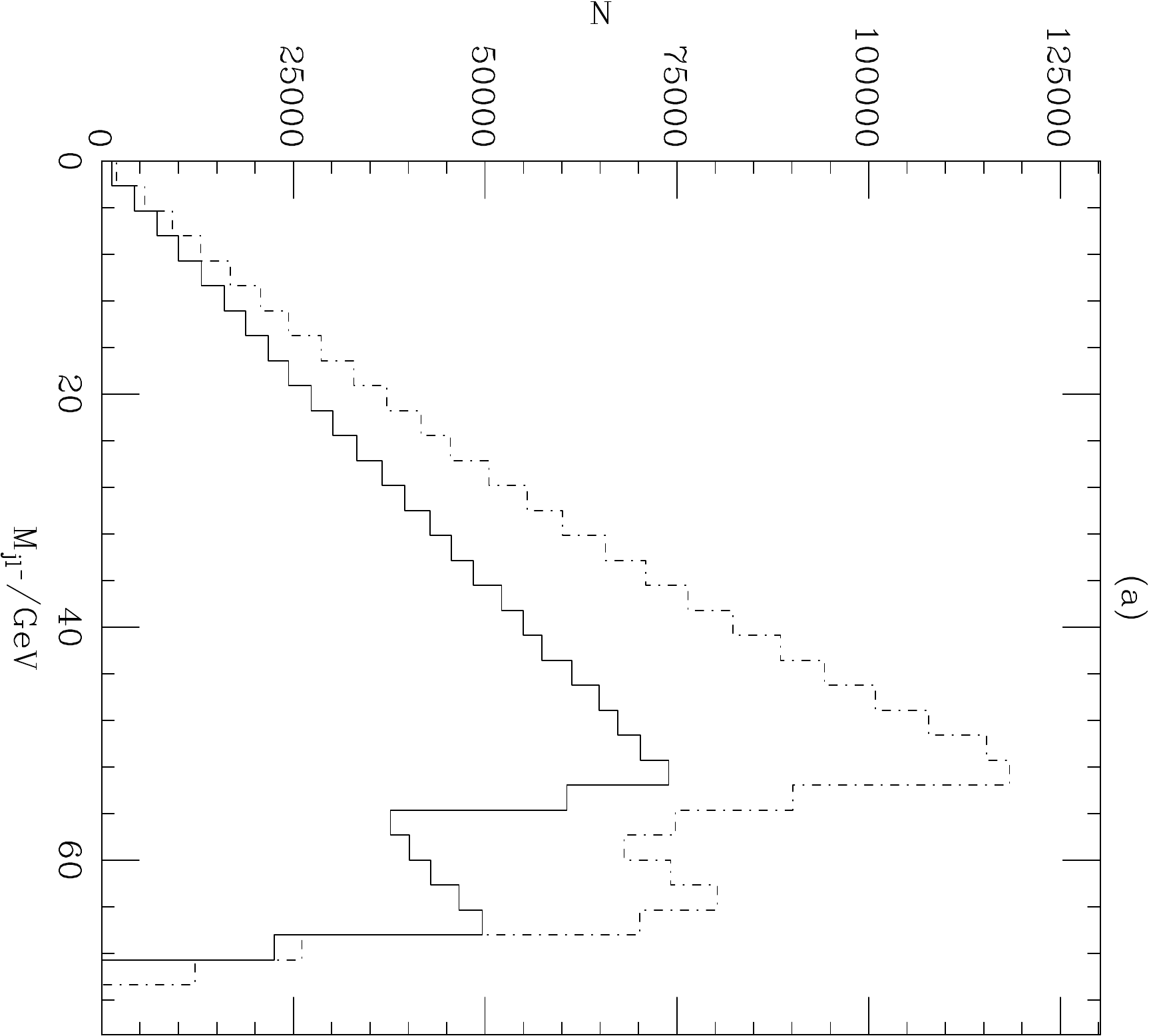} 
\hspace{1.5mm}
\includegraphics[angle=90,width=0.4\textwidth]{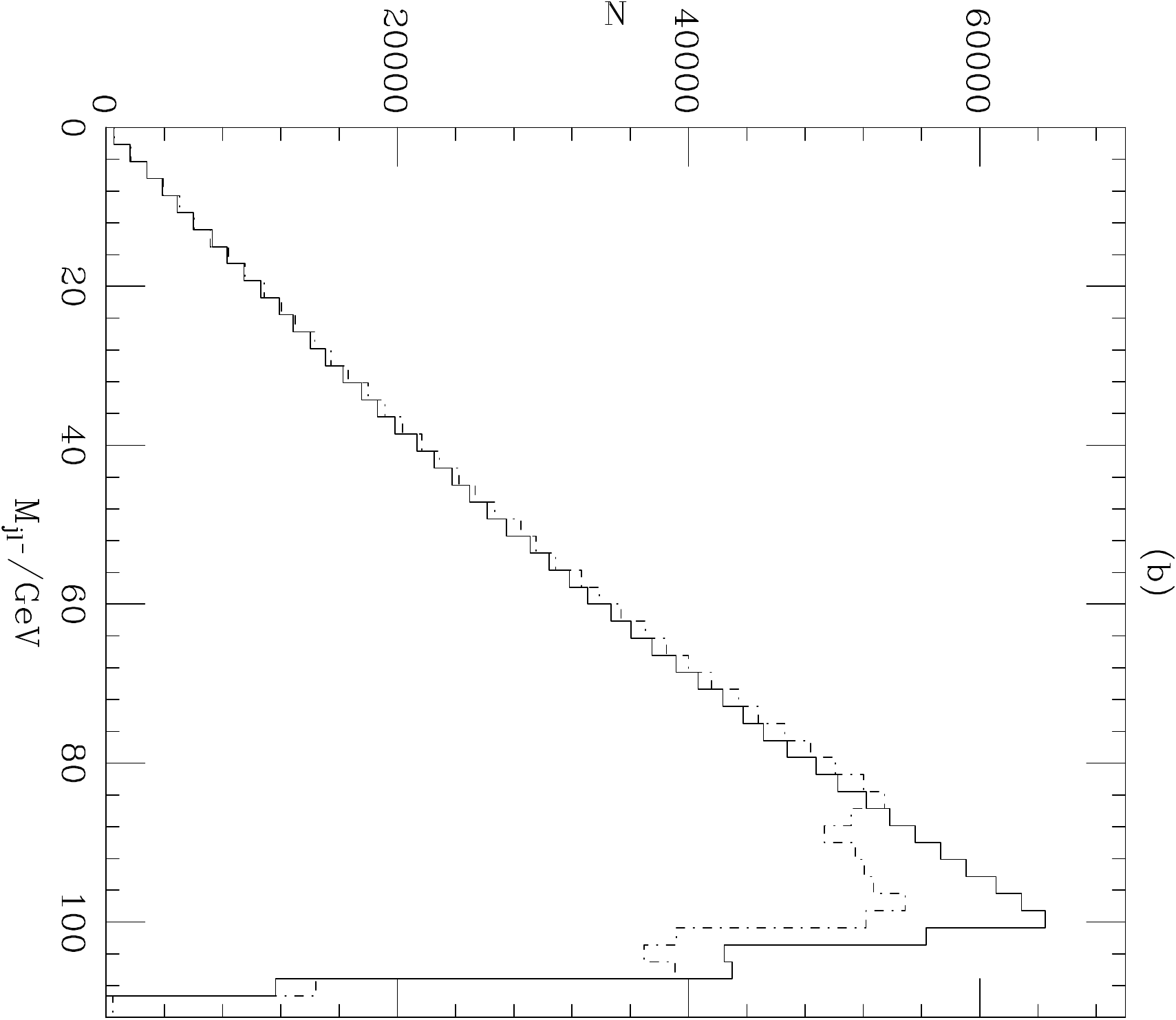}
\end{center}
\vspace{-7mm}
\caption{The invariant mass of a quark or an antiquark with a lepton for
 (a) $\mathrm{R^{-1}=500\,GeV}$ and (b) $\mathrm{R^{-1}=900\,GeV}$. Solid: UED, 
dot-dash: SUSY.}
\label{fig:ued4lepton_jlm}
\end{figure}

\begin{figure}[!htb]
\begin{center}
\includegraphics[angle=90,width=0.4\textwidth]{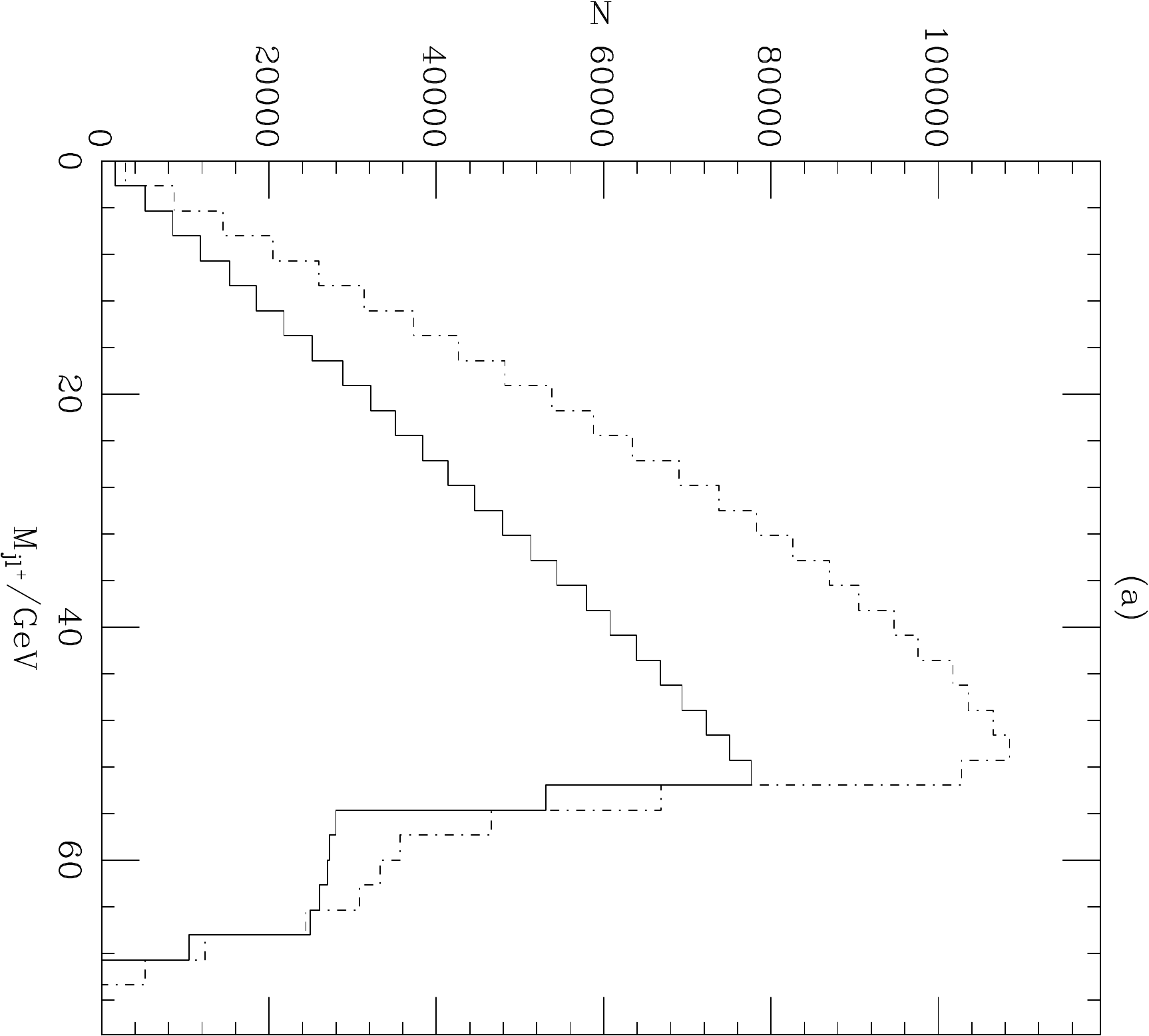}
 \hspace{1.5mm}
\includegraphics[angle=90,width=0.4\textwidth]{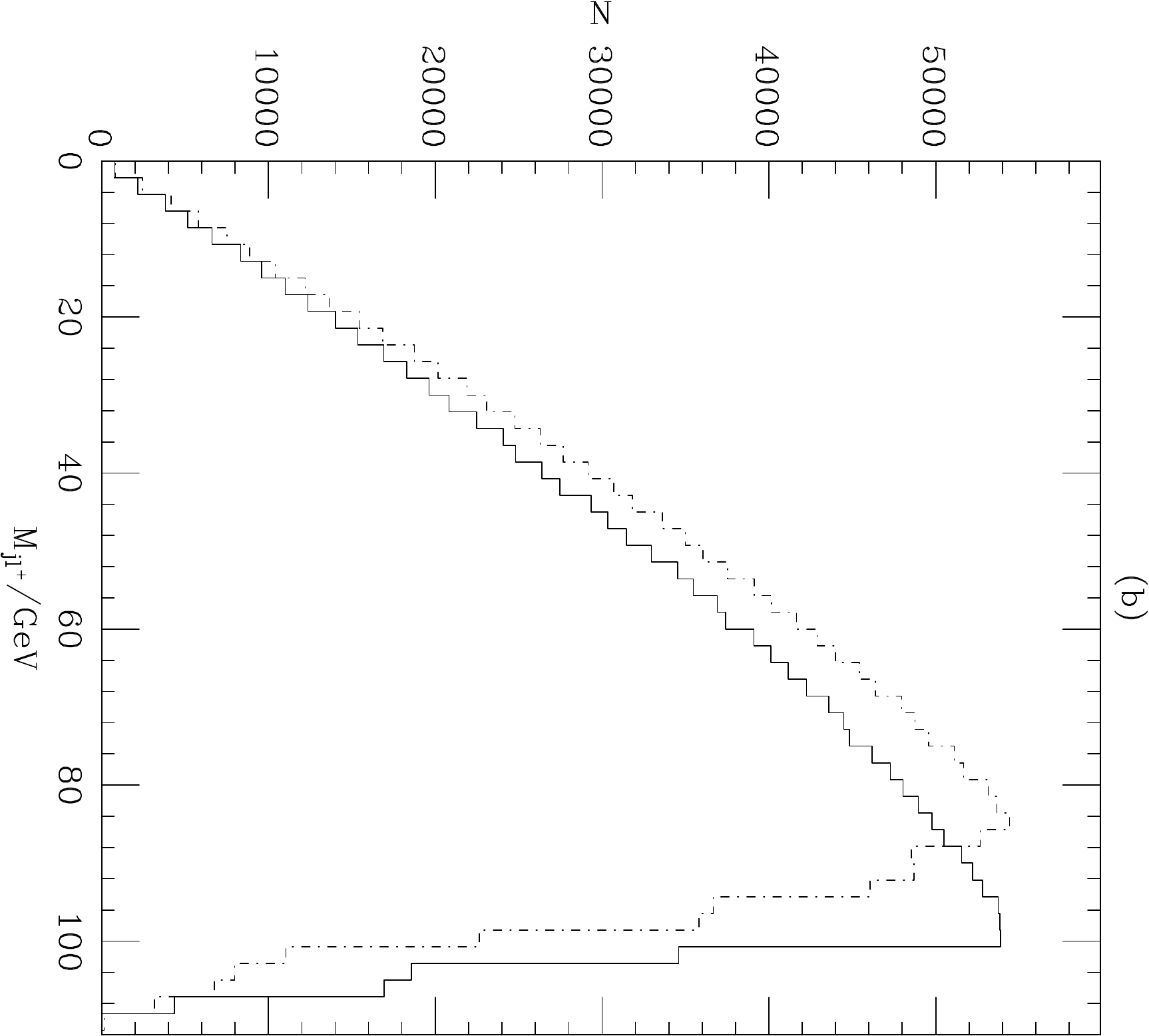}
\end{center}
\vspace{-7mm}
\caption{The invariant mass of a quark or an antiquark with an antilepton for
 (a) $\mathrm{R^{-1}=500\,GeV}$ and (b) $\mathrm{R^{-1}=900\,GeV}$. Solid: UED, 
dot-dash: SUSY.}
\label{fig:ued4lepton_jlp}
\end{figure}

Figure~\ref{fig:ued4lepton_jlm} shows the invariant mass of a jet plus a lepton
while figure~\ref{fig:ued4lepton_jlp} shows the distribution for a jet plus
an antilepton for the two scales under consideration. Again there is a greater
difference at a lower value of the compactification radius where it would
seem that the shapes of the distributions differ more at higher invariant mass
values. The main reason, however, for the similarity in the distributions
under study is that they have combined effects from opposite sets of 
spin correlations. This result can be attributed to firstly
the lack of distinction between quark and antiquark in the jet/lepton
distributions which  means that two sets of data with opposite spin correlations
appear on the same plot thereby cancelling the effect out. Also the run was set 
up so that both left and right-handed partners to the quarks were produced
in the initial hard collision and when these decay they will have, again,
opposite correlations. It is these kinds of effect that will cause the
most trouble in trying to distinguish between the two models. 

\begin{figure}[!htb]
\begin{center}
\includegraphics[angle=90,width=0.48\textwidth]{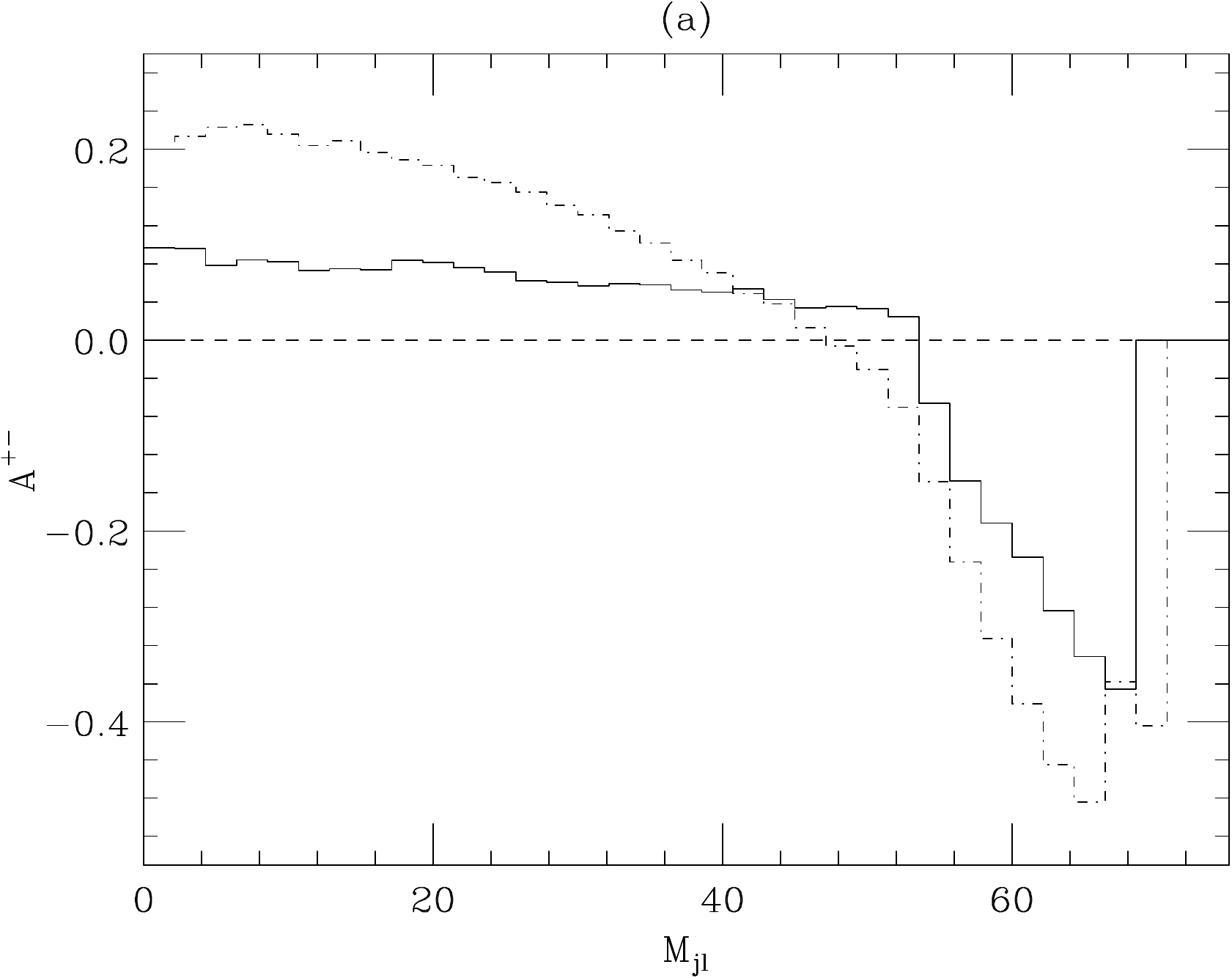} 
\hspace{2mm}
\includegraphics[angle=90,width=0.48\textwidth]{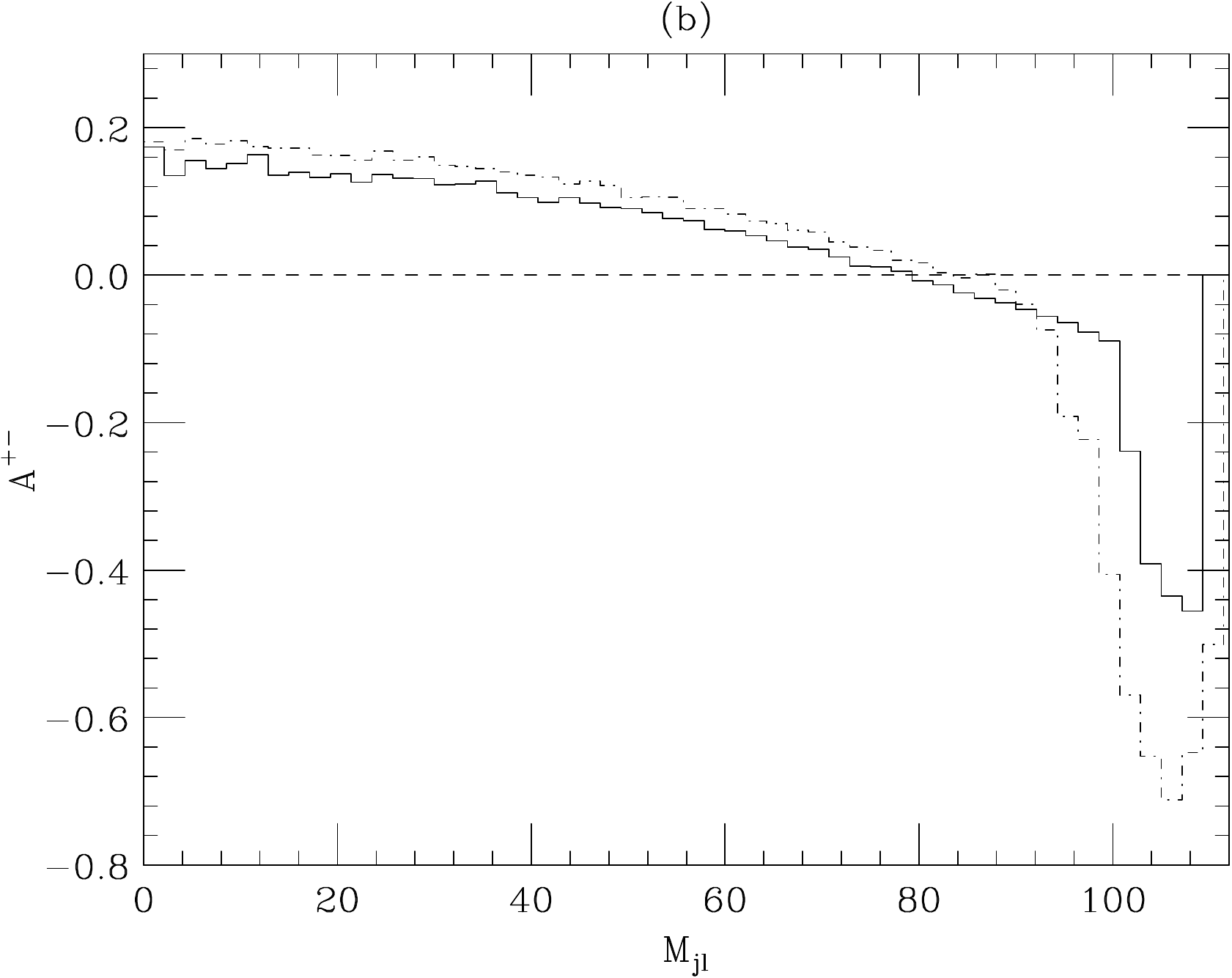}
\end{center}
\vspace{-7mm}
\caption{The asymmetry, as defined in equation~\ref{eqn:ued4lepton_asym}, for
 (a) $\mathrm{R^{-1}=500\,GeV}$ and (b) $\mathrm{R^{-1}=900\,GeV}$. Solid: UED, 
dot-dash: SUSY.}
\label{fig:ued4lepton_asym}
\end{figure}

A useful quantity in trying to achieve this at the LHC will be the asymmetry,
defined as
\begin{equation}\label{eqn:ued4lepton_asym}
  A^{\pm} = \left(\frac{dP}{dm_{jl^+}} - \frac{dP}{dm_{jl^-}}\right)/
  \left(\frac{dP}{dm_{jl^+}} + \frac{dP}{dm_{jl^-}}\right),
\end{equation}
where $dP/dm_{jl^+}$ and $dP/dm_{jl^-}$ are the antilepton and lepton
distributions respectively. Its usefulness stems from the fact that the
LHC is a proton-proton collider and will produce an excess of quarks over 
antiquarks. The result will be a slight favour in one helicity mode 
over the other meaning the asymmetry should be the most sensitive to
the underlying physics model. The distributions from \textsf{Herwig++}\ are shown in
figure~\ref{fig:ued4lepton_asym}.
 
\section{CONCLUSIONS}

The CMS experiment will be able to detect evidence of MUED model 
in the four lepton final state up to $\rm{R^{-1}=900}$ GeV  
with an integrated luminosity of 30 $fb^{-1}$. 
For the purpose of discrimination between the MUED and SUSY scenarios it
is apparent that the analysis of the four lepton signature alone is insufficient.
The best hope is using asymmetry distribution as this is most sensitive to the spin
differences in the underlying physics model.

\section*{ACKNOWLEDGEMENTS}
The work of M.~Gigg was supported by the Science and Technology Facilities 
Council and in part by the European Union Marie Curie Research Training Network
MCnet under contract MRTN-CT-2006-035606. The work of P.~Ribeiro was supported 
by Fundacao para a Ciencia e Tecnologia under grant SFRH/BD/16103/2004.

\AddToContent{M.~Gigg and P.~Ribeiro}
\setcounter{figure}{0}
\setcounter{table}{0}
\setcounter{section}{0}
\setcounter{equation}{0}
\setcounter{footnote}{0}
\clearpage


\part[Fermiophobic $W'$ Bosons]{LHC Events with Three or More Leptons Can Reveal Fermiophobic $W'$ Bosons}

{\it R.S.~Chivukula and E.H.~Simmons}

\section{INTRODUCTION}

Events with three or more leptons plus either jets or missing energy can lead to the discovery of fermiophobic $W'$ bosons associated with the origin of electroweak symmetry breaking.  One possibility is the process $pp \to (W^\ast)\to W'^{(\ast)}Z\to W Z Z \to \,jj\ell^+\ell^-\ell^+\ell^-$ where the $W$ is assumed to decay hadronically and $\ell$ can be an electron or muon.   Another is the process $pp \to W Z jj$ where the $W$ and $Z$ re-scatter through the $W'$ resonance; the final state of interest here includes three leptons, two jets, and missing energy.  This section describes a general class of ``Higgsless" models that include fermiophobic $W'$ bosons, specify the particular model used in our phenomenological studies, and then describe the calculations and results for each multi-lepton channel in turn.

The $pp \to WZ jj$ signal is the classic
$WW$-scattering process studied for a strongly interacting symmetry breaking sector, with
the $W'$ playing an analogous role to the technirho boson. In the three-site higgsless 
model considered below it is possible to calculate this process in a fully gauge-invariant manner,
rather than using the traditional method that involves separately calculating the signal (by using a model of $\pi\pi$ scattering in conjunction with the effective $W$ approximation) and background (usually done by considering
the standard model with a light Higgs boson).

\section{HIGGSLESS MODELS IN GENERAL}

Higgsless models \cite{Csaki:2003dt} provide  electroweak symmetry breaking, including unitarization of 
the scattering of longitudinal $W$ and $Z$ bosons, without 
employing a scalar Higgs boson.   The most extensively studied 
models  \cite{Agashe:2003zs, Csaki:2003zu} are based on a five-dimensional
$SU(2) \times SU(2) \times U(1)$ gauge theory in a slice of Anti-deSitter space, and
electroweak symmetry breaking is encoded in the boundary conditions of the
gauge fields.  Using the AdS/CFT correspondence \cite{Maldacena:1997re}, these theories may be viewed
as ``dual'' descriptions of walking technicolor theories 
\cite{Holdom:1981rm, Holdom:1984sk, Yamawaki:1985zg, Appelquist:1986an, Appelquist:1986tr, Appelquist:1987fc}.
In addition to a massless photon and near-standard $W$ and $Z$ bosons, 
the spectrum includes an infinite tower of  additional 
massive vector bosons (the higher
Kaluza-Klein  or $KK$ excitations), whose exchange is responsible 
for unitarizing longitudinal $W$ and $Z$ boson scattering \cite{SekharChivukula:2001hz}. 
To provide the necessary unitarization, the masses of the lightest $KK$ bosons
must be less than about 1 TeV.
Using deconstruction, it has been shown \cite{SekharChivukula:2004mu} that a 
Higgsless model whose fermions are localized ({\it i.e.}, derive 
their electroweak properties from a single site on the 
deconstructed lattice)  cannot simultaneously satisfy unitarity 
bounds and precision electroweak constraints.

The size of corrections to electroweak processes in Higgsless 
models may be reduced by considering delocalized fermions  
\cite{Cacciapaglia:2004rb, Cacciapaglia:2005pa, Foadi:2004ps}, {\it i.e.},
considering the effect of the distribution
of the wavefunctions of ordinary fermions in the fifth dimension 
(corresponding, in the
deconstruction language, to allowing the fermions to derive 
their electroweak properties from
several sites on the lattice). Higgsless models with delocalized fermions
provide an example of a viable effective theory of a strongly interacting symmetry breaking sector
consistent with precision electroweak tests.

It has been shown \cite{SekharChivukula:2005xm}
that, in an arbitrary Higgsless model, 
if the probability distribution of the delocalized fermions 
is related to the $W$ wavefunction (a condition called ``ideal'' 
delocalization), then deviations in precision electroweak 
parameters are minimized. Ideal delocalization results in the $W'$ resonances
being fermiophobic. Phenomenological limits on delocalized
Higgsless models may be derived \cite{Chivukula:2005ji}
from limits on the 
deviation of the triple-gauge boson ($WWZ$) vertices from their 
standard model value; current constraints allow for the 
lightest $KK$ resonances
to have masses as low as 400 GeV. 

\section{THREE-SITE MODEM IN PARTICULAR}

Many issues of  interest, such as ideal fermion delocalization and the generation of fermion masses (including the top quark mass) can be illustrated in a Higgsless model deconstructed to just three sites \cite{SekharChivukula:2006cg}. The electroweak sector of the three-site Higgsless model 
incorporates an $SU(2)_0 \times SU(2)_1 \times U(1)_2$ gauge group, and $2$ 
nonlinear $(SU(2)\times SU(2))/SU(2)$ sigma models 
responsible for breaking this symmetry down to $U(1)_{em}$.
The extended electroweak gauge sector of the three-site model is that of the Breaking Electroweak Symmetry Strongly (BESS) model \cite{Casalbuoni:1985kq}. The mass-eigenstate vector
bosons are admixtures of the seven gauge-bosons in $SU(2)^2 \times U(1)$, with one
massless photon, three corresponding to the standard model $W^\pm$ and $Z$, and
three nearly-degenerate $W'^{\pm }$ and $Z'$.  For the reasons described above, the masses of the $W'^\pm$ and $Z'$ in
the three-site model (and indeed for the
lightest bosons in any Higgsless model with ideal fermion delocalization) must
be between roughly 400 GeV and 1 TeV.

The left-handed fermions are doublets coupling to the two $SU(2)$ groups,
which may be correspondingly labeled  $\psi_{L0}$ and $\psi_{L1}$. The
right-handed fermions are a doublet coupling to $SU(2)_1$, $\psi_{R1}$, and two singlet
fermions coupled to $U(1)_2$, denoted  $u_{R2}$ and $d_{R2}$ in the case of quarks.
The fermions $\psi_{L0}$, $\psi_{L1}$, and $\psi_{R1}$ have $U(1)$ charges typical
of the left-handed doublets in the standard model, $+1/6$ for quarks and $-1/2$ for leptons.
Similarly, the fermion $u_{R2}$ has $U(1)$ charges typical for the right-handed
up-quarks (+2/3), and $d_{R2}$ has the $U(1)$ charge associated with the right-handed
down-quarks ($-1/3$) or the leptons ($-1$). With these assignments, one may write the
Yukawa couplings and fermion mass term
\begin{equation}
{\cal L}_f = \varepsilon_L M\, \bar{\psi}_{L0} \Sigma_1 \psi_{R1} + M \,\bar{\psi}_{R1}
\psi_{L1} + M \,\bar{\psi}_{L1} \Sigma_2
\begin{pmatrix}
\varepsilon_{Ru} & \\
& \varepsilon_{Rd}
\end{pmatrix}
\begin{pmatrix}
u_{R2} \\
d_{R2}
\end{pmatrix}
+ h.c.
\label{CS_eq:yukawa}
\end{equation}
Here the Dirac mass $M$ is typically large (of order 2 or more TeV), $\varepsilon_L$
is flavor-universal and chosen to be of order $M_W/M_{W'}$ to satisfy the constraint of 
ideal delocalization, and the $\varepsilon_R$ are proportional to the light-fermion
masses (and are therefore small except in the case of the top-quark)  \cite{SekharChivukula:2006cg}.
These couplings yield a  seesaw-like mass matrix,  resulting in
light standard-model-like fermion eigenstates, along with
a set of degenerate vectorial doublets (one of each standard model weak-doublet).

The three-site model is sufficiently simple that it is possible to implement the model
in CompHEP \cite{Pukhov:2004ca} or MADGRAPH \cite{Alwall:2007st}, and  -- by including both
fermion and gauge-boson couplings -- to do so in a way that is fully gauge-invariant.

\section{4 LEPTONS PLUS 2 JETS}

LHC events with 4 charged leptons and 2 jets can reveal \cite{He:2007ge} the presence of a fermiophobic $W'$ boson that is produced through $pp \to W Z Z $ with the $Z$ bosons decaying leptonically and the $W$ decaying hadronically. The signal for the $W'$ boson comes from
associated production $pp \to (W^\ast)\to W^{'}Z$ followed by the decay $W' \to WZ$.
 The backgrounds include the irreducible SM background 
 $pp\to W Z Z \to jj\ell^+\ell^-\ell^+\ell^-$;
 a related, but reducible, SM background  $\,pp\to Z Z Z \to jj\ell^+\ell^-\ell^+\ell^-\,$ in
 which the hadronically-decaying $Z$ is mis-identified as a $W$; and all other SM processes
 leading to the same final state $pp\to jj\ell^+\ell^-\ell^+\ell^-$ through different intermediate steps,
 including processes in which one or more of the jets is gluonic.

 Ref. \cite{He:2007ge} has calculated the full signal and background in the context of the three-site higgsless model \cite{SekharChivukula:2006cg} using the cuts described here.  One set of cuts is used to suppress the SM backgrounds:
 \begin{equation}                            
 M_{jj} \,=\, 80\pm 15\,{\rm GeV},
 ~~~~~~\Delta R(jj)\,<\,1.5\,, ~~~~~~
 \sum_Z p_T^{~}(Z)+\sum_j p_T^{~}(j) \,=\, \pm 15~{\rm GeV}.
 \label{CS_supp-cuts}
 \end{equation}
 The first of these selects dijets arising from on-shell $W$ decay (leaving a margin
 for the experimental resolution\,\cite{unknown:1999fr}); the second reflects the
 dijet separation of the signal events; and the third exploits the 
 conservation of transverse momentum in the signal events.
In addition, a set of transverse momentum and rapidity cuts are imposed on the jets
 and charged leptons
 \begin{equation}                         
 p_{T\ell}^{~}>10\,{\rm GeV},~~~~~~|\eta_\ell^{~}| < 2.5\,,
~~~~~~
 p_{Tj}^{~}>15\,{\rm GeV},~~~~~~ |\eta_j^{~}| < 4.5\,.
\label{CS_det-cuts}
 \end{equation}
for particle identification.

These cuts essentially eliminate the first two sources of background and reduce
the third to a manageable size, with the signal peak standing out cleanly.  Ref. \cite{He:2007ge} concludes that fermiophobic $W'$ bosons will be visible in
this channel at the LHC throughout their entire allowed mass range from $400 - 1200$ GeV.
The integrated luminosity required for detecting the $W'$ in this channel is shown here as a function of $W'$ mass in Fig.\,\ref{CS_IL}.

\section{3 LEPTONS PLUS MISSING ENERGY}

LHC events with 3 charged leptons, missing energy, and forward jets can reveal the presence of a fermiophobic $W'$ boson produced through  the scattering process $pp \to W Z qq'$, where both vector bosons decay leptonically \cite{Bagger:1995mk,He:2002qi,Zhang:2003it}.  In this case, the signal arises from the $W'$ contribution to the vector boson fusion subprocess $W Z \to W Z $.

An initial estimate of the $W'$ signal and related backgrounds was presented in in ref. \cite{Birkedal:2004au} for a 5d higgsless sum rule scenario.  Ref. \cite{He:2007ge} has improved on this by performing the first tree-level calculation that includes both the signal and the full electroweak (EW) and QCD backgrounds for the $2 \to 4$ scattering process \,$pp\to W Z jj'$ in the context of a complete, gauge-invariant higgsless model (the three-site model \cite{SekharChivukula:2006cg}).   A forward-jet tag is used to eliminate the reducible QCD background \cite{Bagger:1995mk} from the annihilation process $qq \to W Z $.  The irreducible QCD backgrounds  $pp \to W Z jj$ with $jj = qg,\,gg$ serving as forward jets are suppressed by the cuts
 \begin{equation}
 E_j^{}    > 300\,{\rm GeV}\,,~~~~~~
 p_{Tj}^{} >  30\,{\rm GeV}\,, ~~~~~~
 |\eta_j^{}| < 4.5\,,~~~~~~
 \left|\Delta\eta_{jj}^{}\right| > 4\,,
 \label{CS_newcuts}
 \end{equation}
 where $E_j^{}$ and $p_{Tj}^{}$ are transverse energy and momentum of each final-state jet,
 $\eta_{j}^{}$ is the forward jet rapidity, and  $\left|\Delta\eta_{jj}^{}\right|$ is
 the difference between the rapidities of the two forward jets.
 The cut on $|\Delta \eta_{jj}^{~}|$ is especially good at suppressing 
 the QCD backgrounds \,$pp \to W Z gg,~W Z qg$\, in the low
 $M_{W Z }$ region \,\cite{HLLHCpaper2}.  The following lepton identification cuts are also employed
 \begin{equation}
 \label{CS_newcuts-lepton}
 p_{T\ell}^{~} \,>\, 10\,{\rm GeV}\,, ~~~~~~~
 |\eta_{\ell}^{~}| \,<\, 2.5\,.
 \end{equation}
 While one must specify a reference value of the SM Higgs boson mass in computing the SM EW backgrounds, the authors of \cite{He:2007ge} found that varying the Higgs mass over the range $M_H=115\,{\rm GeV}-1\,$TeV had little effect.

\begin{figure}[h]
 \vspace*{-10mm}
 \hspace*{-4mm}
 \begin{center}
 \includegraphics[width=8.9cm,height=7.5cm]{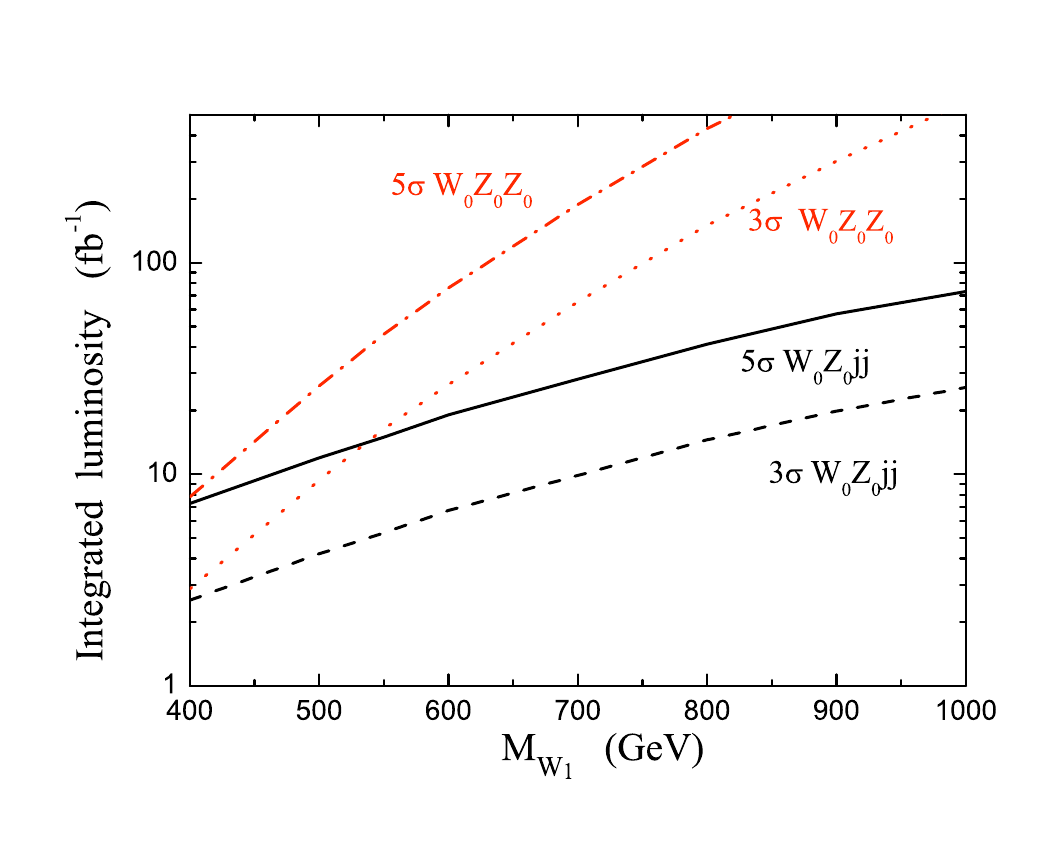}
 \end{center}
 \vspace*{-11mm}
 \caption{Integrated luminosities required
 for $3\sigma$ and $5\sigma$ detection
 of $W'$ signals as a function of $M_{W'}$.
 The dotted and dashed-dotted curves are for the $W Z Z$
 channel, while the dashed and solid curves are
 for the $W Z jj$ channel. From ref. \protect\cite{He:2007ge}.}
 \label{CS_IL}
 \end{figure}

Ultimately, ref. \cite{He:2007ge} reports both the signal and backgrounds
 for the transverse mass distribution of the vector boson pair, where  
 $M_T^2(W Z ) \equiv[\sqrt{M^2(\ell\ell\ell)+p_T^2(\ell\ell\ell)}
    +|p_T^{\rm miss}|]^2-|p_T^{}(\ell\ell\ell)+p_T^{\;\rm miss}|^2$. 
 Counting the signal and background events
 in the range $0.85M_{W'}<M_T<1.05M_{W'}$,
 yields the integrated LHC luminosities required 
 for $3\sigma$ and $5\sigma$ detections of the $W'$ boson in this
 channel, as shown here in Fig.\,\ref{CS_IL}. 

%

\section{CONCLUSIONS}

 Both the $pp \to W' Z \to W Z Z \to jj 4\ell$ and
 $pp \to W' jj\to W Z jj \to  \nu 3\ell\,jj $ channels are promising
 for revealing fermiophobic $W'$ bosons, like those in Higgsless models, at the LHC  \cite{He:2007ge}.
 The $W Z Z $ channel has a distinct signal with a clean resonance peak.
 The $W Z jj'$ channel has a larger cross section when $M_{W'}$ is heavy, but the measurement
 is complicated by the missing $E_T$ of the final-state neutrino.
 Hence, confirming the existence of both signals for the $W'$ boson,
 as well as the {\it absence} of a Higgs-like signal in
 $pp\to Z Z qq\to 4\ell \,qq$, will be strong evidence for
 Higgsless electroweak symmetry breaking \,\cite{He:2007ge}. 
 In particular, as shown here in Fig. \,\ref{CS_IL}, for $M_{W'}=500\,(400)$\,GeV, the $5\sigma$
 discovery of $W'$ requires an integrated luminosity
 of 26\,(7.8)\,fb$^{-1}$
 for $pp \to W Z Z \to jj\,4\ell$, and
 12\,(7)\,fb$^{-1}$ for $pp \to  W Z  jj \to \nu3\ell\,jj$.
 These are within the reach of the first few years' run at the LHC.

\AddToContent{R.S.~Chivukula and E.H.~Simmons}
\setcounter{figure}{0}
\setcounter{table}{0}
\setcounter{section}{0}
\setcounter{equation}{0}
\setcounter{footnote}{0}
\clearpage

\part{Low-Scale Technicolor at the LHC}

{\it G.~Azuelos, K.~Black, T.~Bose, J.~Ferland, Y.~Gershtein, K.~Lane and A.~Martin}\\

%







\begin{abstract}
  If technicolor is responsible for electroweak symmetry breaking, there are
  strong phenomenological arguments that its energy scale is at most a few
  hundred GeV and that the lightest technihadrons are within reach of the
  ATLAS and CMS experiments at the LHC. Furthermore, the spin-one
  technihadrons $\rho_T$, $\omega_T$ and $a_T$ are expected to be very
  narrow, with striking experimental signatures involving decays to pairs of
  electroweak gauge bosons ($\gamma$, $W$, $Z$) or an electroweak boson plus
  a spin-zero $\LSTCtpi$. Preliminary studies of signals and backgrounds for such
  modes are presented. With luminosities of a few to a few tens of
  femtobarns, almost all the spin-one states may be discovered up to masses
  of about 600~GeV. With higher luminosities, one can observe decay angular
  distributions and technipions that establish the underlying technicolor
  origin of the signals. Preliminary ATLAS studies show that, with
  $50$--$100\,\LSTCifb$ and assuming $M_{\LSTCta} \simeq 1.1 M_{\LSTCtro}$, both
  processes $\LSTCtropm,\,a_T^\pm \LSTCra Z^0 W^\pm$ (with $M_{\LSTCtro} \simeq
  500\,\LSTCgev $) may be seen in the $\LSTCellp\LSTCellm\LSTCellpm\nu_\ell$ final state and
  $\LSTCtropm,\,a_T^\pm \LSTCra Z^0 \LSTCtpipm$ (up to $M_{\LSTCtro} \simeq 400\,\LSTCgev $) in
  $\LSTCellp\LSTCellm b\,{\rm jet}$, where $\ell = e,\,\mu$.

\end{abstract}


\section{INTRODUCTION}

Technicolor (TC) is a proposed strong gauge interaction responsible for the
dynamical breakdown of electroweak symmetry~\cite{Weinberg:1979bn,
  Susskind:1978ms}.  Modern technicolor has a slowly-running (``walking'')
gauge coupling~\cite{Holdom:1981rm, Appelquist:1986an, Yamawaki:1986zg,
  Akiba:1986rr}. This feature allows extended technicolor
(ETC)~\cite{Eichten:1979ah} to generate realistic masses for quarks, leptons
and technipions ($\LSTCtpi$) with the very large ETC boson masses
($10^3$--$10^4\,\LSTCtev $) necessary to suppress flavor-changing neutral current
interactions.  (For reviews, see Refs.~\cite{Lane:2002wv, Hill:2002ap}.) The
important phenomenological consequence of walking is that the technicolor
scale is likely to be much lower and the spectrum of this low-scale
technicolor (LSTC) much richer and more experimentally
accessible~\cite{Lane:1989ej, Eichten:1996dx, Eichten:1997yq} than originally
thought~\cite{Eichten:1984eu}. The basic argument is this: (1)~The walking TC
gauge coupling requires either a large number $N_D$ of technifermion doublets
so that $\LSTCLtc \simeq 250\,\LSTCgev /\sqrt{\smash[b]{N_D}} \LSTCsimle 100\,\LSTCgev $, or
two TC scales, one much lower than $250\,\LSTCgev $.\footnote{For an alternate
  view based on a small TC gauge group, $SU(2)$, see
  Refs.~\cite{Appelquist:2002me, Appelquist:2003hn}.}  (2)~Walking enhances
$\LSTCtpi$ masses much more than those of their vector partners, $\LSTCtro$ and
$\LSTCtom$. This effect probably closes the all-$\LSTCtpi$ decay channels of the
lightest techni-vectors. In LSTC, then, we expect that the lightest $\LSTCtro$
and $\LSTCtom$ lie below about $0.5\,\LSTCtev $ and that they decay to an electroweak
boson ($\gamma$, $W$, $Z$) plus $\LSTCtpi$; a pair of electroweak bosons; and $\bar
f f$, especially $\ell^+\ell^-$. These channels have very distinctive
signatures, made all the more so because $\LSTCtro$ and $\LSTCtom$ are very narrow,
$\Gamma(\LSTCtro) \simeq 1$--$5\,\LSTCgev $ and $\Gamma(\LSTCtom) \simeq
0.1$--$0.5\,\LSTCgev $. Technipions are expected to decay via ETC interactions to
the heaviest fermion-antifermion flavors allowed kinematically, providing the
best chance of their being detected.

Many higher-mass states are reasonably expected in addition to $\LSTCtro$ and
$\LSTCtom$. In Refs.~\cite{Lane:1993wz,Lane:1994pg} it was argued that walking TC
invalidates the standard QCD-based calculations of the precision-electroweak
$S$-parameter~\cite{Peskin:1990zt,Golden:1990ig,Holdom:1990tc,Altarelli:1991fk}.
In particular, the spectral functions appearing in $S$ cannot be saturated by
a single $\LSTCtro$ and its axial-vector partner $a_T$. Thus, walking TC
produces something like a tower of vector and axial-vector isovector states
above the lightest $\LSTCtro$ and $\LSTCta$. All (or many) of them may contribute
significantly to the $S$-parameter.\footnote{These higher mass states are
  also important in unitarizing longitudinal gauge boson scattering at high
  energies.} Most important phenomenologically, in models with small $S$, the
lightest $\LSTCta$ and $\LSTCtro$ likely are nearly degenerate and have similar
couplings to their respective weak vector and axial-vector currents; see,
e.g., Refs.~\cite{Appelquist:1998xf, Knecht:1997ts, Hirn:2006nt, Hirn:2006wg,
  Eichten:2007sx}.  The $3\,\LSTCtpi$-decay channels of the $\LSTCat$ are
closed, so these states are also very narrow, $\Gamma(\LSTCta) \LSTCsimle
0.5\,\LSTCgev$.

The $\LSTCtro$, $\LSTCtom$, $\LSTCta$, and $\LSTCtpi$ of low-scale technicolor that we
consider are bound states of the lightest technifermion electroweak doublet,
$(T_U,T_D)$. The phenomenology of these technihadrons is set forth in the
``Technicolor Straw-Man Model'' (TCSM)~\cite{Lane:1999uh, Lane:2002sm,
  Eichten:2007sx}. The TCSM's most important assumptions are: (1) There are
$N_D$ isodoublets of technifermions transforming according to the fundamental
representation of the TC gauge group. The lightest doublet is an
ordinary-color singlet.\footnote{Some of these doublets may be color
  nonsinglets with, e.g., three doublets for each color triplet.
  Technifermions get ``hard'' masses from ETC and ordinary color interactions
  and will have some hierarchy of masses. We expect that the lightest will be
  color-singlets.} We use $N_D = 9$ in calculations; then, the technipion
decay constant $F_T \simeq 246\,\LSTCgev /\sqrt{\smash[b]{N_D}} = 82\,\LSTCgev $. The
technipion isotriplet composed of the lightest technifermions is a simple
two-state admixture,
\LSTCbe\label{eq:LSTC_admix}
|\Pi_T^{\pm,0}\rangle = \sin\chi\,|W_L^{\pm,0}\rangle +
\cos\chi\,|\pi_T^{\pm,0}\rangle\,,
\LSTCee
where $W_L$ is a {\em longitudinally}-polarized weak boson and $\LSTCtpi$ is a
mass eigenstate, the lightest technipion referred to above, and $\sin\chi =
F_T/246\,\LSTCgev  = 1/\sqrt{\smash[b]{N_D}}$. This is why the lightest
spin-one technihadrons are so narrow: all their decay amplitudes are
suppressed by a power of $\sin\chi$ for each $W_L$ emitted and by a power of
$e = g\sin\theta_W$ for each {\em transversely}-polarized $W_\perp$. In
addition, decays to $\LSTCtpi$ are phase-space limited.\footnote{Because the
  interactions of the techni-vectors with electroweak gauge bosons (and
  fermions) are suppressed by $\sin\chi$, they can be light, $\LSTCsimge
  200\,\LSTCgev $, without conflicting with precision electroweak and Tevatron
  data.}  The technihadrons' principal decay modes are listed in Table~1. (2)
The lightest bound-state technihadrons may be treated {\em in isolation},
without significant mixing or other interference from higher-mass states. (3)
Techni-isospin is a good symmetry.\footnote{Also, something like
  topcolor-assisted technicolor~\cite{Hill:1994hp} is needed to keep the top
  quark from decaying copiously into $\LSTCtpip b$ when $M_{\LSTCtpi} \LSTCsimle
  160\,\LSTCgev $. Thus, if $\LSTCtpip$ is heavier than the top, it will not decay
  exclusively to $t \bar b$.} These assumptions allow the TCSM to be
described by a relatively small number of parameters. The ones used for the
present study are, we believe, fairly generic; they are listed below.

{\begin{table}[!ht]
    \begin{center}{
 \begin{tabular}{|c|c|c|}
 \hline
 Process & $V_{V_T/a_T G\LSTCtpi}$ & $A_{V_T/a_T G\LSTCtpi}$\\
 \hline\hline
 $\LSTCtom \LSTCra \gamma \LSTCtpiz$& $\cos\chi$ & 0\\
 $\LSTCts\LSTCts\LSTCts\quad \LSTCra \gamma Z^0_L$ & $\sin\chi$ & 0\\ 
 $\LSTCts\LSTCts\LSTCts\qquad \LSTCra W^\pm \LSTCtpimp$ & $\cos\chi/(2\sin\LSTCthw)$ & 0 \\ 
  $\,\,\,\;\;\qquad \LSTCra W^\pm W_L^\mp$ & $\sin\chi/(2\sin\LSTCthw)$ & 0 \\
 $\qquad \LSTCra Z^0 \LSTCtpiz$ & $\cos\chi\cot 2\LSTCthw$ & 0 \\ 
  $\,\qquad \LSTCra Z^0 Z_L^0$ &$\sin\chi\cot 2\LSTCthw$ & 0\\
 \hline
 $\LSTCtroz \LSTCra W_L^{\pm} \LSTCtpimp$ & $\sin\chi\cos\chi$ & ---  \\
 $\;\;\;\quad \LSTCra W_L^+ W_L^-$  & $\sin^2\chi$ & --- \\
 $\; \LSTCra \gamma \LSTCtpiz$ & $(Q_U + Q_D)\cos\chi$ & 0 \\
 $\; \LSTCra \gamma Z_L^0$ & $(Q_U + Q_D)\sin\chi$ & 0 \\
$\;\;\quad \LSTCra W^\pm \LSTCtpimp$ & 0 & $\pm\cos\chi/(2\sin\LSTCthw)$\\ 
  $\;\;\;\quad \LSTCra W^\pm W_L^\mp$ & 0 & $\pm\sin\chi/(2\sin\LSTCthw)$\\
 $\;\quad\LSTCra Z^0 \LSTCtpiz$ & $-(Q_U+Q_D)\LSTCts \cos\chi\tan\LSTCthw$ & 0\\
  $\;\quad \LSTCra Z^0 Z_L^0$ & $-(Q_U+Q_D)\sin\chi\tan\LSTCthw$ & 0\\
 \hline
 $\LSTCtropm \LSTCra W_L^{\pm} \LSTCtpiz$ & $\sin\chi\cos\chi$ & ---  \\
 $\quad \LSTCra Z_L^0 \LSTCtpipm$  & $\sin\chi\cos\chi$ & ---  \\
 $\,\,\quad \LSTCra W_L^\pm Z_L^0$  & $\sin^2\chi$ & --- \\
 $\; \LSTCra \gamma \LSTCtpipm$ & $(Q_U + Q_D)\cos\chi$ & 0\\ 
 $\;\;\;\LSTCra \gamma W_L^\pm$ & $(Q_U + Q_D)\sin\chi$ & 0\\
 $\;\;\; \LSTCra Z^0 \LSTCtpipm$ & $-(Q_U+Q_D)\cos\chi\tan\LSTCthw$ & $\pm \cos\chi
 /(\sin 2\LSTCthw)$\\  
  $\quad\;\; \LSTCra Z^0 W_L^\pm$ & $-(Q_U+Q_D)\sin\chi\tan\LSTCthw$ &
  $\pm\sin\chi/(\sin2\LSTCthw)$\\
 $\quad\;\; \LSTCra W^\pm \LSTCtpiz$ & 0 & $\mp\cos\chi/(2\sin\LSTCthw)$\\ 
  $\quad\;\; \LSTCra W^\pm Z_L^0$ & 0 & $\mp\sin\chi/(2\sin\LSTCthw)$\\
 \hline
  $a_T^0\LSTCra W^\pm \LSTCtpimp$ & 0 & $\mp\cos\chi/(2\sin\LSTCthw)$ \\ 
  $\qquad \LSTCra W^\pm W_L^\mp$    & 0 & $\mp\sin\chi/(2\sin\LSTCthw)$\\
  \hline
  $a_T^\pm \LSTCra \gamma \LSTCtpipm$ & 0 & $\mp\cos\chi$ \\
  $\qquad \LSTCra \gamma W_L^\pm$ & 0 & $\mp\sin\chi$\\
  $\qquad\;\, \LSTCra W^\pm \LSTCtpiz$ & 0 & $\pm\cos\chi/(2\sin\LSTCthw)$\\ 
  $\quad\;\;\; \LSTCra Z^0\LSTCtpipm$ & 0 & $\mp\cos\chi\cot 2\LSTCthw$\\
  $\,\;\qquad \LSTCra W^\pm Z_L^0$ & 0 & $\pm\sin\chi/(2\sin\LSTCthw)$\\ 
  $\,\;\qquad \LSTCra W_L^\pm Z^0$ & 0 & $\mp\sin\chi\cot 2\LSTCthw$ \\ 
  \hline\hline
\end{tabular}}
\caption{Amplitude factors for the dominant decay modes of $\LSTCtro \LSTCra W_L
  \LSTCtpi$, $W_L W_L$ and $\LSTCtrho,\,\LSTCtom,\,\LSTCta \LSTCra G \LSTCtpi$, $G
  W_L$~\cite{Lane:2002sm, Eichten:2007sx}. Here, $W_L$ is a
  longitudinally-polarized and $G = \gamma, W_\perp, Z_\perp$ a
  transversely-polarized electroweak gauge boson.  Technifermion charges are
  $Q_U = Q_D + 1$, and $\sin\chi = F_T/246\,\LSTCgev  = 1/\sqrt{\smash[b]{N_D}}$.
  Amplitudes for $W_L\LSTCtpi$ and $W_LW_L$ are proportional to $g_{\rho_T} =
  \sqrt{4\pi\alpha_{\rho_T}}$, where $\alpha_{\rho_T} = 2.16(3/N_{TC})$; those
  for emission of a transverse gauge boson are proportional to $e =
  \sqrt{4\pi\alpha}$.}
\end{center}
\end{table}}

The main discovery channel for low-scale technicolor at the Tevatron is
$\LSTCtrho \LSTCra W^\pm \LSTCtpi \LSTCra \ell^\pm \nu_\ell b q$, with at least one tagged
$b$-jet. At the LHC, this channel is swamped by a $\bar t t$ background
100~times larger than at the Tevatron. There the discovery channels will be
$WZ$, $\gamma W$ and $\gamma Z$, with the weak bosons
decaying into charged leptons.\footnote{The channel $\LSTCtropm \LSTCra W^\pm Z^0$
  was studied by P.~Kreuzer, {\em Search for Technicolor at CMS in the
    $\rho_{TC} \LSTCra W + Z$ Channel}, CMS~Note~2006/135, and
  Ref.~\cite{Kreuzer:2007zz}. To consider the decay angular distributions, we
  employ somewhat different cuts than he did.} In the TCSM each of these
modes is dominated (generally $\LSTCsimge 80\%$) by production of a {\em single}
resonance:
\LSTCbe\label{eq:LSTC_VTdecays}
\LSTCtropm \LSTCra W^\pm Z^0,\quad a_T^\pm \LSTCra \gamma W^\pm,\quad \LSTCtom \LSTCra \gamma
Z^0\,.
\LSTCee
In Sects.~2-4, the PGS detector simulator~\cite{PGS} is used for our
preliminary studies of these signals and their backgrounds. None of these LHC
discovery modes involve observation of an actual technipion (other than the
ones already observed, $W_L^\pm$ and $Z_L^0$). There are other
strong-interaction scenarios of electroweak symmetry breaking (e.g.,
so-called Higgsless models in five dimensions~\cite{SekharChivukula:2004mu,
  Chivukula:2005bn} and deconstructed models~\cite{Csaki:2003dt,
  Csaki:2003zu, Agashe:2003zs, Cacciapaglia:2004rb}) which predict narrow
vector and axial-vector resonances, but they do not decay to technipion-like
objects. Therefore, observation of technipions in the final state is
important for confirming LSTC as the mechanism underlying electroweak
symmetry breaking. It is possible to do this at high luminosity with the
decays $\LSTCtropm,\, a_T^\pm \LSTCra Z^0 \LSTCtpipm \LSTCra \ell^+ \ell^- b q$.  This
channel also provides the interesting possibility of observing both $\LSTCtropm$
and $a_T^\pm$ in the same final state. This analysis, using
ATLFAST~\cite{atlfast}, is summarized in Sect.~5~\cite{LSTCazuelos}.

In addition to the discovery of narrow resonances in these channels, the
angular distributions of the two-body final states in the techni-vector rest
frame provide compelling evidence of their underlying technicolor origin.
Because all the modes involve at least one longitudinally-polarized weak
boson, the distributions are
\LSTCbea\label{eq:LSTC_angdist}
&& \frac{d\sigma(\bar q q \LSTCra \LSTCtropm \LSTCra W_L^\pm Z_L^0)}{d\cos\theta},\;\;
\frac{d\sigma(\bar q q \LSTCra \LSTCtropm \LSTCra \LSTCtpipm Z_L^0)}{d\cos\theta} \propto
\sin^2\theta\,;\\
&& \frac{d\sigma(\bar q q \LSTCra a_T^\pm \LSTCra \gamma W_L^\pm)}{d\cos\theta},\;\;
\frac{d\sigma(\bar q q \LSTCra \LSTCtom \LSTCra \gamma Z_L^0)}{d\cos\theta} \propto
1+ \cos^2\theta\,.
\LSTCeea
It is fortunate that each of the two-electroweak-boson final states is dominated
by a single technihadron resonance. Otherwise, because of the resonances'
expected closeness, it would likely be impossible to disentangle the
different forms.  Our simulations include these angular distributions.

For Les Houches, we concentrated on three TCSM mass points that cover most of
the reasonable range of LSTC scales; they are listed in Table~2. In all
cases, we assumed isospin symmetry, with $M_{\LSTCtro} = M_{\LSTCtom}$ and $M_{a_T} =
1.1 M_{\LSTCtro}$; also, the $\LSTCtro$ and $\LSTCta$ constants describing coupling to
their respective weak currents were taken equal; $\sin\chi = 1/3$; $Q_U + Q_D
= 1$; $N_{TC} = 4$ for the TC gauge group $SU(N_{TC})$; and $M_{V_{1,2,3}} =
M_{A_{1,2,3}} = M_{\LSTCtro}$ for the LSTC mass parameters controlling the
strength of $\LSTCtro$, $\LSTCtom$, $\LSTCta$ decays to a transverse electroweak boson
plus $\LSTCtpi/W_L$ or $\LSTCtro$/$\LSTCtom$~\cite{Lane:1999uh, Lane:2002sm,
  Eichten:2007sx}. {\sc Pythia}~\cite{Sjostrand:2006za} has been updated to
include these and other LSTC processes, according to the rules of the TCSM.
The new release and its description may be found at {\tt www.hepforge.org}.

The simulations presented here, especially those using the PGS detector
simulator~\cite{PGS}, are preliminary and in many respects quite
superficial.\footnote{All PGS simulations were done using the ATLAS parameter
  set provided with the PGS extension of MADGRAPHv4.0~\cite{Maltoni:2002qb,
    Alwall:2007st}. The relevant parameters are: calorimeter segmentation
  $\Delta \eta \times \Delta \phi = 0.1 \times 0.1$, jet resolution $\Delta
  E/E = 0.8/\sqrt{E}$, and electromagnetic resolution $\Delta E/E =
  0.1/\sqrt{E} + 0.01$. To model muons more realistically, we changed the
  sagitta resolution to $50\mu m$. With this set of parameters, the PGS
  lepton identification efficiency is $\approx 90$\% in their kinematic
  region of interest, $p_T > 10\,\LSTCgev $ and $|\eta| < 2.5$.} E.g., no attempt
was made to optimize $S/B$ in the PGS studies. Nor did we carry out a serious
analysis of statistical, let alone systematic, errors.\footnote{Potentially
  important sources of systematic error are the higher-order QCD corrections
  to signal and backgrounds. The $K$-factors can be quite large, $\sim 1.5$.}
Still, we believe the simulations establish the LHC's ability to discover, or
rule out, important signatures of low-scale technicolor.  Beyond that, it is
our intent that the present studies will stimulate more thorough ones by
ourselves and by the ATLAS and CMS collaborations.

 \begin{table}[!ht]
     \begin{center}{
  \begin{tabular}{|c|c|c|c|c|c|c|c|c|c|}
  \hline
 Case & $M_{\LSTCtro} = M_{\LSTCtom}$ & $M_{a_T}$ & $M_{\LSTCtpi}$ & $M_{\LSTCtpipr}$  & &
 $\sigma(W^\pm Z^0)$ & $\sigma(\gamma W^\pm)$ & $\sigma(\gamma Z^0)$ &
 $\sigma(Z^0 \LSTCtpipm)$ \\
  \hline\hline
 A & 300 & 330 & 200 & 400 & & 110 & 168 & 19.2 &  158 \\
 B & 400 & 440 & 275 & 500 & & 36.2 & 64.7 & 6.2 & 88.6 \\
 C & 500 & 550 & 350 & 600 & & 16.0 & 30.7 & 2.8 &  45.4 \\
   \hline\hline
 \end{tabular}}
 \caption{Masses (in GeV) and signal cross sections (in fb) for the lightest
   technihadrons for the three TCSM mass points in this study. Isospin
   symmetry is assumed. The $\LSTCtpipr$ is an isosinglet, color-singlet
   technipion expected in TC models; we have assumed it so heavy that the
   $\LSTCtro$, $\LSTCtom$ and $\LSTCta$ cannot decay to it. The cross sections combine
   contributions from $\LSTCtro$, $\LSTCtom$ and $\LSTCta$, but tend to be dominated by a
   single resonance. Branching ratios of the $W$ and $Z$ to electrons and
   muons are included in all cross sections.}
 \end{center}
 \end{table}

\begin{table}
\begin{center}
\begin{tabular}{|c|c|c|} \hline
Background & Cross section ($\LSTCfb$) & Comments \\ \hline
$WZ\rightarrow 3\ell + \nu$ & 430 & \\ \hline
$ZZ\rightarrow 4\ell$ & 52 &  \\ \hline
$Z + \bar b b \rightarrow \ell^+ \ell^-~ \bar b b$ & 7600 & 
$p_T(b) > 15.0\ \LSTCgev ,\ |\eta_b| < 3.5$ \\ \hline
$\bar t t \rightarrow 2\ell~2\nu~ \bar b b$ & 22,800 & {\sc Pythia} generator
\\ \hline\hline
$W\gamma \rightarrow \ell\nu\gamma$ & 2560 & $p_T(\gamma) > 
40\ \LSTCgev ,\ |\eta_{\gamma}|  < 3.5$ \\ \hline
$W\,\LSTCjet \rightarrow \ell\nu\gamma $ (fake) & 3180 & $p_T(\LSTCjet) 
 > 40\ \LSTCgev ,\ |\eta_{\LSTCjet}|  < 3.5$ \\
 & &Includes 0.1\% fake rate \\ \hline \hline
 $Z \gamma  \rightarrow \ell^+ \ell^- \gamma$ & 700 & $p_T(\gamma)
 > 40\ \LSTCgev ,\ |\eta_{\gamma}|  < 3.5$ \\ \hline
 $Z\,\LSTCjet \rightarrow \ell^+ \ell^- \gamma$ (fake) & 315 &  
$p_T(\LSTCjet) > 40\ \LSTCgev ,\ |\eta_{\LSTCjet}|  < 3.5$ \\
 & & Includes 0.1\% fake rate \\ \hline \hline
\end{tabular}
 \caption{Backgrounds to the $W^\pm Z^0$, $\gamma W^\pm$ and $\gamma Z^0$
   signals of low-scale technicolor. The generator is
   ALPGENv13~\cite{Mangano:2002ea} unless indicated otherwise. Branching
   ratios of the $W$ and $Z$ to electrons and muons are included in the cross
   sections.}
\end{center}
\end{table}

\section{$\LSTCtropm  \LSTCra W_L^\pm Z_L^0 \LSTCra \ell^\pm \nu_\ell \ell^+ \ell^-$}

The cross sections for $\LSTCtropm \LSTCra W_L^\pm Z_L^0$, including branching ratios
to electrons and muons, are listed in Table~2.\footnote{For the TCSM
  parameters we use, about 20\% of these $\rho_T^\pm \LSTCra W^\pm
  Z^0$ rates involve one transverse gauge
  boson.} Signal events were generated with the updated {\sc
  Pythia}~\cite{Sjostrand:2006za}. The principal backgrounds are in Table~3,
along with their generators and parton-level cuts. The ALPGEN backgrounds
were passed through {\sc Pythia} for showering and hadronization.

Events were selected which have exactly three leptons, electrons and/or
muons, with two having the same flavor and opposite sign, $|\eta_\ell| <
2.5$, at least one having $p_T > 30\,\LSTCgev $, and the others with $p_T >
10\,\LSTCgev $. No cut on $\LSTCetmiss$ was applied in this analysis, though it may
improve $S/B$ to do so. The $Z$ was reconstructed from two same-flavor,
opposite-sign leptons with the smallest $|M_{\LSTCellp\LSTCellm} - M_Z| < 7.8\,\LSTCgev $.
In reconstructing the $W$, ${\vec p}_T(\nu) = -\sum{\vec {\LSTCetmiss}}$ was
assumed, and the quadratic ambiguity in $p_z(\nu)$ was resolved in favor of
the solution minimizing the opening angle between the neutrino and the
charged lepton assigned to the $W$, as would be expected for a boosted
$W$.\footnote{The efficacy of this procedure, which was adopted at Les
  Houches (``the LH algorithm'') was compared to a ``TeV algorithm'' used at
  the Tevatron. The TeV algorithm chooses the $p_z(\nu)$ solution which gives
  the smaller $W$ energy. In ATLAS~\cite{LSTCblack} and CMS-based
  analyses~\cite{LSTCbose}, it was found that the TeV algorithm does slightly
  better at choosing the correct solution.}

\begin{figure}[!t]
    \begin{center}
      \includegraphics[width=3.50in, height = 6.00in, angle=90]
      {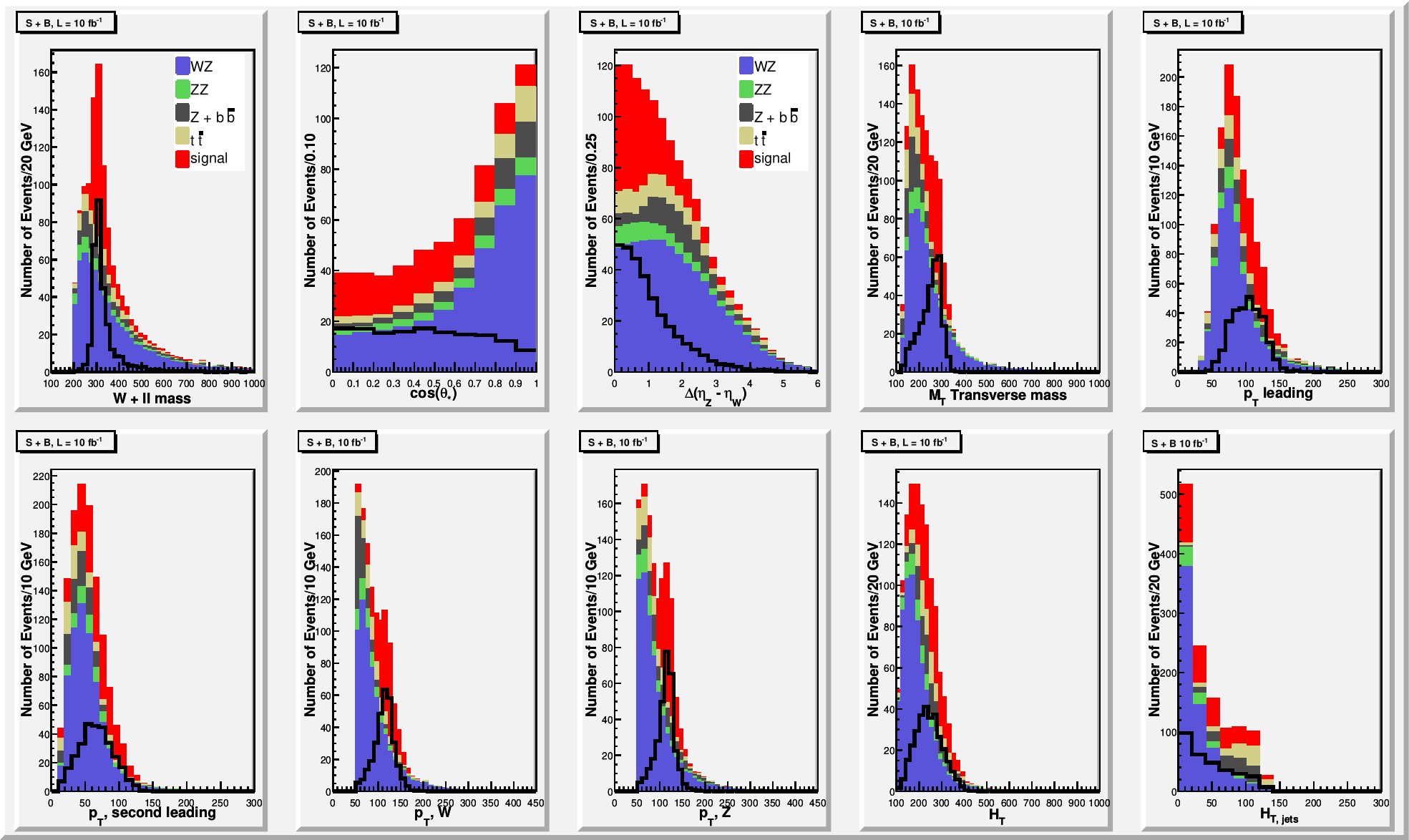}
      \caption{Signal and background distributions of a $300\,\LSTCgev $ $\LSTCtropm \LSTCra
        W^\pm Z^0 \LSTCra \LSTCellpm \nu_\ell \LSTCellp\LSTCellm$ for $10\,\LSTCifb$ at the LHC;
        $p_T(W,Z) > 50\,\LSTCgev $ and $H_{T}({\rm jets}) < 125\,\LSTCgev $. The open
        black histograms are the signal contributions.}
  \label{LSTC_LHC_1}
    \end{center}
  \end{figure}

Figure~\ref{LSTC_LHC_1} shows various distributions for case~A with
$p_T(W),\, p_T(Z) > 50\,\LSTCgev $ and $H_T({\rm jets}) \equiv \sum E_T({\rm
  jet}) < 125\,\LSTCgev $.  The $H_T$ cut significantly reduces the $\bar t
t$ background.  The integrated luminosity is $10\,\LSTCifb$, and a strong
signal peak is clearly visible above background in the first panel. Fitting
the peak to a Gaussian, its mass is $311\,\LSTCgev $. Counting signal and
background within twice the fitted resolution of $25\,\LSTCgev $, only
$\int\LSTCCL dt = 2.4\,\LSTCifb$ is required for a $S/\sqrt{\smash[b]{S+B}} =
5\,\sigma$ discovery of this resonance. Table~4 contains the final-state mass
resolutions and $5\,\sigma$ discovery luminosities for the
two-electroweak-boson modes considered here. The poorer resolution in the
$WZ$ and $\gamma W$ channels is due to the $\LSTCetmiss$
resolution.\footnote{The $p_T$ cuts listed in this table were not optimized
  for discovery; rather they were chosen partly to reveal as much of the
  angular distributions as possible consistent with background reduction.
  Presumably, in a real search, harder cuts would be employed to reveal the
  signal. Once it was found, the $p_T$ cut could be loosened and the
  final-state mass cut tightened to focus on the angular distribution. The
  upward shift of the $\LSTCtropm$ peak mass, evident in their non-Gaussian
  high-mass tails, may be due to $a_T^\pm \LSTCra W^\pm Z^0$ at about 20\%
  the strength of $\LSTCtropm$. These issues are being considered with more
  sophistication using ATLFAST~\cite{atlfast} and CMS Fast Simulation
  ({\tt https://twiki.cern.ch/twiki/bin/view/CMS/WorkBookFastSimulation}) in
  Refs.~\cite{LSTCblack,LSTCbose}. They find discovery luminosities about
  15--30\% lower than estimated here.}

 \begin{table}[!ht]
 \begin{center}
 \begin{tabular}{|c|c|c|c|c|}\hline
 $WZ$ &  $M_{\rm peak}$ $(\LSTCgev )$ & $\sigma\ (\LSTCgev )$ & $\LSTCCL_{\rm min}\,(\LSTCifb)$
 & $p_T$
 cut \\  \hline
 A & 311 & 25.6 & 2.4 & $p_T(W,Z) > 50\ \LSTCgev $ \\ 
 B & 414 & 34.5 & 7.2 & $p_T(W,Z) > 75\ \LSTCgev $ \\
 C & 515 & 41.0 & 14.7 &  $p_T(W,Z) > 75\ \LSTCgev $ \\ \hline 
 $\gamma W$ &  $M_{\rm peak}$ $(\LSTCgev )$ & $\sigma\ (\LSTCgev )$ & $\LSTCCL_{\rm min}\,(\LSTCifb)$ &
 $p_T$ cut \\  \hline 
 A & 328 & 31.2 & 2.3 & $p_T(\gamma,W) > 75\ \LSTCgev $ \\
 B & 439 & 39.1 & 4.5 & $p_T(\gamma,W) > 100\ \LSTCgev $ \\
 C & 547 & 39.3 & 7.8 & $p_T(\gamma,W) > 125\ \LSTCgev $ \\ \hline
  $\gamma Z$ &  $M_{\rm peak}$ $(\LSTCgev )$ & $\sigma\ (\LSTCgev )$ & $\LSTCCL_{\rm min}\,(\LSTCifb)$ &
 $p_T$ cut \\  \hline 
 A & 299 & 7.3 & 16.8 &  $p_T(\gamma,Z) > 80\ \LSTCgev $ \\
 B & 398 & 9.4 & 45.5 &  $p_T(\gamma,Z) > 110\ \LSTCgev $ \\
 C & 498 & 12.0 & 97.2 & $p_T(\gamma,Z) > 150\ \LSTCgev $ \\ \hline
 \end{tabular}
 \caption{PGS simulation data for the spin-one technihadrons decaying to a pair
   of electroweak gauge bosons. A simple Gaussian fit is made to determine
   the mass and width of the resonance; signal and background events are
   counted within $\pm 2\,\sigma$ of the peak value to determine the minimum
   luminosity needed for $S/\sqrt{\smash[b]{S+B}} = 5\,\sigma$.}
 \end{center}
 \end{table} 

\begin{figure}[!t]
    \begin{center}
      \includegraphics[width=3.50in, height = 4.00in, angle=90]
      {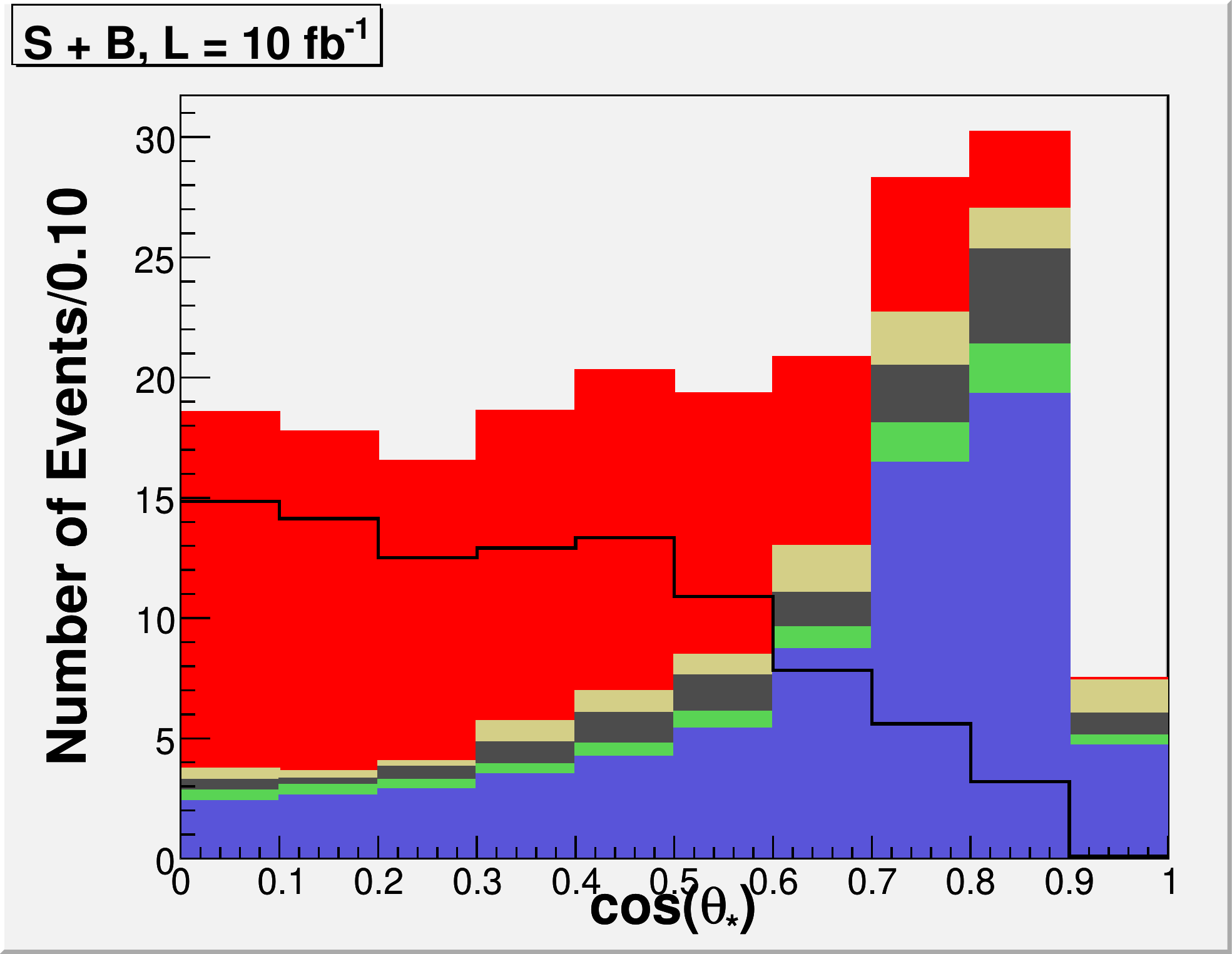}
      \caption{$WZ$ angular distribution of the signal
        and backgrounds for a $300\,\LSTCgev $ $\LSTCtropm \LSTCra W^\pm Z^0 \LSTCra \LSTCellpm
        \nu_\ell \LSTCellp\LSTCellm$ for $10\,\LSTCifb$ at the LHC; $280 < M_{WZ} <
        340\,\LSTCgev $; other cuts are listed in the text. The color code is
        given in Fig.~1.}
  \label{LSTC_LHC_2}
    \end{center}
  \end{figure}

 The second panel in Fig.~\ref{LSTC_LHC_1} shows the total and signal $WZ$
 angular ($|\cos\theta|$) distribution. The distribution is folded since the
 signal and $WZ$ background are even functions of $\cos\theta$. The total
 distribution reflects the forward-backward peaking of the standard $WZ$
 production. The signal distribution (open black histogram) is much flatter
 than the expected $\sin^2\theta$, presumably because of poorly-fit $W$'s and
 their effect on determining $M_{WZ}$ and the $WZ$ rest frame. To remedy
 this, we take advantage of the LSTC technihadrons' very small widths and
 require $280 < M_{WZ} < 340\,\LSTCgev $; see Fig.~\ref{LSTC_LHC_2}. The signal
 distribution now has the expected $\sin^2\theta$ shape. The remaining large
 background at $|\cos\theta| \LSTCsimge 0.7$ can be fit and subtracted by
 measuring the angular distribution in the sidebands $220 \LSTCsimle M_{WZ}
 \LSTCsimle 280\,\LSTCgev $ and $340 \LSTCsimle M_{WZ} \LSTCsimle 400\,\LSTCgev $. We believe that
 $10\,\LSTCifb$ is sufficient to distinguish this angular distribution from
 $1+\cos^2\theta$, but detailed fitting is required to confirm this; see
 Ref.~\cite{LSTCblack}.

  \begin{figure}[!t]
    \begin{center}
      \includegraphics[width=3.00in, height = 3.00in, angle=90]
      {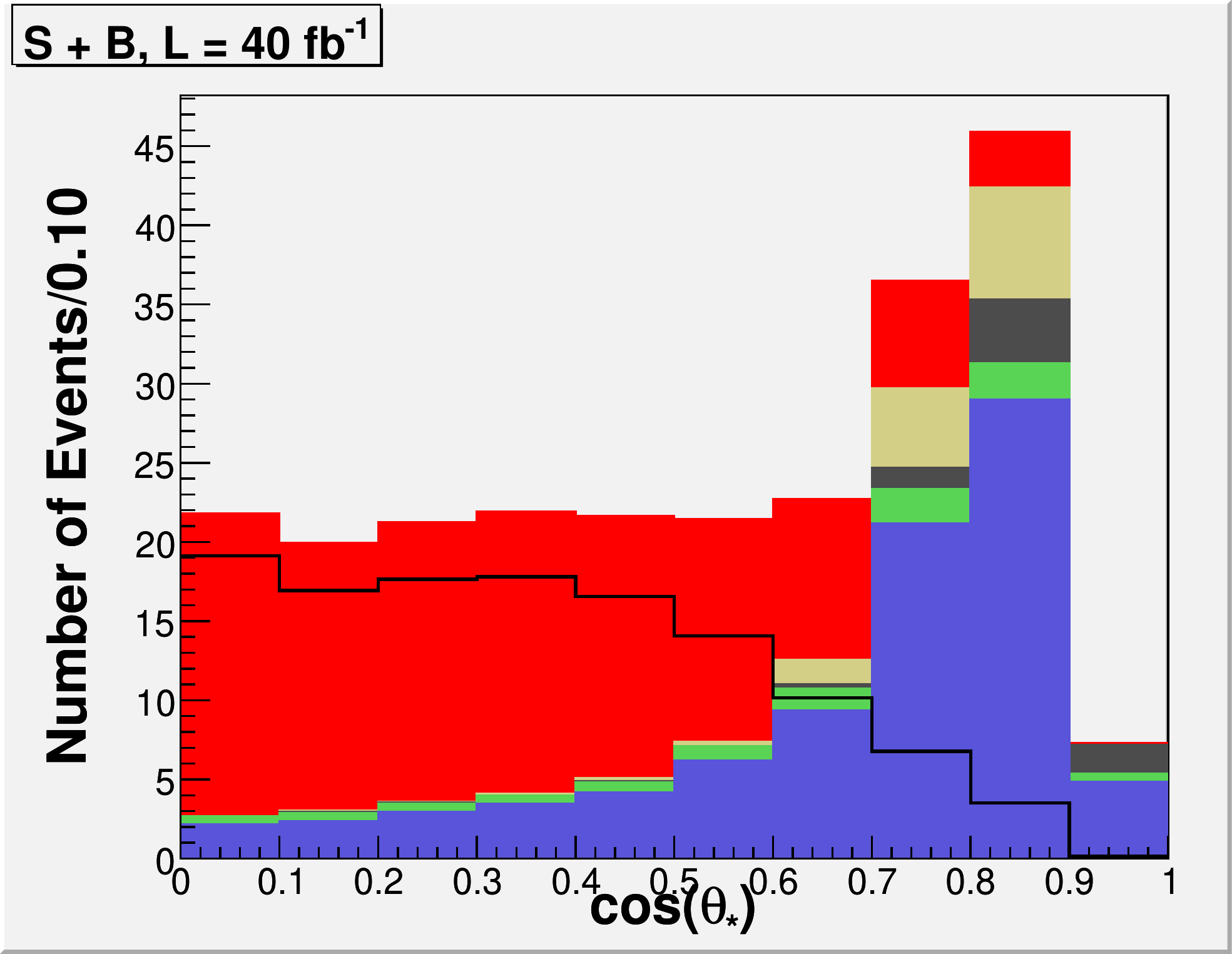}
      \includegraphics[width=3.00in, height = 3.00in, angle=90]
      {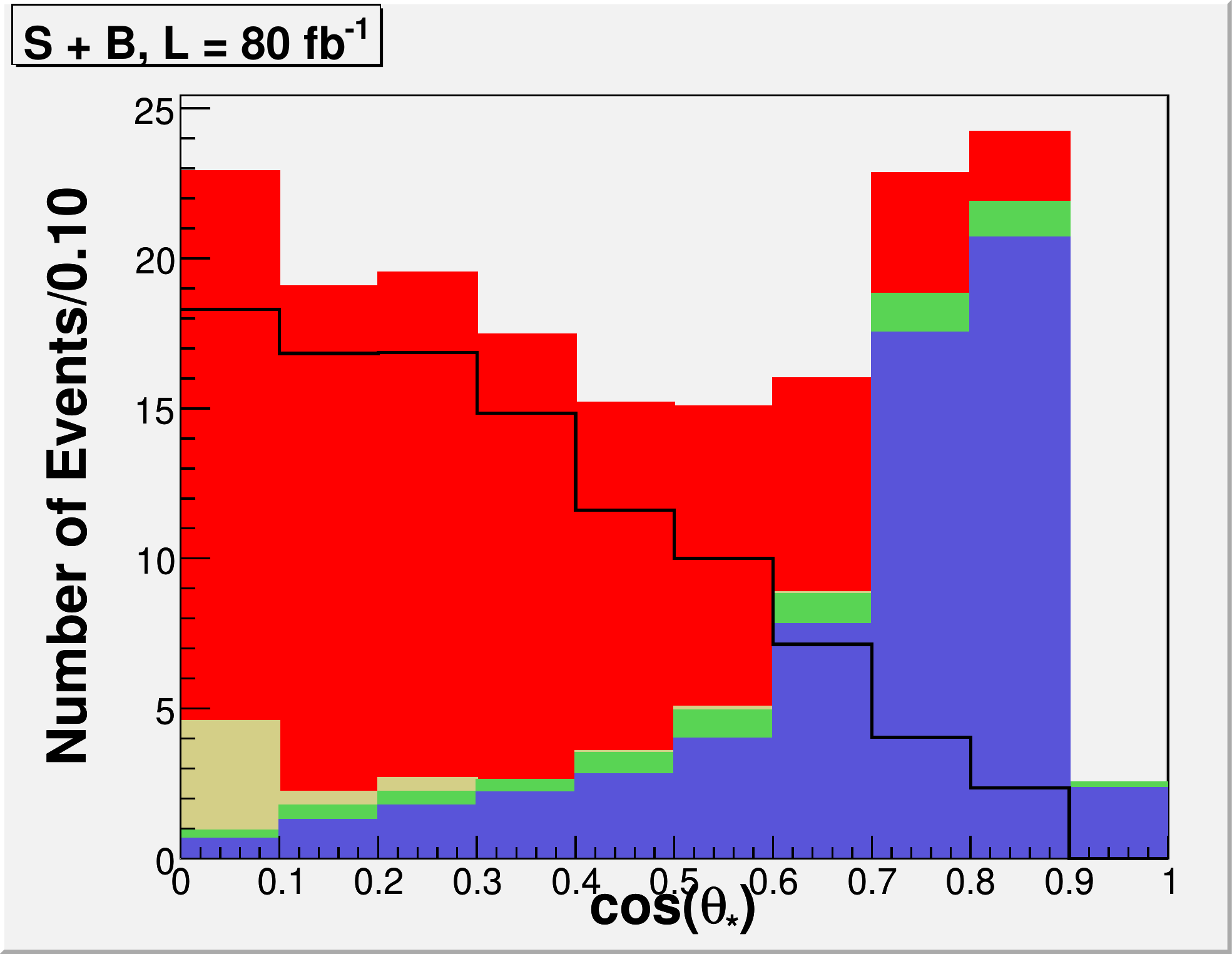}
      \caption{$WZ$ angular distributions of the signal
        and backgrounds for $\LSTCtropm \LSTCra W^\pm Z^0 \LSTCra \LSTCellpm \nu_\ell \LSTCellp
        \LSTCellm$ Left: $M_{\LSTCtro} = 400\,\LSTCgev $ with $380 < M_{WZ} < 440\,\LSTCgev $
        for $40\,\LSTCifb$. Right: $M_{\LSTCtro} = 500\,\LSTCgev $ with $480 < M_{WZ} <
        540\,\LSTCgev $ for $80\,\LSTCifb$. Other cuts are listed in the text. The
        color code is given in Fig.~1.}
  \label{LSTC_LHC_3}
    \end{center}
  \end{figure}
  \begin{figure}[!t]
    \begin{center}
      \includegraphics[width=4.00in, height = 3.50in, angle=0]
      {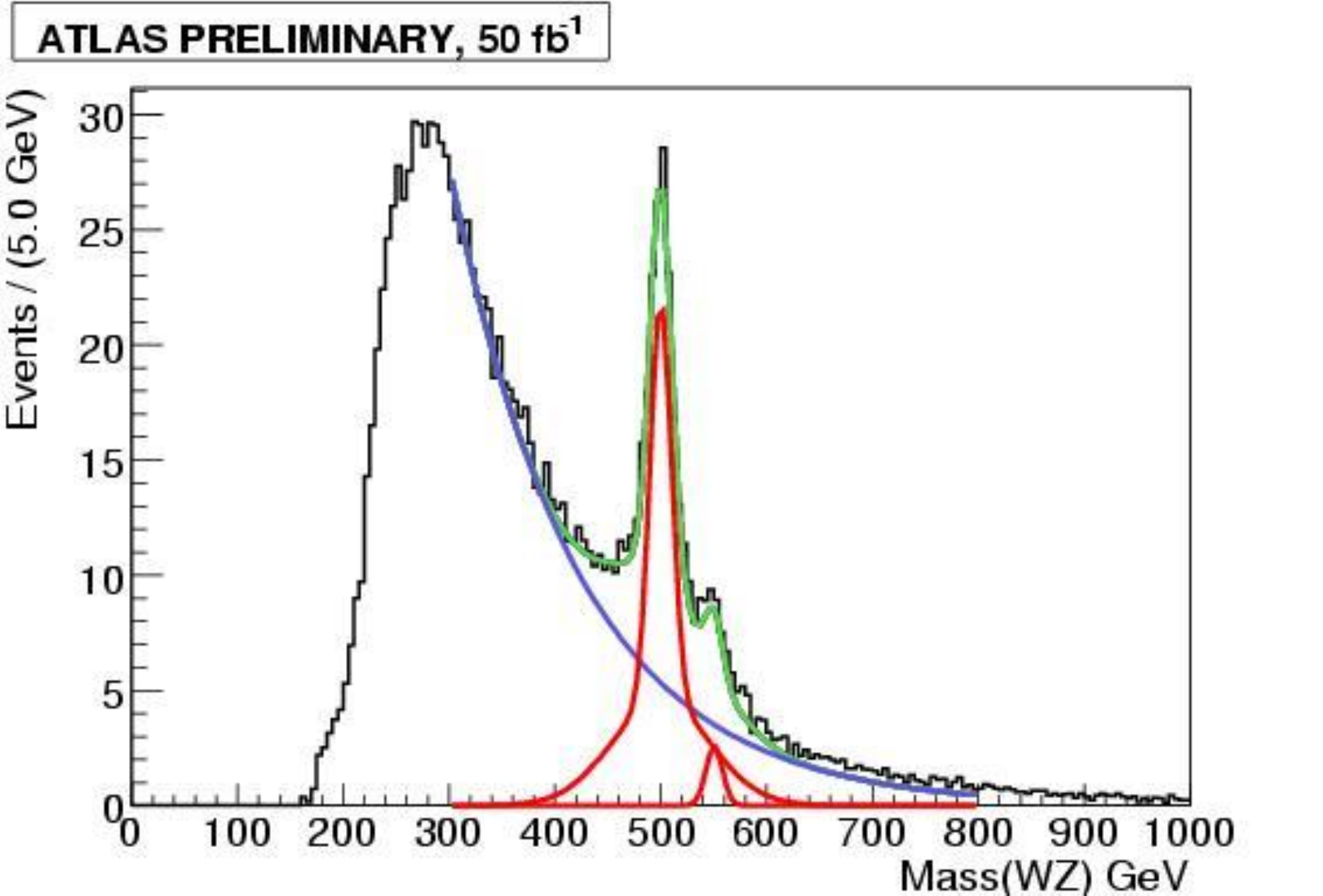}
      \caption{Fit of the signal and background $WZ$ invariant mass
        distribution, for case~C with $50\,\LSTCifb$, in ATLAS. Red curves
        represent the signal resonances, blue the backgrounds, and green the
        total~\cite{LSTCblack}.}
      \label{LSTC_LHC_4}
    \end{center}
  \end{figure}

The $|\cos\theta|$ distribution in the $\LSTCtro$-resonance region is shown in
Fig.~\ref{LSTC_LHC_3} for cases~B ($M_{\LSTCtro} = 400\,\LSTCgev $) and~C
($500\,\LSTCgev $). The $p_T(W,Z)$ cuts of~75 and $100\,\LSTCgev $ were chosen to
accept signal data over the same $\cos\theta$ range, 0.0--0.9, as in Case~A.
The luminosities of 40 and $80\,\LSTCifb$ were chosen to give roughly the same
statistics. For the higher luminosities, the effects of pile-up on
calorimetry and tracking were not considered.

Finally, for the $50\,\LSTCgev$ splitting used in case~C, it appears possible
to see $\rho_T^\pm$ and $a_T^\pm$ as separate peaks in the $M_{WZ}$
distribution. This was studied in Refs.~\cite{LSTCblack,LSTCbose}. With cuts
similar to those used above (except that $p_T(W,Z) > 50\,\LSTCgev$) a
simulation~\cite{LSTCblack} using ATLFAST was performed. The result is seen
in Fig.~\ref{LSTC_LHC_4} where a luminosity of $50\,\LSTCifb$ was assumed.
The $\LSTCtro$ and $\LSTCat$ were modeled as Gaussian distributions and the
background above $300\,\LSTCgev$ as a falling exponential. The $a_T$ appears
as a high-mass shoulder. The CMS analysis finds a similar
result~\cite{LSTCbose}. Both $\rho_T$ and $a_T$ can be observed and a
$5\,\sigma$ (combined) discovery achieved with luminosity $\simeq
9.5\,\LSTCifb$ provided that data is reasonably described by the simulated
$\LSTCetmiss$ resolution.

\section{$a_T^\pm \LSTCra \gamma W_L^\pm$}

    \begin{figure}[!t]
      \begin{center}
        \includegraphics[width=4.50in, height = 6.00in, angle=90]
        {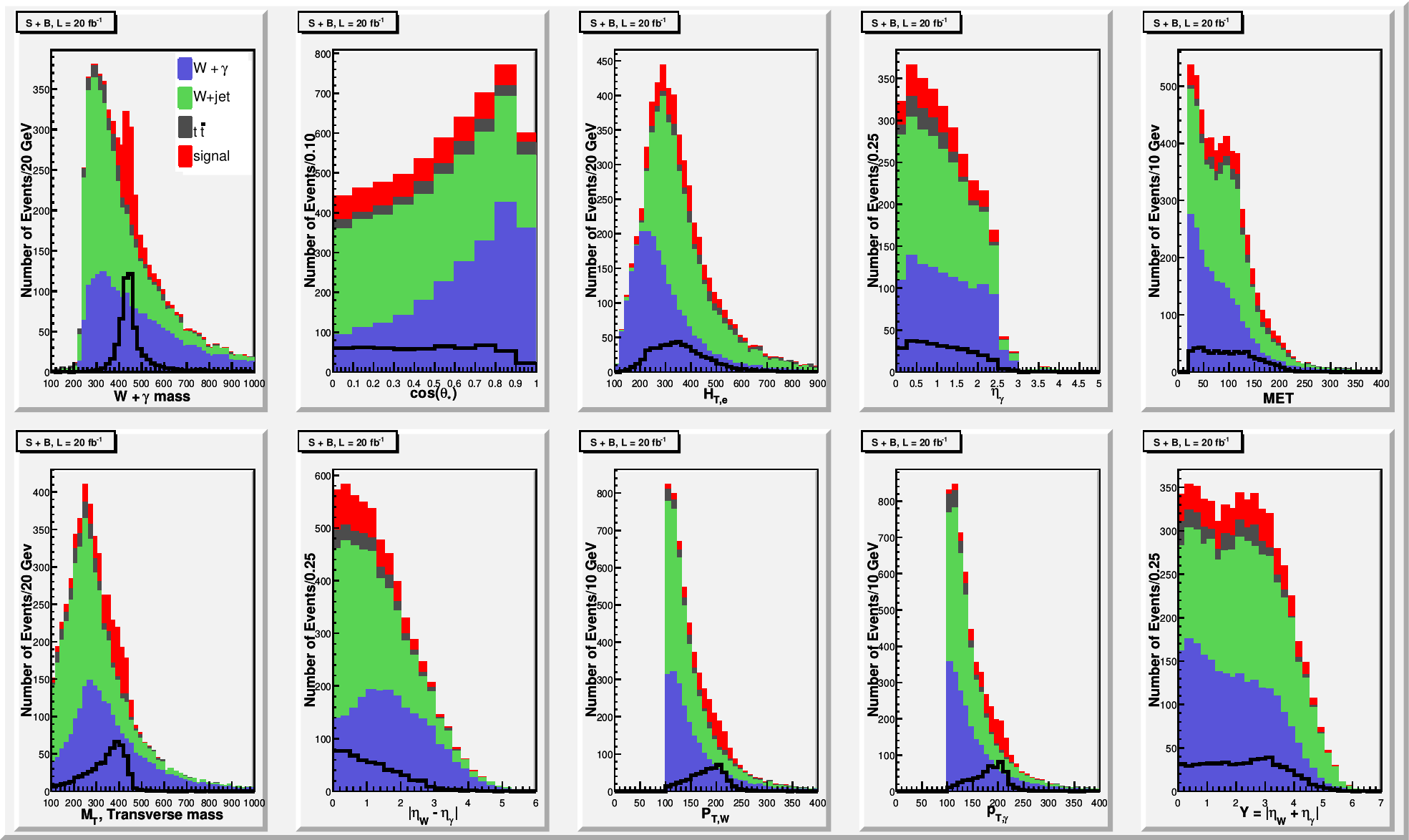}
        \caption{Signal and background distributions of a $440\,\LSTCgev $
        $a_T^\pm \LSTCra \gamma W^\pm \LSTCra \gamma \LSTCellpm \nu_\ell$ for $20\,\LSTCifb$
          at the LHC; $p_T(\gamma,W) > 100\,\LSTCgev $ and $\LSTCetmiss >
          20\,\LSTCgev $. The open black histograms are the signal contributions.}
    \label{LSTC_LHC_5}
      \end{center}
    \end{figure}

The axial-vector isovector $a_T$ is a new addition to the TCSM framework,
motivated by the arguments that the $S$ parameter problem of technicolor is
ameliorated if $\LSTCtro$ and $a_T$ are nearly degenerate and have nearly the
same couplings to the vector and axial-vector weak currents.

On account of space limitation, we show here only the results of PGS
simulation of Case~B, for which $M_{a_T} = 440\,\LSTCgev $. For the decays to a
pair of electroweak bosons considered in this report, $\sigma(a_T \LSTCra \gamma
W) B(W \LSTCra e/\mu\, \nu)$ are the largest; it is $65\,\LSTCfb$ in case~B. Signal
and background events were generated with $p_T(\gamma) > 40\,\LSTCgev $. As noted,
the discovery search could impose a higher threshold. Events were selected
with exactly one lepton, having $p_T > 10\,\LSTCgev $ and $|\eta| < 2.5$.
Distributions are displayed in Fig.~\ref{LSTC_LHC_5}, in which $\LSTCetmiss >
20\,\LSTCgev $ and $p_T(\gamma),\, p_T(W) > 100\,\LSTCgev $. The principal backgrounds
are in Table~3. A ${\rm jet} \LSTCra \gamma$ fake rate of $10^{-3}$ was assumed
for $W+\LSTCjet$~\cite{unknown:1999fr}. Another possible background, $\gamma +
\LSTCjet$ where the jet fakes a lepton, is negligible after the $\LSTCetmiss$ cut.
The luminosity of $20\,\LSTCifb$ was chosen to give reasonable statistics for the
signal's angular distribution. Consequently, the $a_T$ resonant peak has a
significance of $10\,\sigma$.

\begin{figure}[!t]
      \begin{center}
        \includegraphics[width=3.00in, height = 3.00in, angle=90]
        {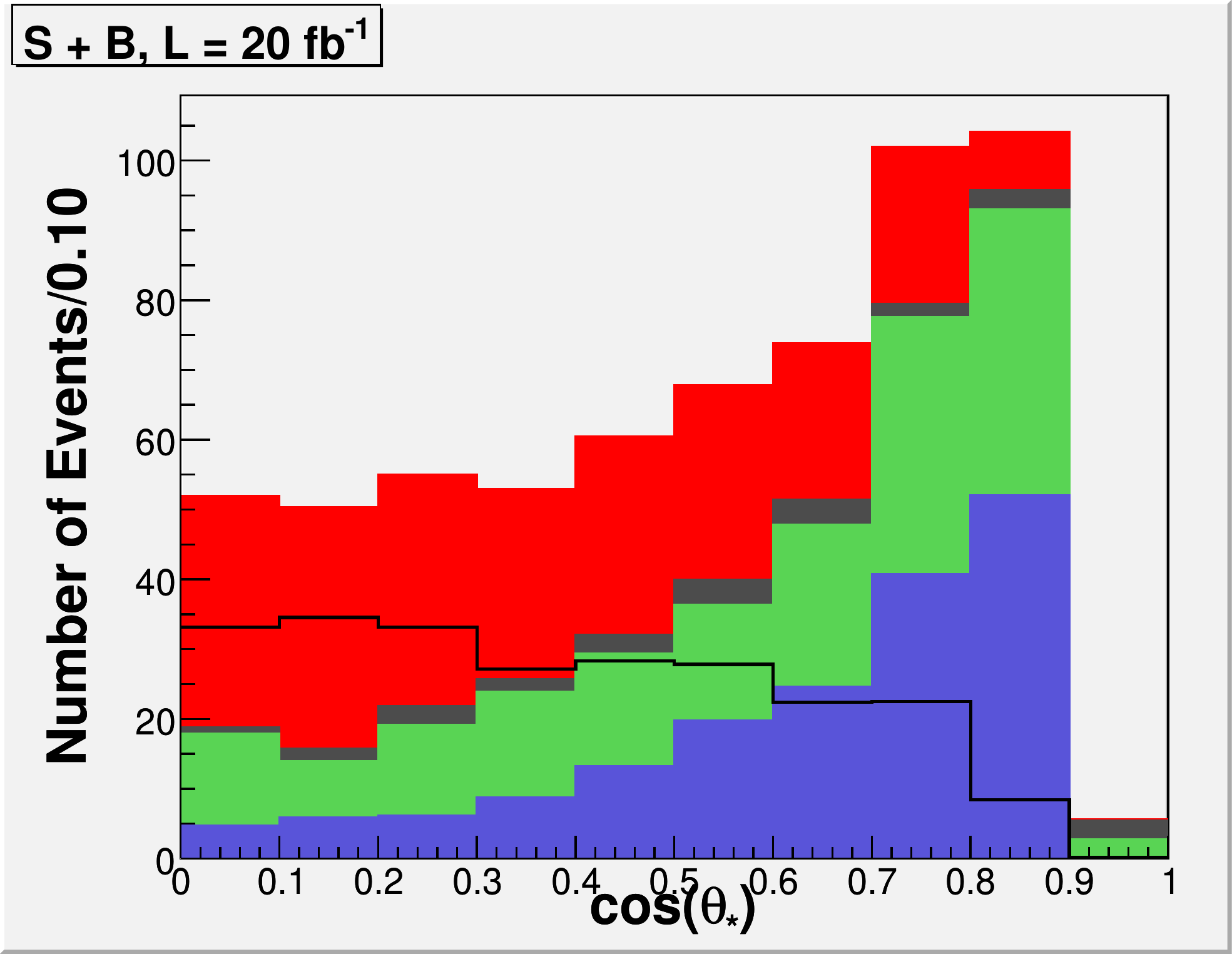}
        \includegraphics[width=3.00in, height = 3.00in, angle=90]
        {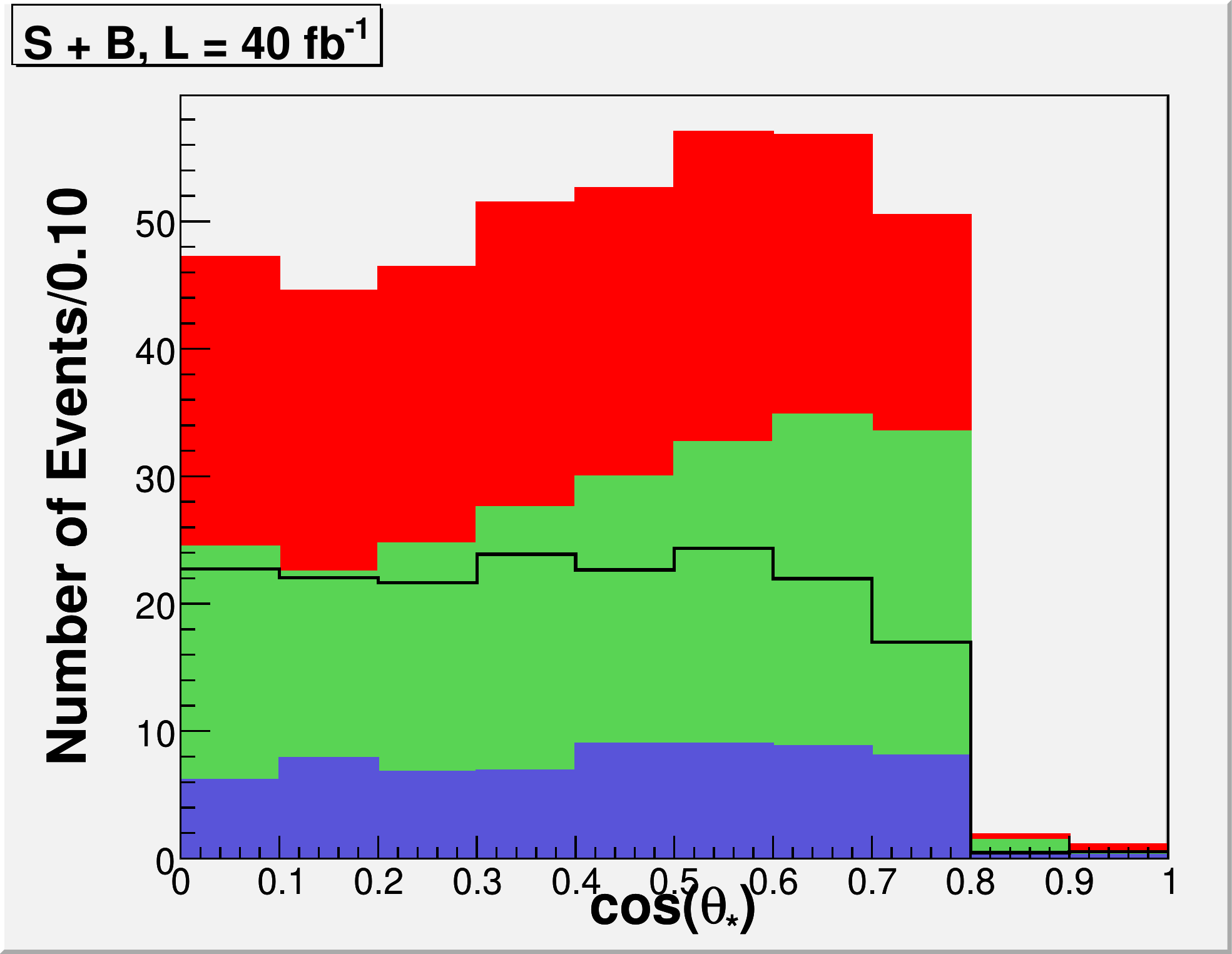}
        \caption{Angular distributions for the signal and backgrounds for
          a $440\,\LSTCgev $ $a_T^\pm \LSTCra \gamma W^\pm$ with $420 < M_{\gamma W} <
          460\,\LSTCgev $ (left) and a $300\,\LSTCgev $ $\LSTCtom \LSTCra \gamma Z^0$ with $290
          < M_{\gamma Z} < 310\,\LSTCgev $ (right) at the LHC; other cuts are
          listed in the text. The luminosities are~20 and $40\,\LSTCifb$,
          respectively. The color codes are given in Figs.~5 and~7.}
    \label{LSTC_LHC_6}
      \end{center}
    \end{figure}

It is clear from these distributions that the backgrounds are a more severe
impediment to observing the signal's angular distribution ($1 +
\cos^2\theta$) than they were in the case of $\LSTCtro \LSTCra WZ$. In particular,
there is no obvious cut to remove them other than one on $M_{\gamma W}$. The
result of requiring $420 < M_{\gamma W} < 460\,\LSTCgev $ is in
Fig.~\ref{LSTC_LHC_6}. Even though the signal's expected forward-backward
excesses are eliminated by the $p_T(\gamma)$ cut, this clearly is a flatter
distribution than the $\sin^2\theta$ ones above. As in that case, subtracting
the background by measuring the sidebands should reveal the signal. Careful
fitting to see that it is consistent with $1+ \cos^2\theta$ after cuts is
work for the future.

\section{$\LSTCtom \LSTCra \gamma Z_L^0$}
 
 The $\LSTCtom$ is as important to find as the $\LSTCtro$, with which it is expected
 to be nearly degenerate, and the $\LSTCta$. Yet it is the most challenging to
 see of the light techni-vectors. At the Tevatron, the primary discovery mode
 is $\LSTCtom \LSTCra \gamma \LSTCtpiz \LSTCra \gamma b \bar b$. Backgrounds to this may make
 this channel difficult at the LHC; studies need to be done! Two other
 channels have much lower branching ratios, but are much cleaner: $\LSTCtom \LSTCra
 \gamma Z_L^0 \LSTCra \gamma \LSTCellp\LSTCellm$ and $\LSTCtom \LSTCra \LSTCellp\LSTCellm$. We discuss
 the first of these in this section.\footnote{Preliminary studies of $\LSTCtom
   \LSTCra \mu^+\mu^-$, including its angular distribution have been carried out
   by J.~Butler and K.~Black. The $\LSTCtom \LSTCra e^+e^-$ mode is ripe for
   picking.}
 
\begin{figure}[!t]
      \begin{center}
        \includegraphics[width=4.50in, height = 6.00in, angle=90]
        {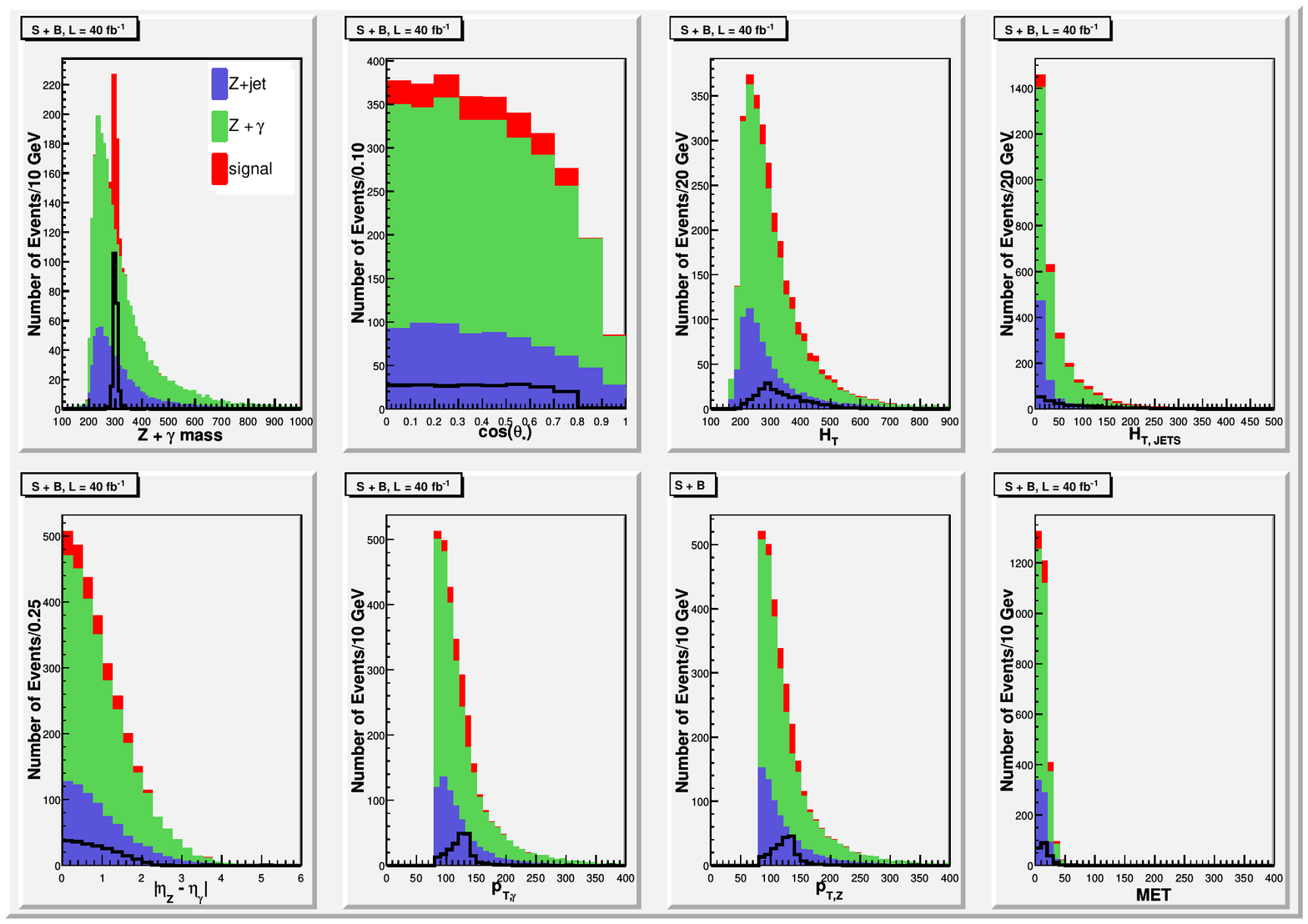}
        \caption{Signal and background distributions of a $300\,\LSTCgev $ $\LSTCtom \LSTCra
          \gamma Z^0 \LSTCra \gamma \LSTCellp\LSTCellm $ for $40\,\LSTCifb$ at the LHC;
          $|M_{\LSTCellp\LSTCellm} - M_Z| < 7.8\,\LSTCgev $ and $p_T(\gamma,Z) >
          80\,\LSTCgev $.  The open black histograms are the signal
          contributions.}
    \label{LSTC_LHC_7}
      \end{center}
    \end{figure}

 As with the $\LSTCta$ search just described, the main backgrounds to $\LSTCtom \LSTCra
 \gamma Z^0$ are standard $Z^0 + \gamma$ and $Z +\,{\rm jet} (\LSTCra \gamma)$
 production. The $\LSTCtom$ signal, however, is about 10~times smaller than the
 $\LSTCta$ one (see Table~2), so considerably higher luminosities are required to
 see a significant signal peak and the characteristic $1 + \cos^2\theta$
 distribution. Here we discuss the PGS simulation for case~A, in which
 $M_{\LSTCtom} = 300\,\LSTCgev $. Events were selected with two same-flavor,
 opposite-sign leptons, each having $p_T > 10\,\LSTCgev $ and rapidity $|\eta| <
 2.5$. The leptons were required to satisfy $|M_{\LSTCellp\LSTCellm} - M_Z| <
 7.8\,\LSTCgev $. Distributions are shown in Fig.~\ref{LSTC_LHC_7} for a
 luminosity of $40\,\LSTCifb$ and for $p_T(\gamma),\, p_T(Z) > 80\,\LSTCgev $. Note
 the much better final-state mass resolution than for $\LSTCtro \LSTCra WZ$ and $\LSTCta
 \LSTCra \gamma W$.  The significance of the signal peak is about $8\,\sigma$;
 the high luminosity is needed to accumulate statistics for the angular
 distribution.\footnote{For case~B, the corresponding luminosity is
   $80\,\LSTCifb$.}
 
 To expose the angular distribution, we take advantage of the superior
 $\gamma Z$ mass resolution and impose a tight cut on $M_{\gamma Z}$ of
 $300\pm 10\,\LSTCgev $. The result is in Fig.~\ref{LSTC_LHC_6}. Because of the
 more stringent $p_T$ cuts, the data above $|\cos\theta| > 0.8$ are lost.
 While quite acceptable, the angular distribution's signal-to-background is
 not as favorable as it was for $\LSTCta \LSTCra \gamma W$. As in that case, detailed
 fitting beyond our scope is needed to determine how well the measured
 distribution fits the expectation. And, as there, the backgrounds can be
 subtracted by measuring the angular distribution in sidebands.

\section{$\LSTCtropm,\,a_T^\pm  \LSTCra Z^0 \LSTCtpipm\LSTCra \ell^+ \ell^- b q$}

Even if narrow resonances in the $WZ$, $\gamma W$ and $\gamma Z$ channels are
found as described above at the LHC, all with nearly the same mass and with
the expected angular distributions, it will remain essential to discover a
technipion to cement the technicolor interpretation of these states. In this
section we present an analysis of $\LSTCtropm \LSTCra Z^0 \LSTCtpipm$ carried
out for the ATLAS detector~\cite{LSTCazuelos}. The large backgrounds to this
signal require large luminosity. On the other hand, for the masses assumed
here, this channel has the extra advantage that the rate for $a_T^\pm \LSTCra
Z_\perp^0 \LSTCtpipm$ is only 2--4 times smaller than for $\LSTCtropm \LSTCra
Z_L^0 \LSTCtpipm$, creating another opportunity for observing both resonant
peaks in the same (well, similar) final state.

As noted, $\LSTCtpi$ are expected to decay into the heaviest
fermion-antifermion flavors kinematically allowed. For the range of
$M_{\LSTCtpi}$ considered here, this implies $\LSTCtpi^+ \LSTCra t \bar b$
and $\LSTCtpiz \LSTCra \bar b b$ or $\bar t t$ are dominant. Actually,
something like topcolor-assisted technicolor~\cite{Hill:1994hp} is needed to
produce $m_t \simeq 175\,\LSTCgev $, and this implies that the coupling of
$\LSTCtpi$ to $t$-quarks is suppressed by a factor of about $m_b/m_t$ from
its naive value. Thus, while the $\LSTCtpi^+$ considered here is massive
enough to decay to $t \bar q$, it should still have an appreciable branching
fraction to $c \bar b$ and $u \bar b$. The latter are the decay channels
considered here. It will be interesting to consider the $t \bar q$ modes.
However, they are not yet included in {\sc Pythia} and, therefore,
$B(\LSTCtpip \LSTCra b\bar q) = 0.87$ in the simulation reported here. This
branching ratio may decrease substantially when the top modes are included.
That would change the search strategy, but we expect that $\LSTCtpip$ can be
seen in $t \bar q$ as well.

    \begin{figure}[!t]
      \begin{center}
        \includegraphics[width=3.00in, height = 3.00in, angle=90]
        {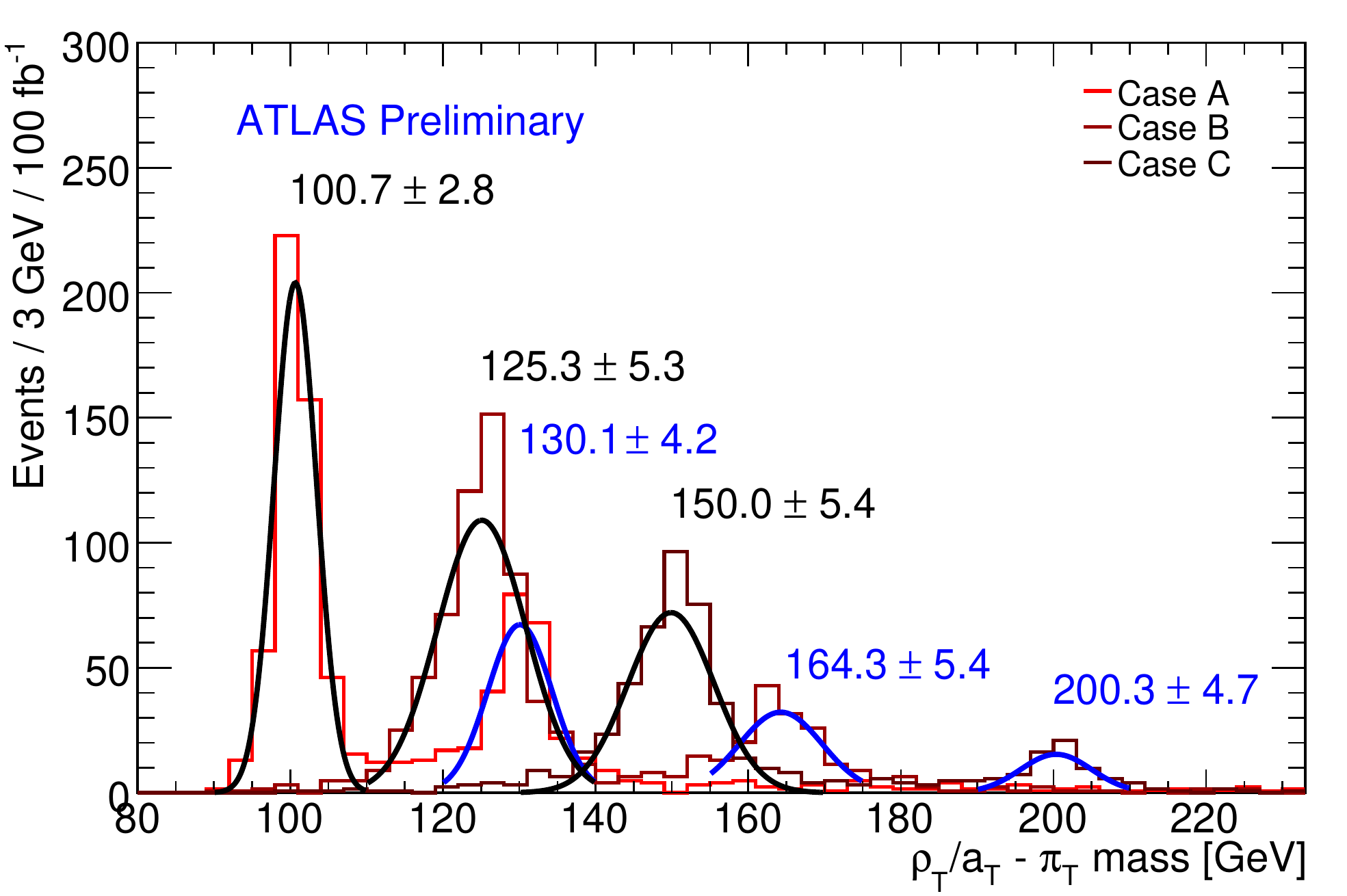}
        \includegraphics[width=3.00in, height = 3.00in, angle=90]
        {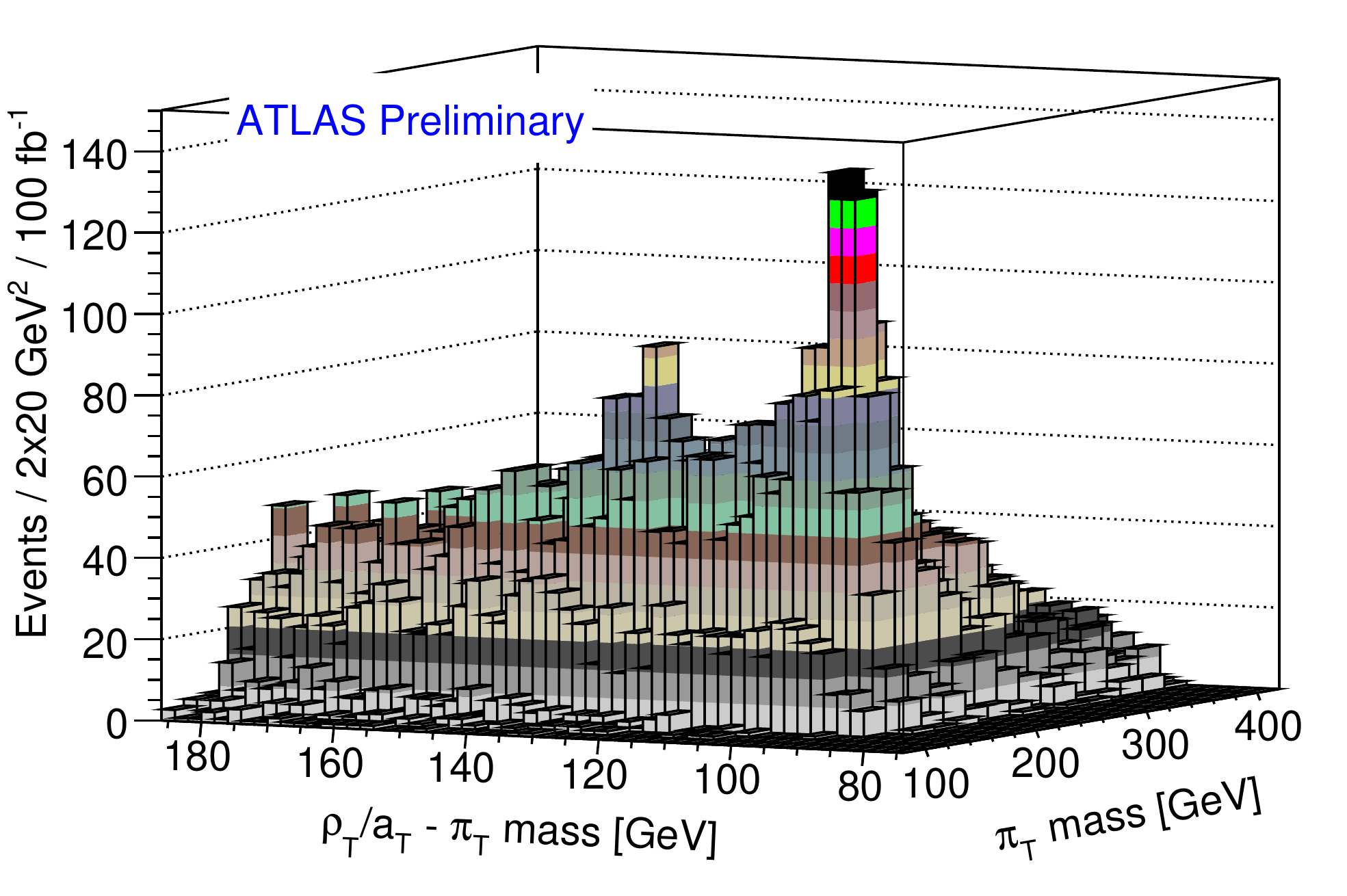}
        \caption{Left: $M_{\LSTCtro}-M_{\LSTCtpi}$ and $M_{\LSTCta}-M_{\LSTCtpi}$ for cases~A,
          B, C in ATLAS. Right: $M_{\LSTCtro,\,\LSTCta} - M_{\LSTCtpi}$ vs. $M_{\LSTCtpi}$
          signal and background events per $2\times 10\,\LSTCgev ^2$ per
          $100\,\LSTCifb$ for case~A~\cite{LSTCazuelos}.}
    \label{LSTC_LHC_8}
      \end{center}
    \end{figure}

    The signal cross sections are $\sigma(pp \LSTCra \LSTCtropm,\, a_T^\pm \LSTCra Z^0
    \LSTCtpipm) B(Z^0 \LSTCra \LSTCellp\LSTCellm) = (99,\,59)$ (A), $(71,\,17)$ (B), and
    $(37,\,9)\,\LSTCfb$ (C), where the two numbers are approximately the $\LSTCtropm$
    and $a_T^\pm$ contributions. The principal backgrounds and their
    leading-order cross sections are: $\bar t t$ ($500\,\LSTCpb$) and, including
    the branching ratio of $Z^0$ to $e^+e^-$ and $\mu^+\mu^-$, $Z^0 jj$
    ($344\,\LSTCpb$), $Z^0 b \bar b$ ($56\,\LSTCpb$) and $Z^0 bj$
    ($11\,\LSTCpb$).\footnote{Recall footnote~8 regarding systematic errors on
      such backgrounds. Background contributions from processes with even
      more jets are possible; they are partly accounted for by the
      leading-log parton showering approximation and initial and final state
      QCD radiation in Pythia. The $Z^0 b \bar b$ rate here is much larger
      than in Table~3 because it was generated with greater $b$-jet
      acceptance.}  See Ref.~\cite{LSTCazuelos} for generation details.  The
    ATLAS detector simulation used ATLFAST~\cite{atlfast}. An additional
    factor of 90\% was applied to the simulation for lepton identification
    efficiency.  The $b$-jet tag efficiency used was 50\%; this corresponds
    to a light-jet mistag rate of~1\% and a $c$-jet mistag rate of 10\%.

To satisfy ATLAS trigger and high-luminosity ($100\,\LSTCifb$ per year) running
conditions, events were preselected with (1) two same-flavor opposite-sign
leptons with $p_T > 20\,\LSTCgev $ and (2) at least one $b$-tagged jet and one
non-$b$-tagged jet, both with $p_T > 20\,\LSTCgev $; the two highest-$p_T$ jets
satisfying these conditions are the $\LSTCtpi$-candidate jets. For the $\LSTCtropm$,
these selections resulted in 548~(A), 382~(B) and 184~(C) signal events per
$100\,\LSTCifb$. For the $a_T^\pm$, there were 297~(A), 117~(B) and 34~(C)
events. The total background event numbers, dominated by $Zjj$ and $Z b \bar
b$, were: (6930, 10670)~(A), (7505, 6285)~(B) and (3015, 2550)~(C). Here, the
first number in each pair refers to $\LSTCtro$ and the second to $\LSTCta$; the
background is the number of events in an elliptical region in
$M_{\LSTCtpi}$--$(M_{\LSTCtro,\,\LSTCta} - M_{\LSTCtpi})$ space centered at the mean and with
widths corresponding to $1.5\,\sigma$.

The following cuts were then applied to optimize the signal significances:
(1) $\LSTCetmiss < 35\,\LSTCgev $ to suppress $t\bar t$; (2)~the highest-$p_T$ jet had
$p_T > 80$~(A), 115~(B), $150\,\LSTCgev $~(C); (3)~the second highest-$p_T$ jet
had $p_T > 65$~(A), 80~(B), $100\,\LSTCgev $~(C); (4) there is exactly one
$b$-tagged jet; and (5) $M_{\LSTCellp\LSTCellm} = 91\pm 5\,\LSTCgev $. After these cuts,
the number of remaining signal events is (344, 215)~(A), (242,75)~(B) and
(126,21)~(C) for ($\LSTCtropm,\,a_T^\pm$).The backgrounds under these signals are
(403,900)~(A), (346,242)~(B) and (96,69)~(C). For the parameters used in
this simulation, then, only in case~C is the $a_T$ not observable in the
$Z\LSTCtpi$ channel in $100\,\LSTCifb$.

The resolution in $M_{\LSTCtpi}$ varies from 16 to $23\,\LSTCgev $ and in
$M_{\LSTCtro,\,\LSTCta}$ from~19 to $30\,\LSTCgev $. Most of the error comes from the
$\LSTCtpi$ jets' energy measurements. Therefore, much of it cancels in $Q =
M_{\LSTCtro,\,\LSTCta} - M_{\LSTCtpi}$. This is shown on the left in
Fig.~\ref{LSTC_LHC_8}, where the resolution in this difference ranges from~3
to $5\,\LSTCgev $. This sharpness will facilitate the discovery of $\LSTCtro,\,a_T \LSTCra
Z \LSTCtpi$ and other technivector-to-technipion decays.\footnote{The $Q$-value
  was used to advantage in the most recent CDF search for $\LSTCtro \LSTCra W^\pm
  \LSTCtpi \LSTCra \LSTCellpm \nu_\ell b\,\LSTCjet$; see CDF/ANAL/EXOTIC/PUBLIC/8566, {\tt
    http://www-cdf.fnal.gov/physics/exotic/r2a/20061025.techcolor/}, and its
  importance was emphasized in Ref.~\cite{Eichten:2007sx}.} The signals and
background for case~A are on the right in Fig.~\ref{LSTC_LHC_8}. The twin
peaks stand out dramatically (looking rather like Boston's Back Bay).

In summary: For the TCSM parameters used here, there should be no difficulty
seeing $\LSTCtropm$ and $a_T^\pm$ in the $Z^0\LSTCtpipm$ channel in case~A, and the
$\LSTCtro$ and a strong indication of the $\LSTCta$ in case~B. In case~C, only
$\LSTCtropm \LSTCra Z^0 \LSTCtpipm$ can be seen in $100\,\LSTCifb$. The minimal cross
sections (times $B(Z \LSTCra e^+e^-\,/\mu^+\mu^-)$) and luminosities required to
see the $\LSTCtropm$ and $a_T^\pm$ signals at $5\,\sigma$ significance are in
Table~5.

{\begin{table}[!ht]
\begin{center}{
    \begin{tabular}{|c|c|c|c|c|c|c|c|}\hline
       peak & A  & B  & C  & & A  & B  & C\\
       \hline\hline
       $\rho_T^\pm$ & 29 & 28  & 14 & & 8.3  & 15  & 15 \\
       $a_T^\pm$    & 41& 18 & 18 & & 48 & 106 & 390 \\
      \hline\hline
    \end{tabular}}
    \caption[limits]{Minimal cross-section times branching fractions (in fb,
      left) and minimal luminosities (in $\LSTCifb$, right) required for
      $5\,\sigma$ significance in cases~A, B, C.}
    \label{tab:LSTC_limits}
\end{center}
\end{table}}

\section{CONCLUSIONS AND OUTLOOK}

Low-scale technicolor (with $N_D = \LSTCCO(10)$ isodoublets transforming as
$SU(N_{TC})$ fundamentals) is a well-motivated scenario for strong
electroweak symmetry breaking with a walking TC gauge coupling. The
Technicolor Straw-Man framework provides the simplest phenomenology of this
scenario by assuming that the lightest technihadrons --- $\LSTCtro$, $\LSTCtom$,
$\LSTCta$ and $\LSTCtpi$ --- and the electroweak gauge bosons can be treated in
isolation. This framework is now implemented in {\sc Pythia}.

We used {\sc Pythia} and (mainly) the generic detector simulator PGS to study
the final-state mass peaks and angular distributions for the LSTC discovery
channels at the LHC: $\LSTCtropm \LSTCra W^\pm Z^0$, $a_T^\pm \LSTCra \gamma W^\pm$ and
$\LSTCtom \LSTCra \gamma Z^0$, with leptonic decays of the weak bosons. We also
carried out an ATLFAST simulation for $\LSTCtropm,\,a_T^\pm \LSTCra Z^0 \LSTCtpipm \LSTCra
\LSTCellp\LSTCellm b\,\LSTCjet$. The results are very promising. For the fairly
generic TCSM parameters chosen, the technivector mesons can be discovered up
to about 500--$600\,\LSTCgev $ in the two-gauge boson modes, usually with a few to
a few tens of~$\LSTCifb$. The angular distributions, dispositive of the
underlying technicolor dynamics, can be discerned with a few tens to
$100\,\LSTCifb$ (except for a higher mass $\LSTCtom \LSTCra \gamma Z^0$). Taking
advantage of the superb resolutions in $Q = M_{\LSTCtro,\LSTCta}-M_{\LSTCtpi}$ and
$M_{\LSTCellp\LSTCellm}$ for $\LSTCtro,\,\LSTCta \LSTCra Z\LSTCtpi \LSTCra \LSTCellp\LSTCellm b\,\LSTCjet$, both
resonances and the technipion can be seen for $M_{\LSTCtro} \LSTCsimle 500\,\LSTCgev $ and
$M_{\LSTCta} \LSTCsimle 400\,\LSTCgev $.

Still, these studies just scratch the surface of what can and needs to be
done to gauge the potential of the ATLAS and CMS detectors for discovering
and probing low-scale technicolor. Simulating detector response to the
signals and backgrounds of the relatively simple processes we considered
requires considerably more sophistication, in both depth and breadth, than we
have been able to deploy. Issues such as the accuracy with which technivector
masses and decay angular distributions can be determined as a function of
luminosity are especially important. While we believe that the TCSM
parameters --- $\sin\chi$, $Q_U + Q_D$, $M_{\LSTCtpi}$, $M_{V_i}$ and
$M_{A_i}$, $N_{TC}$ --- we chose are reasonable, relative branching fractions
can be fairly sensitive to them, as Table~1 indicates~\cite{Lane:2002sm}. It
would be valuable to reconsider the processes examined here for a range of
these parameters.  Finally, there are other modes we have not been able to
consider but which are nevertheless of considerable interest. Two outstanding
examples are $\LSTCtroz,\,\LSTCtom,\, a_T^0 \LSTCra \LSTCellp\LSTCellm$ and
$\LSTCtom,\,\LSTCtroz \LSTCra \gamma \LSTCtpiz \LSTCra \gamma \bar b b$.
Thus, the main goal of our Les Houches studies, as it is for the other
``Beyond the Standard Model'' ones started at Les Houches, is to motivate the
ATLAS and CMS collaborations to broaden the scope of their searches for the
origin and dynamics of electroweak symmetry breaking.

\vskip0.15truein
\hskip0.44truein ``Faith'' is a fine invention

\hskip0.5truein    When Gentlemen can see ---

\hskip0.5truein    But  {\em Microscopes}  are prudent

\hskip0.5truein    In an Emergency.\hfil\break


\hskip1.0truein --- Emily Dickinson, 1860

\bigskip

\section*{ACKNOWLEDGEMENTS}

We thank the organizers and conveners of the Les Houches workshop, ``Physics
at TeV Colliders'', for a most stimulating meeting and for their
encouragement in preparing this work. We are especially grateful to Steve
Mrenna for updating {\sc Pythia} to include all the new TCSM processes. We
also thank participants, too many to name, for many spirited discussions.
Lane and Martin are indebted to Laboratoire d'Annecy-le-Vieux de Physique des
Particules (LAPP) and Laboratoire d'Annecy-le-Vieux de Physique Theorique
(LAPTH) for generous hospitality and support throughout the course of this
work. Part of this work has been performed within the ATLAS Collaboration
(Azuelos and Ferland; Black) and the CMS Collaboration (Bose), and we thank
members of both collaboration for helpful discussions. We have made use of
their physics analysis framework and tools which are the result of
collaboration-wide efforts. This research was supported in part by NSERC,
Canada (Azuelos and Ferland) and the U.S.~Department of Energy under Grants
DE-FG02-91ER40654 (Black), DE-FG02-91ER40688 (Bose), DE-FG02-97ER41022
(Gershtein), DE-FG02-91ER40676 (Lane), and DE-FG02-92ER40704 (Martin).


\AddToContent{G.~Azuelos, K.~Black, T.~Bose, J.~Ferland, Y.~Gershtein, K.~Lane and A.~Martin}
\setcounter{figure}{0}
\setcounter{table}{0}
\setcounter{section}{0}
\setcounter{equation}{0}
\setcounter{footnote}{0}
\clearpage


\part{Technivectors at the LHC}

{\it J.~Hirn, A.~Martin and V.~Sanz}


\begin{abstract}
 Assuming composite spin-1 states to be the most relevant particles produced by EW scale strong interactions, we model them with a simple parametrization inspired by extra dimensions. Our flexible framework accommodates deviations from a QCD-like spectrum and interactions, as required by precision electroweak measurements. 
\end{abstract}

\section{INTRODUCTION:}
As was emphasized in the Les Houches non-SUSY BSM working group, very
few LHC simulations of dynamical electroweak symmetry breaking (DEWSB) scenarios are
available. In order to remedy this situation, Les
Houches 2007 called for an effective description of strong
interactions, flexible enough to interpolate between
some known models of resonance interactions in 4D or 5D, yet
economical enough to have a tractable parameter space. Ultimately,
such a framework could play the same role for strong interactions as was played by minimal
Supergravity (mSUGRA) for the case of SUSY: simplifying assumptions reduce the number of parameters from $\mathcal{O}(100)$  down to
a few, enabling a slew of phenomenological studies.

As a step towards an effective DEWSB description, we present a flexible yet manageable
  model of interactions between spin-1 resonances and the Standard Model (without Higgs) using the
  framework of Holographic Technicolor (HTC)~\cite{Hirn:2006nt}. 
  In HTC the number of
  parameters in the effective lagrangian of resonance interactions is reduced by
  deriving the interactions from a precursor 5D
  lagrangian, as suggested by the AdS/CFT correspondence {\cite{Gubser:1998bc, Witten:1998qj}}.
 In practice,
  we see no reason to be constrained by a strict 5D
  formulation~\cite{Agashe:2007mc}: we simply model interactions between 4D
  resonances, but resort to 5D techniques to compute the parameters. 
  
 HTC is similar to Higgsless~\cite{Csaki:2003zu} models, but contains deviations from pure AdS 5D geometry in the form of
  effective warp factors that differ for the various
  fields. These warp factors are a departure from true 5D modelling~\cite{Agashe:2007mc} but they are motivated by the  requirement of
  small deviations from the SM in the gauge sector (oblique
  corrections \cite{Hirn:2006nt}, and cubic couplings, see below).  With nonstandard 5D geometry we achieve a different resonance spectrum from rescaled QCD, confirming and building upon previous 4D results \cite{Sundrum:1991rf, Lane:1993wz, Appelquist:1998xf}. 
    
  We also refrain from modelling the fermions in
  the extra-dimension, as this would not reduce the number of
  parameters: in the present study, the couplings of fermions to
  resonances  $g_{f fV}$ are free parameters, set to pass experimental
  constraints, while the couplings of fermions to $W, Z$ are assumed to
  exactly obey SM relations. This can be relaxed in the future.
 
 In sections 2 and 3 we present the basics of HTC parameterization and the constraints on its parameter space.  In sections 4 and 5 we describe some LHC signals using a MadGraph~\cite{Alwall:2007st}/BRIDGE~\cite{Meade:2007js} implementation of HTC. In the future we hope to provide a more complete package to allow further study.
     
  \section{HOLOGRAPHIC TECHNICOLOR:}
      
    In HTC, as in Higgsless models, the SM $SU(2)_{\rm w}$ and $U(1)_{{\rm em}}$ gauge fields are the lightest Kaluza-Klein (KK) states of 5D $SU(2)_L\otimes SU(2)_R$ gauge fields. The higher KK excitations of the same 5D fields are interpreted as new spin-1 resonances.  Provided they are light enough, these spin-1 resonances assist in unitarizing $W W$ scattering in the absence of a Higgs.
    
    The masses and interactions of the resonances are dictated by the geometry of the 5th dimension, which is set by a warp factor $w(z)$. The warp factor appears in the 5D metric  $ds^2 = w(z)^2 (dx^2 - dz^2)$, where the extra coordinate $z$ is restricted to the interval $l_0
  \leqslant z \leqslant l_1$. Instead of a single warp factor, in HTC we allow the axial ($A$) and vector ($V$) combinations of 5D gauge fields to feel different backgrounds. Specifically  we define $w_X = (l_0 / z) \exp \left( \frac{o_X}{2}  \left(
  \frac{z - l_0}{l_1} \right)^4 \right)$, $X = A, V$. The power $4$ in $(z /
  l_1)^4$ was based on walking technicolor arguments
  {\cite{Appelquist:1986an}}, but is irrelevant for LHC phenomenology:
  one can absorb the effect of a different power in the $o_X$ value.  Pure AdS geometry corresponds to
  $o_{V,A} = 0$.  Choosing boundary conditions that preserve only $U (1)_{{\rm em}}$ leads to a
  massless photon, and light $W, Z$ compared to the resonances. Although 5D provides a tower of  resonances, we restrict our study to the lightest two
  triplets of resonances $(W_{1,
  2}^{\pm}, Z_{1, 2})$. These resonances are narrow, $\Gamma \sim {\rm GeV}$~\footnote{The tower of narrow resonances is also expected in 4D large $N$ gauge theories.}.
  
  While we assume the strong interactions
  themselves to be parity symmetric~\footnote{Meaning $g_{5L} = g_{5R}$, where $g_{5L} ,g_{5R}$ are the 5D gauge couplings of $SU(2)_L \otimes SU(2)_R$.}, the coupling to the EW sector (set by boundary conditions)
  leads to physical mass eigenstates that
  are an
  admixture of axial and vector components.  The mass splitting between resonances is directly affected by nonzero $o_V, o_A$, as are their couplings. Specifically, the permutation symmetry among triboson couplings does not hold: for $B,C,D$ representing three different HTC spin-1 particles (including $\gamma, W^{\pm}, Z$)
  \begin{equation}
  \label{technivector_eq:coupl}
  g_{BCD}(\partial_{[ \mu }B^-_{\nu ]} C^{+\mu} D^{0\nu}) +  g_{BCD2}(\partial_{[ \mu }C^+_{\nu ]} D^{0\mu} B^{-\nu})  + g_{BCD3}(\partial_{[ \mu }D^0_{\nu ]} B^{-\mu} C^{+\nu}),
  \end{equation}
 we find $g_{BCD} \ne g_{BCD2} \ne g_{BCD3}$. We modified both MadGraph and BRIDGE accordingly.
  
  In summary, the HTC description is very economical. The remaining free parameters are: the size of the ED ($l_1$), which sets the overall mass scale for the new resonances, $M \sim 1/l_1$, the amount of departure from AdS geometry ($o_{V, A}$) and the coupling of the resonances to SM fermions ($g_{f f V}$).     
  
  \section{PARAMETER CONSTRAINTS:}  
    
\begin{figure}[!htb]
  \begin{center}
    \includegraphics[width=5.0in, height = 1.5in]{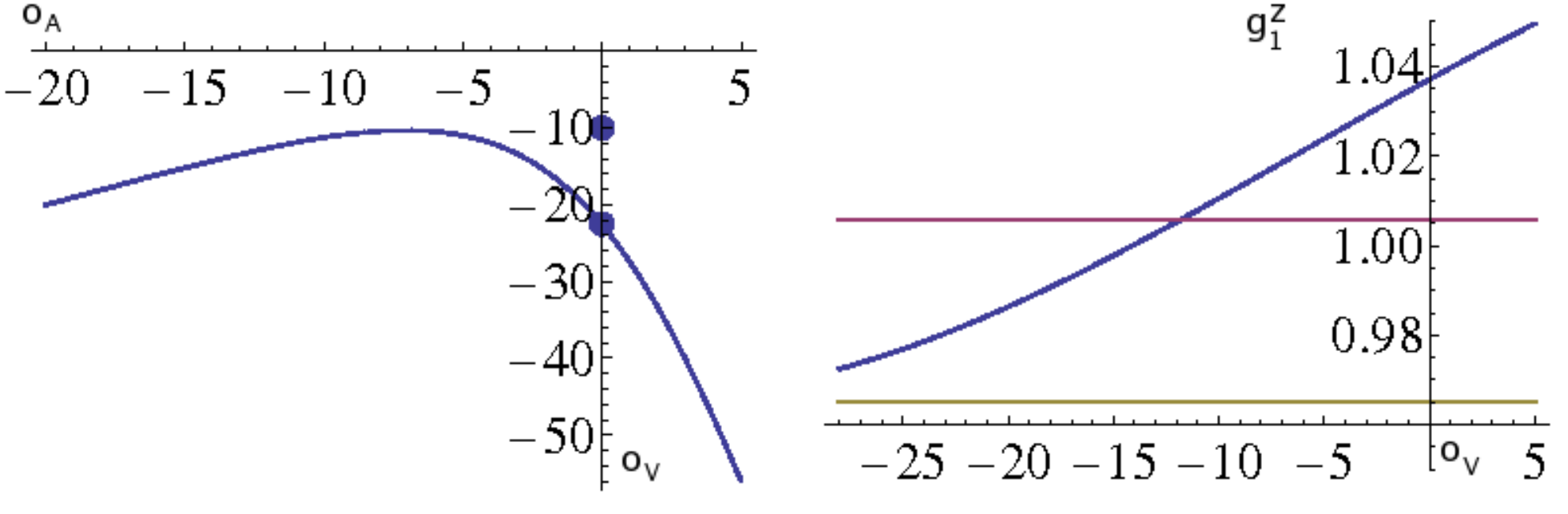} \\
    \caption{ Lefthand side: line of $S=0$ in the $o_A, o_V$ plane with the two sample points on the $o_A = 0$ axis. Righthand side: setting $o_A = 0$ and $M_{W_{1}}=500\ {\rm GeV}$, value of the trilinear gauge coupling $g_1^Z$ as a function of $o_V$. The two horizontal lines correspond to the 1$\sigma$ bounds {\cite{Yao:2006px}}.}
   \label{technivector_fig:S0}
\end{center}
  \end{figure}

     In a pure-AdS model ($o_V = o_A = 0$), consistency with precision electroweak measurements (especially the $S$ parameter) requires fermiophobic resonance couplings, $g_{ffV} \approx 0$~\cite{Cacciapaglia:2004rb}. This is not true in HTC, where we find regions of parameter space in which $S$ is small due to cancellations {\em between} resonance multiplets. In the lefthand side of Fig. \ref{technivector_fig:S0}, we show the line along which oblique corrections cancel. Along that line, the lightest  two resonances are separated by
  only $\gtrsim 100\ {\rm GeV}$, though the exact spacing depends on the full set of 5D parameters. The mass separation between the $W_1$ and $W_2$ greatly impacts the phenomenology, as we see in section 4.
  
   Having narrowed down the $o_V, o_A$ region of interest, for the present study we simply set
  the couplings of fermions to the $W, Z$ to follow the SM relations,
  and take the couplings $g_{f f V}$ of fermions to resonances as free parameters. The $g_{f fV}$ are still constrained by direct $Z', W'$
  Tevatron cross section bounds {\cite{:2007sb, Abulencia:2006kh}} and
  by contact interaction limits {\cite{Cheung:2001wx, Yao:2006px}}. In any particular HTC model, one must also check that the resonances do not disrupt the measured
  Tevatron diboson cross sections
  {\cite{Abazov:2005ni, Abulencia:2007tu}} and high $p_{T, Z}, p_{T, \gamma}$ distributions
  {\cite{Abazov:2005ni,CDFnote}}.
  
  For a given resonance mass, the geometry parameters $o_V, o_A$ are constrained by LEP limits on anomalous triboson couplings {\cite{Yao:2006px}}, as depicted in the righthand side of  Fig. \ref{technivector_fig:S0}. As a first application of the HTC framework we now summarize the LHC signals of the two HTC points indicated in Fig. \ref{technivector_fig:S0}. For the details of the analysis, see Ref.~\cite{Hirn:2007we}.

  \section{S-CHANNEL PRODUCTION:} 
  
  Because HTC resonances need not be fermiophobic they can be produced as $s$-channel resonances. In our setup, $W Z$ is the dominant decay mode for charged resonances~\footnote{Because our 5D setup mimics a strong sector with minimal chiral
   symmetry, $SU(2)_L\otimes SU(2)_R$ the spectrum doesn't contain any
   uneaten pseudo-Goldstone bosons (technipions).}, therefore in Ref.~\cite{Hirn:2007we} we considered the mode: 
  \begin{equation}
  {\rm pp} \rightarrow W^{\pm}_{1,2} \rightarrow W^{\pm} Z, W^{\pm}Z \rightarrow 3 \ell + \nu
  \end{equation}

\begin{figure}[!htb]
  \begin{center}
    \includegraphics[width=2.2in, height = 2.0in]{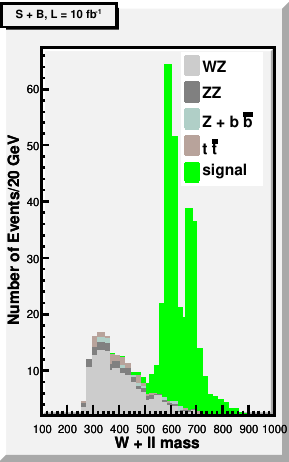}\includegraphics[width=2.2in, height=2.0in]{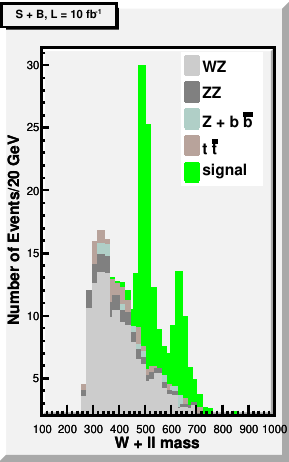} \\
    \caption{ Invariant mass distributions for $W_{1,2}$ signal and background  in the WZ channel for two HTC points and assuming $\mathcal{L}= 10\ {\rm fb}^{- 1}$.  In the left plot, $o_V = -10, o_A = 0, l_1 =  6.3\ {\rm TeV}^{-1}, g_{ffV} = 0.1~g_2$~($g_2$ is the SM $SU(2)_{\rm w}$ gauge coupling), while in the right plot $o_V = -22.5, o_A = 0, l_1 = 8\ {\rm TeV}^{-1}, g_{ffV} = 0.05~g_2$. The dominant background is SM $WZ$ production. All backgrounds are greatly reduced by imposing cuts on the $p_T$  of the $W$ and $Z$. The $\bar t t$ background is  suppressed further by cutting on the maximum $p_T$ carried in jets.}
      \label{technivector_fig:DYfig}
\end{center}
  \end{figure}

   In figure (\ref{technivector_fig:DYfig}) we show the  invariant mass distributions in the $W^{\pm} Z$ channel for the two sample HTC points in~\cite{Hirn:2007we}. For these points we have set  the values of $g_{ffV}$  to be compatible with Tevatron-LEP limits and yet both resonances could be discovered within the first few ${\rm fb}^{- 1}$
  at the LHC.  These points are just an example, chosen because they have large signals at the LHC. 

 Qualitatively, the overall size of the  signal is set by the
 fermion-resonance coupling $\sigma \propto g_{ffV}^2$ and the mass scale of
 the new resonances ($M_{W_1} \sim 1/l_1$), while the relative height of the
 peaks is determined by the relative strengths of the couplings $g_{W_1 W Z},
 g_{W_2 W Z}$ and the mass separation $M_{W_2} - M_{W_1}$, both of which
 depend (primarily) on the geometry parameters $o_V, o_A$. Preliminary
 studies of the $o_V, o_A$ and $l_1$ dependence of  $g_{W_1 W Z}, g_{W_2 W
   Z}$ are being carried out \cite{ustheory}, however a more thorough analysis of the HTC parameter space remains to be done. 

  The large s-channel signals to $WZ$ look similar to signals of Low-Scale Technicolor (LSTC)~\cite{Lane:1999uh, Lane:2002sm, Eichten:2007sx}.   However, in LSTC the interactions are carefully chosen to preserve an approximate techni-parity symmetry. With this symmetry only the vector
  resonance couples to longitudinal $W$ and $Z$ polarizations.  In HTC, we are not free to tune the interactions. All interactions, including vector-axial mixing, are determined by the 5D parameters and boundary conditions.  In the region of interest (viable with
  electroweak constraints) we find that techni-parity is not a good
  approximation for HTC, so both low-lying resonances couple to $W_L, Z_L$. Since the resonance contribution to ${\rm pp} \rightarrow WZ$ is dominated by $W_{1,2} \rightarrow W^{\pm}_L Z_L$ , LSTC predicts only one peak in figure~(\ref{technivector_fig:DYfig}), while in HTC we see two.
  
  Another interesting s-channel production mode is
  \begin{equation}
  {\rm p p} \rightarrow W^{\pm}_{1,2} \rightarrow W^{\pm} \gamma~,~W\rightarrow \ell \nu
  \end{equation}
  Of the conventional three vector boson terms, the only permutation consistent with
  $U (1)_{em}$ gauge invariance is $g_{\gamma W_{1, 2} W} (\partial_{[\mu}
  A_{\nu]} (W_{1, 2[\mu}^- W^+_{\nu]}) + h.c.),$ i.e. where the derivative
  acts on the photon field. A nonzero value for only one triboson coupling
  permutation is not possible in traditional, AdS-based Higgsless models.
  However, this final state as been considered recently~{\cite{LH:2007db}} in
  the context of LSTC, exhibiting only one resonance.
  
\begin{figure}[!htb]
  \begin{center}
    \includegraphics[width=2.2in, height = 2.0in]{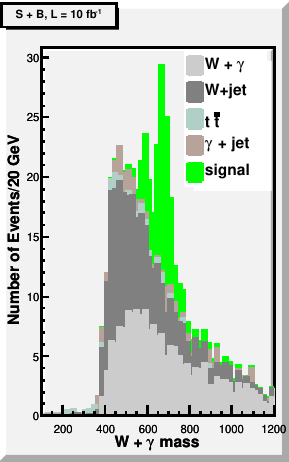}\includegraphics[width=2.2in, height=2.0in]{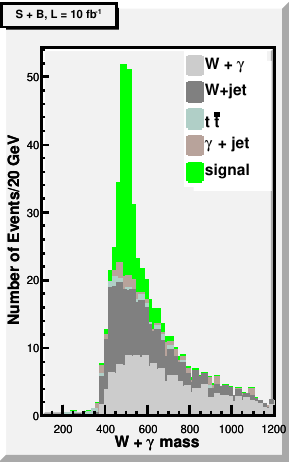} 
      \caption{Invariant mass for $W_{1,2}$ signal and background in the $W\gamma$ channel assuming $\mathcal{L}= 10\ {\rm fb}^{- 1}$. The HTC points are the same as in Figure~(\ref{technivector_fig:DYfig}). The dominant backgrounds are SM $W + \gamma$, and $W + {\rm jet}$ where the jet fakes a photon. The background is suppressed through hard cuts on $p_{T,\gamma}, p_{T,W} $.  }
        \label{technivector_fig:DYfig2}
\end{center}
  \end{figure}

   This channel was also investigated in~\cite{Hirn:2007we}, and in figure (\ref{technivector_fig:DYfig2}) we plot the invariant $W+\gamma$ mass for the same sample HTC points used in figure (\ref{technivector_fig:DYfig}).
  
  As in the $W Z$ case, the signals for both these HTC points are dramatic and could be seen within the first few
  ${\rm fb}^{_{^{- 1}}}$.  However, the difference
  between HTC parameters sets is more evident here than in the $W Z$ case. When $o_V, o_A$
  are such that the separation between the resonances is $\gtrsim 100\
  {\rm GeV}$, as in the righthand plot, only the lightest resonance
  is visible, whereas in the $W Z$ case the second peak is still visible. The main reason for this is  that the decay modes  $W_2
  \rightarrow W_1Z, Z_1 W$ are open, thus suppressing the
  branching ratio to $W \gamma$.   Since the BR to $W \gamma$ is smaller than
  to $W Z$, only when the resonances are very degenerate, as in the lefthand example, are both resonances visible.
  
 Finally, neutral resonances $Z_{1,2}$ can also be produced in the s-channel,
 but the most promising final state is into leptons rather than gauge bosons $W
 W$.  Despite the smaller cross section, the cleaner dilepton channel may reveal both resonances within $\mathcal L \sim {\rm few\ }{\rm fb}^{-1}$ for the two points presented here.
    
 \section{VECTOR BOSON FUSION: }  Although $s$-channel processes will be the
 most important in the early years of the LHC, alternative channels do exist
 and can expose different aspects of the resonance theory. One example process is vector boson fusion (VBF): ${\rm pp} \rightarrow W^{\pm}_{1,2}jj, Z_{1,2} jj$. For resonances decaying to a pair of gauge bosons, VBF directly probes $W_LW_L$ scattering and thus it is important to study regardless of the fermion-resonance coupling~\footnote{Also, in
 fermiophobic models where $g_{ffV} \approx 0$, channels such as VBF are the only way to discover the resonances.   
}. One VBF channel, ${\rm pp} \rightarrow W^{\pm}_{1,2} jj \rightarrow W^{\pm}Z jj $, was studied in~\cite{Hirn:2007we}. For the two HTC points, the existence of two nearby resonances could be seen as two edges in the transverse mass~\footnote{The transverse mass is defined as: $M_T^2 = ( \sqrt{M^2 (\ell \ell \ell) + p_T^2 (\ell
  \ell \ell)} + |E_{miss, T} |)^2 - |p_T (\ell \ell \ell) +  E_{miss,T} |^2$.} $M_T$, though only with luminosity $ \mathcal L \gtrsim O (100)\ {\rm fb}^{-1}$. VBF signals which require the  $g_{W_{1,2} W \gamma}$ coupling, such as ${\rm pp} \rightarrow W^{\pm}\gamma jj$, may also be interesting.

  \section{CONCLUSIONS: } An effective lagrangian description of two new triplets of vector resonances would introduce ${\cal O}(100)$ new
  parameters. In this paper we perform a first step towards an
  economical parametrization of models of Dynamical EWSB: Holographic Technicolor (HTC). Although a departure from 5D
  modelling, HTC uses 5D techniques to parameterize a wide class of models in
  terms of 4 parameters: $l_1, o_V, o_A, g_{ffV}$. After imposing current experimental
  constraints, we identified the relevant region in the HTC parameter
  space. We have chosen two sample HTC points and discussed
  the early discovery (${\cal L}\sim 10\ {\rm  fb}^{-1}$) of two nearby resonances in
  the s-channel $pp\rightarrow W_{1,2} \rightarrow  W
  Z, W \gamma$. The framework
  presented here can be extended to add new particles,
  e.g. techni-pions, techni-omegas and composite Higgs. 

   \section*{ACKNOWLEDGEMENTS: } We thank T. Appelquist, G. Azuelos, G. Brooijmans,
  K. Lane, S. Chivukula, N. Christensen, M. Perelstein and W. Skiba for helpful comments. The work of JH and AM is supported by DOE grant DE-FG02-92ER-40704
  and VS is supported by DE-FG02-91ER40676.

\AddToContent{J.~Hirn, A.~Martin and V.~Sanz}
\setcounter{figure}{0}
\setcounter{table}{0}
\setcounter{section}{0}
\setcounter{equation}{0}
\setcounter{footnote}{0}
\clearpage
%

\superpart{Dilepton Final States}

\part[Searches for New Physics in the Dilepton Channel]{Generic Searches for New Physics in the Dilepton Channel at the Large Hadron Collider}

{\it G.~Landsberg}

\begin{abstract}
We propose a model-independent framework applicable to searches for new physics in the dilepton channel at the Large Hadron Collider. The feasibility of this framework has been demonstrated by the D\O\ searches for large extra dimensions. The proposed framework has a potential to distinguish between various types of models and determine most favorable parameters within a particular model, or set limits on their values.
\end{abstract}

\section{INTRODUCTION}

This letter is devoted to searches for signals for new physics at the Large Hadron
Collider (LHC) in the dilepton channel\footnote{In what follows by ``dileptons'' we will imply the dielectron and dimuon channels, and is focussed on the early discovery potential in this channel. The discussed formalism also applies to the ditau channel, but since this is a more challenging channel experimentally, we are not pursuing it here for the purpose of early searches at the LHC.} This channel has been historically fruitful for discoveries: $J/psi$ and $\Upsilon$ mesons, as well as the $Z$ boson were all discovered using dileptons. The LHC may not be an exception! 

\noindent
The advantages of the dilepton channels for searches for new physics are numerous:
\begin{itemize}
\item Easy triggering;
\item Relatively low instrumental and standard model (Drell-Yan) backgrounds;
\item Well known (NNLO) standard model (SM) cross section;
\item Number of theoretical models that predicts relatively narrow resonances with non-vanishing decay branching fraction to dileptons; a typical example is a generic type of models with an extra $U(1)$ group, which leads to the existence of a $Z'$ boson, often with non-zero couplings to dileptons.
\end{itemize}

\section{THE MODEL}

In the SM, the dilepton final state at hadron colliders is produced via the $s$ and $t$-channel exchange of virtual photons or $Z$ bosons. While the $s$ and $t$-channel diagrams interfere, their main contributions are well separated in the phase space of the dilepton system: high-$p_T$ dileptons are dominantly produced via the $s$-channel exchange, while the $t$-channel process mainly results in very forward leptons. While this doesn't really matter for the generic formalism discussed below, we will focus on the high-$p_T$ dileptons in the $s$-channel, as the most promising signature for new physics at the LHC, with the exception of one case~-- $t$-channel exchange of a leptoquark (LQ) or a SUSY particle in the models with $R$-parity violation (RPV).

In the presence of additional diagrams contributing to the dilepton final state via exchange of new particles, the overall cross section for the dilepton production is given by the interference of the SM diagrams with the new ones coming from new physics. Consequently, it makes sense to parameterize the double-differential cross section, $d^2\sigma/dM_{ll}/d\cos\theta^*$, where $M_{ll}$ is the dilepton invariant mass and $\cos\theta^*$ is the cosine of the scattering angle in the dilepton c.o.m. frame, in the following form:
\begin{equation}
\frac{d^2\sigma}{dM_{ll}d\cos\theta^*}  =  \left(\frac{d^2\sigma}{dM_{ll}d\cos\theta^*}\right)_{\rm SM} + 
\left(\frac{d^2\sigma}{dM_{ll}d\cos\theta^*}\right)_{\rm int} + 
\left(\frac{d^2\sigma}{dM_{ll}d\cos\theta^*}\right)_{\rm NP}. \label{eqGL:1}
\end{equation}
Here the first term describes the SM contribution, the second term corresponds to the interference between the SM and new physics contributions, and the third one describes direct contribution from new physics. In terms of matrix elements, the first three terms are proportional to the appropriate derivatives of $|{\cal M}_{\rm SM}|^2$, $|{\cal M}_{\rm SM}^* {\cal M}_{\rm NP} + {\cal M}_{\rm SM} {\cal M}_{\rm NP}^*|$, and $|{\cal M}_{\rm NP}|^2$, respectively.

Eq. (\ref{eqGL:1}) describes well general case of new physics due to, e.g., compositeness-like operator. However, in case new physics appears in a form of a narrow resonance, the corresponding matrix element has a Breit-Wigner pole, and therefore it makes sense to explicitly specify it. Moreover, in the case of relatively narrow resonance, the interference effect nearly cancels out when integrating over the width of the resonance, so it simply could be added to the above equation:
\begin{eqnarray}
\frac{d^2\sigma}{dM_{ll}d\cos\theta^*} & = & \left(\frac{d^2\sigma}{dM_{ll}d\cos\theta^*}\right)_{\rm SM} + 
\left(\frac{d^2\sigma}{dM_{ll}d\cos\theta^*}\right)_{\rm int} + 
\left(\frac{d^2\sigma}{dM_{ll}d\cos\theta^*}\right)_{\rm NP} + \nonumber \\
 & & {\rm BW}(M_{ll}, M_0, \Gamma_0) \frac{d\Omega}{d\cos\theta^*},\label{eqGL:2}
\end{eqnarray}
where BW is the line-shape for a Breit-Wigner resonance with the mass $M_0$ and width $\Gamma_0$, and $\Omega$ is the angular distribution of its decay products. The remaining NP contribution described by the second two terms now does not include the resonance, which has been explicitly treated separately. In case of more than one resonance (e.g., Kaluza-Klein tower of resonances), additional terms similar to the last one can be added. However, since e focus our attention on early searches for new physics, chances are that only the lowest mass resonance is going to be visible.

Note that the advantage of using double-differential cross section for description of the process is that the two variables, $M_{ll}$ and $\cos\theta^*$ define the tree-level $2 \to 2$ process completely. Thus, at leading order, they contain entire information about both the SM and new physics contributions, i.e. offer the most powerful separation between the NP signal and SM background possible in the entire phase space.

An additional advantage of using full double-differential cross section is that the ``standard candle'' -- the $Z$ peak~-- is contained in the data. That allows {\it in situ\/} calibration of the search sample: in particular the limits on new physics cross section, or the measurement of its cross section can be expressed in terms of the $Z$ production cross section, which is well-known theoretically. Moreover, normalization to the $Z$-peak results in the reduced systematic error, as many important uncertainties, such as the uncertainty of the luminosity measurements, signal acceptance, and lepton identification efficiency would largely cancel out.

\section{THE METHOD}

The proposed method to look for generic deviations due to NP effects is based on the idea outlined in Ref.~\cite{Cheung:1999wt}, generalized to a more complete case of non-resonant and resonant NP contributions.

The fist step is to generate templates corresponding to each of the four terms of Eq. (\ref{eqGL:2}) for a particular model. The simplest one would have a flat resonance decay angular distributions, i.e. $d\Omega/d\cos\theta^* = $const, and a particular non-resonant NP model, e.g., compositeness. These templates are result of parton-level Monte-Carlo (MC), followed by the detector simulation, which includes acceptance requirements as well as smearing of particle momenta. Once the templates are generated, one would fit experimental $d^2\sigma/dM_{ll}d\cos\theta ^*$ spectrum (typically in a form of a 2-dimensional histogram) to the sum of four terms. The results of the fit can then be used to evaluate the relative weights of the SM-only and SM $+$ NP hypotheses and distinguish between various types of new physics. The MC templates also include next-to-leading order corrections in the form of non-zero transverse momentum of the dilepton system. Since the final state is colorless, the new particles contributing to the final state must be color-singlet as well. That implies that the only source of next-to-leading order corrections is initial-state radiation, which is expected to be the same for the Drell-Yan dilepton production and for contributions from new physics. Thus the $p_T$ spectrum of the dilepton system can be reliably modeled by using the well-known $p_T$ spectrum of Drell-Yan pairs.

Since for a heavy narrow resonance experimental mass resolution is typically worse than the internal width, $\Gamma$, after the detector effects the Breit-Wigner in Eq. (\ref{eqGL:2}) will be replaced with a Gaussian with the r.m.s. $\sigma$ determined by the experimental mass resolution. (In case the resonance is not sufficiently narrow, a convolution of the Breit-Wigner and a Gaussian should be used instead.)

\begin{figure}[hbt]
\begin{center}
\includegraphics[width=0.9\textwidth]{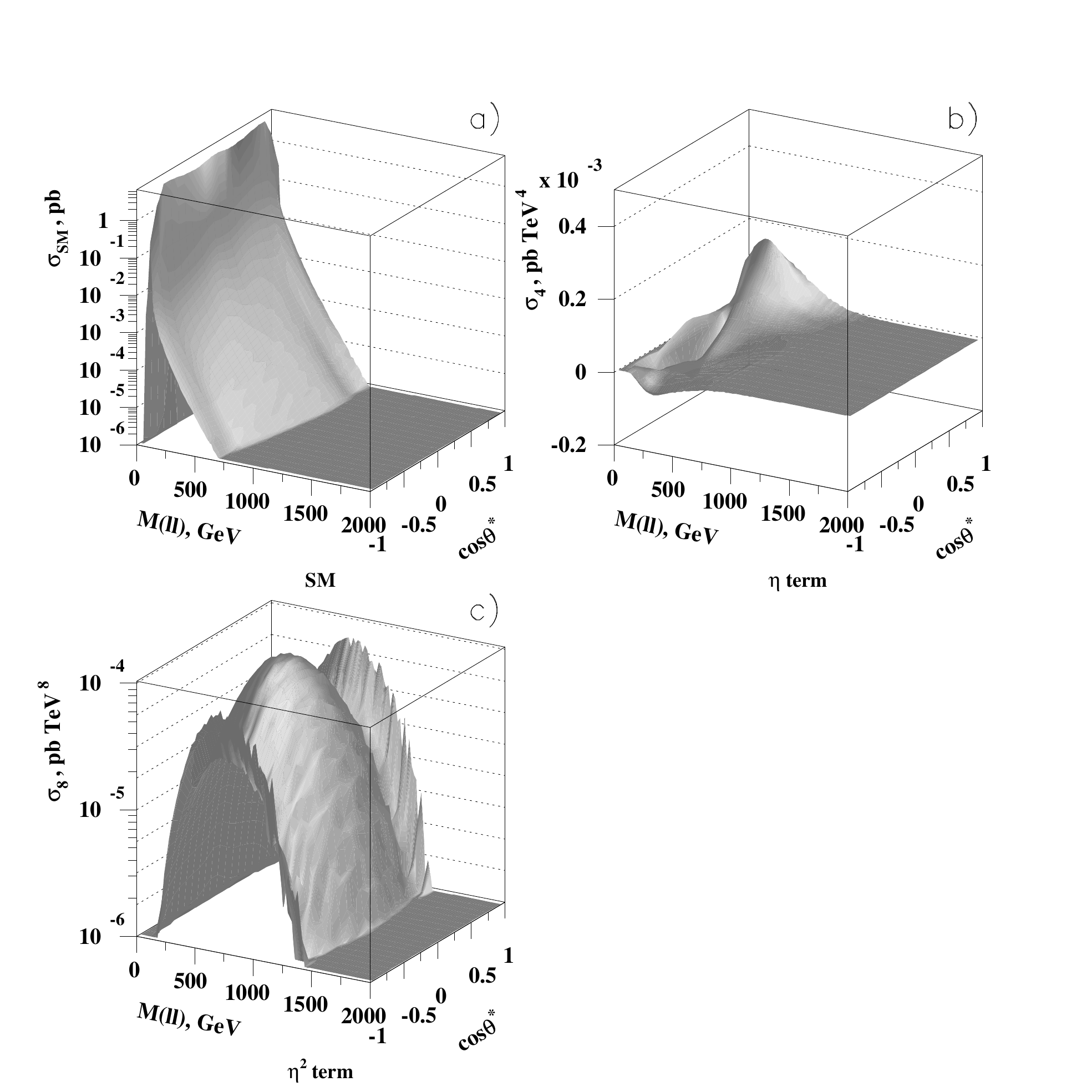}
 \caption{Examples of the templates for Drell-Yan production at the Tevatron: a) the SM ($f_0$); b) $f_1$ for models with large extra dimensions; c)   $f_2$ for models with large extra dimensions. Here $\eta$ is related to the ultraviolet cutoff for the Kaluza-Klein graviton tower, $M_S$ as $\eta = 1/M_S^2$. From Ref.~\protect\cite{Cheung:1999wt}.
}
\label{figGL:1}
\end{center}
\end{figure}

We can further quantify effects of non-resonant new physics via a parameter $\eta$, which shows the relative strength of the NP contribution. For example, if the new physics has characteristics of compositeness with the scale $\Lambda$, a good choice for this parameter is $\eta = 1/\Lambda^2$. Putting it all together, we obtain a modified Eq. (\ref{eqGL:2}):
\begin{equation}
\frac{d^2\sigma}{dM_{ll}d\cos\theta^*} =  f_0(M_{ll},\cos\theta^*) + 
\eta f_1(M_{ll},\cos\theta^*) + \eta^2 f_2(M_{ll},\cos\theta^*) + N \times G(M_0,\sigma) f_3(\cos\theta^*),
\label{eqGL:3}
\end{equation}
where $f_0$, $f_1$, $f_2$ and $f_3$ are the templates discussed above, $G(M_0,\sigma)$ is the Gaussian, and $N$ is the normalization. An example of templates from Ref.~\cite{Cheung:1999wt} is shown in Fig.~\ref{figGL:1}.

The fit of the double-differential distribution in data with the function described in Eq. (\ref{eqGL:3}) yields the values and the uncertainties on the two free parameters of the fit: $\eta$ and $N$. Depending on the values of these parameters, the following broad classes of new physics can be identified and possibly distinguished from one the other by doing steps outlined in the last column of the Table:

\begin{table}
\begin{center} 
\caption{Contributions of various types of new physics in the dilepton final state.}
\begin{tabular}{l|c|c|l}
\hline\hline
New physics model & $\eta$ & $N$ & Next steps \\
\hline
{\small Standard Model} & $= 0$ & $= 0$ & {\small Look elsewhere} \\
{\small Compositeness} & $\neq 0$ & $=0$ & {\small Look at diphotons, dijets; correlate dielectrons and dimuons.} \\
{\small Large extra dimensions} & $\neq 0$ & $=0$ & {\small Look at diphotons, search for black holes and monojets.}\\
{\small LQ's, RPV SUSY} & $\neq 0$ & $=0$ & {\small Look for pair-produced leptoquarks in the dilepton} $+$ {\small dijet}\\
&&& {\small channel; focus on a dedicated search for RPV SUSY.}\\
$Z'$ & $= 0$ & $\neq 0$ & {\small Confirm absence in the} $\gamma\gamma$ {\small channel, determine couplings.} \\
{\small Randall-Sundrum model} & $= 0$ & $\neq 0$ & {\small Confirm presence in the} $\gamma\gamma$ {\small channel, look for black holes.} \\
{\small Technirho/Tecniomega} & $= 0$ & $\neq 0$ & {\small Confirm in the} $W + $ {\small dijet channel.}\\
$Z_{KK}$ & $\neq 0$ & $\neq 0$ & {\small Correlate destructive interference with the peak height;} \\
&&& {\small look for the next excitation.} \\
\hline\hline
\end{tabular}
\end{center}
\end{table}

A simpler version of this method has been successfully used by the D\O\ experiment in searches for large extra dimensions~\cite{Abbott:2000zb,Abazov:2005tk}. It is therefore expected that the above method would work well at the LHC.

\section{CONCLUSIONS}

A framework for generic searches for new physics in the dilepton channel at the LHC is discussed. This framework would allow to statistically distinguish between various models of new physics and to determine parameters (or set limits on their values) within a particular model.

\section*{ACKNOWLEDGEMENTS}
I am grateful to my co-author on the original paper introducing this idea in the context of searches for large extra dimensions, Kingman Cheung. I would like to thank the Les Houches 2007 workshop organizers for an excellent and exciting  venue, and the participants of the dilepton group for a number of stimulating discussions and suggestions. This work is partially supported by the U.S.~Department of Energy under Grant No. DE-FG02-91ER40688 and by the National Science Foundation under the CAREER Award PHY-0239367.

\AddToContent{G.~Landsberg}
\setcounter{figure}{0}
\setcounter{table}{0}
\setcounter{section}{0}
\setcounter{equation}{0}
\clearpage

\part{$Z^\prime$ Rapidity and Couplings to Quarks}

{\it T.M.P.~Tait}


\begin{abstract}
The rapidity distribution of a $Z^\prime$ produced at the LHC encodes information about the
relative sizes of the couplings to up quarks and down quarks which is different from the inclusive
cross section.  Thus, by measuring the $Z^\prime$ production rate at different rapidities, we can
help pin down the coupling to up quarks independently from the coupling to down quarks.
\end{abstract}

\section{INTRODUCTION}

A massive neutral vector $Z^\prime$  decaying into a pair of leptons would be a fascinating discovery of physics beyond the Standard Model, one the LHC could hope to identify with a relatively small sample of data provided it couples to up- or down-type quarks and decays into charged leptons, 
$\ell^+ \ell^- = e^+ e^-$ or 
$\mu^+ \mu^-$.  After the initial excitement of discovery, it will be important to measure as many of the $Z^\prime$ couplings independently as possible, to unravel the underlying model from which it arose.

The primary initial observables will be the mass, given by the position of the excess in the lepton
invariant mass distribution, and the cross section, given by the magnitude of the excess.  Assuming the $Z^\prime$ has flavor universal couplings (motivated so that it does not introduce large FCNCs), the cross section can be parameterized at a hadron collider by \cite{Carena:2004xs},
\begin{eqnarray}
\sigma (p p \rightarrow Z^\prime \rightarrow \ell^+ \ell^- X) & = & c_u w_u + c_d w_d
\label{eq:sigcucd}
\end{eqnarray}
where the $w_u$ and $w_d$ factors contain the parton distribution functions and higher order
QCD corrections (exactly at NLO and to good approximation to NNLO) and thus contain only SM
inputs, and $c_u$ and $c_d$ contain the $Z^\prime$-dependent quantities,
\begin{eqnarray}
c_u & = & \left( g_Q^2 + g_u^2 \right) BR \left( Z^\prime \rightarrow \ell^+ \ell^- \right) \nonumber \\
c_d & = & \left( g_Q^2 + g_d^2 \right) BR \left( Z^\prime \rightarrow \ell^+ \ell^- \right)
\end{eqnarray}
where $g_Q$, $g_u$, and $g_d$ are the $Z^\prime$ coupling to left-handed quarks (assumed equal by $SU(2)_L$ invariance), right-handed up-type quarks, and right-handed down-type quarks, and
$BR ( Z^\prime \rightarrow \ell^+ \ell^- )$ is the branching ratio into $\ell^+ \ell^-$.

\section{$c_u$ VERSUS $c_d$}

A measurement of the cross section becomes, through, Eq.~(\ref{eq:sigcucd}), a determination of a combination of couplings times the branching ratio into leptons.  A single measurement constrains only a combination weighted by $w_u$ and $w_d$.  As a first step toward separating out $c_u$ and $c_d$
individually, we consider the cross section differential in the rapidity distribution of the $Z^\prime$,
$d\sigma / dy$.  We expect that because there are more up than down quarks in the proton, that the
contribution to forward $Z^\prime$ rapdities will depend more strongly on the coupling to up quarks than to down quarks.  This can be captured by introducing $y$-dependent $w_{u,d}(y)$,
\begin{eqnarray}
\frac{d \sigma}{dy_{Z^\prime}} & = & c_u w_u ( y_{Z^\prime}) + c_d w_d (y_{Z^\prime})
\end{eqnarray}
where the integral of $w_{u(d)} (y)$ over $y$ reproduces the original $w_{u(d)}$ 
in Eq.~(\ref{eq:sigcucd}) for the inclusive $Z^\prime$ cross section.

\begin{figure}[!thb]
\begin{center}
\includegraphics[width=0.4\textwidth]{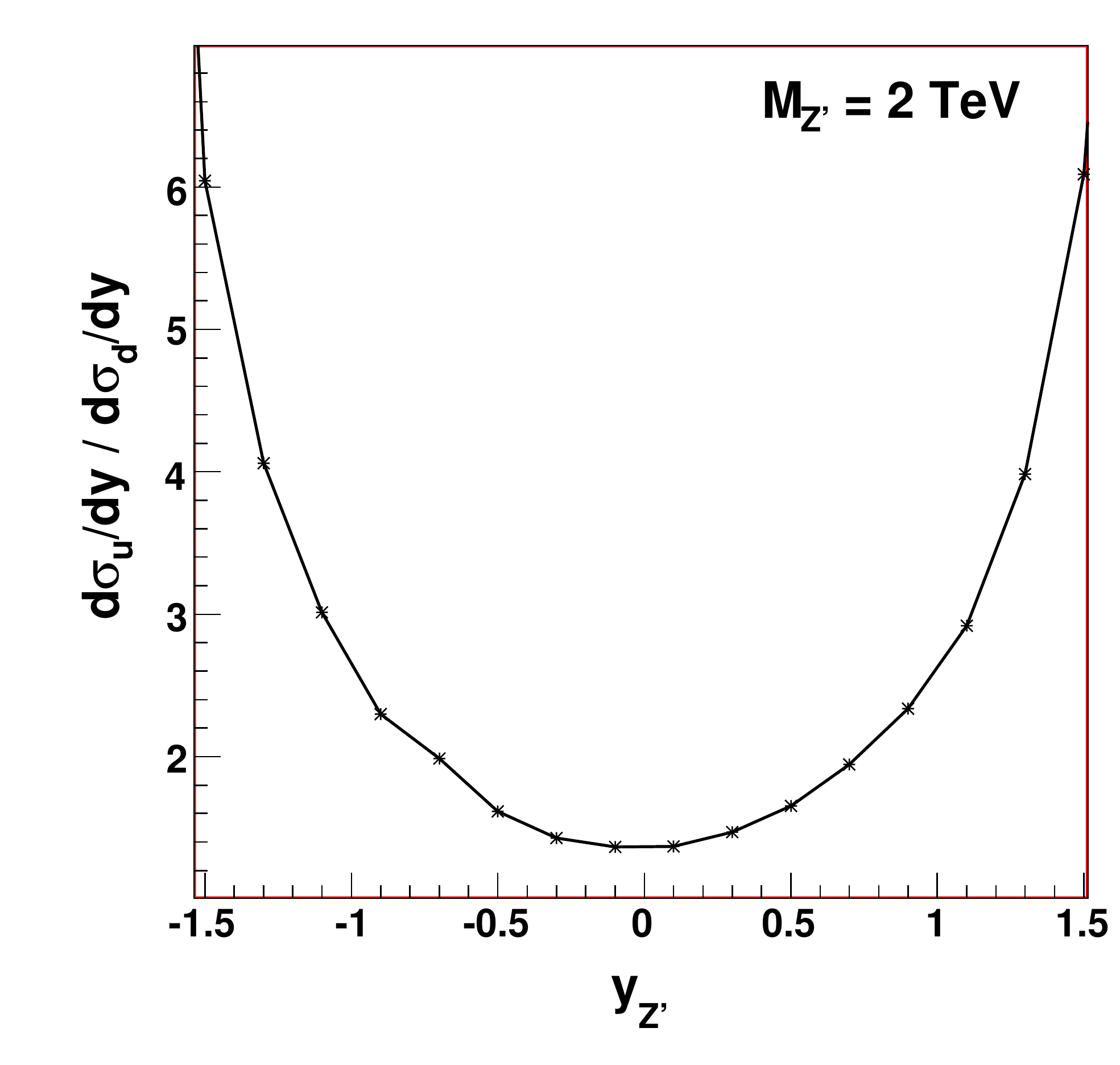}
\includegraphics[width=0.4\textwidth]{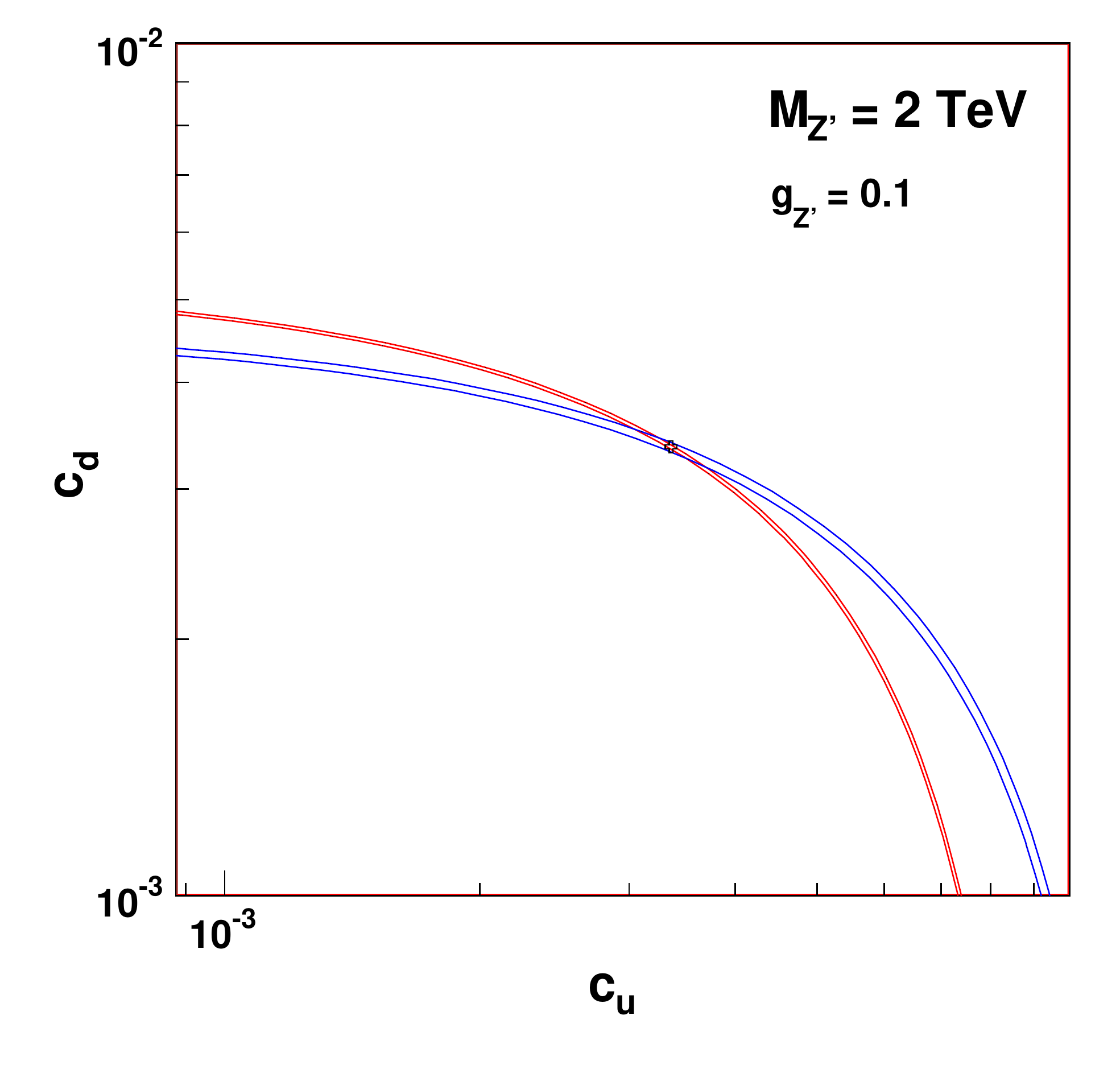}
\caption{Left: The ratio of $w_u (y) / w_d (y)$ as a function of $y$.
Right: The uncertainties on the combination of $c_u$ and $c_d$ obtained from inclusive (red) and
differential (blue) $Z^\prime$ cross section measurements, for the $Z^\prime$ considered in the text.}
\label{fig:cucdratio}
\end{center}
\end{figure}

To illustrate how this works, in Figure~\ref{fig:cucdratio}, we present the ratio of $w_u (y) / w_d (y)$
as a function of $y$ for a $Z^\prime$ of mass 2 TeV, computed at tree level.  A more precise analysis
would want to use higher order corrections, which are known to NNLO for this distribution
\cite{Anastasiou:2003ds}, 
and are straight-forward to include.  For this study which aims to examine a proof of
principle, tree level is sufficient to estimate the utility of the measurement.
(This quantity is equivalent to the ratio of
cross sections $d\sigma / dy$ for two $Z^\prime$s of equal masses and couplings, one coupling only to $u$-quarks
and the other coupling only to $d$-quarks).
We see that at high rapidities 
$w_u (y) \gg w_d (y)$, and the sensitivity to the up-type quark couplings is enhanced.

As an example to illustrate how this could work, we consider an example $Z^\prime$ with 
$M_{Z^\prime} = 2$~TeV, $g_{Z^\prime} = 0.1$, for which all quarks and leptons have the same
charge (taken to be unity).  This $Z^\prime$ has $c_u = c_d \simeq 3.2 \times 10^{-3}$ and
is compatible with LEP-II data.  We imagine two measurements - one of the inclusive 
$Z^\prime \rightarrow e^+ e^-, \mu^+ \mu^-$ and another with a cut of $|y_{Z^\prime}| > 1$,
selecting only $Z^\prime$s in the forward region.  No detector effects or efficiencies are included.
We apply statistical errors assuming $100~{\rm fb}^{-1}$ of collected data.  These turn out to
be on the order of $1\%$ ($3\%$) for the inclusive (cut) $Z^\prime$ cross sections we consider, and thus are probably similar in magnitude to the
the residual theoretical error in an NNLO calculation, and coming from the PDFs.

We unfold the $1\sigma$ uncertainties in the combination of $c_u$ and $c_d$ for both measurements
and plot the results in Figure~\ref{fig:cucdratio}.  Also shown on the plot is the correct value of
$c_u$ and $c_d$ for the example $Z^\prime$.   The plot illustrates that adding the measurement
of the cut $Z^\prime$ cross section does help break the degeneracy in $c_u$ and $c_d$ which
is left after measuring the inclusive cross section.

\section{CONCLUSIONS}

We examine the possibility that one can use the $Z^\prime$ rapidity distribution to help disentangle the coupling of the $Z^\prime$ to up- and down-type quarks.  The result is encouraging, and would
motivate further investigation, including realistic detector simulations, a better treatment of uncertainties, and a more sophisticated analysis such as comparison with the full $Z^\prime$ rapidity distribution (and not just a cut rate measurement).  

Between the time this study was performed at Les Houches 07 and the writing of this report, a more comprehensive theoretical treatment of $Z^\prime$ observables at the LHC has appeared
\cite{Petriello:2008zr}, 
including discussion of this (and other) observables.   The results of that study seem
to continue to suggest that the $Z^\prime$ rapidity distribution is a useful quantity to unravel the
$Z^\prime$ couplings to quarks.

\section*{ACKNOWLEDGEMENTS}

Research is supported at Argonne National Lab by the US Department of Energy
under contract DE-AC02-06CH11357.  It is a pleasure to acknowledge the organizers of Les Houches
for a stimulating physics environment, and the LAPTH theory group for their hospitality while some
of the work herein was completed.


\AddToContent{T.M.P.~Tait}
\setcounter{figure}{0}
\setcounter{table}{0}
\setcounter{section}{0}
\setcounter{equation}{0}
\setcounter{footnote}{0}
\clearpage


\part[A search for top partners using same-sign dilepton final states]{A search for top partners at the LHC using same-sign dilepton final states}

{\it T.~Bose, R.~Contino, M.~Narain and  G.~Servant}


\begin{abstract}
A natural, non-supersymmetric solution to the hierarchy problem generically requires
fermionic partners of the top quark with masses not much heavier than $500\, \text{GeV}$.
We study the pair production and detection at the LHC of the top partners
with electric charge $Q_e=5/3$ ($T_{5/3}$) and $Q_e=-1/3$ ($B$), that are
predicted  in models where the Higgs is a pseudo-Goldstone boson.
Both kinds of new fermions decay to $Wt$, leading to a $t\bar{t}WW$ final state.
We focus on the golden channel with two same-sign leptons, that offers the best chances of discovery  in the very early phase of LHC and permits a full mass reconstruction of 
the $T_{5/3}$. 
Samples are processed with the CMS Fast Simulation.
\end{abstract}

\section{INTRODUCTION}

The most notorious example of symmetry protection for the light Higgs
is Supersymmetry: according to its paradigm, the radiative correction
of each SM field to the Higgs mass is fine tuned against that of a superpartner
of opposite statistics. The top quark contribution, in particular, is balanced by
the contribution of its scalar partners, the stops.
Another kind of symmetry protection, however, could be at work: the light Higgs could be
the pseudo-Goldstone boson of a spontaneously broken global 
symmetry~\cite{Weinberg:1972fn,Georgi:1975tz,Kaplan:1983fs}.
In this case the radiative correction of the top quark to the Higgs mass is balanced 
by the contribution of new partners of the same spin. 
The naturalness criterium suggests that these new heavy fermions 
should have masses below, or not much heavier than, 1 TeV.
The production and detection of these top partners at the LHC was studied in 
\cite{Contino:2008hi} focussing on final states with two same-sign leptons. 
This note reports those findings  and complements them with a preliminary analysis using 
the CMS Fast Simulation~\cite{Bose:2008cms}.

Particularly motivated  is the possibility that
the spontaneous breaking of the global symmetry and the new states originate
from a strongly-coupled dynamics. This would allow for a complete 
resolution of the Hierarchy Problem without the need of fundamental scalar fields,
and would make it possible to generate a large enough quartic coupling for the Higgs
via radiative effects. 

The LEP precision data are  crucial in guiding our theoretical investigation, 
as they seem to be compatible only with a specific kind of strong dynamics: the new
sector must possess a custodial symmetry $G_{C}=$SU(2)$_C$ to avoid large 
tree-level corrections to the $\rho$ parameter~\cite{Sikivie:1980hm}. This in turn implies an unbroken
SU(2)$_L \times$SU(2)$_R\times$U(1)$_X$ invariance of the strong dynamics before EWSB,
meaning that its resonances, in particular the heavy partners of the top quark, will
fill multiplets of such symmetry.
It has been recently pointed out~\cite{Agashe:2006at} that the LEP constraint on 
the $Z\bar b_Lb_L$ coupling  
is more easily satisfied if the custodial symmetry of the strong sector includes a 
$LR$ parity, $G_{C}=$SU(2)$_C\times P_{LR}$, 
and 
$b_L$ couples linearly to a composite fermionic operator transforming as a 
$(\mathbf{2},\mathbf{2})_{2/3}$ under SU(2)$_L \times$SU(2)$_R\times$U(1)$_X$ (hypercharge being
defined as $Y=T^3_R+X$).
In this case, 
the heavy partners of 
$(t_L,b_L)$ can themselves fill a $(\mathbf{2},\mathbf{2})_{2/3}$ representation. 
The latter consists of two SU(2)$_L$ doublets: the first, $(T,B)$, has the quantum numbers
of $(t_L,b_L)$; the second -- its ``custodian'' -- is made of one fermion with exotic electric charge 
$Q_e = +5/3$, $T_{5/3}$, and one with charge $Q_e = +2/3$, $T_{2/3}$.
Since the Higgs transforms like a $(\mathbf{2},\mathbf{2})_{0}$, the partners of $t_R$, if any, will form a 
$(\mathbf{1},\mathbf{1})_{2/3}$ or a $[(\mathbf{1},\mathbf{3})\oplus (\mathbf{3},\mathbf{1})]_{2/3}$
of SU(2)$_L \times$SU(2)$_R\times$U(1)$_X$~\cite{Agashe:2006at}.

\begin{figure}[htb!]
\begin{center}
\includegraphics[width=7.5cm,height=6.cm]{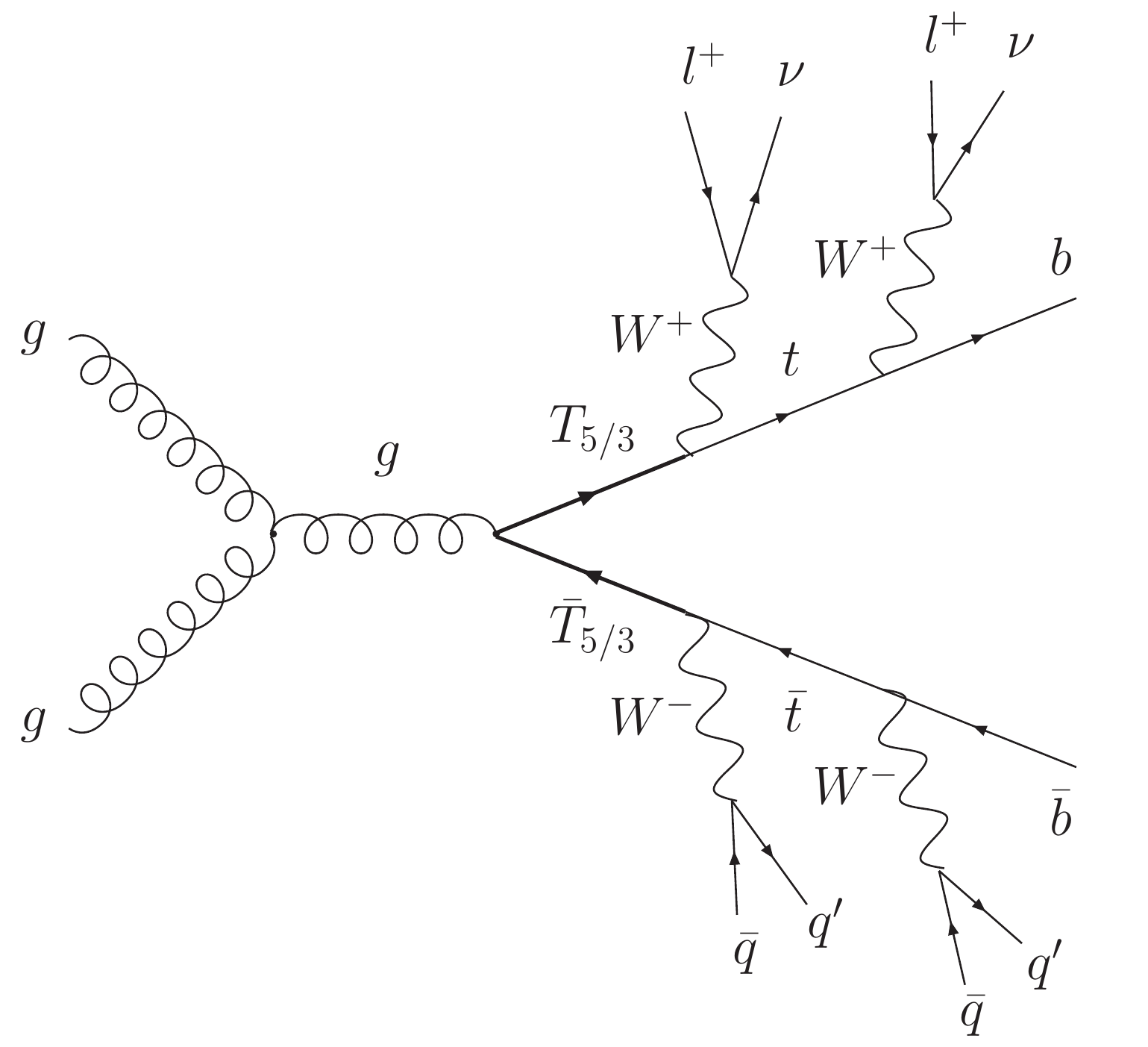}
\hspace{0.5cm}
\includegraphics[width=7.5cm,height=6.cm]{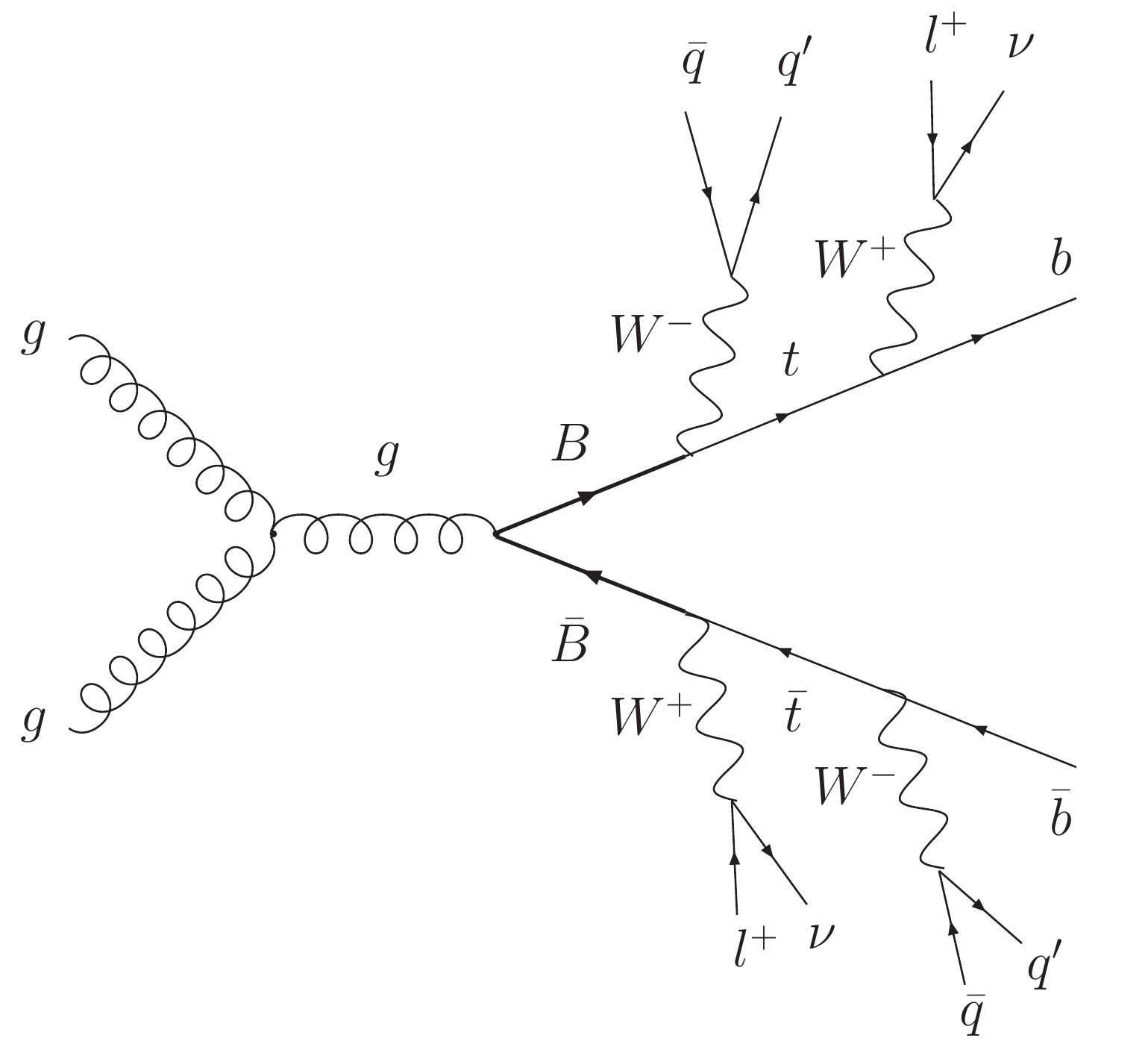}
\caption{\label{BCNS_fig1}  \small
Pair production of $T_{5/3}$ and $B$ to same-sign dilepton final states.
}
\end{center}
\end{figure}

As discussed in more details in Ref.~\cite{Contino:2008hi}, these heavy partners  couple strongly to 
the third generation SM quarks plus one longitudinal $W$, $Z$ gauge boson or the Higgs.
As in \cite{Contino:2008hi}, we focus on the pair production of the $B$ and of its custodial partner $T_{5/3}$.
Once pair produced, both the heavy bottom $B$ and the exotic $T_{5/3}$ decay exclusively 
to $W^+W^+W^-W^-b\bar b$, although with different
spatial configurations as dictated by their different electric charges, see Fig.~\ref{BCNS_fig1}.
In the case of the $T_{5/3}$ the same-sign $W$'s  
come from the decay of the same heavy fermion, 
while in the case of the heavy bottom they come from different $B$'s.
Selecting events with two same-sign leptons thus allows one to
fully reconstruct the hadronically-decaying $T_{5/3}$.
Even though  
a full reconstruction of the $B$ is not possible, it was found in \cite{Contino:2008hi} that 
the same-sign dilepton channel is probably the most promising one for its 
discovery (see \cite{Dennis:2007tv,Skiba:2007fw} for other recent studies of pair-production of $B$-type quarks).

In this work, we report some of the results derived in Ref.~\cite{Contino:2008hi} and we improve on them
by adding the CMS Fast Simulation~\cite{Bose:2008cms} to the analysis.
Section~2 presents our Monte Carlo simulations.
In Section~3, we discuss our cuts and derive the discovery potential,
while  in Section~4 we reconstruct the $W$ and $t$ candidates and pair them to 
reconstruct the $T_{5/3}$ invariant mass.

\section{FAST SIMULATION AND EVENT SELECTION}
\label{BCNS_sec2}

We focus on the pair production at the LHC of $B$ and $T_{5/3}$ to two same-sign leptons :
\begin{equation}
gg ,\,  q\bar q \to B\bar B ,\, T_{5/3} \bar T_{5/3} \to t \overline{t} W^+ W^-  \to l^{\pm}\nu\, l^{\pm}\nu\, b\bar b\, q\bar q^\prime q\bar q^\prime \, .
\end{equation}
The physical, observed final state is of the form
\begin{equation} 
pp\to l^{\pm} l^{\pm} + n\; jets \, + \not\!\! E_T \, , \qquad\quad l = e,\mu \, ,
\label{BCNS_Eq1}
\end{equation}
where the number of jets depends on the adopted jet algorithm and on its parameters.  
We take $M=500$ GeV for the mass of the heavy fermions (we refer the reader to ~\cite{Contino:2008hi} 
for the related study of the $M=1 $ TeV case) and set the coupling to $tW$ to be 
$\lambda_{T_{5/3}} = \lambda_B = 3$. Such large values of the couplings 
are naturally expected if the heavy fermions are bound states of a strongly coupled sector, and $t_R$ is mainly 
composite.
\footnote{
Notice, however, that our final results are largely independent of the specific values of $\lambda_{T_{5/3}}$, 
$\lambda_B$, since the latter determine only the decay width of the heavy fermions. 
For our choice of couplings and $M=500$ GeV,  $\Gamma = 31$ GeV.}

The most important SM backgrounds to the process of eq.(\ref{BCNS_Eq1}) are $t\bar t W+jets$, $t\bar t WW+jets$
(including the $t\bar t h+jets$ resonant contribution for $m_h\ge 2 m_W$), $WWW+jets$
(including the $W h+jets$ resonant contribution for $m_h\ge 2 m_W$), $W^{\pm}W^{\pm}+jets$
and $Wl^+l^-+jets$ (including the $WZ+jets$ contribution) where one lepton is missed.
To be conservative and consider the case in which the background is largest, we have
set the Higgs mass to $m_h = 180\, \text{GeV}$. This greatly enhances the $t\bar tWW$ and $WWW$ backgrounds.
The production cross sections for the signal  
and for the various backgrounds are reported in Table~\ref{BCNS_table1}.
No K-factors have been included, since those for the backgrounds are not all available
(the K-factor for the signal is $\simeq 1.8$ for $M=500$ GeV~\cite{Bonciani:1998vc}).
\begin{table}[h!]
\begin{center}
\begin{tabular}{|p{6.5cm}|c|c|}
\hline
 & $\sigma$ [fb] & $\sigma \times BR(l^\pm l^\pm)$ [fb] \\[0.10cm] \hline 
$T_{5/3}\overline{T}_{5/3} / B {\overline B}+ jets$  \ $(M=500$ GeV)   & $2.5\times 10^3$ & 104 \\[0cm]
$T_{5/3}\overline{T}_{5/3} / B {\overline B}+jets$  \ $(M=1$ TeV)      & 37   & 1.6 \\[0.35cm]
$t {\overline t} W^+W^- + jets$ \ ($\supset t\bar t h + jets$)        & 121  & 5.1 \\[0cm]
$t {\overline t} W^{\pm}+ jets$                                        & 595  & 18.4\\[0cm]
$ W^{+}W^{-} W^{\pm} + jets$ ($\supset h W^{\pm}+jets$)                 & 603  & 18.7 \\[0cm]
$ W^{\pm}W^{\pm} + jets$                                               & 340  & 15.5 \\[0.10cm]
\hline
\end{tabular}
\caption{ \small \label{BCNS_table1}
Signal and background cross sections at leading order (left column). The right column
reports the cross section times the branching ratio to two
same-sign leptons final states ($e$ or $\mu$). }
\end{center}
\end{table}

Given its complexity, we were not able to 
fully simulate the $Wl^+l^-+jets$ background,
and for that reason we have not included it in our 
analysis. 
However, we estimated that after the main cuts presented below, 
the $Wl^+l^-+jets$ background is expected to be smaller than $\sim 30\%$ of the sum of the
other backgrounds~\cite{Contino:2008hi}. Even though this is not entirely negligible, the error due to 
its exclusion is within the uncertainty of our leading-order analysis.
Moreover, the $Wl^+l^-+jets$ cross section is expected to be strongly suppressed after
requiring the reconstruction of one $W$ and one top as done in the last section. 

Another potential source of background are $t\bar t + jets$ events where the charge of one of the two
leptons from the top decays is misidentified. 
Given the large $t\bar t+jets$ cross section, even a charge misidentification probability
$\BCNSeps_{mis}\sim \text{a few}\times 10^{-3}$ would result into a same-sign dilepton background of the same
order of $t\bar t W+jets$.
~\footnote{Requiring the reconstruction of one $W$ and one top
as in Section~3 is however expected to reduce significantly 
more the $t\bar t + jets$ events background than $t\bar t W+jets$ or $t\bar t WW+jets$.}
The hardest lepton in the
$t\bar t+jets$ events has a $p_T$ distribution peaked at values smaller than $100\, \text{GeV}$
(the second hardest lepton has instead a significantly softer $p_T$ distribution).
From the latest ATLAS and CMS TDRs, probabilities as low as $\sim 10^{-4}$ seem to be realistic in the case
of 100 GeV muons, while slightly larger values are expected for electrons~\cite{CMS_TDR1,ATLAS_TDR1}.
In such case the $t\bar t+jets$ background would be safely negligible. 
In the absence of a realistic estimate of $\BCNSeps_{mis}$ as a function of the lepton's $p_T$ and pseudo-rapidity,
we decided not to include the $t\bar t+jets$ background events in our analysis.
It is however clear that a specific and accurate estimate of this background is required to validate 
our results.

In these proceedings, we review part of the findings obtained with the simulation and analysis of Ref.~\cite{Contino:2008hi}, that we will refer to as {\it generic} analysis, and compare with the results obtained by adding the CMS Fast simulation of the detector.
Both the signal and the SM background events were generated at the partonic level
with MadGraph/MadEvent~\cite{Alwall:2007st,Maltoni:2002qb,Stelzer:1994ta},~\footnote{The factorization and renormalization
scales have been chosen as follows: $\mu = M_{T,B} $ for the signal; 
$\mu = 2 m_t+m_W$ for $t\bar tW+jets$; $\mu = 2 m_t+m_h$ for $t\bar tWW+jets$; 
$\mu = m_W + m_h$ for $WWW+jets$; $\mu = 2 m_W$ for $W^{\pm}W^{\pm}+jets$.}.

In the {\it generic} analysis, we have used Pythia~\cite{Sjostrand:2006za} for showering 
and 
to include the initial and final-state radiation. Hadronization and underlying event
have been switched off for simplicity. Jets have been reconstructed using the 
GetJet~\cite{getjet} cone algorithm with $E_T^{min}=30\, \text{GeV}$ and a cone size $\Delta R=0.4$.
The parton-jet matching has been performed following the MLM 
prescription \cite{Alwall:2007fs, Mangano:2006rw}.
No detector effects were taken into account,
except for a simple gaussian smearing on the jets. Both the jet energy and momentum absolute
value were smeared by $\Delta E/E=100\%/\sqrt{E/\text{GeV}}$, and the jet momentum direction using an angle resolution 
$\Delta \phi=0.05$ radians and $\Delta\eta=0.04$.

 On the other hand, in the CMS Fast Simulation analysis~\cite{Bose:2008cms}, 
the  samples are first 
processed via {\sc{PYTHIA}}  for showering, to include 
initial and final-state radiation and to fragment and 
hadronize quarks and gluons. In addition, the  underlying event is switched on. A jet-matching algorithm, 
following the MLM  prescription~\cite{mlmmatch:2006} is employed to  ensure 
that there is no double  counting due to the parton showering in {\sc{PYTHIA}}. 
These samples are  then processed via a dedicated fast simulation
processor (FAMOS) of the CMS detector and the event reconstruction
software. Jets with cone size $\Delta R=0.5$ are reconstructed with 
the  iterative cone algorithm.  Generic jet energy corrections
are applied to the cone jets. The jet resolution used in the CMS Fast Simulation  
is $\Delta E_T/E_T=1.25/\sqrt{E_T/\text{GeV}} \oplus 5.6/{E_T/\text{GeV}} 
\oplus 0.033$~\cite{CMS_TDR1}. This jet energy resolution will be the main source of different results 
with the {\it generic} analysis.

\begin{figure}[!htb]
\begin{center}
\includegraphics[height=0.43\textwidth,width=0.49\textwidth]{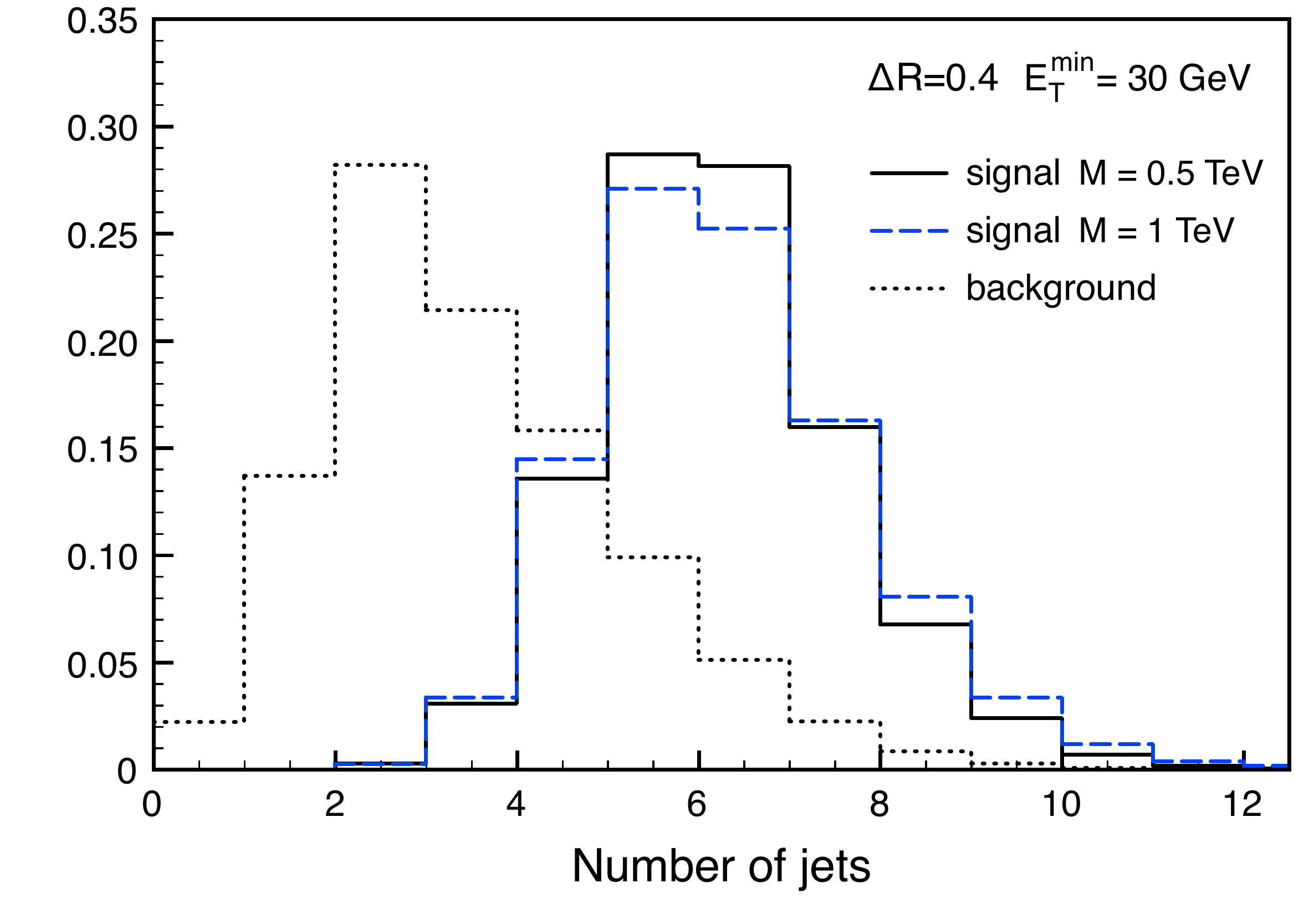}
\includegraphics[height=0.46\textwidth,width=0.5\textwidth]{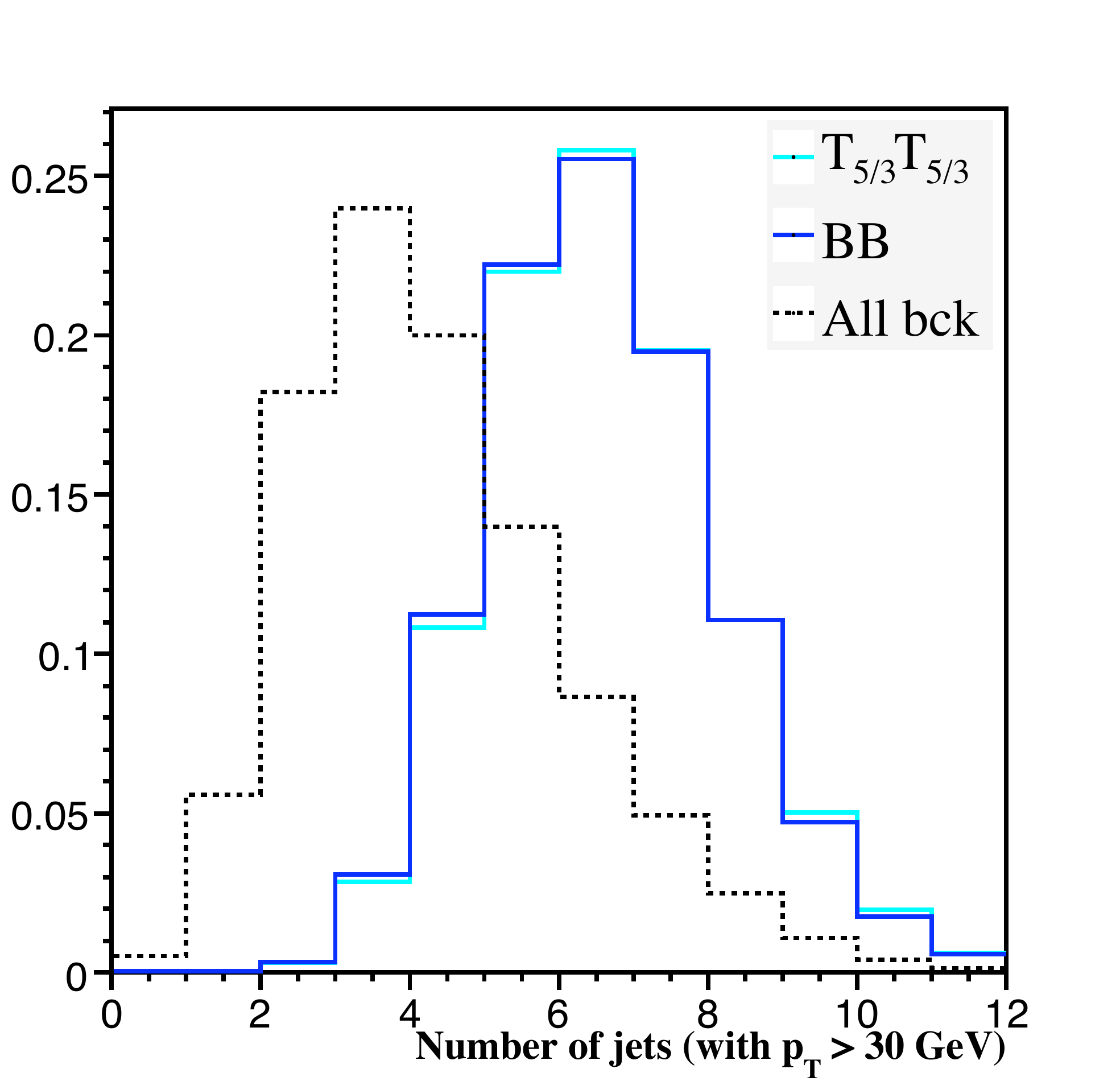}
\caption{Fractions of signal and background events with a given number of jets 
for $E^{min}_T=30\,\text{GeV}$. Left plot is from Ref.~\cite{Contino:2008hi}  
(using the GetJet cone algorithm with $\Delta R= 0.4$)  and right plot is
from the CMS Fast Simulation (using iterative cone algorithm with $\Delta R= 0.5$).}
\label{BCNS_fig3}
\end{center}
\end{figure}

A first important information on the kinematics of signal and background events comes 
from the number of reconstructed jets.
Fig.~\ref{BCNS_fig3}  compares the distributions for the number of
jets  in both analysis. The background and signal values are peaked at 2 and
5 (3 and 6) respectively in the {\it generic} (CMS Fast
Simulation) analysis~\cite{Bose:2008cms}. By signal here we mean either $T_{5/3}\bar T_{5/3}$ or
$B\bar B$ events. 
  The CMS Fast Simulation leads to a slightly larger number of 
reconstruted jets above $p_T >$ 30 GeV due to a combination of  the underlying event settings, jet
reconstruction  algorithm, and energy corrections used.
 The total background distribution is peaked at smaller values. This is mainly
due to the low jet multiplicity in the $WWW+jets$ and $W^{\pm}W^{\pm}+jets$ backgrounds.
In the case of the signal, the hard scattering process produces $6$ quarks, after the
decay of the top and of the $W$. It turns out that for $M=500\,\text{GeV}$
the 5-jet bin is mostly populated by events where the 6th jet is lost because it is too soft
(i.e. it does not meet the minimum transverse energy requirement, $E_T\ge 30\,\text{GeV}$).

\section{DISCOVERY POTENTIAL}
\label{BCNS_sec3}

In this section, we focus on the discovery of the top partners, proposing a simple strategy that does not
rely on any sophisticated reconstruction, nor does it require $b$-tagging. 
\begin{figure}[!htb]
\begin{center}
\includegraphics[height=0.51\textwidth,width=0.495\textwidth]{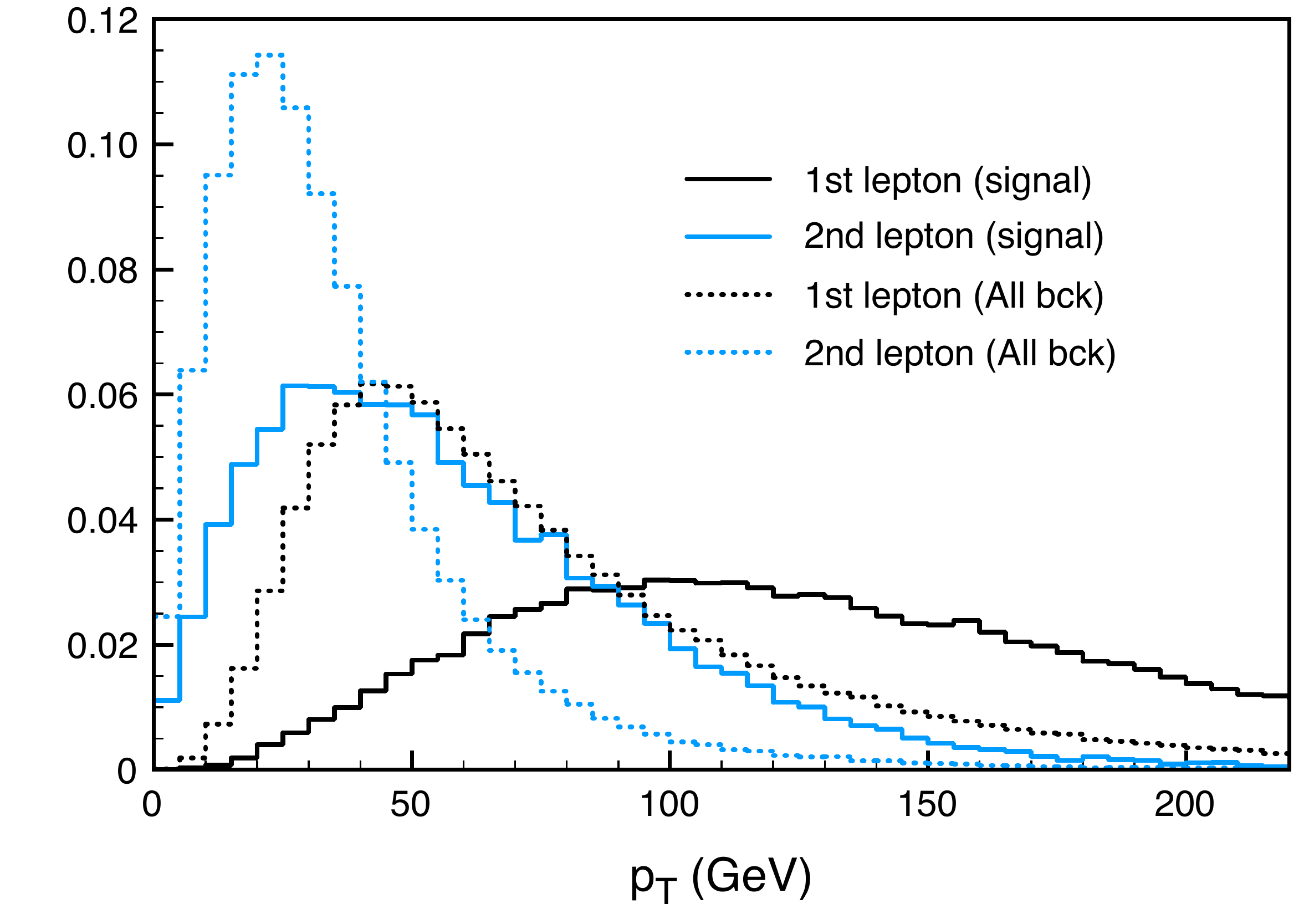}
\includegraphics[height=0.51\textwidth,width=0.495\textwidth]{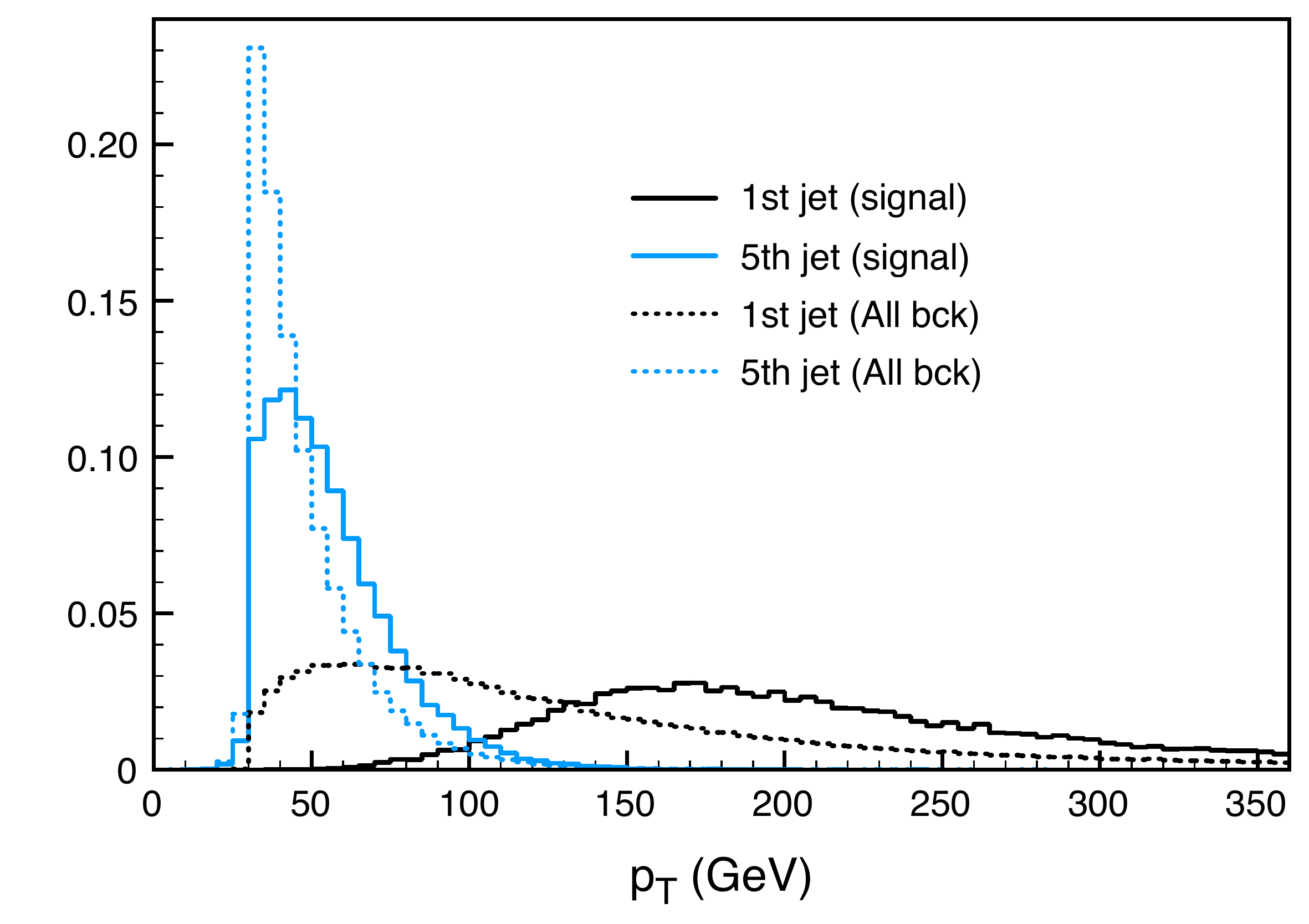}
\caption{$p_T$ distributions of the first and fifth hardest jets (left) and of the first and second hardest leptons (right) before any cut (from the {\it generic} simulation).}
\label{BCNS_fig5}
\end{center}
\end{figure}
To isolate the signal, the following cuts are applied (both for the {\it generic} and the CMS Fast simulations):
\begin{equation} \label{BCNS_Eq2}
\text{\underline{leptons}}: 
\begin{cases}
p_T(\text{1st})  \geq 50 \; \text{GeV} \\
p_T(\text{2nd})  \geq 25 \; \text{GeV} \\
|\eta_l| \leq 2.4 \, ,  \ \  \Delta R_{lj} \geq 0.4
\end{cases} \quad
\text{\underline{jets}}: 
\begin{cases}
p_T(\text{1st})  \geq 100 \; \text{GeV} \\
p_T(\text{2nd})  \geq 80 \; \text{GeV} \\
n_{jet} \geq 5 ,  \ \ |\eta_j| \leq 5
\end{cases} \quad
\not\!\! E_T  \geq 20 \; \text{GeV} \, ,
\end{equation}
where 1st and 2nd refer respectively to the first 
and second hardest jet or lepton (electron or muon). These cuts were motivated by the $p_T$ distributions 
shown in Fig.~\ref{BCNS_fig5}.
The relative efficiencies are reported in Table~\ref{table:epsmain}.
\begin{table}[htb!]
\begin{center}
\begin{tabular}{|l|c|c|c|c|c|}
\hline
& signal  & \multirow{2}{*}{$t\bar t W$}
& \multirow{2}{*}{$t\bar t WW$} & \multirow{2}{*}{$WWW$} & \multirow{2}{*}{$W^{\pm}W^\pm$} \\
& ($M=500$ GeV)  & & & & \\[0.05cm]
\hline 
Efficiencies ($\BCNSeps_{main}$) & 0.42 \ \ (0.55) & 0.074 \ \ (0.15)& 0.12 \ \ (0.06)& 0.008 \ \ (0.02)& 0.01 \ \ (0.03) \\[0.01cm]
$\sigma\, \text{[fb]} \times BR \times \BCNSeps_{main}$ & 44.2  \ \ (57.2)& 1.4 \ \ (2.80) & 0.62 \ \ (0.30) & 0.15 \ \ (0.37) & 0.16 \ \ (0.48) \\[0.05cm]
\hline 
\end{tabular}
\caption{\label{table:epsmain} \small
Efficiencies of the main cuts of eq.(\ref{BCNS_Eq2}) for the {\it generic} analysis. Here signal means either
$T_{5/3}\bar T_{5/3}$ or $B\bar B$ events. Numbers in parenthesis refer to the CMS Fast Simulation.~\cite{Bose:2008cms}
}
\end{center}
\end{table}

\begin{figure}[!htb]
\begin{center}
\includegraphics[width=0.55\textwidth]{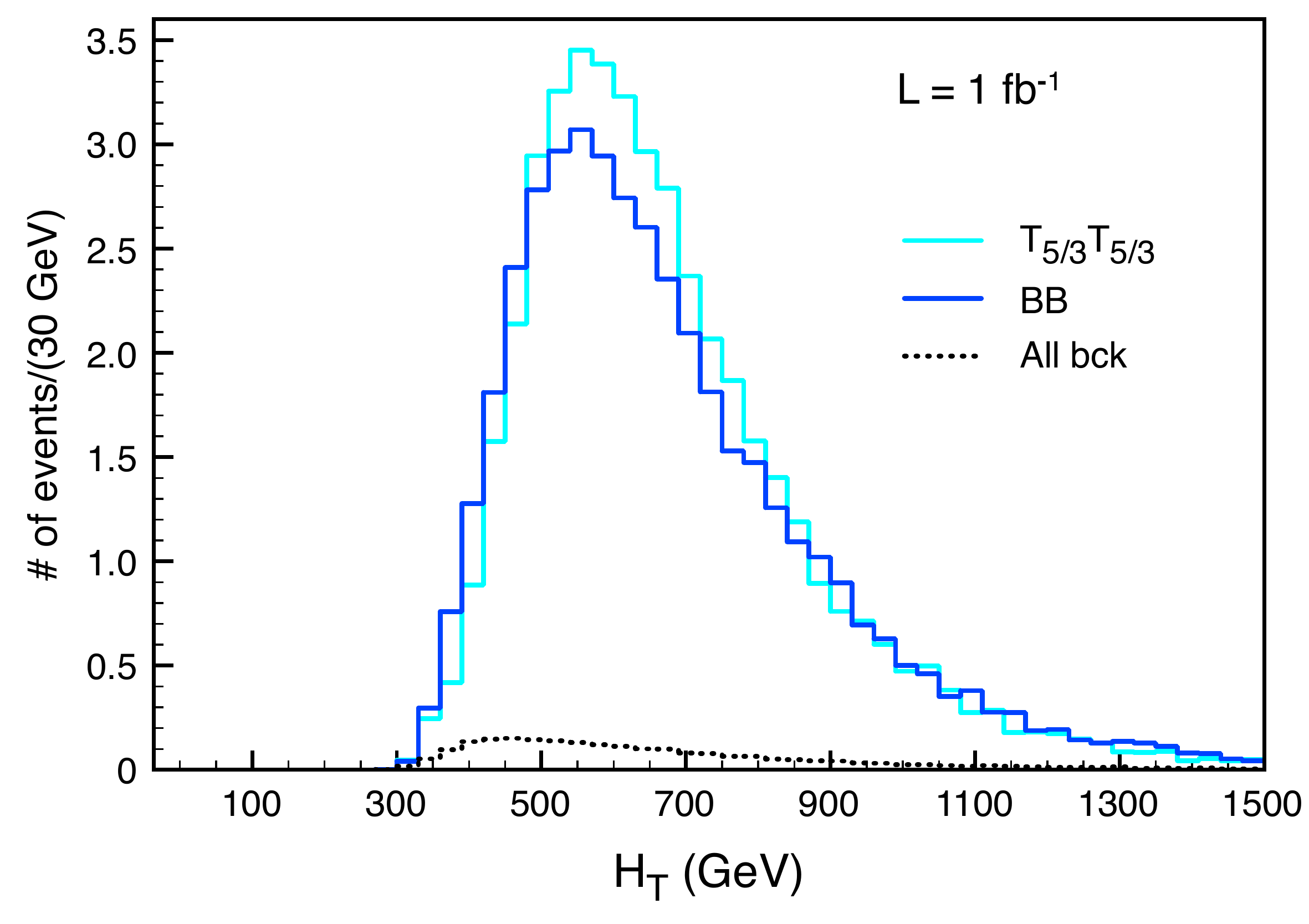}
\includegraphics[width=0.44\textwidth]{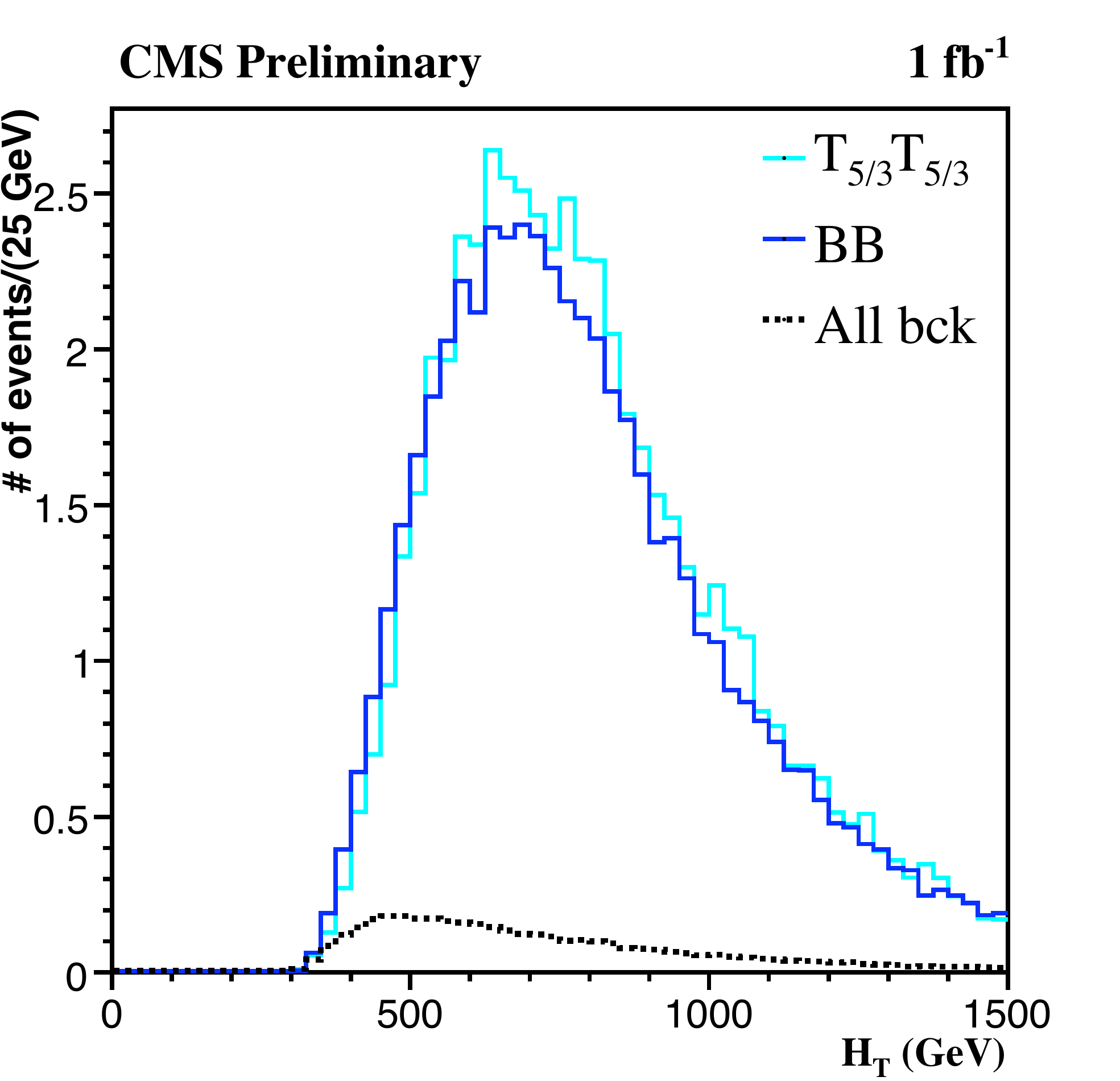}
\caption{$H_T$ distribution (scalar sum of $p_T$ of all jets in the
  event). Comparison between the simulation of Ref.~\cite{Contino:2008hi}
  (left) and the output of the CMS Fast Simulation for an integrated
luminosity corresponding to 1 $\text{fb}^{-1}$ (right). }
\label{BCNS_fig7}
\end{center}
\end{figure}

After the cuts, the signal is much larger than the background. 
A rough indication on the mass of the heavy fermions can be extracted from the 
distribution for the scalar sum of $p_T$ of all jets in the event ($H_T$), see Fig.~\ref{BCNS_fig7}.
This is a quantity that is not much affected by a poor jet energy resolution, and as such it is particularly
appropriate in the first LHC phase.
Further evidence on the  production of a pair of  heavy fermions comes from the distributions of the total
invariant mass, the invariant mass of the hardest $5$ jets, and transverse mass of the $(ll\nu\nu j)$ system.
We refer the interested reader to Ref.~\cite{Contino:2008hi} for the relative plots.

By counting the number of signal and background events that 
pass the main cuts of eq.(\ref{BCNS_Eq2}),  
one can estimate the statistical significance of the signal over 
the background, as well as the minimum integrated luminosity required for a discovery. In the {\it generic} analysis, 
we defined the latter to be the integrated luminosity for which a goodness-of-fit test of the SM-only hypothesis
with Poisson distribution gives a p-value = $2.85\times 10^{-7}$, 
(see for instance ~\cite{Yao:2006px}).~\footnote{This p-value 
corresponds to a $5\sigma$ significance in the limit of a gaussian distribution.}
Results  are reported in Table~\ref{table:significances}. 
The numbers in parenthesis are those from the CMS Fast simulation, were we used a different and more conservative 
estimate for the significance: $S/\sqrt{S+2B}$.
\begin{table}[htb!]
\begin{center}
\begin{tabular}{|l|l|c|c||r @{} l|}
\hline
& & ${\cal S}$ & ${\cal B}$ & \multicolumn{2}{c|}{$L_{disc}$}\\
\hline
\multirow{2}{*}{$M=500$ GeV} & $T_{5/3}+B$ & 864 \ (1124) & 23 \ (42)& $56$ & $\,\text{pb}^{-1}$ \ (240 pb$^{-1}$) \\
                             & $B$ only & 424 \ (548) & 23 \ (42) & $147$ & $\,\text{pb}^{-1}$ \ (530 pb$^{-1}$) \\ [0.05cm]
\hline
\end{tabular}
\end{center}
\caption{\label{table:significances} \small
Number of signal (${\cal S}$) and background (${\cal B}$) events
that pass the main cuts of eq.(\ref{BCNS_Eq2}) with $L=10\,\text{fb}^{-1}$ for $M = 500\,\text{GeV}$.
The last column reports the corresponding integrated luminosity needed for the
discovery ($L_{disc}$), 
as computed in the {\it generic} analysis by means of a goodness-of-fit test with Poisson distribution
and p-value = $2.85\times 10^{-7}$. Values in parenthesis are from the CMS Fast Simulation and in the last column $L_{disc}$ is computed using  $S/\sqrt{S+2B}$ as an estimate of the significance.
}
\end{table}
%
In the most favorable case where both $T_{5/3}$ and $B$ partners exist and have mass $M = 500\,\text{GeV}$,
a discovery will need less than $\sim 250\,\text{pb}^{-1}$.
Our estimates should be conservative, as 
we did not include any K-factor in our
analysis, although it is known that next-to-leading order corrections enhance the signal cross section
by $\sim 80\%$ for $M = 500\,\text{GeV}$ ~\cite{Bonciani:1998vc}.
Even a common K-factor $\kappa$ for both the signal and the background would imply a statistical significance
larger by a factor $\sim\sqrt{\kappa}$, as well as a discovery luminosity smaller by the same factor.

\section{$T_{5/3}$ MASS RECONSTRUCTION}
\label{BCNS_sec4}

More direct evidence for the production of a pair of $T_{5/3}$ or $B$ comes from
reconstructing the hadronically decayed top quark and $W$ boson, as well as 
from the distribution of 
the invariant mass of their system. 
The mass reconstruction of $T_{5/3}$ crucially depends on the jet energy
resolution when using the full detector simulation.
 We find that the width of the reconstructed top quark  mass as obtained
in the {\it generic} analysis is 12.5 GeV, while that obtained
 from the CMS
  Fast Simulation is about  19.5 GeV. This latter value more realistically reproduces the expected resolution of in the early phase of the LHC.  It will improve as the flavor dependent jet energy
  corrections and use of high purity $b$-tagging for correct assignment of
  the $b$-jets are incorporated in the analysis. 

\begin{figure}[!htb]
\begin{center}
\includegraphics[width=0.48\textwidth]{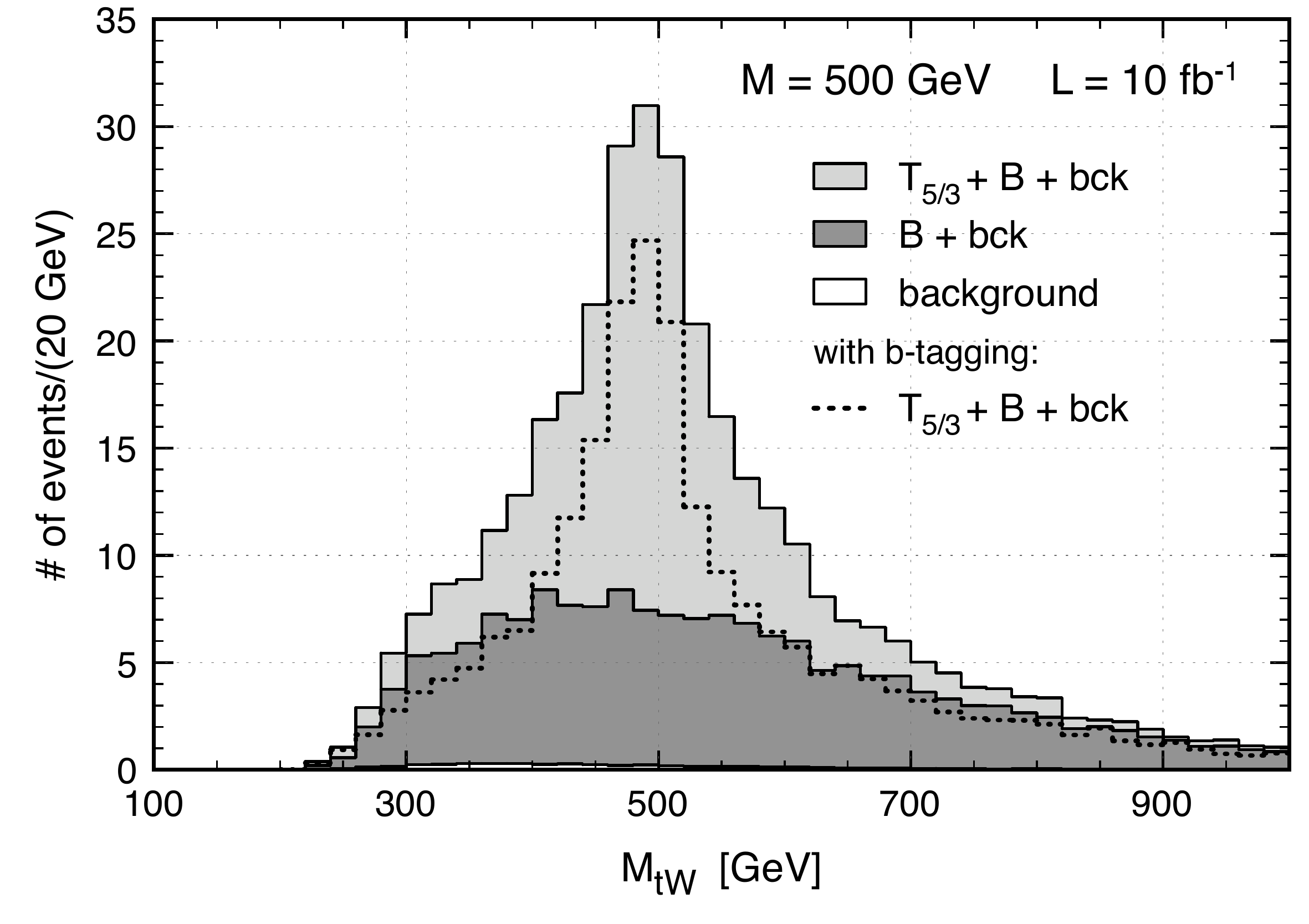}
\caption{Invariant mass of the $Wt$ system for $M=500\,\text{GeV}$ with $L=10\,\text{fb}^{-1}$.
The dotted curve refers to
the case in which $b$-tagging is performed in the reconstruction.
It assumes two $b$ tags, though no $b$-tagging
efficiency has been included.}
\label{BCNS_fig9}
\end{center}
\end{figure}

We report here the results from the {\it generic} analysis using the energy
resolution defined in Section~2, which is probably
optimistic for the early LHC phase. 
We 
first select the events where two $W$'s can be 
simultaneously reconstructed, each $W$ candidate being formed by a pair of jets
with invariant mass in the window $|M(jj)-m_W|\leq 20\,\text{GeV}$. To avoid wrong pairings
and reduce the fake ones from the background, we impose the following cuts: 
\begin{align}
& \Delta R_{jj}\leq 1.5\, , \quad |\vec p(W)|\geq 100\,\text{GeV} & &
 \text{on the first $W$ candidate}\, ; \\[0.2cm]
& \Delta R_{jj}\leq 2.0\, , \quad |\vec p(W)|\geq 30\,\text{GeV} & &
 \text{on the second $W$ candidate}\, .
\end{align}
The $p_T$ cuts, in particular, have been optimized using the $p_T$ distributions of the $W$ and top from the signal events (see \cite{Contino:2008hi}).
If more than one pair of $W$ candidates exists which satisfies the above cuts, we select that
with the smallest $\chi^2 = \Delta R_{jj}^2(\text{1st pair})+\Delta R_{jj}^2(\text{2nd pair})$.
We then reconstruct the top by forming $Wj$ pairs, made of one $W$ and one of the remaining jets, 
with invariant mass in the window $|M(Wj)-m_t|\leq 25\,\text{GeV}$.
If more than one top candidate exists, we select that with invariant mass closest to $m_t$.
We discard events where no top can be reconstructed. 
The efficiencies of this reconstruction algorithm are reported in Table~\ref{table:effrec500GeV}.
\begin{table}[htb!]
\begin{center}
\begin{tabular}{|l|c|c|c|c|c|}
\hline
& signal & \multirow{2}{*}{$t\bar t W$}
& \multirow{2}{*}{$t\bar t WW$} & \multirow{2}{*}{$WWW$} & \multirow{2}{*}{$WW$} \\
& ($M=500$ GeV) & & & & \\
\hline 
$\BCNSeps_{2W}$    & 0.62 & 0.36 & 0.49 & 0.29 & 0.15 \\[0cm]
$\BCNSeps_{top}$   & 0.65 & 0.56 & 0.64 & 0.35 & 0.35 \\[0.05cm]
\hline 
\end{tabular}
\end{center}
\caption{\label{table:effrec500GeV} \small
Efficiencies for the reconstruction of two $W$'s ($\BCNSeps_{2W}$) and one top ($\BCNSeps_{top}$) 
using the algorithm  and the cuts described in the text for the case $M=500\,\text{GeV}$.
}
\end{table}
The distribution of the $Wt$ invariant mass is plotted in 
Fig.~\ref{BCNS_fig9}.
As expected, in the scenario with $T_{5/3}$ partners there is a 
resonant peak centered at $M_{T_{5/3}}=500\,\text{GeV}$, while 
the distribution has a non-resonant, continuous shape if only $Q_{e}=-1/3$ heavy fermions exist.
The dotted curve refers to the case in which $b$-tagging is performed in the
reconstruction algorithm. More in detail, we have selected events with two $b$ tags
and we have reconstructed the top from $Wb$ pairs, excluding at the same time the $b$ jets
when selecting the $W$ jet pair candidates. No $b$-tagging efficiency has been included,
in order not to commit to any specific value (hence, for a $b$-tagging efficiency $\epsilon_b$
the final distribution will be rescaled by a factor $\epsilon_b^2$).

\section{CONCLUSIONS}

Our results  show
that the analysis of final states with two same-sign leptons  at the LHC
is an extremely promising
method to discover the top partners $B$ and $T_{5/3}$.
By requiring two same-sign leptons one avoids the large $t\bar t$ background and selects
a particularly clean channel where evidence for 
the existence of the heavy fermions
could come in the early phase of the LHC.
If both $B$ and $T_{5/3}$ exist,
a discovery will require only $\sim$ 60 -- 250 pb$^{-1}$,
depending on how the statistical significance is estimated.
Even without $b$-tagging, and before reconstructing the hadronically decayed $W$ and top,
one can have a first crucial indication on the value of the mass of the 
heavy fermions from the distributions of the total invariant mass, the 
invariant mass of the hardest $5$ jets or the $H_T$ distribution (scalar sum of jet $p_T$'s).

Ultimately,
a crucial information to understand the origin and the role of the heavy fermions would
come from the measurement of their decay width, which will in turn lead to a determination of
their couplings $\lambda_{T_{5/3},B}$.
A large value of $\lambda_{T_{5/3},B}$
will be strong circumstantial evidence for the compositeness of the heavy fermions.
Extracting the decay width from the invariant mass distribution will be challenging,
as one will have to cope with the issue of jet energy resolution.
Most likely, a measurement will be possible only with large statistics and will require
sophisticated $W$ and $t$ reconstruction techniques.

Given the strong theoretical motivations for a search of the heavy partners of the top,
we think that our explorative study would also deserve to be followed by a 
more detailed
investigation. Our results suggest that the same-sign dilepton channel might 
be one of the golden modes to discover the top partners $B$ and $T_{5/3}$, but only a complete
analysis with a full simulation of the detector effects,
an exact calculation of the $Wl^+ l^-+jets$ and $t\bar t+jets$ backgrounds,
and the use of fully realistic reconstruction techniques
will eventually establish its ultimate potentialities.

\section*{ACKNOWLEDGEMENTS}

We thank the organizers of the Les Houches workshop for their invitation to participate to a very fruitful meeting.

\AddToContent{T.~Bose, R.~Contino, M.~Narain and  G.~Servant}
\setcounter{figure}{0}
\setcounter{table}{0}
\setcounter{section}{0}
\setcounter{equation}{0}
\clearpage

\superpart{Single Lepton + X Final States}

\part[Top Quark Pairs]{Top Quark Pairs as a Model-Independent Discriminator of New Physics at the LHC}

{\it D.G.E.~Walker}

%

\begin{abstract}
We review top quark pair production as a way to probe new physics at the LHC.  Our scheme requires identifying integer-spin resonances from $t\bar{t}$ semileptonic decays in order to favor/disfavor new models of electroweak physics.  The spin of each resonance can be determined by the angular distribution of top quarks in their c.m. frame.  In addition, forward-backward asymmetry and CP-odd variables can be constructed to further distinguish the new physics.  We parametrize the new resonances with a few generic parameters and show high invariant mass top pair production may provide a framework to distinguish models of new physics beyond the Standard Model.  
\end{abstract}

\section{INTRODUCTION}

A considerable number of experimentally viable extensions to the Standard Model (SM) have been proposed to describe electroweak symmetry breaking.  A sample of popular scenarios include:  the Minimal Supersymmetric Standard Model (MSSM) \cite{Dimopoulos:1981zb}, models with new strong dynamics \cite{Weinberg:1975gm,Susskind:1978ms,Hill:1980sq,Hill:1991at,Chivukula:1998wd,Dobrescu:1997nm,Hill:2002ap}, composite Higgs models at the TeV scale \cite{Kaplan:1983sm}, Little Higgs models \cite{ArkaniHamed:2001nc,ArkaniHamed:2002qy,Schmaltz:2005ky}, and models with extra dimensions at the electroweak scale \cite{ArkaniHamed:1998rs,Randall:1999ee}.  In addition, some string-inspired extensions \cite{Hewett:1988xc,Cvetic:1995rj} can also lead to new signatures.  In this review, we describe a scheme to sort out the particle content in a model independent fashion.

The LHC will be a ``top factory":  About 80 million $t\bar{t}$ events will be produced via QCD production for integrated luminosity of 100 fb$^{-1}$.  With such a large number of events (and knowing natural models have a preferential coupling to top quarks), it seems worthwhile to study top pairs.  If the new physics contributes to $t\bar{t}$ production as an $s$-channel resonance, we want to identify the signal as a bump \cite{Lane:1995xx} on the smoothly falling $t\bar t$ invariant mass  distribution.  We want to then reconstruct the  $t \bar t$ c.m.~frame so the integer spin of the resonance can be determined from the polar angular  distribution of the top quark.  Further, any asymmetry of this distribution could probe possible chiral couplings.  In addition, any CP properties of the couplings could be elucidated with the help of CP-odd kinematical variables constructed from the final state particle momenta. 

We look at only the $t \bar{t}$ semileptonic decay mode:  $t \bar t$ $\to b j_1j _2\ \bar b\ell^- \bar\nu + \mathrm{c.c.}$ where $\ell=e$ or $\mu$.  The purely hadronic decay mode not only suffers from a much larger QCD background, but also loses the identification of $t$ from $\bar t$.  For the purely leptonic mode, with a small  branching fraction of about $4/81$, one cannot reconstruct the $\bar t t$ invariant mass with two missing neutrinos.  Thus, our signal will be an isolated charged lepton plus missing energy ($\walkeretmiss$), 2 $b$-jets plus 2 light jets. Note:  The branching ratio of  the semileptonic to the hadronic channel is 2/3.

In this article, we summarize the analysis of~\cite{Barger:2006hm}.  Since \cite{Barger:2006hm} appeared, many papers have been published which look at the effectiveness of using top quarks to probe new physics.  A brief sample of papers is~\cite{Fitzpatrick:2007qr,Lillie:2007yh,Shiu:2007tn,Baur:2007ck,Agashe:2007ki,Mohapatra:2007af,Gerbush:2007fe,Cavicchia:2007dp,Frederix:2007gi,Agashe:2006hk}.  Please note:  \cite{Barger:2006hm} focused on early indications of new physics by restricting to only 10 fb$^{-1}$ of data.  When necessary, for emphasis, we include the relevant analysis for 100 fb$^{-1}$.  In all, the exposition of this review follows \cite{Barger:2006hm} closely.
 
\section{EVENT RECONSTRUCTION}

In order to search for new physics, we first have to reliably reconstruction $t\bar{t}$ semileptonic decays at high invariant mass on an event-by-event basis.  $t\bar{t}$ semileptonic decays contain a neutrino in the final state -- thereby complicating the reconstruction. The transverse momentum  of the neutrino is identified with the observed $\walkeretmiss$.  The neutrino longitudinal momentum  is subject to a two-fold ambiguity from solving the kinematic quadratic equation. 

We choose a reconstruction method useful for high invariant mass $\walkerttb$ events~\cite{Barger:2006hm}.  Several top reconstruction methods have been used at the Tevatron~\cite{Abbott:1998dc,Abe:1997vq}.  There, however, the top quarks are produced near threshold and the kinematics of the subsequent decay products are very complicated.  High invariant mass $\walkerttb$ events tremendously simplify the kinematics, especially by distinguishing the $b$ quark from the $\bar b$.  Throughout this review, we use a $2 \to 6$ partonic level monte-carlo simulation that incorporates full spin correlations from production through decay \cite{Barger:1988jj,Arai:2004yd}.  \cite{Barger:2006hm} made a Pythia simulation, including gluon radiation and hadronization, that confirmed the results presented here.  

To ensure only high invariant mass $\walkerttb$ events, we impose a cluster transverse mass cut on the $\walkerttb$ system
\begin{eqnarray*}
M_T= \sqrt{(p_{b} + p_{\bar b} + p_{j_1} + p_{j_2} + p_\ell)^2 + \walkeretmiss^2} + \walkeretmiss > 600\ {\rm GeV}.
\label{walker_mtc}
\end{eqnarray*}
Following ATLAS and CMS, we adopt the following kinematical cuts:
\begin{eqnarray*}
E_T^j > 20 \,\,\, \mathrm{GeV} && |\eta_j | < 2.5 \\
p_T^\ell > 20 \,\,\,\mathrm{GeV} && |\eta_\ell| < 2.5 \nonumber. 
\label{walker_kinematicalcuts1}
\end{eqnarray*}
In addition, we require a lepton-jet and jet-jet isolation cut and a $\walkeretmiss$ cut of
\begin{eqnarray*}
\Delta R > 0.4 && \walkeretmiss >20 \,\,\, \mathrm{GeV}.
\label{walker_kinematicalcuts2}
\end{eqnarray*}
The hadronic energy is smeared according to a Gaussian error given by $\Delta E_j/E_j=0.5/\sqrt {E_j/\textrm{GeV}} \oplus 0.03$.  Additionally, the lepton momentum is smeared by $\Delta p_T^\ell /p_T^\ell= 0.36 (p_T^\ell/\textrm{TeV}) \oplus 0.013/\sqrt{\sin\theta}$.  Here $\theta$ is the polar angle of the lepton with respect to the beam direction in the lab frame.  
\begin{figure}[!htb]
\centering
\label{walker_fig:mwmt}
\vspace{-1.75in}
\includegraphics[width=15truecm,height=15truecm,clip=true]{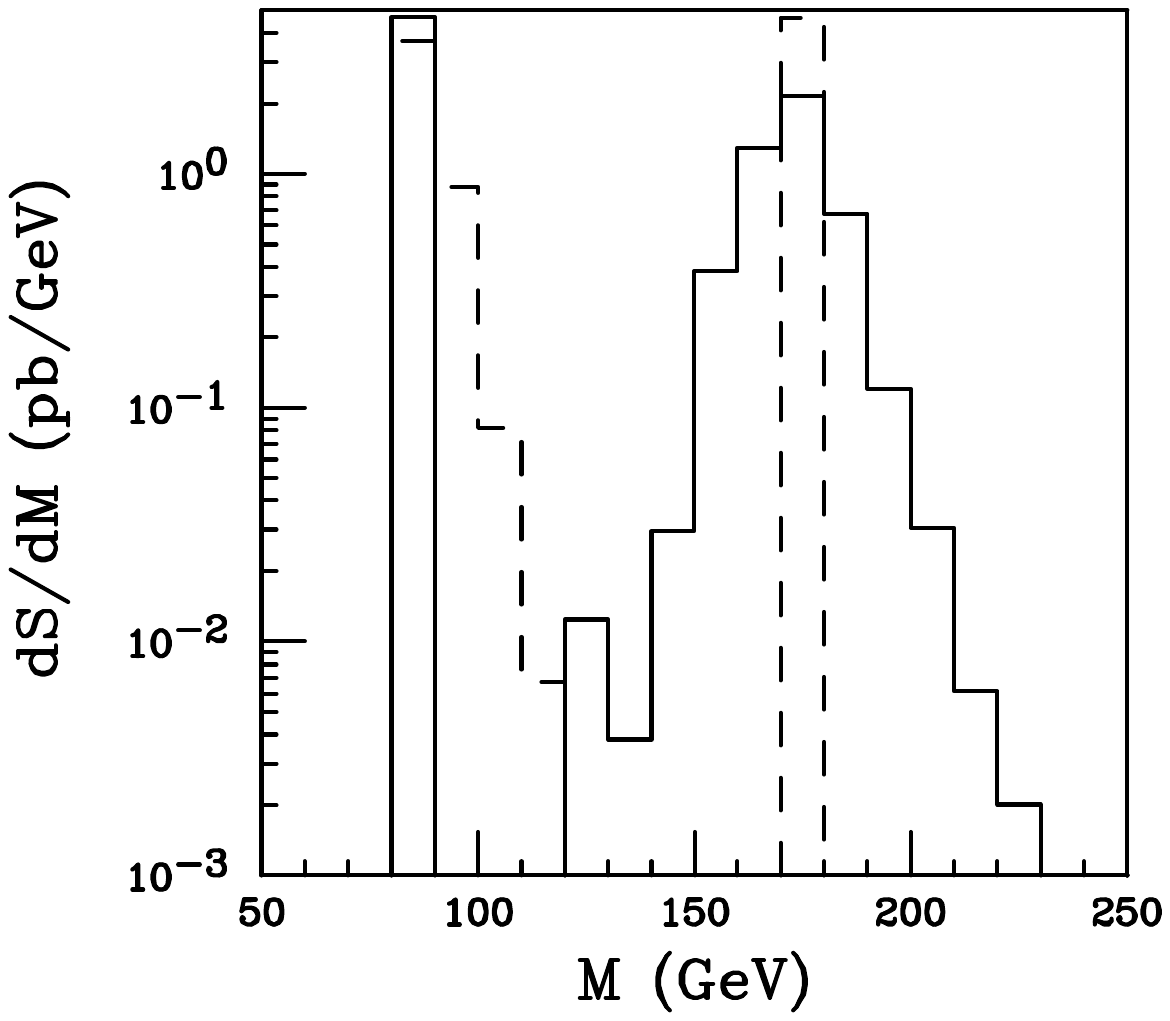}
\vspace{-2in}
\caption{The $W$ and top mass reconstructions from our reconstruction scheme. Step I (Step II) is solid (dashed), respectively.}
\end{figure}

We choose a reconstruction scheme that takes $M_W$ and $m_t$ as inputs for their on-shell production and decays.  We break the scheme into the following steps:
\\ \\
    {\bf Step I:} 
 Demand $m_{l \nu}^2 = M_W^2$.  The longitudinal momentum of the neutrino is formally expressed as
\begin{displaymath}
p_{\nu L}  =\frac{1}{2\, p_{e T}^2}
 \left( {A\, p_{e L} \pm E_e \sqrt{A^2  - 4\,{p}^{\,2}_{e T}\walkeretmiss^2}} \right),
\end{displaymath}
where $A = M_W^2 + 2 \,\vec{p}_{e T} \cdot \,\vec{\walkeretmiss} $.  If $A^2 - 4 \, {p}^{\,2}_{e T}\,\walkeretmiss^{\,2} \geq 0$, 
the value of $p_{\nu L}$ that best yields the known top mass via $m_{l\nu b}^2 = m_t^2$ is selected.  This ideal situation may not always hold when taking into account the detector resolutions.  
For cases with no real solutions, we then  proceed to the next step.
\\ \\
 {\bf Step II:} To better recover the correct kinematics, 
we instead first reconstruct the top quark directly by demanding $m_{l\nu b}^2 = m_t^2$.  
The longitudinal momentum of the neutrino is expressed as
\begin{eqnarray*}
p_{\nu L} & = &  {A'\, p_{bl L}}/ 2(E_{bl}^2 - p_{bl L}^2) \  \pm 
\frac{1}{2(E_{bl}^2 - p_{bl L}^2)}\\
&\times & ( {{p_{bl L}^2 A^{'\,2}  + (E_{bl}^2 - p_{bl L}^2) \,
( A^{'\,2} - 4 E_{bl}^2 E\!\!\!\!\slash_{T}^2  )} })^{1/2} ,
\end{eqnarray*}
where $A' = m_t^2 - M_{bl}^2 + 2 \,\vec{p}_{bl T} \cdot \,\vec{\walkeretmiss} $.
The two-fold ambiguity is broken by choosing the value that best reconstructs 
$M_W^2 = m_{l \nu}^2 $. 
A plot of the top and $W$ mass distributions is  shown in  Fig.~1.  
The  solid histogram is from the procedure {\it Step I},  and the dashed histogram from {\it Step II}.  With these two steps, there could still be some events that do not lead to a real solution.  Because we want accurate reconstruction of $\walkerttb$ at high invariant mass, we thus discard them in our event collection. 
The discard rate is about 16\%.

\section{BACKGROUNDS}

The major backgrounds to our $t\bar t$ events include the processes $W+$ jets, $Z+$ jets, $WW$, $WZ$ and $ZZ$.  Both the ATLAS and CMS Technical Design Reports~\cite{unknown:1999fr, Ball:942733} performed detailed studies of the selection efficiencies for these background processes in comparison to a reconstructed $t\bar{t}$ semileptonic signal.  The ATLAS (CMS) group found for an integrated luminosity of $10$ fb$^{-1}$ ($1$ fb$^{-1}$)  a signal to background ratio of $S/B = 65$ ($S/B = 26$).  This ratio has been obtained using the kinematical cuts listed in the previous section.  Because of the expected high $S/B$ ratio,  our analysis is concentrated solely on the $\tt$ events without including the small background contamination.  Our analysis does not include misidentification of faked leptons from jets in $t\bar{t}$ total hadronic decays.

Although the $t$ ($\bar{t}$) is primarily identified by the charged lepton, $\ell^+$ ($\ell^-$), a concern is the matching of the b-jet associated with this top quark decay.  Both ATLAS and CMS studies~\cite{unknown:1999fr, Ball:942733} show a combination of kinematic fits, designed to properly reconstruct the $W$ boson and the hadronically decaying top significantly reduces misidentification.  The cut on $M_T$ helps significantly in this regard. 

\section{SEARCHING FOR NEW PHYSICS}

\begin{figure}[!htb]
\centering
\label{walker_Fig:LHC}
\vspace{-1in}
\includegraphics[width=15truecm,height=15truecm,clip=true]{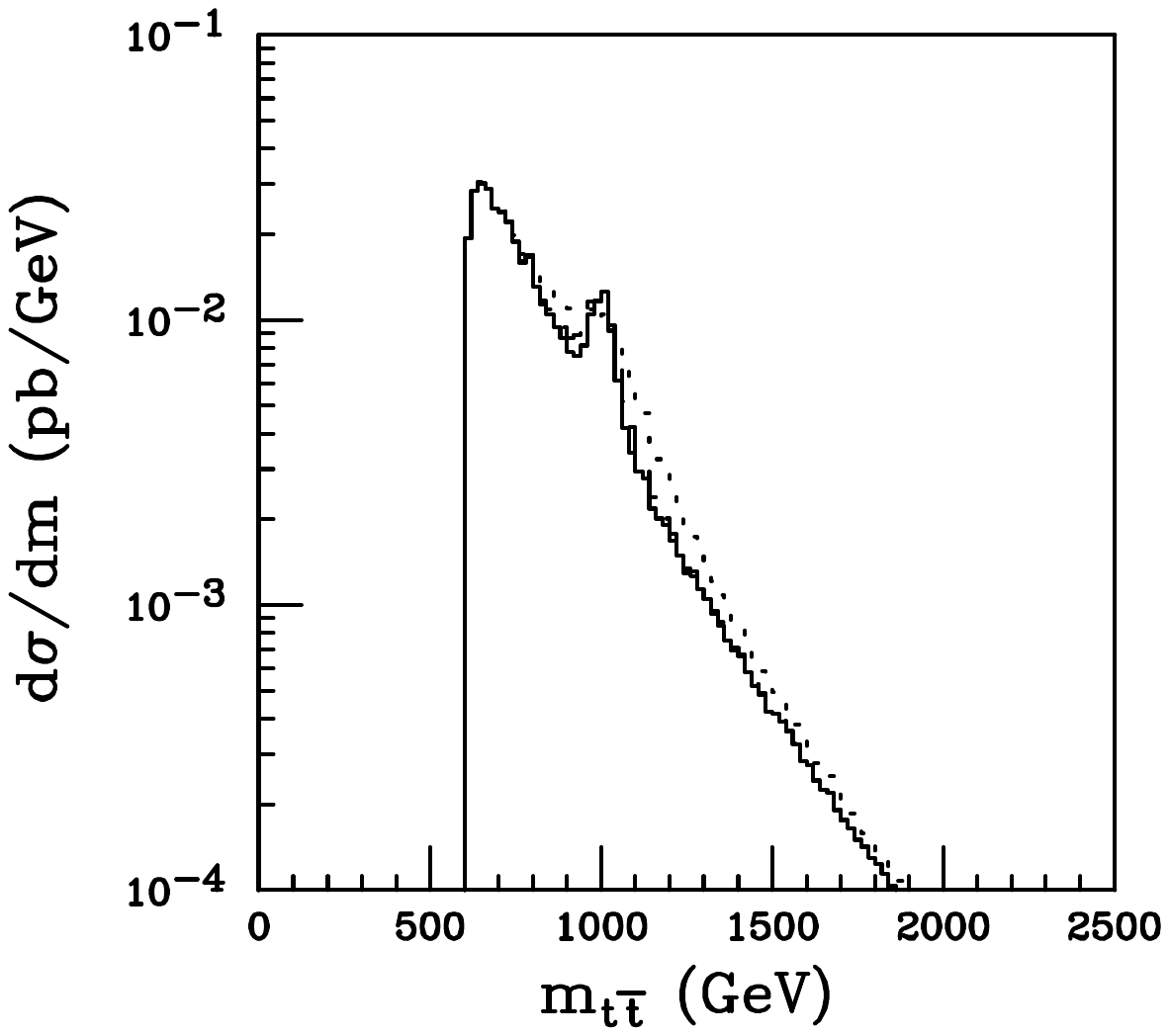}
\vspace{-1.75in}
\caption{Reconstructed $t\bar{t}$ invariant mass distributions.  The plot features a 1 TeV resonance with a total width of 2\% (solid), 5\% (dashed), and 20\% (dotted) of the resonance's mass. }
\end{figure}
\noindent
We want to search for new resonances that couple to $t\bar{t}$ in a model independent manner. We consider $t\bar{t}$ production via
\begin{eqnarray*}
g  g \to \,\,\phi \to t \bar{t},    \quad
q  \bar q \to \,\,V \to t \bar{t},   \quad
q \bar q,\  gg  \to \,\,\tilde{h} \to   t \bar{t},
\end{eqnarray*}
where $\phi$, $V$ and $\tilde{h}$ are the spin-0, spin-1, and spin-2 resonances.  We characterize the effects on the invariant mass spectrum with three parameters:  mass, total width, and the signal cross section normalization  ($\omega^2$).  The normalization $\omega=1$ defines our benchmark for the spin 0, 1 and 2 resonances.  They correspond to the SM-like Higgs boson, a $Z'$ with electroweak coupling strength and left (L) or right (R) chiral couplings to SM fermions, and the Randall-Sundrum graviton $\tilde h$ with the couplings scaled as $\Lambda^{-1}$ for $\tilde h q\bar q$, and $(\Lambda \ln(M^*_{pl}/\Lambda))^{-1}$ for $\tilde h gg$, respectively.\footnote{More precisely, we use the Feynman rules given in \cite{Han:1998sg} and include the additional warp correction factors from \cite{Davoudiasl:2000wi}.}  Numerically, we take $\Lambda = 2$ TeV.

\begin{figure}[!htb]
\centering
\label{walker_fig:parameterscan}
\includegraphics[width=0.4\textwidth,viewport=60 260 400 650]{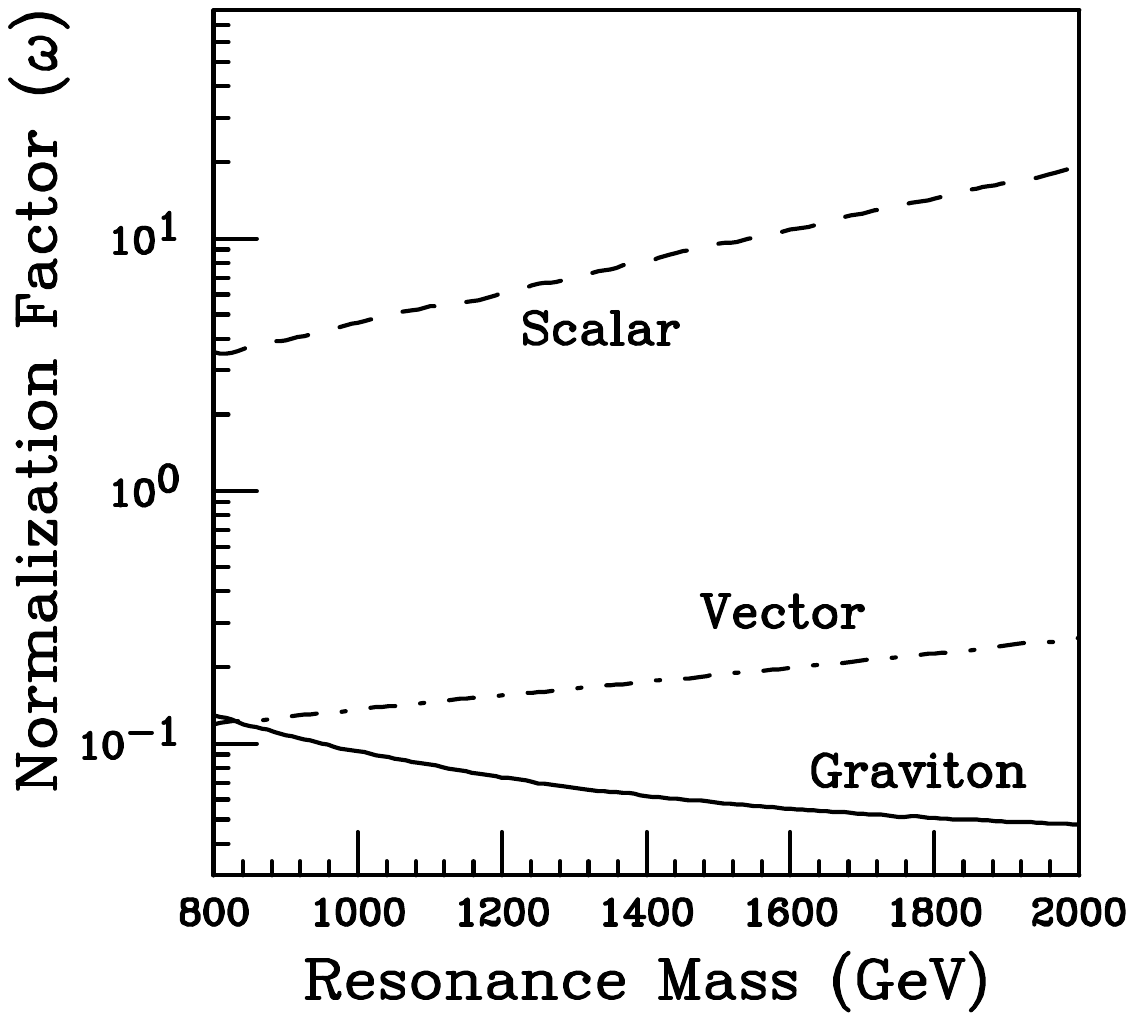}
\hspace{0.5in}
\includegraphics[width=0.4\textwidth,viewport=60 260 400 650]{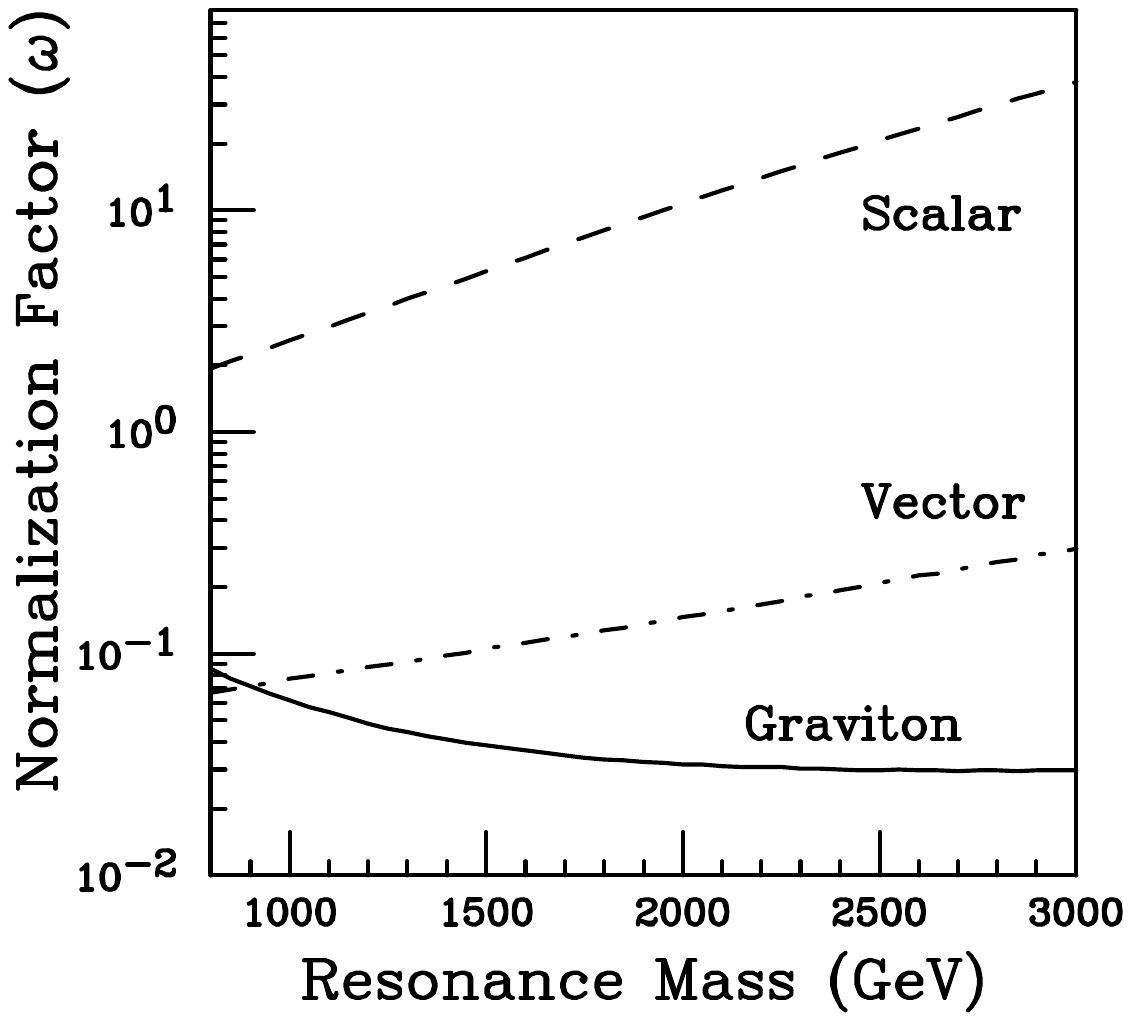}\\
\caption{Normalization factor versus the resonance mass  for the scalar (dashed) with a width-mass ratio of $20\%$, 
vector (dot-dashed) with 5\%,  and graviton (solid) 2\%, respectively.   The region above each curve represents values of $\omega$ that give 5$\sigma$  or greater statistical  significance with 10 fb$^{-1}$ (left panel) and 100 fb$^{-1}$ (right panel) integrated luminosity.}
\end{figure}
In Fig.~2 we show the reconstructed $t\bar t$ invariant mass distribution for our reconstruction scheme.  The SM $t\bar{t}$ total cross section is theoretically known beyond the leading order in QCD \cite{Laenen:1991af,Berger:1996ad,Cacciari:2003fi}.  We thus expect to have a good control of this distribution even at high invariant masses.  As for new physics,  we include the contribution of  a 1 TeV vector resonance for illustration, for $\omega_\mathrm{v} = 1$, with total widths specified in the caption of Fig~2.  

We maximize the signal observability by isolating the resonance within an invariant mass window of $\pm 100$~GeV, $\pm 30$~GeV and $\pm 25$~GeV for the scalar, vector and graviton resonance, respectively.  Given a resonance mass and total width, we can  quantify how large $\omega$ needs to be for a $5 \sigma$ discovery.  With the number of events for a signal (S) and background (B), we require $S/\sqrt{B+S} > 5$.  This translates to a bound $\omega^2 > ( 25 + 5 \sqrt{25 + 4 B}) / 2S_1$ where $S_1$ is the benchmark signal rate for $\omega=1$.  This is illustrated by Fig.~3 versus the mass for a scalar, vector and graviton resonance for total widths of 20\%, 5\%, and 2\% of its mass,  respectively, for an integrated luminosity of 10 fb$^{-1}$ (Fig.~3, left panel) and 100 fb$^{-1}$ (Fig~3, right panel).

It is of critical importance to reconstruct the c.m.~frame of the resonant particle, where the fundamental properties of the particle can be best studied.  In Fig.~4, we show the top quark angular distribution, $\cos\theta^*$, with $\theta^*$ defined as the angle  in the $t\bar t$ c.m.~frame between the top-quark momentum and the incident quark momentum, with the latter determined by the longitudinal boost direction of the c.m.~system.  Although events in the forward and backward regions are suppressed due to the stringent kinematical cuts, we still see the impressive features of the $d$-function distributions\footnote{For the definition and convention of the $d$-functions, we follow the PDG.\cite{Yao:2006px}} in Fig.~4:  a flat distribution for a scalar resonance (dashed), $d^1_{11}$ distribution for the left/right chiral couplings of a vector (dotted), and $d^2_{1\pm1}$ from $q\bar q$ (solid) and $d^2_{2\pm 1}$ from $gg$ (dot-dashed) for a spin-2 resonance.
\begin{figure}[!htb]
\centering
\vspace{-0.5in}
\includegraphics[width=0.4\textwidth,viewport=60 260 400 650]{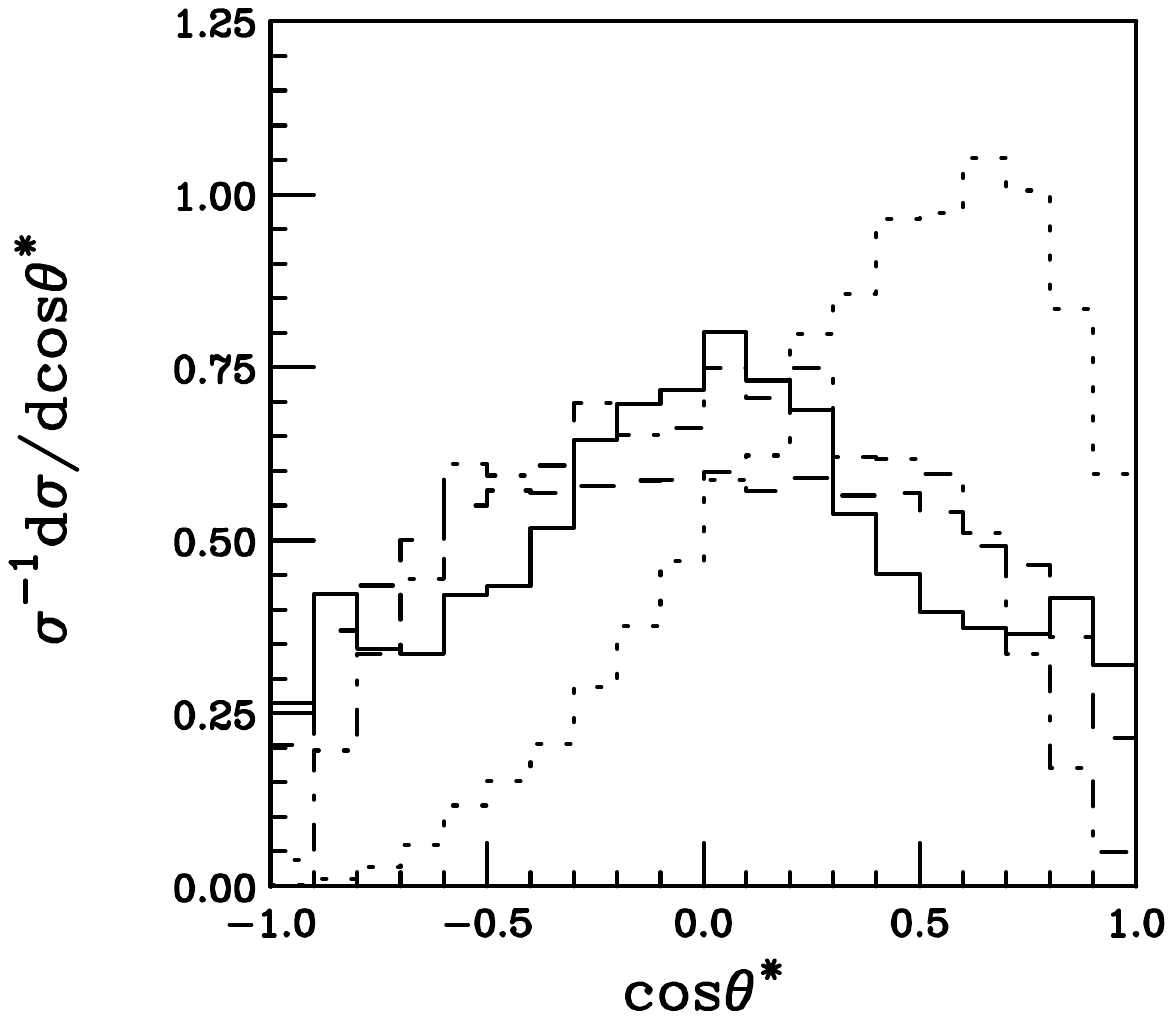}\\
\caption{Polar angular distributions for the top quark in the c.m.~frame, 
(a) Signal only by the for a scalar (dashed), a vector (dots), and a graviton from $q\bar q$ (solid)
or from $gg$ (dot-dashed).}
\label{walker_fig:angledist}
\end{figure}
\begin{figure}[!htb]
\begin{center}
\includegraphics[width=0.4\textwidth,viewport=60 260 400 650]{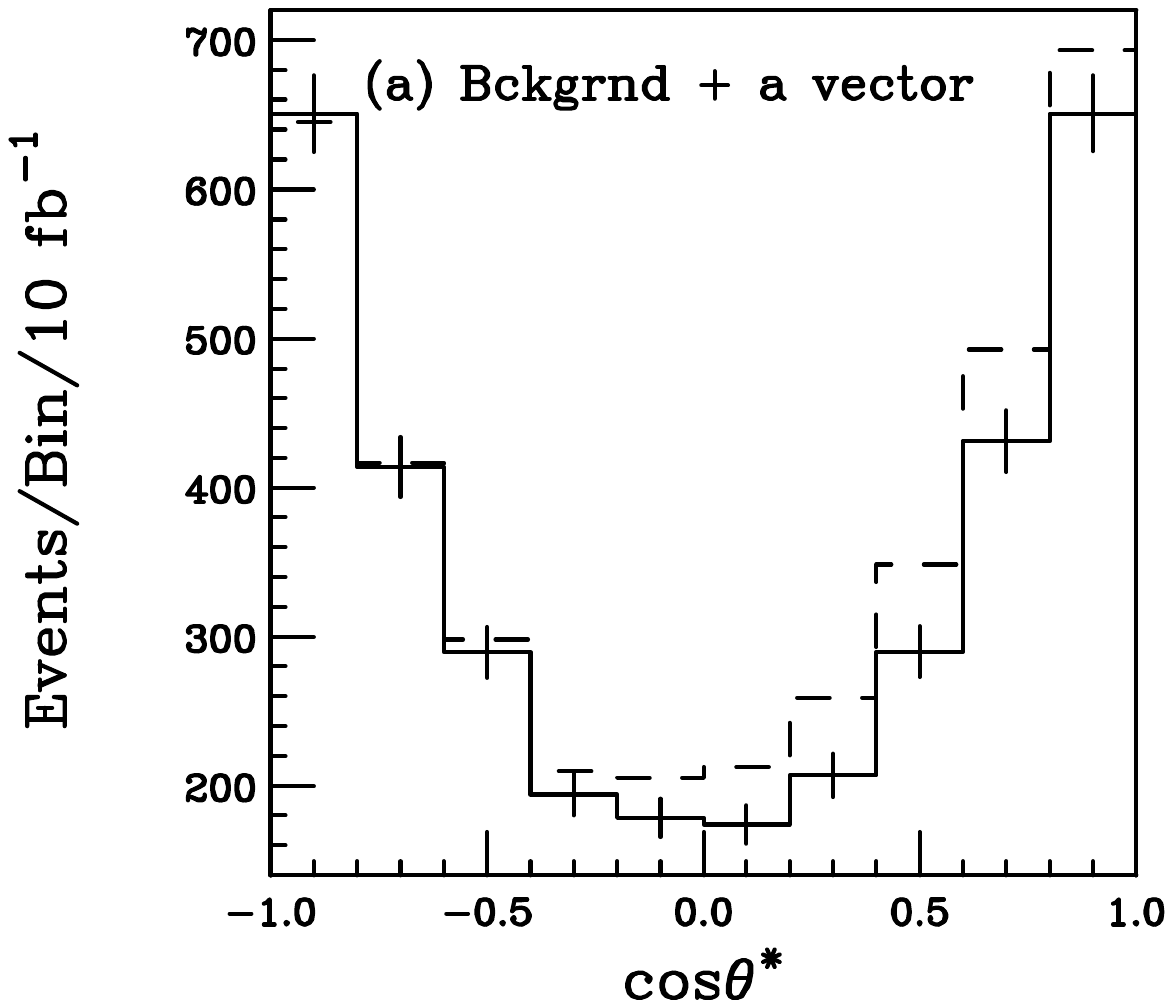}
\hspace{0.5in}
\includegraphics[width=0.4\textwidth,viewport=60 260 400 650]{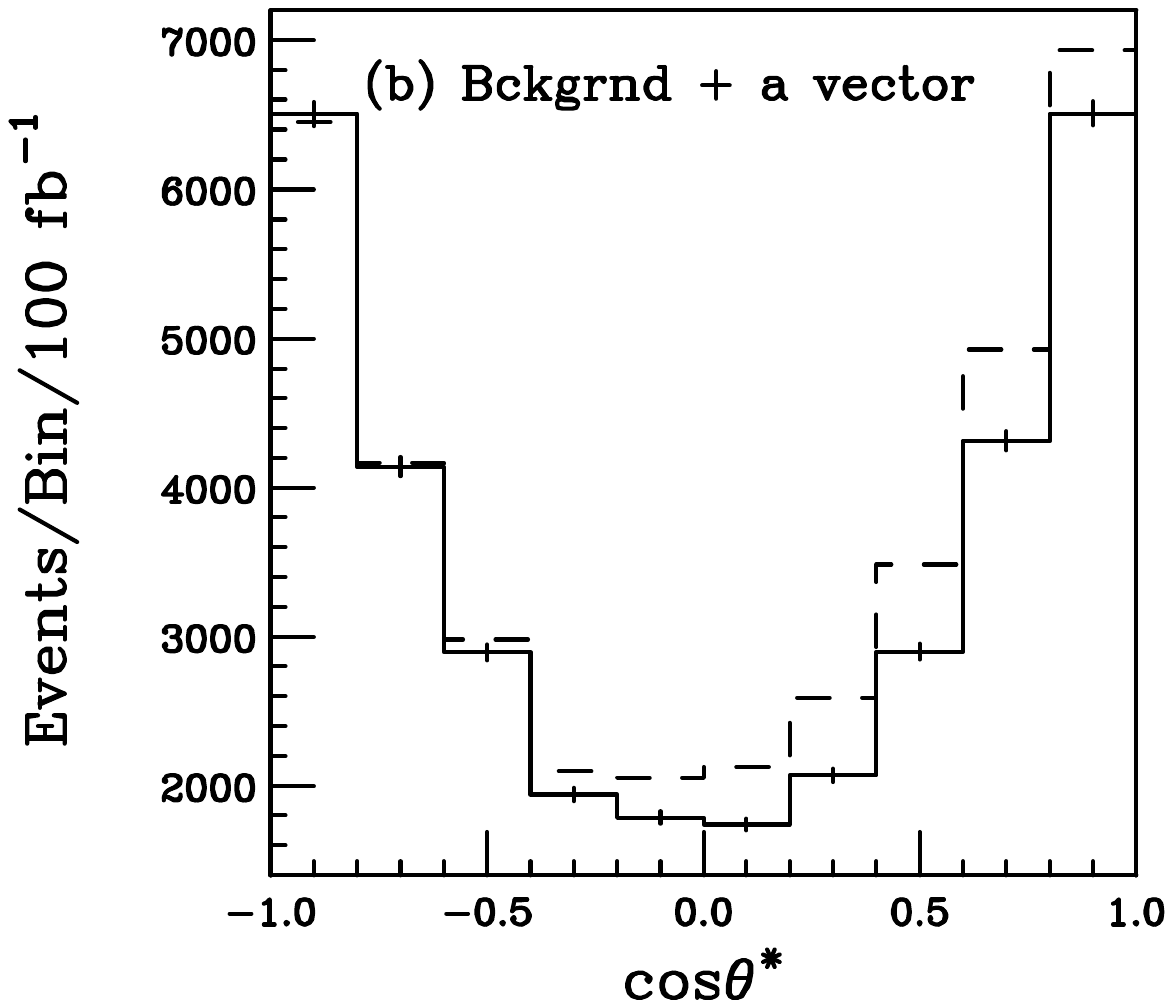}\\
\caption{Polar angular distributions for the top quark in the c.m.~frame.  Number of events for the SM $t\bar t$ background (solid) with 1$\sigma$  statistical error bars, and the background plus a vector resonance (dashed) for (a) 10 fb$^{-1}$ and (b) 100 fb$^{-1}$}
\end{center}
\end{figure}
To illustrate the statistical sensitivity for observing the characteristic distributions, we show in Fig.~5 the expected SM $t\bar t$ events (solid)  with 1$\sigma$ statistical error bars in each bin for a 10 fb$^{-1}$ and 100 fb$^{-1}$ integrated luminosity, along with a $5\sigma$ signal of a chirally coupled vector summed with the $t\bar t$  background in the resonant region (dashed). Due to the large event sample, the statistical significance is evident in the central and forward region. 
The forward-backward asymmetry in $\cos\theta^*$ can thus be constructed to probe the chiral couplings of the particle to the top quark. 
With the identification of the charged leptons, one may even form kinematical triple products to test the CP properties of the couplings 
\cite{Chang:1992tu,Valencia:2005cx}.

\section{CONCLUSIONS}

In summary, we reviewed the use for top quarks in discovering new physics in the form of integer-spin resonances.  We showed the use of angular distributions of the top in the reconstructed CM frame to reveal the spin of the resonance, and the relative contribution from the initial states $q\bar q$ or $gg$.  The forward-backward asymmetry and CP-odd variables can be constructed to further differentiate models.  Because SM top quark physics is well predicted, high invariant mass top pair production may provide an early indicator for new physics beyond the Standard Model at the LHC.

\section*{ACKNOWLEDGEMENTS}

The author thanks Vernon Barger and Tao Han for being part of a productive collaboration.

\AddToContent{D.G.E.~Walker}
\setcounter{figure}{0}
\setcounter{table}{0}
\setcounter{section}{0}
\setcounter{equation}{0}
\setcounter{footnote}{0}
\clearpage

\part[Production of KK gauge bosons at the LHC]{Production of Kaluza--Klein  excitations of gauge bosons at the LHC}

{\it A.~Djouadi, G.~Moreau and R.K.~Singh}


\begin{abstract}    
We consider the Randall--Sundrum model with fields propagating  in the bulk and
study  the production of the strongly and weakly interacting gauge boson
Kaluza--Klein excitations at the LHC. These states have  masses of order of a
few TeV and can dominantly decay into top quark pairs.   We perform a Monte
Carlo  study of the  production process  $pp  \to t\bar t$ in  which the
Kaluza--Klein excitations  are exchanged and find that the latter can lead to a
significant excess of events  with respect to the Standard Model prediction.
\end{abstract}

\section{INTRODUCTION}   
   
If the extra--dimensional model suggested by Randall and Sundrum (RS) \cite{Randall:1999ee,Gogberashvili:1999ad}
is to solve the gauge hierarchy problem of the Standard Model (SM), the  masses
of the first KK excitations of gauge bosons, $\KKgluonmkk$, must be in  the vicinity of
the TeV scale.  Direct experimental searches for KK gluon excitations at the
Tevatron   lead to $\KKgluonmkk \gtrsim 800$ GeV \cite{Guchait:2007ux, Lillie:2007ve}, while   
high--precision measurements  impose stronger bounds as the exchanges of the KK
excitations  lead to unacceptably large contributions to the electroweak
observables~\cite{Burdman:2002gr, Kim:2002kkb, Hewett:2002fe, Huber:2000fh, Huber:2001gw, Csaki:2002gy}. Nevertheless, it was shown \cite{Agashe:2003zs} that if the
SM symmetry is enhanced to the left--right  structure ${\rm SU(2)_L\! \times\! 
SU(2)_R\! \times\! U(1)_{\rm B-L}}$, with $B$ and $L$ the  baryon and lepton
numbers, electroweak precision data can be fitted while  keeping  the KK masses
down to the acceptable value  of $\KKgluonmkk \simeq 3$ TeV.

In the RS model with SM fields  in the bulk, one can generate through a  simple
geometrical mechanism the large mass hierarchies  prevailing  among SM fermions 
\cite{Gherghetta:2000qt} by placing them differently   along the extra dimension:  their
different wave  functions overlap with the Higgs boson, which remains  confined
on the  so--called TeV--brane for its mass to be protected, generate
hierarchical  patterns among the effective four--dimensional Yukawa couplings.
In this case, the KK gauge  bosons dominantly couple to heavy SM fermions as 
they are localized toward the TeV--brane (this typical feature can only be
avoided  in some particular situations \cite{Ledroit:2007ik}). In this case, the
processes involving  the third generation $b$ and $t$ quarks are those which are
expected to be   significantly affected  by the presence of the new vector
states.
 
In the extended gauge symmetry   originally proposed in Ref.~\cite{Agashe:2003zs},  a
${\rm U(1)_{\rm B-L}}$ group was included; other ${\rm U(1)_{X}}$ groups  with
different  fermion charges can be considered \cite{Agashe:2006at,Djouadi:2006rk}.  An 
important motivation would be  that  with specific charges of the new Abelian
group and  for specific fermion localizations, the three standard discrepancy
between the forward--backward  asymmetry $A_{FB}^b$ in $Z\to b\bar b$ decays 
measured  at LEP  and the SM prediction \cite{Djouadi:1989uk} is  naturally resolved, 
while keeping all the other observables in agreement with data.  There are also
constraints on the $t,b$ couplings from flavour changing neutral current
processes,  but those can be satisfied for $M_{KK}$ values around the TeV scale
\cite{Huber:2003tu, Moreau:2006np, Agashe:2004ay, Agashe:2004cp, Ligeti:2006pm, Agashe:2005hk, Agashe:2006iy, Agashe:2006wa}.

At the LHC, one can produce directly the KK excitations of the gluons and the
electroweak gauge  bosons with masses in the multi--TeV range \cite{Djouadi:2007eg,Davoudiasl:2000wi, Lillie:2007yh, Agashe:2006hk, Agashe:2007ki}. 
In this note,  we study the main production mechanism, the Drell--Yan  channel 
$pp \to \KKgluonvkk=g^{(1)},$ $\gamma^{(1)},$ $Z^{(1)},$  $Z^{'(1)}$, and  perform a Monte
Carlo simulation of the process $q\bar q \to t \bar t$.  We base our analysis on
the framework which  resolves the $A_{FB}^b$ anomaly \cite{Djouadi:2006rk}, but the
results that we obtain can  be easily generalized to other scenarios.   
Including only the dominant QCD backgrounds, we show that for a set of 
characteristic points of the parameter space,  the exchange of KK gauge bosons 
can lead to visible deviations with respect to the SM production  rates.

\section{PHYSICAL FRAMEWORK} 
 
We consider the RS model in which SM fields propagate along the  extra spatial
dimension, like gravity, but the Higgs boson remains confined on the 
TeV--brane.   In the RS scenario, the warped extra   dimension is compactified
over a $S^{1}/\mathbb{Z}_{2}$   orbifold.   While the gravity scale on the
Planck--brane is $M_{P}= 2.44\times  10^{18}$ GeV, the effective scale on the
TeV--brane, $M_{\star}=e^{-\pi  kR_{c}} M_{P}$, is suppressed by a warp factor
which depends on the curvature  radius of the anti--de Sitter space $1/k$ and
the compactification radius $R_c$.  The product $k R_{c} \simeq 11$ leads to
$M_{\star}\!=\!{\cal O}(1)$ TeV, thus  addressing the gauge hierarchy problem.  
The values for the fermion masses are dictated by their wave function
localization.  In  order to control these localizations, the five--dimensional
fermion fields  $\Psi_{i}$, with $i=1,2,3$ being the generation index, are
usually coupled to  distinct masses $m_{i}$ in the fundamental theory. If
$m_{i}= {\rm sign}(y)  c_{i} k$, where $y$ parameterizes the fifth dimension and
$c_{i}$ are  dimensionless parameters, the fields decompose as $\Psi _{i}(x^{\mu
},y)=  \sum_{n=0}^{\infty }\psi_{i}^{(n)}(x^{\mu }) f_{n}^{i}(y)$, where $n$
labels  the tower of KK excitations and $f_{0}^{i}(y)=e^{(2-c_{i})k|y|} /
N_{0}^{i}$  with $N_{0}^{i}$ being a normalization factor. Hence, as $c_i$
increases, the  wave function $f_{0}^{i}(y)$ tends to approach the Planck--brane
at $y=0$.\smallskip

We consider the scenario `RSb' developed in Ref.~\cite{Djouadi:2006rk} where, in order
to protect the electroweak observables against large deviations and,   at the
same time, resolve the anomaly in $A_{FB}^b$, the electroweak gauge symmetry is
enhanced to  ${\rm SU(2)_L \times SU(2)_R \times U(1)_{X}}$ with some specific
fermion  representations/charges under the group gauge.  The usual symmetry of
the SM is recovered after the breaking  of both ${\rm SU(2)_{R}}$ and ${\rm
U(1)_{X}}$ on the Planck--brane, with  possibly a small breaking of the ${\rm
SU(2)_R}$ group in the bulk. Note the  appearance of a new $Z'$ boson (but
without a zero--mode) which is a  superposition of the state $\widetilde W^3$
associated to the ${\rm SU(2)_R}$  group and $\widetilde B$ associated to the
${\rm U(1)_{X}}$ factor; the  orthogonal state is the SM hypercharge $B$ boson. 
Solving the gauge hierarchy problem forces  the masses of the first KK
excitations of the SM gauge bosons, $\KKgluonmkk=M_{  \gamma^{(1)}}=M_{g^{(1)}} \simeq
M_{Z^{(1)}} \simeq M_{Z^{\prime (1)}}$, to be ${\cal O}$(TeV)  and we will fix
the common mass value to  $\KKgluonmkk= 3$ TeV in the present study.

The light SM fermions [leptons and first/second generation quarks]  are
characterized by $c_{\rm light}>0.5$, $c$ being the parameter which determines 
the fermion localization to cope with high--precision data. The large value of
the top quark mass requires $c_{t_R}<0.5$ and  $c_{Q^3_L}<0.  5$, with
$c_{Q^3_L}\!=\!c_{t_L}\!=\!c_{b_L}$ [as the states $b_L$  and $t_L$ belong to
the same ${\rm SU(2)_L}$ multiplet] so that the top and  bottom quarks have to
be treated separately. In the framework of Ref.~\cite{Djouadi:2006rk}, the precision 
data in the $b$ sector, that is $A_{FB}^b$ and $R_b$, are correctly 
reproduced with $\KKgluonmkk=3$ TeV and e.g. 
$g_{Z'}=0.3 \sqrt{4 \pi}$ for the coupling of the new $Z'$ boson. Then, the best 
fit of $R_b$ and $A_{FB}^b(\sqrt{s})$ (which also lead to the correct range for 
the top and bottom quark masses) is obtained for $c_{Q^3_L} \simeq 0.36$ and 
$c_{b_R} \simeq 0.135$. One gets the values $Q(c_{b_R}) \simeq 3.04$ and 
$Q'(c_{b_R})\simeq 3.19$, where $Q(c)$ ($Q'(c)$) is the ratio of the 
four--dimensional effective coupling between the $g^{(1)}/\gamma^{(1)}/Z^{(1)}$ 
($Z^{\prime (1)}$) boson and the SM fermions, over the coupling of the  
gluon/photon/$Z$ (would be $Z'$) boson zero--mode. A small $Q$ charge holds
for the light
fermions, $Q(c_{\rm light}) \simeq -0.2$, as well as zero $Z'$ charge, $Q'(c_{\rm
light})=0$. This means that the couplings of the KK excitations of the gluon,
photon and $Z$ boson are an order of magnitude smaller to  light fermions
compared to the couplings to top and bottom quarks, while the KK excitations of
the $Z'$ boson do not couple to  these  light fermions at all.

From the previous discussion, one concludes that in the chosen scenario, the sum
of the branching ratios  BR$(\KKgluonvkk  \to t\bar t)$ and BR$(\KKgluonvkk  \to  b\bar b)$,
with $\KKgluonvkk =\gamma^{(1)}$, $Z^{(1)}$, $Z'^{(1)}$ and $g^{(1)}$,  is close to
unity which means that the KK excitations decay almost exclusively into the
heavy $t,b$ quarks and that  little room is left for decays into light quarks
and  leptons. In the case of $g^{(1)}$ for instance, one has  BR$(g^{(1)}  \to
t\bar t)=0.69$ and BR$(g^{(1)}  \to  b\bar b)=30$. Because of the large
couplings to fermions, the total decay widths of the KK excitations $\KKgluonvkk $
[which grow proportionally to the mass $M_{\KKgluonvkk } \KKgluongsim 3$ TeV] are very large. 
For instance, the decay width of $g^{(1)}$ is of the order of a few hundred GeV
and is between 10\% and 20\%  of its mass; the KK state can be thus considered
as a relatively narrow resonance.

\section{PRODUCTION CROSS SECTIONS AT THE LHC}

 The most straightforward way to produce the KK excitations of the gauge  bosons
$\KKgluonvkk $  at the LHC is via the Drell--Yan process, $pp \to q\bar q \to \KKgluonvkk  
\to Q\bar Q \, , \, Q=t,b$ with $\KKgluonvkk $ subsequently decaying into top and 
bottom quarks.  The relatively  small couplings of the initial quarks $q \equiv
u,d,s,c$ to  $\KKgluonvkk $ lead to  smaller production rates compared to, for
instance, the production of $Z'$  bosons from GUT's. Since the KK gauge bosons
have different  couplings to left-- and right--handed fermions, one expects the
produced $t/b$   quarks to be polarized and to have a forward--backward
asymmetry. Because $\KKgluonvkk$ have substantial total decay widths, the narrow width
approximation in which the production and  decay processes are factorized is not
sufficient and one needs to consider the virtual exchange of $\KKgluonvkk$ in which the
total width is included in a Breit--Wigner form, together with the exchange  of
the zero modes; the full interference   should be taken into account. For $b
\bar b$ production, the subprocesses are also initiated by bottom  partons and
one should also consider the channel in which $\KKgluonvkk$ are exchanged in the
$t$--channel.

The signal  $q\bar q \to \KKgluonvkk  \to  Q\bar Q$ and  the main SM background $q \bar
q \to  Q\bar Q$  and $gg  \to  Q\bar Q$ [which  gives a much more substantial
contribution] have to be considered simultaneously.   For the signal reaction,
we have calculated the matrix element squared of the  process $pp \to Q\bar Q$
with {\it polarized} final state quarks and incorporated  the exchange
[including the $t$--channel contributions] of all the SM gauge bosons as well as
their KK excitations and  those of the  $Z'$ boson; we use the CTEQ5M1 set of
parton  distributions  with the factorization and a renormalization  scales set
to the invariant mass of the $Q\bar Q$ system, $\mu_F=\mu_R= m_{Q\bar Q}$.\smallskip  
The significance ${\cal S}$ of the signal in the RS model can be then defined 
as ${\cal S}_{\cal L}  = (\sigma^{\rm RS+SM} - \sigma^{\rm SM})/ (  \sigma^{\rm
SM} )^{1/2} \times  {\cal L}^{1/2}$  with ${\cal L}$  the total  LHC 
luminosity. 

In order to enhance the signal, which is peaked at $m_{Q\bar Q}$ and to suppress
the continuum background, one needs to select events  near the KK resonance. We
thus impose the cut $|m_{Q\bar Q}-\KKgluonmkk| \le \Gamma_{\KKgluonvkk}$.   To further reduce
the backgrounds, we  also impose the following cuts on the transverse momenta of
the two final jets and their rapidity $ p_T^{Q,\bar Q} \ge 200\, {\rm GeV}$ and
$|\eta_{Q,\bar Q}| \le 2$,  as in the signal, the $p_T$ of the jets is peaked
close to $\frac12 M_{\KKgluonvkk }$ and the production is central, while in the
background, the jets are peaked in the forward and backward directions and the
bulk of the cross section is for low $p_T$ jets. These cuts can certainly be
optimized but in this preliminary and simple parton--level investigation, we
will simply compare the signal and the main corresponding physical background to
determine if, grossly, one can have a detectable signal. More efficient and 
realistic cuts and detection efficiencies will not be discussed.

The invariant mass distribution d$\sigma$/d$m_{t\bar t}$ of the process $pp \to
t\bar t$ is shown in the left--hand side of Fig.~1 for the chosen scenario with 
$\KKgluonmkk =3$ TeV, including the cuts mentioned above. As can be seen, there is a
substantial contribution of the KK excitations to the invariant mass
distribution, in particular around the peak $m_{t\bar t} \sim 3$ TeV. At higher
$m_{t\bar t}$, the KK contribution becomes small, while at lower $m_{t\bar t}$, 
it is significant even for $m_{t\bar t} \sim 2$ TeV; only for $m_{t\bar t} \KKgluonlsim
1$ TeV the KK contribution becomes negligible. Outside the KK mass peak, the RS
effect is mostly due to the interference between the excited state and SM 
contributions; this interference is positive below and negative above the peak. 
The dominant contribution compared to the SM case is by far due to the exchange
of the excitation of $g^{(1)}$ which has the largest (QCD versus EW) couplings
to the initial state partons; the contributions of $\gamma^{(1)}$ and $Z^{(1)}$
increase the peak only slightly. In turn, $Z'^{(1)}$ has a negligible impact as
it does not couple to the initial light quarks and the parton density of the
heavier bottom quark in the proton is small.

The significance of the excess of events in the RS scenarios when all KK
excitations are  included is large, ${\cal S}_{10}^{\rm RS} \sim 30$ for a 
moderate luminosity  and ${\cal S}_{100}^{\rm RS}\sim 95$ for a high
luminosity.   Since the excess over the SM background is mainly due to 
$g^{(1)}$ exchange, the significance ${\cal S}_{100}^{g^{(1)}} \sim 90$, is
almost the same as when the full signal is considered, ${\cal S}_{100}^{\rm
RS}$. In the case where only the first KK excitation of the photon or the $Z$
boson  is considered (assuming that the peaks can be disentangled, which is not
obvious), the significance  is much smaller ${\cal S}_{100}^{\gamma^{(1)}} \sim
5$ and ${\cal S}_{100}^{Z^{(1)}} \sim 7$.  This is a mere consequence of the
fact that the EW  $\gamma^{(1)}, Z^{(1)}$ couplings are much smaller than the 
$g^{(1)}$ QCD couplings,  leading to limited production cross sections. The
smaller rates are, however,  partly compensated by the smaller total decay
widths . 

\begin{figure}[!ht]
\begin{center}
\epsfig{file=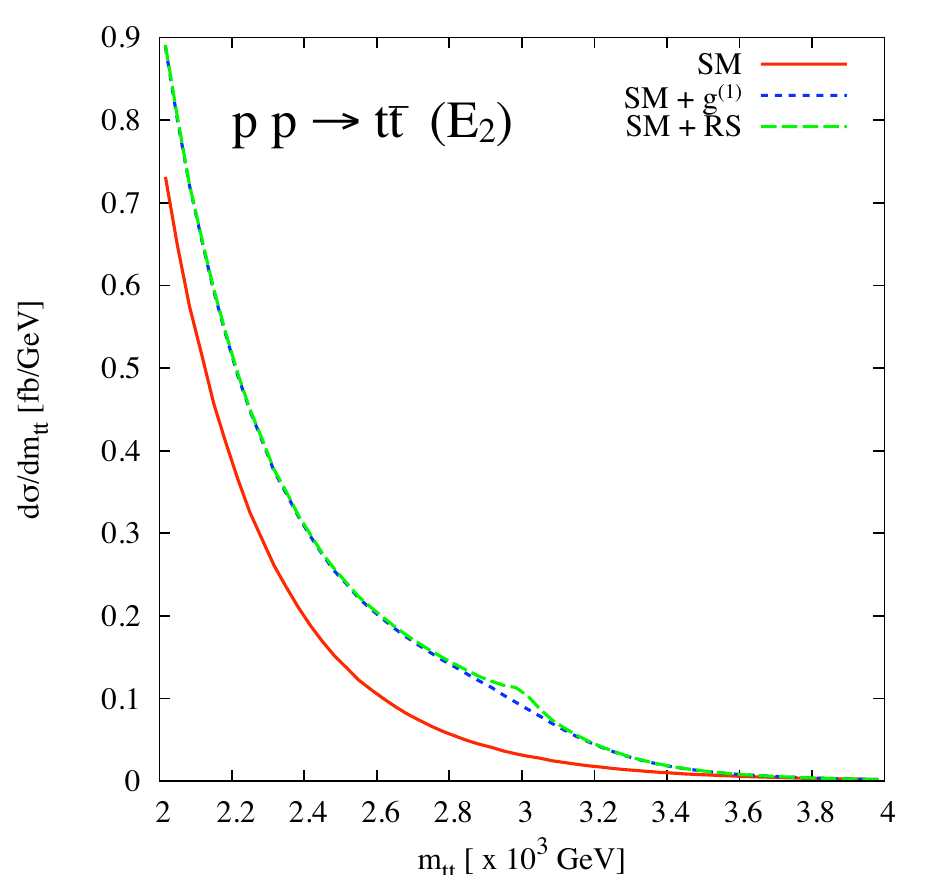,width=7.0cm}
\epsfig{file=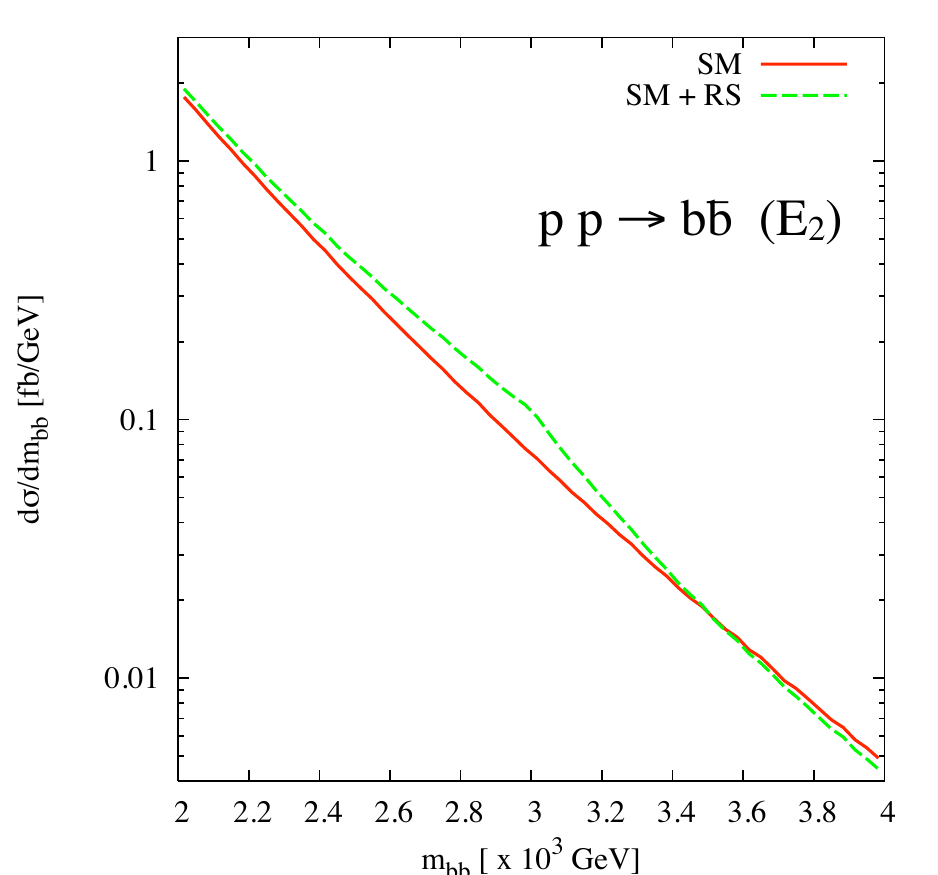,width=7.0cm}\vspace*{5mm}
\end{center}
\vspace*{-5mm}
\caption{The invariant mass distribution of the cross sections for the 
$pp\to t\bar t$ (left) and $b\bar b$ (right) processes for the scenario E2 
discussed in the text including the RS signals and the SM backgrounds with 
the relevant cuts } 
\vspace*{-5mm}
\end{figure}

The discussion for the  $pp \to b\bar b$ process is quite similar to the one of
$pp \to t\bar t$ except for the fact that the small and not peaking $t$--channel
$b\bar b \to b\bar b$ contribution. The invariant mass distribution d$\sigma$/d$
m_{b \bar b}$ for bottom quark pair production is shown in the right--hand side
of Fig.~1 for $m_{b\bar b}$ between 2 and 4 TeV with the relevant cuts. Here, we
simply show the SM background and the signal excess  in the case where the
contributions of all KK excitations are simultaneously included; again, this
excess is largely dominated by the exchange of $g^{(1)}$. The signals are less 
striking than in the $pp \to t\bar t$ case, the main reason being that
BR($g^{(1)} \to b\bar b)$ is smaller than BR($g^{(1)} \to t\bar t)$. However, 
the significances are large enough, ${\cal S}_{10}^{\rm RS}\sim 15$ and  ${\cal
S}_{100}^{\rm RS}\sim 50$, to allow for detection if no other background or
experimental problem is included. Note that when combining the $pp \to t\bar t$
and $pp \to b\bar b$ processes, one  would in principle be able to have access
to the couplings of the KK  states (at least $g ^{(1)}$ to $t,b$ quarks.  

Note that  we have also calculated QCD higher order processes to the heavy quark
production involving exclusively KK gauge couplings to these heavy quarks, which
are the favored couplings in  the present framework. First, we have studied the
one loop level reaction $gg \to \KKgluonvkk \to t\bar t$  where the anomalous $gg \KKgluonvkk
$ vertex was regulated via the St\"uckelberg mixing term.  Secondly, we have
analyzed the four--body reactions $pp \to t\bar t b\bar b$,  $t\bar t t\bar t$,
$b\bar b b\bar b$ and three--body reactions $gb \to b t \bar t$,  $b b\bar b$.
Our result there is that the RS effects will be difficult to test at the LHC, 
due to the small amplitudes involved relatively to the SM background. 

Each of the KK excitations of the gauge bosons, $g^{(1)}, \gamma^{(1)}$ and
$Z^{(1)}$, has a different coupling to the right-- and left--handed top quarks 
[which are themselves different from the SM ones]. These couplings  appear in
the forward--backward asymmetry as well as in the polarization of the  produced
top quarks. While the enhancement in the production rate due to a single KK
gauge boson is proportional to the sum of squared couplings,  the polarization
and forward--backward asymmetries are proportional to their difference. Thus, a
combined measurement of the cross section together with asymmetries would
determine the couplings of  the vector boson $\KKgluonvkk $. However, since the process
is mediated by the exchange of several KK gauge bosons with differing right--
and left--handed couplings, it  will be challenging to measure precisely these
couplings for each of the KK  excitations. This is particularly true as the
major contribution to the total rate for $\sigma(pp \to Q\bar Q)$ is coming from
$g^{(1)}$, as the contribution from the electroweak excitations $\gamma^{(1)}$,
$Z^{(1)}$ and $Z^{\prime (1)}$  is relatively small. Nevertheless, the
measurement of the polarization and forward--backward asymmetries for top
quarks  can be instrumental in establishing the presence of parity violating KK
gauge bosons.
 
\section{CONCLUSION} 
 
 We have considered the version of the RS model with SM fields in the bulk,
concentrating on quark geometrical localizations and gauge quantum numbers that
allow to solve the    the LEP anomaly on the forward--backward asymmetry
$A_{FB}^b$.  We have studied the main potential effects from KK excitations of
gauge bosons at LHC, which come from their exchange in the pair production of
third generation quarks constituting new contributions with respect to the pure
SM ones. Based on the computation of cross sections and estimations of the SM
backgrounds,  it has been shown that simple kinematical cuts permit to detect
the excesses of  events in $t \bar t$ production due to the KK resonances
(essentially the KK gluon)  for some characteristic points of parameter space
and $\KKgluonmkk \KKgluongsim 3$ TeV (to satisfy EW constraints).  For the case of $b \bar b$
production,    large significances are also obtained.  Furthermore, relevant top
quark polarization and angular asymmetries were computed  and turn out to
provide a good test of the chiral structure for top quark couplings  to KK gauge
modes. In this simple parton--level analysis, we did not take into account the
non--leading and non--physical backgrounds as well as as detector efficiencies,
etc. An implementation of this process in a Monte Carlo generator, interfaced
with a detector response simulator, have started at this Les Houches workshop
\cite{KKgluon_atlas}. 

\section*{ACKNOWLEDGEMENTS}
This work is supported by the  Indo--French CEFIPRA
project no.\,3004-B and by the French ANR project PHYS@COL\&COS.

\AddToContent{A.~Djouadi, G.~Moreau and R.K.~Singh}
\setcounter{figure}{0}
\setcounter{table}{0}
\setcounter{section}{0}
\setcounter{equation}{0}
\setcounter{footnote}{0}
\clearpage

\part{LHC studies of the left-right twin Higgs model}

{\it X.~Miao, S.~Su, K.~Black, L.~March, S.~Gonzalez de la Hoz, E.~Ros and M.~Vos}\\

\begin{abstract}
The twin Higgs mechanism has recently been proposed to solve the little hierarchy problem. We study the LHC collider phenomenology of the left-right twin Higgs model. We focus on the cascade decay of the heavy top partner, with a signature of multiple $b$ jets $+$ lepton $+$ missing energy. We also present the results for the decays of heavy gauge bosons: $W_H\rightarrow t \bar{b}$, $W_H\rightarrow \phi^{\pm} \phi^0 $ and $Z_H\rightarrow e^+ e^-$.
\end{abstract}

\section{INTRODUCTION}
Naturalness requires the stabilization of the Higgs mass against large radiative corrections. The scale of new physics needs to be around the electroweak scale to avoid the fine-tuning of the Higgs potential. On the other hand, electroweak precision measurements push the cutoff scale for new physics to be likely above 5$-$10 TeV. This conflict in the energy scale of new physics is the so-called `little hierarchy' problem. 

Recently, the twin Higgs mechanism has been proposed as a solution to the little hierarchy problem \cite{Chacko:2005pe, Chacko:2005vw, Foot:2006ru, Chacko:2005un,Falkowski:2006qq,Chang:2006ra}. Higgses emerge as pseudo-Goldstone bosons once a global symmetry is spontaneously broken. Gauge and Yukawa interactions that break the global symmetry give masses to the Higgses, with the leading order being quadratically divergent. Once an additional discrete symmetry (twin symmetry) is imposed, the leading one-loop quadratically divergent terms respect the global symmetry. Thus they do not contribute to the Higgs masses. The Higgs masses do obtain one-loop logarithmically divergent contributions, resulting in masses around the electroweak scale when the cutoff is around 5$-$10 TeV.

The twin Higgs mechanism can be implemented in left-right models with the discrete symmetry being identified with left-right symmetry\cite{Chacko:2005un}. Many new particles which have order of one interaction strength with the Standard Model (SM) sector are predicted and rich phenomenology is expected at the Large Hadron Collider (LHC). 

This paper is organized as follows. 
In Sec.~2, we describe the left-right twin Higgs (LRTH) model briefly. In Sec.~3, we present the results on the LHC studies of the cascade decay of the heavy top partner. In Sec.~4, we present the results on two channels for the heavy counterpart of the charged gauge boson: $W_H\rightarrow t \bar{b}$ and $W_H\rightarrow \phi^{\pm} \phi^0 $ and the decay $Z_H\rightarrow e^+ e^-$ of the neutral heavy gauge boson. Finally, in Sec.5, our conclusions are presented.

\section{THE LEFT-RIGHT TWIN HIGGS MODEL}
\label{lrthsec:model}

The LRTH model was first proposed in Ref.~\cite{Chacko:2005un} and the details of the model as well as the Feynman rules, particle spectrum, and collider phenomenology have been studied in Ref.~\cite{Goh:2006wj, Goh:2006pt}. Here we briefly introduce the model and focus our attention on the heavy top partner and heavy gauge bosons. 

In the LRTH model proposed in Ref.~\cite{Chacko:2005un},  the global symmetry is U(4)$\times$U(4), with the diagonal subgroup of ${\rm SU}(2)_L \times  {\rm SU}(2)_R \times  {\rm U}(1)_{B -L}$ gauged. The twin symmetry which is required to control the quadratic divergences of the Higgs mass is identified with the left-right symmetry which interchanges L and R.  For the gauge couplings $g_{2L}$ and $g_{2R}$ of ${\rm SU}(2)_L$ and ${\rm SU}(2)_R$, the left-right symmetry implies that $g_{2L} = g_{2R} = g_2$. 

Two Higgs fields, $H = (H_L, H_R )$ and $\hat{H} = (\hat{H}_L, \hat{H}_R )$\footnote{The introduction of $\hat{H}$ and the requirements that it couples to gauge boson sector only and has a vev $\hat{f}\gg f$ are due to the electroweak precision constraints.},  are introduced and each transforms as
({\bf 4}, {\bf 1}) and ({\bf 1}, {\bf 4}) respectively under the global symmetry.  $H_{L,R}$ ($\hat{H}_{L,R}$ ) are two component objects which are charged under ${\rm SU}(2)_L$ and ${\rm SU}(2)_R$, respectively. Each Higgs obtains a vacuum expectation value  (vev): $\langle H  \rangle = (0, 0, 0, f )$, $\langle \hat{H} \rangle = (0, 0, 0, \hat{f} )$ with $\hat{f} \gg   f$, breaking one of the U(4) to U(3), respectively.  The Higgs vevs also break ${\rm SU}(2)_R \times  {\rm U}(1)_{B -L}$ down to the SM ${\rm U}(1)_Y$.

Below the cutoff scale $\Lambda$, the effective theory can be described by a nonlinear sigma model of the 14 Goldstone bosons. After spontaneous global symmetry breaking by $f$ and $\hat{f}$, three Goldstone bosons are eaten by the massive gauge bosons $W_H^\pm$ and $Z_H$, and become their longitudinal components. The masses of the heavy gauge bosons are given approximately by 
\begin{equation}
m_{W_H} \sim \frac{e}{\sqrt{2}\sin\theta_W} \hat{f},\ \ \ 
m_{Z_H}\sim \frac{\cos\theta_W}{\sqrt{\cos 2\theta_W}} m_{W_H}.
\end{equation}
After the SM electroweak symmetry breaking, three additional Goldstone bosons are eaten by the SM gauge bosons $W^\pm$  and $Z$. With certain re-parametrizations of the fields, we are left with four Higgses that couple to both the fermion sector and the gauge boson sector: one neutral pseudoscalar $\phi^0$, a pair of charged scalars $\phi^\pm$, and the SM physical Higgs $h_{SM}$.  In addition, there is an ${\rm SU}(2)_L$ doublet $\hat{h}$ that  couples to the gauge boson sector only.  It could contain a good dark matter candidate\cite{Dolle:2007ce}.

The fermion sector of the LRTH model is similar to that of the SM, with the right handed quarks ($u_R, d_R$) and leptons ($l_R, \nu_R$) form fundamental representations of ${\rm SU}(2)_R$. In order to give the top quark a mass of the order of the electroweak scale, a pair of vector-like quarks $q_L$ and $q_R$ are introduced. The mass eigenstates, which contain one SM-like top $t$ and a heavy top $T$, are mixtures of the gauge eigenstates. Their masses are given by  
\begin{equation}
m_t \sim   y v /\sqrt{2},\ \ \  m_T\sim y f. 
\end{equation}
The mixing angle $\alpha_L$ and $\alpha_R$ are controlled by the mass mixing term 
$M \bar{q}_L q_R$.  The collider phenomenology differs significantly for a very small value of $M\leq 1$ GeV or for not so small values of $M$. In our analysis below, we assume that $M = 150$ GeV.

\begin{table}
\begin{center}
\begin{tabular}{c|cccccc|ccc} \hline
&\multicolumn{6}{|c|}{mass spectrum (GeV)}&
\multicolumn{3}{|c}{LHC production cross sections (fb)} \\ \hline
$f$ (GeV)&$m_T$&$m_{W_H}$&$m_{Z_H}$&$m_{\phi^0}$&$m_{\phi^{\pm}}$
&$m_{h_{SM}}$&$\sigma(Tj)$&$\sigma(W_H)$&$\sigma(Z_H)$ \\ \hline
555 & 567  & 1250 & 1495 & 109 & 184 & 173   & 3768 & 13200 & 2290 \\
600 & 614  & 1393 & 1665 & 111 & 199 &173    & 2409 & 9598  & 1418\\
800 & 812  & 2000 & 2407 & 116 & 260 & 175   & 501  & 1570  & 224 \\ 
1000 &1007 & 2605 & 3115 & 118 & 321 &175    & 134  & 473   & 49.5 \\
1134 &1144 & 3000 & 3589 & 119 & 362 & 176   & 60   & 179   & 19.5   \\
1500 &1504 & 4053 & 4846 & 120 & 476 &179    & 7.3  & 28    & 2.3 \\ \hline
\end{tabular}
\caption{Mass spectrum and the LHC production cross sections for $T j$, $W_H$ and $Z_H$ in the LRTH model for several benchmark values of $f$. The other parameters in the model are chosen as $\Lambda = 4\pi f$, $M = 150$ GeV, and $\mu_r = $50 GeV.}
\label{lrthtable:spectrum}
\end{center}
\end{table}

The new particles in the LRTH model are: heavy gauge bosons $Z_H$, $W_H$, heavy top quark $T$, neutral Higgs $\phi^0$, a pair of charged Higgses $\phi^\pm$, and an ${\rm SU}(2)_L$ complex Higgs doublet $\hat{h}$.  The free parameters in the model that are relevant for the collider studies are ($f$, $\Lambda$, $M$, $\mu_r$), where $f$ is the vev for Higgs boson\footnote{The vev $\hat{f}$ can be determined by minimizing the Coleman-Weinberg potential for the SM Higgs and requiring that  the SM Higgs obtains an electroweak symmetry breaking vev of 246 GeV.} $ H $, $\Lambda$ is the cutoff scale, $M$ is the top quark vector singlet mass mixing parameter,  and $\mu_r$ is the mass parameter for $\phi^0$. Table.~\ref{lrthtable:spectrum} shows the masses for the new particles in the model for several  benchmark values of  $f$.

\section{CASCADE DECAY OF HEAVY TOP PARTNER}
\label{lrthsec:heavytop} 

The dominant production mode for the heavy top $T$ at the LHC is a single heavy top quark produced in associated with a jet (most likely a $b$ jet). For a heavy top mass of 500$-$1500 GeV, the cross section is in the range of $7\times 10^3 \ {\rm fb} - 10 \ {\rm fb}$, (predominantly via the on-shell decay of $W_H$). The LHC cross section for $T j$ associated production for several benchmark points are given in Table.~\ref{lrthtable:spectrum}. The cross section for QCD pair production is about a factor of five smaller due to the heavy top quark mass. 

The dominant decay channel for the heavy top is $T\rightarrow \phi^+  + b$.   Considering the subsequent decay of 
\begin{equation}
\phi^+ \rightarrow t\bar{b},\ \ \ t \rightarrow W^+ b \rightarrow l^+ \nu b, 
\end{equation}
and taking into account the additional energetic jet (most likely a $b$-jet) that accompanies $T$ from single heavy top production, the signal is typically four $b$-jets + one charged lepton ($e$ or $\mu$) + missing $E_T$. The SM backgrounds dominantly come from $t\bar{t}$, $W j j j j$, $Wcjjj$, $Wccjj$, and $W bbj j$.  Two independent studies have been performed to identify this process at the LHC. Their procedures, cuts and results are  summarized below. 

The study by Miao and Su used MadGraph \cite{Alwall:2007st} and BRIDGE \cite{Meade:2007js} to generate the signal processes.   Background $t\bar{t}$ events are generated using Madgraph while $W+$ jets events are generated using Alpgen \cite{Mangano:2002ea}. Both the signal and background events are passed through PYTHIA \cite{Sjostrand:2006za} and PGS4 \cite{PGS} for hadronization and detector simulations. For $f$=600 GeV, they have adopted the following cuts:
\begin{itemize}
\item{} At least three jets with $p_T >$ 30 GeV, with leading jet has $p_T >$ 400 GeV, and the second leading jet has $p_T >$ 250 GeV.
\item{} One energetic lepton ($e$ or $\mu$) with $p_T >$ 30 GeV and $|\eta | < $2.5.
\item{} Missing $E_T >$  15 GeV.
\item{} Reconstructed transverse top mass (from $b l \nu$ ) within $m_t\pm 20$ GeV.
\item{}Reconstructed transverse $\phi^\pm$ mass within $m_{\phi^\pm} \pm $ 30 GeV.
\item{} Reconstructed transverse heavy top mass within $m_T \pm$ 50 GeV.
\item{} At least one  jet is tagged as $b$ jet.
\end{itemize}
Similar cuts are imposed for $f=$ 1000 GeV and $f=$ 1500 GeV.  The results for the signal and background cross sections before and after the cuts are summarized in Table.~\ref{lrthtable:heavytop}.   For LHC integrated luminosity of 30 ${\rm fb}^{-1}$, the significance for heavy top discovery is more than 10 $\sigma$ for $f=$ 600 and 1000 GeV.  For $f$=1500 GeV, more luminosity is required to reach a significant discovery.  The significance level could be further increased with higher luminosity or when both detectors at the LHC are taken into account.

\begin{table}[h]
\begin{center}
\begin{tabular}{ccccc|ccccc} \hline
\multicolumn{5}{c|}{Miao/Su}&\multicolumn{4}{|c}{Vos et al.} \\  \hline
$f$ (GeV) &$\sigma_S^{\rm before}$&$\sigma_S^{\rm after}$&$\sigma_B^{\rm after}$
&$S/\sqrt{B}$& $f$ (GeV) & $\sigma_S^{\rm before}$ & $\sigma_S^{\rm after}$&$\sigma_B^{\rm after}$
&$S/\sqrt{B}$  \\ \hline
600 &419&4.08&4.64&10.4& 555 & 547 & 8.7 & 5.0 & 21 \\   
1000 &26.5&1.23&0.15&17.1 & 800 & 75 & 4.0 & 0.16 & 55 \\ 
1500 &1.54&0.21&0.24&2.4 & 1134 & 9.5 & 0.89 & 0.09  & 16 \\ \hline 
\end{tabular}
\caption{ Results for the cascade decay of heavy top $T$ from two studies. The LHC integrated luminosity is taken to be $L = 30 \ {\rm fb}^{-1}$. In the analysis of Miao/Su all contributions to the $ T $ cross-sections are included, while in the analysis of Vos et al. only the $ pp \rightarrow W_H \rightarrow T \bar{b} $ is considered.}
\label{lrthtable:heavytop}
\end{center}
\end{table}

In the study by Vos et al. the signal production processes and some key decay modes are implemented in Pythia~\cite{Sjostrand:2006za}. The response of the ATLAS detector is simulated using the ATLAS fast simulation package ATLFAST~\cite{atlfast}. The decay is reconstructed step by step, starting from the $W$ decay into lepton neutrino. At each step kinematical constraints - on the mass and transverse momentum of the reconstructed particles - are applied.  The values of the cuts employed in the selection vary with model parameter $f$ that governs the masses of the involved particles. In the following, the selection criteria for $ f = $ 555 GeV (i.e. a reconstruction aimed at a 1.25 TeV $ W_H $-boson) are given. 

\begin{itemize} 
\item{} A lepton ($ e^{\pm} $, $ \mu^{\pm} $) with transverse momentum greater than 25 GeV. The presence of the lepton ensures that the events can be triggered efficiently.
\item{} A minimum missing transverse energy of 25 GeV. A $W$ candidate is reconstructed from the missing transverse energy and the lepton momentum using the collinear approximation (i.e. assuming $ p_z^{\nu} = p_z^l $ ). 
\item{} The $W$ candidate is combined with all jets with $ 25 < p_T (j) < 200 $ GeV. The combination that gives the best match with the top mass is selected. If none of the combinations yields a mass $ m_t < $ 250 GeV, the event is discarded. 
\item{} A second jet with $ 25 < p_T < 100 $ GeV is added to reconstruct the charged Higgs boson $ \phi^{\pm} $. Again, events with a reconstructed $ \phi^{\pm} $ mass greater than 250 GeV are discarded. 
\item{} A third jet with $ p_T (j) > $ 100 GeV is required to reconstruct the heavy top quark $ T $. The $ T $-candidate is required to satisfy the following constraints: $ m_T < $ 700 GeV and $ p_T (T) > $ 150 GeV. This latter cut, that
takes advantage of the Jacobean peak in the signal, is particularly useful to  reduce the dominant $ t \bar{t} $ background. 
\item{} Finally, a fourth jet with $ p_T (j) > $ 150 GeV is used to form the $ W_H $ candidate. As the dominant $ T $ production process is through $ W_H \rightarrow T \bar{b} $, the explicit reconstruction of the $ W_H $ is instrumental in reducing the background. 
\end{itemize}
The width of the reconstructed mass peaks is dominated by the experimental resolution for jet energy and missing transverse energy. For the lightest $ W_H $ boson of 1.25 TeV, the Gaussian width of the peak is approximately 100 GeV, large compared to the natural width of 30 GeV. The total efficiency for the kinematical reconstruction is 9 \%. 

The $ t \bar{t} $ and $ W + $ jets backgrounds are generated using Pythia. The former is found to be the dominant contribution to the background after the kinematical reconstruction is performed. 

At this stage, a significance (when estimated as $ S / \sqrt{B} $ ) well above 5 is found for all mass points considered. The shape of signal and background mass distributions are, however, quite similar. Experimental methods to normalize the background, like the sideband estimate, cannot readily be employed. Taking into account the large uncertainty on the background production cross section and selection efficiency, the significance of the signal is greatly reduced. 

To further reduce the background, the lifetime signature of the multiple $b$-jets is used. The signal topology contains very high $ p_T $ $b$-jets. To correctly describe the ATLAS performance, jets with transverse momenta up to 1 TeV have been studied using a detailed GEANT4 simulation of the detector response~\cite{atlas_csc_btagging}. Several detector effects are found to lead to a quite significant degradation of the performance for the highest $ p_T $ bins. In this study, a parameterization of the full simulation results are used. 

To discriminate the signal topology, with four $b$-jets, against the dominant $ t \bar{t} $ background a four jet likelihood is constructed by summing the tag likelihood of the four leading jets. The four-jet likelihood allows for a significant reduction of the background. For the studied mass points, the signficance ($ S  / \sqrt{B}$) improves slightly or remains unaltered, but the $ S/B $ is greatly improved, thus rendering the analysis much more robust against uncertainties in the number of background events. The results  - listed in Table~\ref{lrthtable:heavytop} - indicate that the discovery potential reaches a $ W_H $ mass of 3 TeV, even for a relatively small integrated luminosity (30 $\rm fb^{-1}$). For larger $ W_H $ masses, the discovery potential is rapidly degraded by the small absolute number of signal events.

A closer look into the dominant $ t \bar{t} $ background for different versions of the generator yields significantly different results for the high-$p_T $ tail of the top quark spectrum, and therefore of the number of background events that pass the kinematical reconstruction and selection cuts. Therefore, the ATLAS study is being repeated using the MC@NLO~\cite{Frixione:2002ik,Frixione:2003ei} generator. These results will be published at a later time~\cite{atlas_twinhiggs}.

The heavy top could also decays into $bW$, $t h_{SM}$, $tZ$, and $t\phi^0$. The branching ratios for those channels quick drop for larger $m_T$ and smaller $M$. Therefore, we will not discuss those channels further here.

\section{HEAVY GAUGE BOSONS}
\label{lrthsec:heavygaugeboson}

The dominant production channels for heavy gauge bosons at the LHC are the Drell-Yan processes: $pp\rightarrow  W_H X$ and $pp \rightarrow Z_H X$. The production cross sections for several benchmark points are given in Table.~\ref{lrthtable:spectrum}. 

The dominant decay modes for $ W_H $ are into two jets, with a branching ratio of about 60\%. Such modes suffer from the overwhelming QCD di-jets background for large $p_T$ jets. $W_H$ could also decay into a heavy top plus a $b$-jet, with a branching ratio of about 20\%$-$30\%. This is  the main channel for $T j$ associated production as discussed earlier in Sec.~3. 

The decay $ W_H \rightarrow \phi^{\pm} \phi^{0} $ has a branching fraction of 3\%. The subsequent decay of the charged and neutral Higgs bosons $ \phi^{\pm} \rightarrow t b $ and $ \phi^0  \rightarrow b \bar{b} $ have large branching ratios. Thus, the same final state with four $b$-jets and a lepton and neutrino as for the decay $ W_H \rightarrow T \bar{b} $ is obtained. A kinematical reconstruction of the decay chain along the lines of the previous analysis described in Sec.~3 is quite successful for small values of $ f $. For larger $ W_H $ mass the boost of the relatively light Higgs bosons increasingly leads to difficulties in the reconstruction of the jets. The reduction in reconstruction efficiency, in combination with the sharply dropping production cross section, limits the discovery range of this signature to $ W_H $ masses below 1.5 $-$ 2 TeV.

The branching ratio for  $W_H\rightarrow  t\bar{b}$  is of the order of 4\%. Due to the much reduced branching ratio this channel is a priori less promising than in the Littlest Higgs model, studied by ATLAS~\cite{hadronic_littlehiggs}. Recent work~\cite{atlas_twinhiggs} investigates the reduction of the dominant $ t \bar{t} $ background by exploiting the presence of additional jets or leptons in the background sample. Preliminary results indicate that isolation of this signal for $ W_H $ masses up to 1.5 $-$ 2 TeV may well be possible. 

Although the dominant decay modes of $Z_H$ are into dijets, the discovery modes for $Z_H$ would be $Z_H\rightarrow l^+l^-$ (with a branching ratio of 2.5\% for $e^+ e^-$, $\mu^+ \mu^-$ and  $\tau^+\tau^-$ individually). The natural width of the heavy $ Z_H $ ranges from 25 to 75 GeV for a $ Z_H $ boson from 1.2 to 3.6 TeV. The di-lepton modes $e^+ e^-$ and $\mu^+ \mu^-$ therefore provide clean signatures, which can be separated from the SM background by studying the invariant dilepton mass distribution. 

A study of the di-lepton signature has been performed by ATLAS. The excellent momentum resolution for high $ p_T $ electrons yields an error in the invariant mass of the parent boson that is inferior to the natural width throughout the studied mass range. For the di-muon final state the invariant mass resolution is limited by the resolution of the combined measurement of inner tracker and muon spectrometer. Therefore, this first exploration only considers the di-electron signature. The discovery potential of this channel is evaluated using a classical analysis counting the (small) number of signal and background events in a narrow mass window. This signature may give rise to very early discovery, with only a few inverse $\rm fb$ of data, provided the $ Z_H $ mass is less than 2.5 TeV. With an integrated luminosity of 75 $\rm fb^{-1} $ the discovery reach is extended up to 3.5 TeV. Potentially, the LHC experiments are sensitive to much larger masses through the interference of the heavy neutral gauge boson with the Standard Model Z and photon. 

$Z_H$ could also decay into $t\bar{t}$ final states, with a branching ratio of about 2 $-$ 5\%. Searches of $t\bar{t}$ resonance have been studied in Refs.~\cite{unknown:1999fr, AguilarSaavedra:2005pv}. The reach for $Z_H \rightarrow t\bar{t}$ at the LHC is very limited, due to the small decay branching ratio into the $t\bar{t}$ final states. 

\section{CONCLUSIONS}
\label{lrthsec:conclusion}
The twin Higgs mechanism provides an alternative method to solve the little hierarchy problem. The LRTH model has rich collider phenomenology that could be studied at the upcoming LHC. In this paper, we presented LHC studies on the searches of the heavy top partner and the heavy gauge bosons in the LRTH models. For the heavy top quark partner, a significance level of over 3 $\sigma$ could be reached for almost the entire interesting parameter regions of the LRTH model with an integrated luminosity of 30 ${\rm fb}^{-1}$. An independent study in the ATLAS fast simulation framework has used a parameterization of the full simulation results for high $ p_T $ $b$-tagging. This study finds a significant signal up to $ f =$ 1100 GeV with an integrated luminosity of 30 $\rm fb^{-1} $. Several $ W_H $ decays have been studied. The decays $W_H \rightarrow \phi^{\pm} \phi^0 $ and $W_H \rightarrow t\bar{b}$ give rise to a final state with a lepton and a neutrino and four, respectively two, $b$-jets. The search for both channels is expected to yield a significant signal only for small values (of the order of 1 TeV) of the $ W_H $ mass. $Z_H \rightarrow  e^+ e^-$ is likely to be the discovery channel for the LRTH model. For relatively light $ Z_H $ (up to 2.5 TeV) this signature could be observed with only a few inverse $\rm fb$ of data. The discovery potential for the LRTH model at the LHC is very promising and further studies towards the identification of the twin Higgs mechanism and the distinction between various electroweak models are currently under investigation. 

\section*{ACKNOWLEDGMENTS}
XM and SS  would like to thank Johan Alwall and Matt Reece for helps on using MadGraph and BRIDGE package. XM and SS would also like to thank Michelangelo Mangano for help using Alpgen. MV would like to thank Manouk Rijpstra and Marcel Vreeswijk for their analysis of the $ W_H \rightarrow t b $ channel.
SS and MV would like to thank the organizers for the Les Houches 2007 program: "Physics at TeV Colliders". We had a lot of fruitful discussions during the workshop. XM and SS are supported under U.S. Department of Energy contract\# DE-FG02-04ER-41298.

\AddToContent{X.~Miao, S.~Su, K.~Black, L.~March, S.~Gonzalez de la Hoz, E.~Ros and M.~Vos}
\setcounter{figure}{0}
\setcounter{table}{0}
\setcounter{section}{0}
\setcounter{equation}{0}
\setcounter{footnote}{0}
\clearpage


\part{$W_L W_L$ Scattering}

{\it A.~Delgado, C.~Grojean, E.~Maina and R.~Rosenfeld}

\begin{abstract}
In this report we intend first to review the main models
where strong dynamics are responsible for EWSB.
An overview of tests of new models through the production of new 
resonant states at the LHC is presented.
We illustrate how different models can be related by
looking at two general models with resonances.
\end{abstract}

\section{INTRODUCTION}

One of the main goals of the LHC is to find the mechanism
responsible for electroweak symmetry breaking (EWSB) at the TeV 
scale. In the SM of electroweak interactions, this is accomplished
by postulating the existence of a complex scalar Higgs field with a potential crafted
in such a way as to result in the breaking of $SU(2)_L \times U(1)_Y$ into the residual 
electromagnetic $U(1)$ symmetry. 
The couplings of the Higgs field with gauge bosons and fermions 
generate the masses we observe.
This rosy picture has its thorns: a scalar sector is unstable under
radiative corrections and the Higgs sector has to be understood as an
effective theory valid up to some energy scale $\Lambda$.  

One of the ideas that has been intensely studied in the past is that there
are new strong interactions responsible for EWSB and
the Higgs sector effective description will break down at a scale $\Lambda \simeq $ TeV.
Some of these ideas have resurfaced recently in the form
of various models.
The longitudinal components of the gauge bosons have their origin
in the EWSB mechanism and hence provide a window to study these models.

The charge of our working group is to review the large amount of work
that has been done in the past and to identify improvements that can be made
in the light of these novel models of strongly interacting
EWSB sector and of the new
tools available to study possible signatures at the LHC.
In particular, we want to concentrate on the more 
model-independent features of these models.

There are three main themes we would like to address in this report:\\
$\bullet$ $WW$ scattering and unitarity; \\
$\bullet$ Drell-Yan versus WW fusion as discovery processes at the LHC;\\
$\bullet$ constraints from electroweak precision measurements.

\section{MODELS}

Heavy, relatively wide resonances are the hallmark of strong interactions.
They may or may not be at the LHC reach for direct detection, depending on 
their masses. Hence it is convenient to classify different models of strong interactions 
in terms of having light or heavy resonances, where light means within LHC reach.
It is also convenient to separate out the scalar from the vector resonances, since they 
are expected to have a very different phenomenology.
An incomplete list of models is classified in Table \ref{models}.
This classification is of course arbitrary. In fact, many of these models can actually
move among different classes for a different set of parameters.
For an alternative classification of models, see {\it e.g.} Cheng~\cite{Cheng:2007bu}.

\vspace{0.5cm}
\begin{table}[h]
\flushleft
\begin{tabular}{|l||l|l|}
\hline 
 & No (light) vector resonances & Light vector resonances  \\
\hline \hline
  No (light) Higgs &  Chiral lagrangians~\cite{Appelquist:1980vg, Longhitano:1980iz}  &  LSTC~\cite{Eichten:1984eu}, Higgsless~\cite{Csaki:2003dt, Csaki:2003zu} \\
&& (D)BESS~\cite{Casalbuoni:1985kq}\\ 
\hline Light Higgs  &SM, SILH~\cite{Giudice:2007fh}   & Warped/Composite~\cite{Contino:2006nn}\\
&& Holographic~\cite{Agashe:2004rs}, Little~\cite{ArkaniHamed:2001nc, ArkaniHamed:2002qy}, \\ 
&& Gaugephobic~\cite{Cacciapaglia:2006mz}, Twin~\cite{Chacko:2005pe, Chacko:2005un}\\
&& LDBESS~\cite{Casalbuoni:1995qt}, Gauge-Higgs~\cite{Manton:1979kb, Antoniadis:2001cv, Csaki:2002ur, Scrucca:2003ra} \\
\hline
\end{tabular} 
\caption{Attempt to classify different models of strong dynamics}
\label{models}
\end{table}
\vspace{0.5cm}

Many of these non-SUSY models share the same low-energy phenomenology. Models based on
extra dimensions, such as Gauge-Higgs unification in either flat or warped extra dimension,  
may resemble conventional 4D models with new particles such as Little Higgs models since only the
lightest KK modes will be accessible at the LHC. In fact, they can both be described by a so-called
three-site moose model~\cite{Cheng:2006ht}. Also, warped Higgs models and Higgsless models can be smoothly
interpolated in the so-called gaugephobic Higgs model~\cite{Cacciapaglia:2006mz}.
In the latter Sections we will compare 2 general models with new scalar and vector particles, namely LDBESS and
Composite/Warped Higgs models. 

We will focus primarily on the optimistic view that new resonances are within the LHC reach and
construct a general model that captures some general features of the phenomenology of 
these resonances. In particular, we will not consider here the more model dependent possibility
of composite fermions, since we will be interested mostly in the interaction of these resonances to light
fermions, which is important for Drell-Yan production, for example.
If the resonances are heavy and outside the LHC reach one will have to test their indirect effects in anomalous gauge boson couplings
coming from higher dimensional operators.

We will denote generically by $V$ these new vector resonances. They will in general mix with the SM gauge bosons. 
They may or may not have direct couplings to light fermions. In the latter case, the couplings will arise solely from the mixings with
gauge bosons.
The models will then be mainly characterized by:\\
$\bullet$ the mass eigenstates $M_W$ and $M_V$ of the gauge bosons and resonances;\\
$\bullet$  the couplings $g_{VWW}$ and $g_{hWW}$ between gauge bosons and resonances and the Higgs 
boson arising from the mixings.

In the SM the $W_LW_L$ scattering amplitudes (here $W$ generically denotes massive gauge boson, the $W^\pm$ as well as the $Z$) are unitarized by the Higgs boson contribution. If the Higgs coupling $g_{hWW}$ is modified, unitarity is not exactly restored by the Higgs boson alone and the contribution from other resonances may be relevant.
In models with no Higgs at all, the contribution arising from a tower of resonances can help to either unitarize or
delay unitarity violation to higher energies.

\subsection{Electroweak constraints}
We will study the conditions imposed on these models by the electroweak observables. This goes as far as the early nineties when the oblique parameters $S,T$ and 
$U$~\cite{Peskin:1991sw} and the $\epsilon$'s~\cite{Altarelli:1990zd} where introduced, it was shown that minimal technicolor had difficulties accommodating the results from the different accelerators running at that time, specially SLC and LEP. 

The effects of new physics can be classified as ``oblique" or universal with respect to the fermionic currents and those which affect differently to different flavors. The second ones tend to be more restrictive because they imply FCNC and they are also more model dependent so we will only describe here in some detail the first ones because, in general any model predicting FCNC will have great difficulties accommodating actual data . 

These universal effects can only affect the self energies of the electroweak bosons since anything that affect equally all fermionic currents can be move into the kinetic term for the gauge bosons via field redefinition. Those self energies can be expanded in powers of the momentum, following the notation of~\cite{Barbieri:2004qk}:
\begin{equation}
\Pi_{AB}(p^2)=\Pi_{AB}(0)+p^2 \Pi'_{AB}(0)+\frac{1}{2} (p^2)^2\Pi''_{AB}(0)+\dots
\end{equation}
\noindent
where $AB={W^+,W^-,W_3W_3,BB,BW_3}$. There are twelve coefficients, three of them are reabsorbed into $g$, $g'$ and $v$ and other two are fixed to ensure that the photon is massless and couples to the correct current. We are left with seven parameters, three of them correspond to the usual $\epsilon$'s or $S,T$ and $U$ and four new ones only relevant for higher order corrections. 

Let as focus to the most important ones that are $S$ or $\epsilon_3$ and $T$ or $\epsilon_1$, their definition are as follows:
\begin{eqnarray}
S,\epsilon_3&\sim&\Pi'_{W_3 B}(0)\nonumber\\
T,\epsilon_1&\sim&\Pi_{W_3W_3}(0)-\Pi_{W^+W^-}(0)
\end{eqnarray}
The latter has a direct connection to the $\rho$-parameter which measures the breaking of the custodial symmetry. In the SM that breaking is very weak leading to very small deviations from 1 in the $\rho$-parameter. The way to avoid large contributions to the $T$ parameter is to ensure that  the strong interaction describing the EW breaking sector has some kind of custodial symmetry embedded on it~\cite{Agashe:2003zs}. This, in turn, implies that the resonances fall into representation of the custodial group, and also that there are vectorial right-handed resonances.

The second big issue is that technicolor or Higgsless theories tend to predict very large contributions to the $S$ parameter. In that sense having a light scalar resonance, i.e. Higgs, is preferred from the electroweak constraints. We can rewrite the $S$ parameter in the following suggesting way:
\begin{equation}
S=4\pi\left[\frac{F^2_V}{M^2_V}-\frac{F^2_A}{M^2_A}\right]
\end{equation}
\noindent
where $F_i$ and $M_i$ are the decay constants and masses for vectorial (V)  and axial (A) resonances, in minimal technicolor inspired by QCD the value is $0.3$, too large compare to the experimental data, so a way to suppress big contribution to $S$ is to have vector and axial resonances almost degenerate which in turn will mean that experimentally these resonances should be discovered with similar masses.

For theories where such resonances do not follow that rule, like in Ref.~\cite{Cacciapaglia:2004rb}, another way to cancel those large contribution to the $S$ parameter is to suppose that light fermions do have suppressed couplings to heavy resonances, this will  make Vector Boson Fusion (VBF) the favorite channel to produced these resonances. 

To summarize, the electroweak data impose two main conditions into theories with strongly coupled sectors:
\begin{enumerate}
\item  Some implementation of custodial symmetry is needed which means extra right-handed states in the spectrum.
\item  A particular choice of the spectrum of the resonances or cancelation with fermionic contributions is granted to reduce dangerous contributions to $S$. This will also have particular implications in the spectrum.
\end{enumerate}

Before finishing this discussion let us mention that the third family may potentially deviate from this picture since getting a big enough top mass can be in conflict with the requirement that the bottom quark couples to the $Z$ the same way the rest of the quarks do but since we are just dealing with $W_L W_L$ scattering those potential problems will not affect this scattering process.

\subsection{Strong interactions from mixings}

In this subsection we show the origin of a possible strong interaction between the vector resonance composite fields and the
SM gauge fields. Let us consider the decay of a vector resonance to two gauge bosons. Due to the gauge structure of the 
vertices before mixing one would have:
\begin{eqnarray}
&&{\cal M}^{\lambda \lambda_1 \lambda_2} 
\left( V^\lambda_\sigma (p) \rightarrow  
W^{\lambda_1}_\nu (q_1)  W^{\lambda_2}_\mu (q_2) \right)  =
g_{VWW}  \left[ (p-q_1)_\mu g_{\sigma \nu} + \right. \\ \nonumber
&& \left.  + (q_1-q_2)_{\sigma} g_{\mu \nu} + (q_2 - p)_{\nu} g_{\sigma \mu} \right] 
\varepsilon^{\lambda,\sigma}(p) \; \varepsilon^{\lambda_1,\nu}(q_1) \; \varepsilon^{\lambda_2,\mu}(q_2) 
\end{eqnarray}
Using the rest frame of the decaying vector resonance and the usual representation for the polarization vector one finds
for the decay amplitude into longitudinal polarizations of the gauge bosons:
\begin{equation}
{\cal M}^{\lambda L L} = g_{VWW} \frac{M_V^2}{2 M_W^2} (q_1 - q_2) \cdot \varepsilon^{\lambda}.
\end{equation} 

Comparing this result with the one following from a simple QCD inspired effective lagrangian describing the couplings of $\rho$'s to pions:
\begin{equation}
{\cal L} \sim g_{\rho \pi \pi} \rho_\mu \left( \pi \stackrel{\leftrightarrow}{\partial^\mu} \pi \right)
\end{equation} 
one immediately obtains the correspondence
\begin{equation}
g_{\rho \pi \pi} = g_{VWW} \frac{M_V^2}{2 M_W^2}
\end{equation}
and strong interactions can arise from mixing when
\begin{equation}
 g_{VWW} \frac{M_V^2}{2 M_W^2} \WWgsim 1
\end{equation}

\subsection{Drell-Yan versus Vector Boson Fusion}

In general there are two competing processes for the production of resonances at hadron colliders: Drell-Yan (DY) and vector boson fusion (VBF) processes. Let us for the moment focus on the production of a generic neutral resonance $R$ at the LHC which couples to first generation fermions and to longitudinally polarized $W_L$'s.
The DY cross section can be written as:
\begin{equation}
\sigma^{DY} (pp \rightarrow W_L W_L X) = \sum_{i,j} \;
\int_{\tau_{min}}^1 \; d\tau \; \left( dL/d\tau \right)_{pp/q_i q_j}
\hat{\sigma}(q_i q_j \rightarrow R \rightarrow W_L W_L) \;,
\end{equation} 
where the sum is over quarks flavors that can produce the pair of $W_L$'s, $\tau = M_{WW}^2/s$, $\tau_{min} = 4 M_W^2/s$,
$\hat{\sigma}$ is the partonic cross section at a center-of-mass energy of $\sqrt{\tau s}$ and the partonic luminosity
function is given in terms of the parton distribution function by:
\begin{equation}
\left( dL/d\tau \right)_{pp/q_i q_j} = \int_{\tau}^1 \; \frac{dx}{x} \; \left[ u^{(a)}(x) \bar{u}^{(b)}(\tau/x) +  d^{(a)}(x) \bar{d}^{(b)}(\tau/x)
+ (a) \leftrightarrow (b) \right]
\end{equation}
In the narrow-width approximation of the Breit-Wigner form for the partonic cross section one has
\begin{equation}
\hat{\sigma}(q_i q_j \rightarrow R \rightarrow W_L W_L) = 
\frac{8 \pi^2}{M_R^2} \frac{\Gamma(R \rightarrow q_i q_j) \Gamma(R \rightarrow W_L W_L)}{M_R \Gamma(R)}
\tau \; \delta(\tau - M_R^2/s)
\end{equation}
which results in
\begin{equation}
\sigma^{DY} (pp \rightarrow W_L W_L X)  = \sum_{i,j} \;
 \frac{8 \pi^2}{M_R^3} \Gamma(R \rightarrow q_i q_j) BR(R \rightarrow W_L W_L)
\left( \tau dL/d\tau \right)_{pp/q_i q_j} 
\end{equation}

Analogously, the VBF cross section can be estimated as
\begin{equation}
\sigma^{VBF} (pp \rightarrow W_L W_L X)  = \sum_{i,j} \;
 \frac{8 \pi^2}{M_R^3} \Gamma(R \rightarrow W_L W_L) BR(R \rightarrow W_L W_L)
\left( \tau dL/d\tau \right)_{pp/W_L W_L} 
\end{equation}
where $\left( dL/d\tau \right)_{pp/W_L W_L} $ is the luminosity of a pair of $W_L$'s inside
the proton and is approximately given by:
\begin{equation}
\left( dL/d\tau \right)_{pp/W_L W_L}  = 
\int_{\tau}^1 \; \frac{d \tau^\prime}{\tau^\prime} \;  \int_{\tau^\prime}^1 \; \frac{dx}{x} \;
 \left[ u^{(a)}(x) d ^{(b)}(\tau^\prime/x) + (a) \leftrightarrow (b) \right] \; 
\left( dL/d\xi \right)_{ud/W^+_L W^-_L} ,
\end{equation}
where $\xi = \tau/\tau^\prime$ and
\begin{equation}
\left( dL/d\xi \right)_{ud/W^+_L W^-_L} = 
\left( \frac{\alpha}{\pi} \right)^2 \frac{1}{\xi} \left[ (1+\xi) \ln (1/\xi) + 2 ( \xi-1) \right]
\end{equation}

The relative DY and VBF contributions can be easily estimated in the context of these approximations:
\begin{equation}
\frac{\sigma^{VBF} ( pp \rightarrow W_L W_L X)}{\sigma^{DY} ( pp \rightarrow W_L W_L X)} = 
\frac{ \Gamma(R \rightarrow W_L W_L)}{\Gamma(R \rightarrow q_i q_j)} 
\frac{ \left( dL/d\tau \right)_{pp/W_L W_L}}{\left( dL/d\tau \right)_{pp/q_i q_j} }
\end{equation}

Estimating the ratio of luminosities one finds that in order for the VBF process to be competitive, 
the coupling of the resonance to light quarks must be suppressed so that the ratio of the 
partial widths can compensate for the ${\cal O} (10^{-6})$ smaller luminosity. 
This is usually the case for scalar resonances but it can also be the case for some models with vector resonances.

\section{A BRIEF HISTORY OF VECTOR BOSON FUSION}

\begin{figure}[h,t,b]
\begin{center}
\includegraphics[width=14cm,height=10cm]{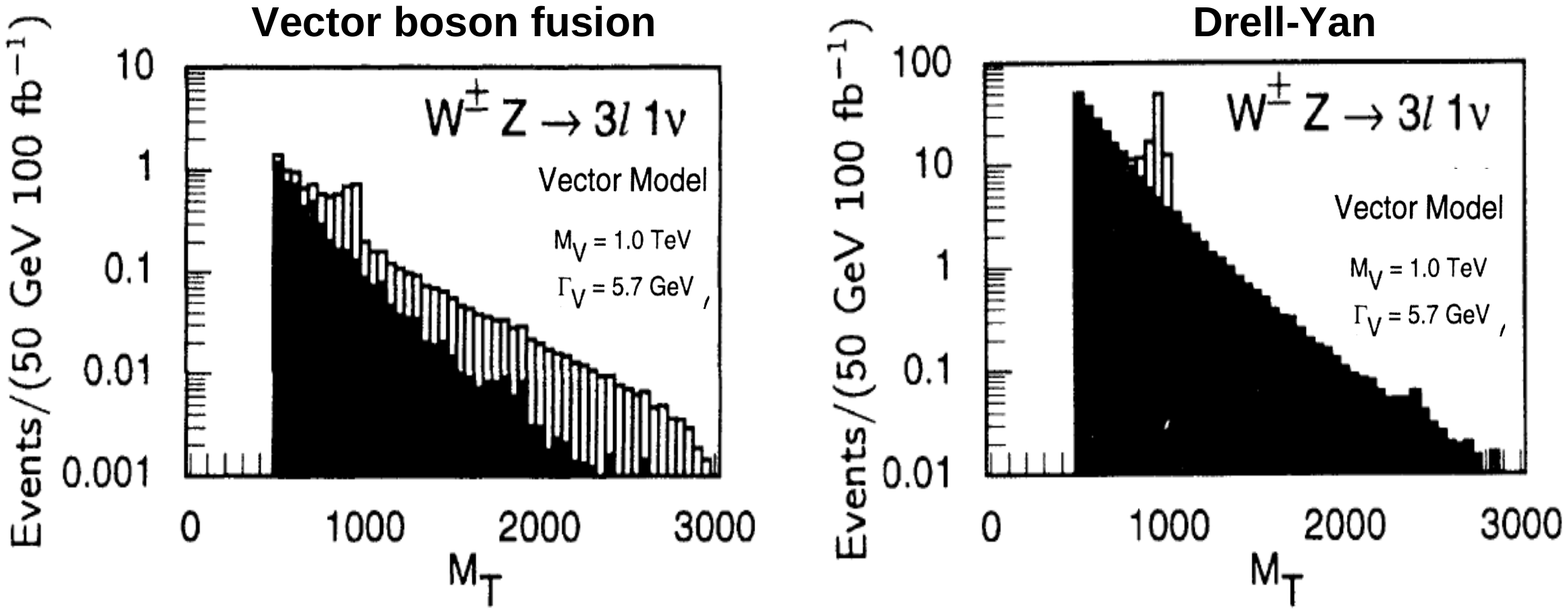}%
\vspace{-3cm}
\caption{Signal and background for a vector resonance in $WW$ (left panel)
and Drell-Yan (right panel) (from~\cite{Bagger:1995mk}).
 \label{WW_fig1} }
\end{center}
\end{figure}

Vector boson fusion has come a long way since first proposed by Cahn and Dawson~\cite{Cahn:1983ip}.
The development of the effective W approximation (EWA)~\cite{Dawson:1984gx}, used in conjunction with the equivalence theorem (ET)~\cite{Cornwall:1974km,Lee:1977eg,Chanowitz:1985hj}, 
provided an important tool for estimations  of the signals for these processes.
A first realistic study of signatures of strong interactions in $WW$ scattering at the LHC was performed for different
models in the context of EWA/ET in~\cite{Bagger:1995mk}, see Fig. \ref{WW_fig1}. For a model with vector resonances, 
it was recognized that the Drell-Yan production
is more important than VBF for resonances lighter than ${\cal O}$(TeV). 

Similar conclusions were reached in models of chiral electroweak lagrangians with resonances generated via unitarization procedures, again in the EWA/ET 
context~\cite{Dobado:1999xb,Butterworth:2002tt}. However, a detailed simulation for the ATLAS detector of this type of models
 with a $1.2$ TeV vector resonance using VBF with EWA/ET implemented in Pythia has been reported~\cite{Stefanidis:2007zz,AllwoodSpiers:2007mi}, 
where it is claimed that $100$ ($300$) fb$^{-1}$ is needed to detect this resonance in $qq \rightarrow q' q' W Z$ using the 
$WZ \rightarrow  l \nu jj$ ($WZ \rightarrow  l \nu ll$) mode.

Realistic simulations using the CMS detector have been recently performed for the case of a vector $\rho_{TC}$ in the context of 
LSTC implemented in Pythia, using the dominant Drell-Yan process~\cite{Kreuzer:2007zz}. A 5$\sigma$ discovery is reported
for a integrated luminosity of $30$ fb$^{-1}$ for $M_{\rho_{TC}}$ up to $700$ GeV. 

\begin{figure}[h,t,b]
\begin{center}
\includegraphics[width=14cm,height=10cm]{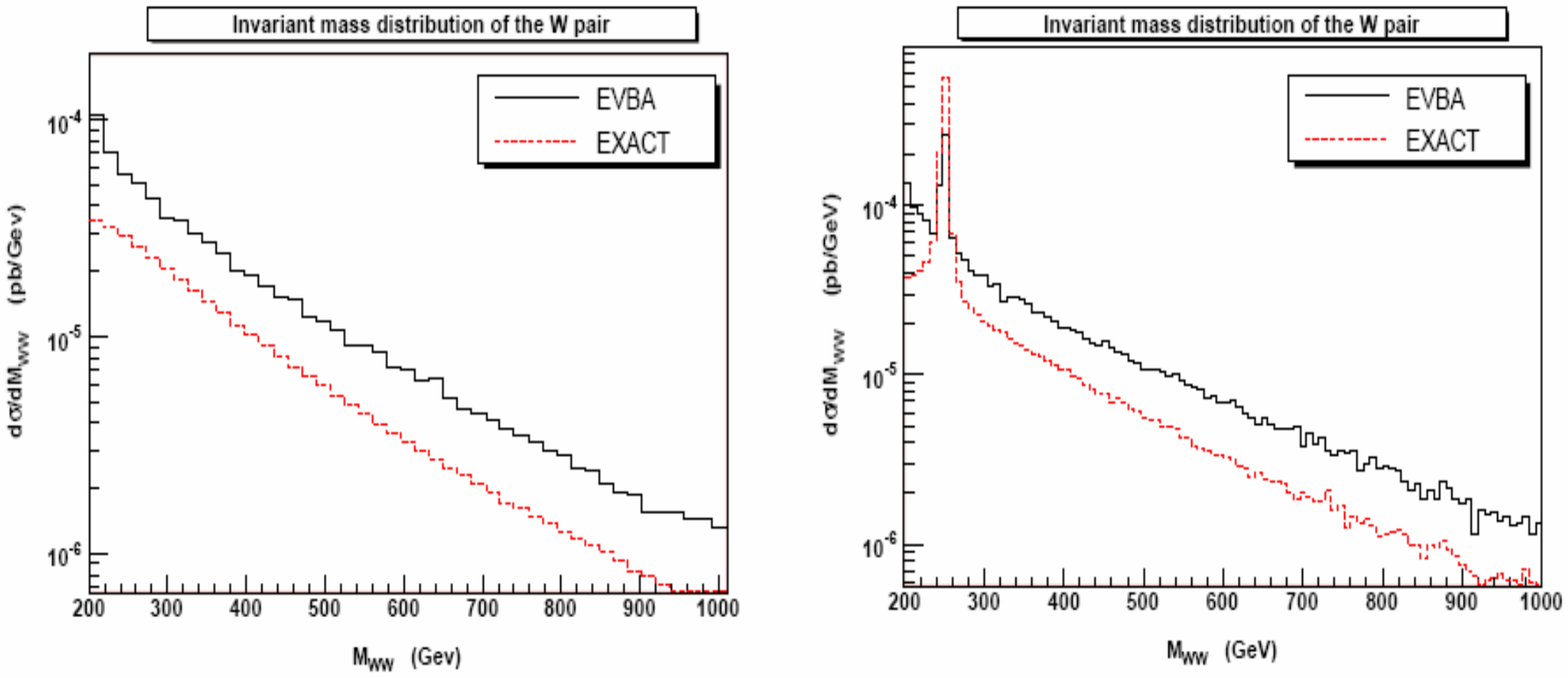}%
\vspace{-3cm}
\caption{ $WW$ invariant mass distribution for the process $us \rightarrow d c W^+ W^-$
with EWA (black solid curve) and the exact result (red dashed curve) for infinite Higgs boson mass
(left panel) and $M_H = 250$ GeV (right panel)
 (from~\cite{Accomando:2006mc}).
 \label{WW_fig2} }
\end{center}
\end{figure}

The signatures for a strongly interacting sector becomes even more challenging when there are no resonances at the LHC reach.
They manifest at low energies in anomalous gauge boson interactions generated by integrating out the heavy resonances.
An exact tree-level study of anomalous quartic gauge boson couplings in VBF was performed, where bounds on some coefficients 
of the chiral lagrangian were determined~\cite{Belyaev:1998ih} and refined recently in the Ref.~\cite{Eboli:2006wa}. The VBF production of a 
Higgs boson with non-standard couplings to electroweak gauge bosons was analysed in~\cite{Zhang:2003it}. 
Significant differences between a complete calculation of VBF and the EWA/ET approximation were found in 
the large $M_{WW}$ invariant mass distribution region,
see Fig.~\ref{WW_fig2}, where a full calculation of $2 \rightarrow 6$ amplitudes was also performed~\cite{Accomando:2006mc, Ballestrero:2007xq}.

\begin{figure}[h,t,b]
\begin{center}
\includegraphics[width=14cm,height=10cm]{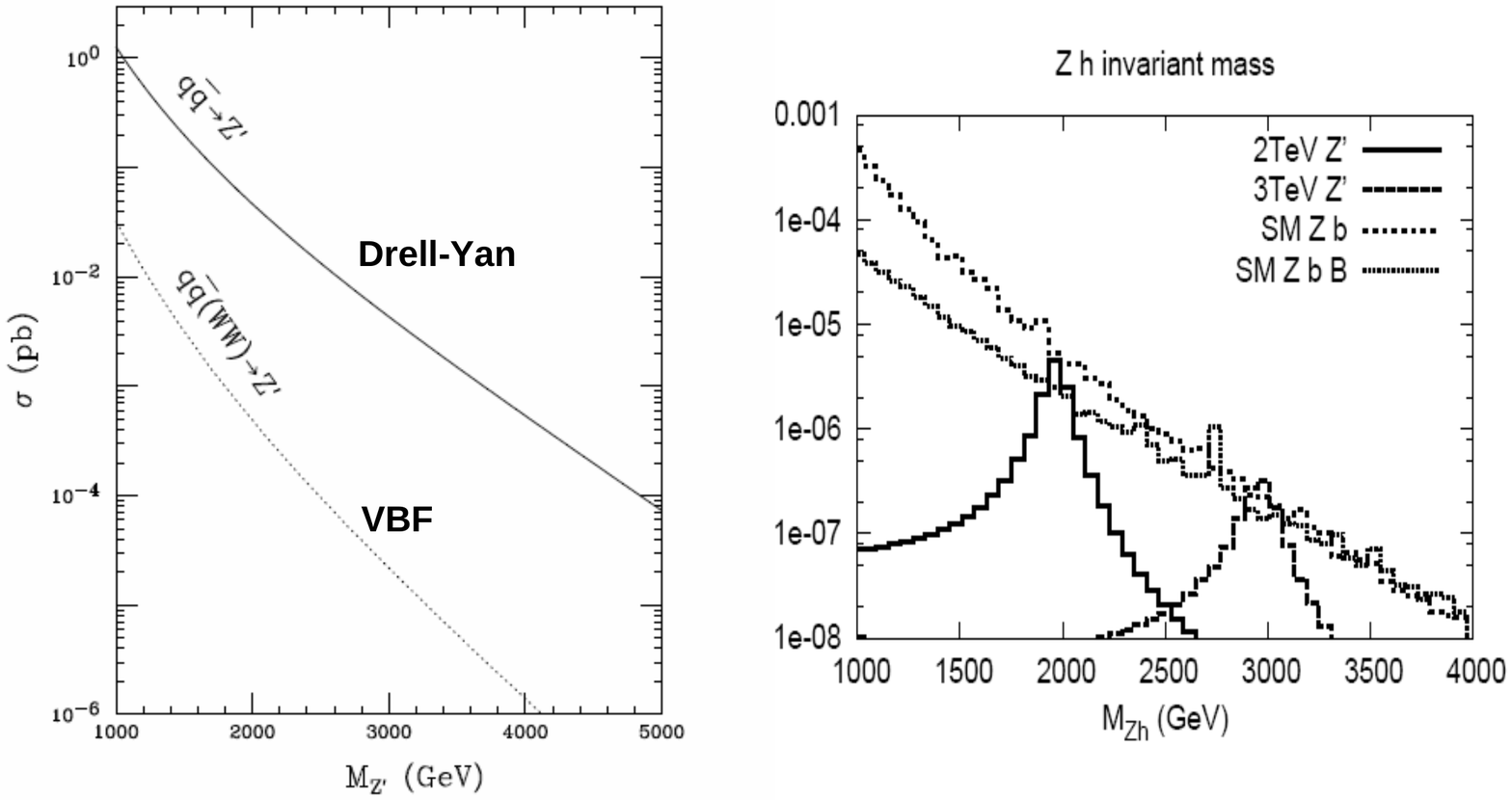}%
\vspace{-3cm}
\caption{Cross section for $Z'$ production at the LHC for different $Z'$ masses, with the Drell-Yan and VBF processes shown separately (left panel). $Z'$ mass reconstruction from  $p p \rightarrow Z' \rightarrow H Z \rightarrow b \bar{b} l^+ l^-$  (right panel) (from~\cite{Agashe:2007ki}).
 \label{WW_fig3} }
\end{center}
\end{figure}

Turning now to the production of resonances in recent developments, the production at the LHC of a neutral KK excitation 
of the electroweak gauge bosons, generically called $Z'$, in the context of a warped extra dimension model was studied by
Agashe {\it et al.}\cite{Agashe:2007ki}. They obtained that the Drell-Yan process is dominant over the VBF. The best discovery mode is
$Z' \rightarrow H Z \rightarrow b \bar{b} l^+ l^-$, see Fig.~\ref{WW_fig3}. They conclude that a 2 (3) TeV $Z'$ can be discovered with
approximately 100 fb$^{-1}$ (1 ab$^{-1}$) integrated luminosity. Production of KK excitations of gauge bosons and its contribution to top quark pair production at the LHC was discussed in~\cite{Djouadi:2007eg}.

\begin{figure}[h,t,b]
\begin{center}
\includegraphics[width=14cm,height=10cm]{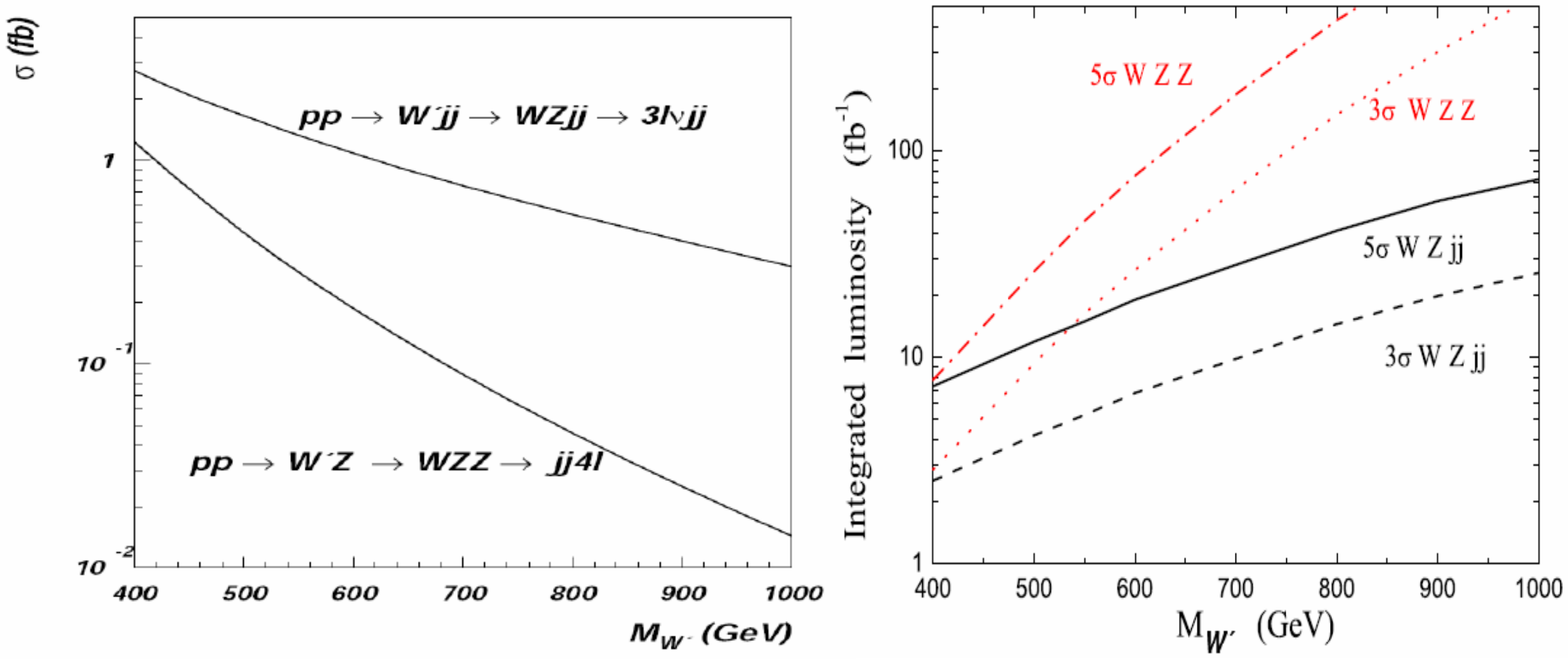}%
\vspace{-3cm}
\caption{Cross section for $W'$ production at the LHC for different $W'$ masses, with the associated $W'Z$ and VBF processes shown separately (left panel).Significance for $W'$ detection for the different modes (right panel) (from~\cite{Belyaev:2007ss}).
 \label{WW_fig4} }
\end{center}
\end{figure}

In the case of Higgsless models, an opposite situation occurs for the production of $W'$~\cite{He:2007ge}. 
Due to the vanishing of fermionic couplings to the new vector resonance states, which guarantees a null correction
to the electroweak precision parameters, the VBF is the dominant production process. A detailed study, including 
a complete leading order computation of  the signal and relevant background, shows that 
with a 100 fb$^{-1}$ integrated luminosity the LHC will completely cover the parameter space of this model, 
with a 5 $\sigma$ discovery of $W'$ up to a mass of 1.2 TeV, see Fig.~\ref{WW_fig4}.

It is also interesting to point out that it could be possible to use gluon-gluon fusion Higgs production rate 
(and the Higgs decay rate to 2 photons) to 
discriminate among different models, such as UED and Gauge-Higgs unification models~\cite{Djouadi:2007fm,Falkowski:2007hz,Maru:2007xn}.

The next frontier in VBF is certainly the next-to-leading order (NLO) computations.  A parton level Monte Carlo 
implementation of NLO QCD
corrections to vector boson pair production via VBF was recently reported~\cite{Bozzi:2007te}. 
The implementation of new models in the same framework may be available in the near future.

 See also~\cite{Hirn:2007we, Foadi:2008ci} for recent studies of strong $WW$ scattering.
 
\section{RESONANCES IN A STRONGLY COUPLED EWSB SECTOR}

\subsection{Inspiration from low energy QCD}

Low energy QCD (by low energy we mean $\sqrt{s} \ll m_\rho$) with two flavors is a 
theory of strongly coupled pions
described by a chiral lagrangian (non-linear $\sigma$ model):
\begin{equation}
{\cal L} = \frac{f_\pi^2}{4} Tr \left[ \partial_\mu U \partial^\mu U^\dagger \right]
\label{chiral1}
\end{equation}
with
$U = e^{i 2 \pi^a T^a/f_\pi}$ transforming as a $(2,2)$ under a global chiral symmetry $G=SU(2)_L \times SU(2)_R$:
\begin{equation}
U \rightarrow g_R U g_L^\dagger; \;\; g_R \in SU(2)_R, \;\; g_L \in SU(2)_L. 
\end{equation}
This lagrangian is the lowest order term of an infinite expansion in an increasing number of derivatives of the field $U$.
The pions are the Nambu-Goldstone bosons resulting from the spontaneous symmetry breaking
$SU(2)_L \times SU(2)_R \rightarrow SU(2)_V$ generated by a non-zero vaccum expectation value $\langle U \rangle = 1$.

It follows from the lagrangian (\ref{chiral1}) that $\pi \pi$ scattering amplitudes grow as $s/f_\pi^2$, as predicted by current algebra arguments. They violate $s-$wave perturbative unitarity when $\sqrt{s} \approx 4 \pi f_\pi$. Higher derivative terms in the chiral lagrangian
can delay the energy scale for unitarity violation. 

There are several {\it ad-hoc} methods to unitarize $\pi \pi$ scattering but we will not dwell on them. In QCD, violation of perturbative unitarity points to the existence of $\pi \pi$ resonances, like the $\rho$ and $a_1$. Hence, one must extend the formalism of chiral
perturbation theory to include these resonances in order to describe physics at energy scales $\sqrt{s} \approx m_\rho$.
Possibly one of the most successful attempts to introduce resonances consists in describing them as dynamical gauge bosons of a local symmetry, as first proposed by Sakurai~\cite{Sakurai:1960ju}. 
This idea led to the so-called hidden symmetry approach~\cite{Bando:1987br},
where the masses of the resonances are generated via a Higgs mechanism.
The hidden symmetry model can nicely accomodate features such as vector meson dominance and gauge invariance.
It can also be used to describe resonances in a strongly coupled EWSB sector.

\subsection{Strongly coupled EWSB sector}

The EWSB sector in the SM is a gauged linear $\sigma$-model which can be nicely written 
as~\cite{Appelquist:1980vg,Longhitano:1980iz}:
 \begin{equation}
{\cal L}_{SB} = \frac{1}{4} Tr \left[ (D_\mu U)^\dagger D^\mu U \right] - 
\frac{1}{4} \lambda \left(\frac{1}{2} Tr\left[ U^\dagger U \right] - v^2  \right)^2
\end{equation}
where 
\begin{equation}
D_\mu U = \partial_\mu U + \frac{1}{2} i g W_\mu^i \tau_i U - \frac{1}{2} i g' Y_\mu U \tau_3.
\end{equation} 
In the limit $g' \rightarrow 0$ this lagrangian has an additional global symmetry $G=SU(2)_L \times SU(2)_R$, with $U$ transforming
as a $(2,2)$:
\begin{equation}
U \rightarrow e^{i \epsilon_L^i  \tau_i/2} U e^{-i \epsilon_R^i  \tau_i/2}. 
\end{equation}
Electroweak symmetry is spontaneously broken by a non-zero
vacuum expectation value $\langle U \rangle = v$, which generates masses for the electroweak gauge bosons. 
A so-called custodial global $SU(2)_V$ symmetry ($\epsilon_L=\epsilon_R$) survives, which is responsible for 
keeping the parameter $\rho=M_W^2/M_Z^2 \cos \theta_W = 1$ at tree level.

One way to introduce resonances from a strongly coupled sector in an effective lagrangian is to extend the SM
global symmetry to a linearly realized $[SU(2)_L \times SU(2)_R]^{el} \times [SU(2)_L \times SU(2)_R]^{comp}$ symmetry.
Here we are using the superscripts $(el)$ and $(comp)$ to denote the elementary and
composite sectors respectively.
The subgroup $[SU(2)_L \times U(1)_X]^{el} \times [SU(2)_L \times SU(2)_R]^{comp}$ will be gauged.
The masses of all gauge bosons are generated via an extended Higgs mechanism.
The SM Higgs can be part of the elementary sector (LDBESS model by Casalbuoni et al.~\cite{Casalbuoni:1985kq, Casalbuoni:1986vq, Casalbuoni:1997rs}) or part of the composite sector (2-site composite Higgs model by Contino et al.~\cite{Contino:2006nn}).

\subsection{LDBESS model}

\begin{figure}[h,t,b]
\begin{center}
\includegraphics[width=12cm]{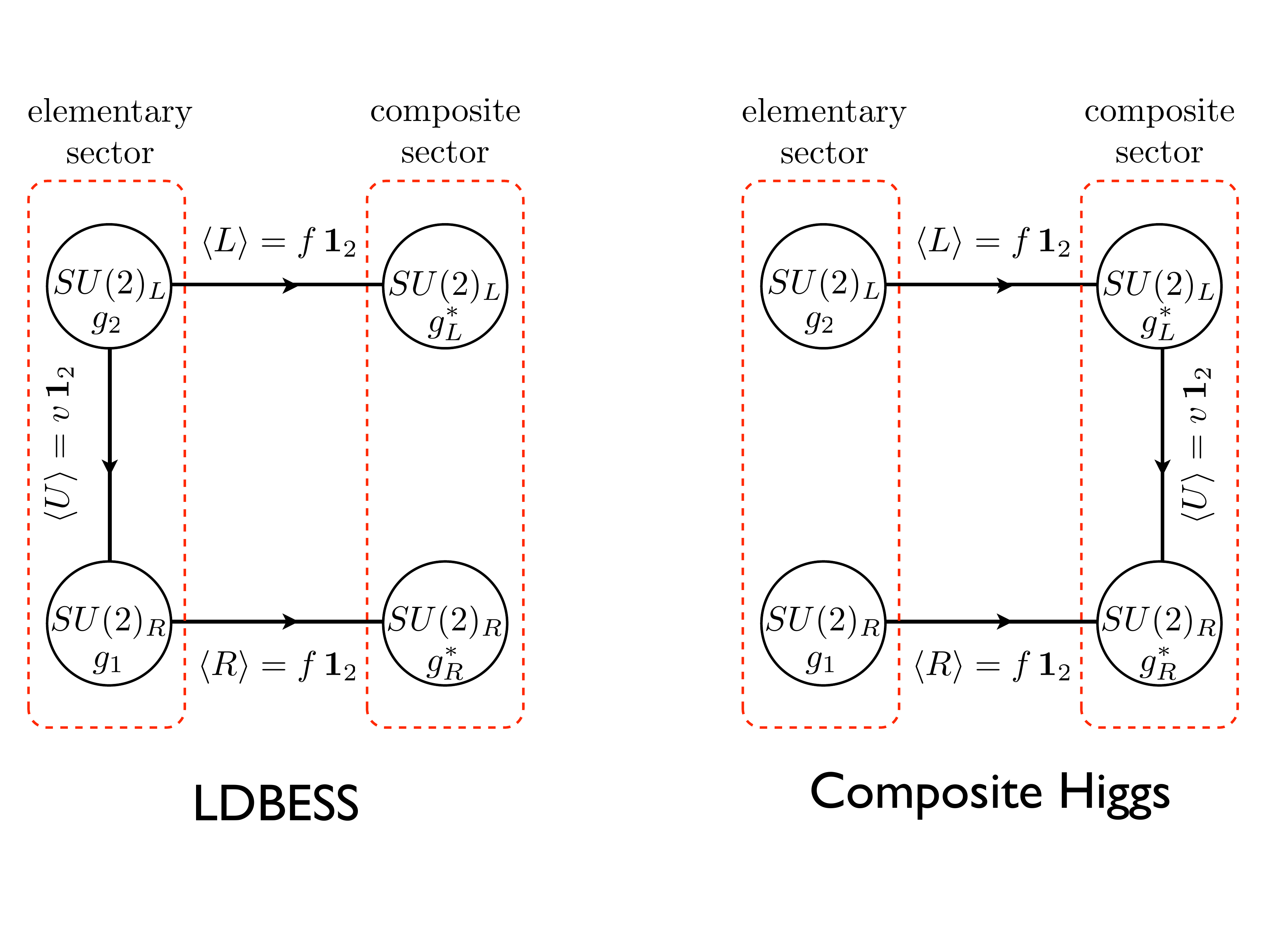}%
\vspace{-.2cm}
\caption{Moose diagrams of the LDBESS model (left panel) and a 2-site composite Higgs model (right panel).
\label{WW_mooseLDBESS}}
\end{center}
\end{figure}

The scalar sector of the LDBESS model is given, as depicted on the moose diagram of Fig.\ref{WW_mooseLDBESS},  by 3 sets of fields transforming under $[SU(2)_L \times SU(2)_R]^{el} \times [SU(2)_L \times SU(2)_R]^{comp}$  as
\begin{eqnarray} 
L \in (2,0,{\bar 2},0) \; & \Longrightarrow & \; L \rightarrow g_L L h^\dagger_L \\
R \in (0,2,0,{\bar 2}) \; & \Longrightarrow & \; R \rightarrow g_R R h^\dagger_R \\
U \in (2,{\bar 2},0,0) \; & \Longrightarrow & \; U \rightarrow g_L U g_R^\dagger, 
\end{eqnarray}
where $g_{L,R} \in SU(2)_{L,R}^{el}$ and $h_{L,R} \in SU(2)_{L,R}^{comp}$.
Notice that the Higgs field $U$ transforms under the elementary group. 

The covariant derivatives acting on the scalar fields are (suppressing Lorentz indices):
\begin{eqnarray} 
D L &=& \partial L + i g_2 \frac{\tau_i}{2} W^i L - i g^*_L L \frac{\tau_i}{2} V_L^i \\
D R &=& \partial R + i g_1 \frac{\tau_3}{2} Y R - i g^*_R R \frac{\tau_i}{2} V_R^i \\
D U &=& \partial U + i g_2 \frac{\tau_i}{2} W^i U - i g_1 U \frac{\tau_3}{2} Y ,
\end{eqnarray}
where $W,Y,V_L$ and $V_R$ are the gauge bosons of the local 
$[SU(2)_L \times U(1)_Y]^{el} \times [SU(2)_L \times SU(2)_R]^{comp}$ symmetry. The couplings
of the elementary sector are $g_1$ and $g_2$, whereas $g^*_L$ and $g^*_R$ are the couplings of the
composite sector.

The scalar potential is chosen in such a way that $\langle L \rangle = \langle R \rangle = f \mathbf{1}_2$ and
$\langle U \rangle = v\mathbf{1}_2$, with $f\gg v$, resulting in the symmetry breaking pattern
\begin{eqnarray}
[2_L \times 1_X]^{el}& \times& [2_L \times 2_R]^{comp} 
\stackrel{\langle \tilde{L} \rangle = \langle \tilde{R} \rangle}{\longrightarrow}
[SU(2)_L \times U(1)_Y]^{SM}  \ \
\stackrel{\langle \tilde{U} \rangle}{\longrightarrow} U(1)_{em}
\end{eqnarray}

As usual, the scalar kinetic terms generate masses and mass-mixing terms for the gauge bosons:
\begin{equation}
{\cal L} = \frac{1}{4} \textrm{Tr} \left[ (D_\mu \tilde{U})^\dagger D^\mu \tilde{U} +
(D_\mu \tilde{L})^\dagger D^\mu \tilde{L} + (D_\mu \tilde{R})^\dagger D^\mu \tilde{R}
 \right]
\end{equation}
with the result:
\begin{eqnarray*}
{\cal L}_{mass}
&=& \frac{1}{8} \left[ v^2 \left( g_2^2 W_i^2 + g_1^2 Y^2 - 2 g_2 g_1 W_3 Y \right) + 
f^2 \left( g_2^2 W_i^2 + g^{*\,2}_L V_{L,i}^2 - 2 g_2 g^*_L W_i V_L^i \right) \right. \\ 
&+&  \left. f^2 \left( g_1^2 Y^2 + g^{*\,2}_R V_{R,i}^2 - 2 g_1 g^*_R Y V_{R}^3 \right) \right]
\end{eqnarray*} 

Even in absence of EWSB ($v=0$), the $L$ link field induces a mixing between the elementary and the composite fields.
The mass eigenstates are a mixture of elementary and composite fields. Diagonalizing the mass matrices after EWSB,  we obtain the following spectrum
\begin{itemize}
\item Electrically charged sector
\vspace{.3cm}
\begin{itemize}
\item a composite massive $W_{h1}^\pm=V_R^\pm$ with a mass $m_{W_{h1}^\pm}=\frac{1}{2} g^*_R f$
\item a mostly composite massive $W_{h2}^\pm$ which is a combinaison of $W^\pm$ and $V_L^\pm$:
\begin{equation}
W_{h2}^\pm= \sin\theta (1+ r \cos^2 \theta) W^\pm - \cos\theta ( 1 - r \sin^2\theta) V_L^\pm,
\end{equation} 
with a mass given by
\begin{equation}
M_{W_{h2}}^2 = \frac{1}{4}  g^{*\,2}_L f^2 \left( \frac{1}{\cos^2 \theta} + r \tan^2\theta +  {\cal O}(v^4/f^4) \right).
\end{equation}
\item a mostly elementary light $W_l^\pm$ which is the linear combination orthogonal to $W_{h2}^\pm$
\begin{equation}
W_{l}^\pm= \cos\theta (1- r \sin^2 \theta) W^\pm + \sin\theta ( 1 + r \cos^2\theta) V_L^\pm,
\end{equation} 
with a mass proportional the SM Higgs vev
\begin{equation}
M_{W_l}^2 = \frac{1}{4}  g^2 v^2 \left( 1 - r \sin^2\theta +  {\cal O}(v^4/f^4) \right).
\end{equation}
\end{itemize}
\item Electrically neutral sector
\vspace{.3cm}
\begin{itemize}
\item a massless photon $\gamma= Y$.
\item a mostly elementary light $Z$ which is a linear combination of $W^3$ and $V_L^3$
\begin{equation}
Z_{l}= \cos\theta (1- r \sin^2 \theta) W^3 + \sin\theta ( 1 + r \cos^2\theta) V_L^3,
\end{equation} 
with a mass proportional to the SM Higgs vev and equal to $M_{W_l}$
\item a mostly composite heavy $Z_{h2}$ corresponding to the linear combination orthogonal to $Z_l$. Its mass is equal to $M_{W_{h2}}$
\item a composite heavy $Z_{h1}=V_R^3$ with a mass $m_{Z_{h1}^\pm}=\frac{1}{2} g^*_R f$.
\end{itemize}
\end{itemize}
\hspace{.5cm}
The mixing angle $\theta$ corresponds to the ratio of the gauge couplings of the elementary and composite sectors
\begin{equation}
\sin \theta = \frac{g_2}{\sqrt{g_2^2+g_L^{*\,2}}}.
\end{equation}
It measures the amount of compositeness in the massless gauge bosons before EWSB.
$g$ is the gauge coupling of the low-energy SM $SU(2)$ gauge coupling
\begin{equation}
g^2 = \frac{g_2^2 g_L^{*\, 2}}{g_2^2 + g_L^{*\, 2}},
\end{equation}
after the breaking $SU(2)^{el}\times SU(2)^{comp}$ to $SU(2)_L^{SM}$.
Note that $g\simeq g_2$ if the composite sector is strongly coupled ($g^*_L \gg g_2$).
Finally, the parameter $r$ measures the mass gap between the light and the heavy states
\begin{equation}
r= \frac{g^2 v^2}{g_L^{*\,2} f^2} \simeq \frac{M_{W_l}^2}{M_{W_{h2}}^2}.
\end{equation}
We have neglected here the coupling $g_1$ of the elementary $U(1)$ gauge group. Turning it on will induce a mixing of $V^3_L, V^3_R, W^3$ and $Y$.

Trilinear couplings between the light and the heavy states arise from the gauge kinetic terms,
\begin{equation}
{\cal L}_{gauge} = \sum_{W,Y,V_L,V_R} \frac{1}{2} \textrm{Tr} \left[ F_{\mu \nu}^2 \right].
\end{equation}
For instance, we obtain the coupling
\begin{equation}
g_{Z_{h1} W^+_l W^-_l} = g \, r \sin\theta \cos\theta.
\end{equation}
Notice that the coupling is suppressed by a factor $r$ and that, in the absence of symmetry breaking,
this coupling vanishes.

This model can also be extended to include more resonances by simply introducing more copies of  $[SU(2)_L \times SU(2)_R]^{comp}$  or  $[SU(3)_c \times SU(2)_L \times SU(2)_R]^{comp}$ if color octet vector bosons (gluon resonances) are present in the spectrum. However, the basic idea is the one shown above. Notice that in the case of gluon resonances, their coupling to gluons would vanish since
the color symmetry is unbroken, as pointed out in~\cite{Zerwekh:2001uq}. 

\subsection{A 2-site composite Higgs model}

The basic difference between the LDBESS model and a 2-site composite Higgs is that the Higgs is now part of the strong sector and interacts directly with the composite gauge bosons. So
\begin{equation}
U \in (0,0,2,\bar{2}) \;  \Longrightarrow  \; U \rightarrow h_L U h_R^\dagger
\end{equation}
and
\begin{equation}
D U = \partial U + i g^*_L \frac{\tau_i}{2} V_L^i U - i g^*_R U \frac{\tau_i}{2} V_R^i ,
\end{equation}
in the notation adopted in the previous section. One still keeps $\langle L \rangle =\langle R \rangle = f$
and $\langle U \rangle = v$.
Consequently we now have the mass-mixing lagrangian given by:
\begin{eqnarray*}
{\cal L}_{mass} 
&=& \frac{1}{8} \left[ v^2 \left( g^{*\,2}_L V_{L,i}^2 + g^{*\, 2}_R V_{R,i}^2 - 2 g^*_L g^*_R V^i_L V^i_R \right) + 
f^2 \left( g_2^2 W_i^2 + g^{*\,2}_L V_{L,i}^2 - 2 g_2 g^*_L W_i V_L^i \right) \right. \\ 
&+&  \left. f^2 \left( g_1^2 Y^2 + g^{*\,2}_R V_{R,i}^2 - 2 g_1 g^*_R Y V_{R}^3 \right) \right]
\end{eqnarray*} 

As before, the mass eigenstates are linear combinations of elementary and composite states.
Diagonalizing the mass mixing terms, we easily obtain the spectrum after EWSB. For example, in the electrically charged sector, we get two massive $W^\pm_{h,1-2}$ with masses given by
\begin{eqnarray}
m_{W^\pm_{h1}}^2=\frac{1}{4} g^{*\,2}_R f^2 + \frac{1}{4} g^{*\,2}_{R} v^2 + \mathcal{O}(v^4/f^2)\\
m_{W^\pm_{h2}}^2= \frac{1}{4}  g^{*\,2}_L f^2 \left( \frac{1}{\cos^2 \theta} + r\, \textrm{cotan}^2\theta +  {\cal O}(v^4/f^4) \right).
\end{eqnarray}
and one light $W_l^\pm$ whose mass is proportional to the Higgs vev and that can be identified with the SM $W^\pm$
\begin{equation}
M_{W_l}^2 = \frac{1}{4}  g^2 v^2 +  {\cal O}(v^4/f^2).
\end{equation}
As for the LDBESS model, $g$ is the gauge coupling of the low-energy SM $SU(2)$ gauge coupling
\begin{equation}
g^2 = \frac{g_2^2 g_L^{*\, 2}}{g_2^2 + g_L^{*\, 2}},
\end{equation}
after the breaking $SU(2)^{el}\times SU(2)^{comp}$ to $SU(2)_L^{SM}$ at the scale $f$ and the mixing angle $\theta$ is
the ratio of the gauge couplings of the elementary and composite sectors
\begin{equation}
\sin \theta = \frac{g_2}{\sqrt{g_2^2+g_L^{*\,2}}}.
\end{equation}
Finally, $r$ is the mass gap between the light and the heavy states
\begin{equation}
r= \frac{g^2 v^2}{g_L^{*\,2} f^2} \simeq \frac{M_{W_l}^2}{M_{W_{h2}}^2}.
\end{equation}
The spectrum is the same as in the LDBESS model up to differences of  $\mathcal{O}(v^2/f^2)$ in the heavy sector.
Still, these differences have big effects in the couplings of the Higgs boson to the heavy resonances. These couplings can be simply obtained by substituting
$v$ by $v+ h$ in the lagrangian: therefore, in the LDBESS model the Higgs coupling to $W^\pm_{h2}$ are reduced two powers of the compositeness angle $\theta$, while in the composite Higgs model, the coupling is enhanced by the same factor.

\AddToContent{A.~Delgado, C.~Grojean, E.~Maina and R.~Rosenfeld}
\setcounter{figure}{0}
\setcounter{table}{0}
\setcounter{section}{0}
\setcounter{equation}{0}
\setcounter{footnote}{0}
\clearpage


\part{Vector-like quarks: a toolkit for experimenters}

{\it J.~Santiago}


\begin{abstract}
We review the motivation and main features of vector-like quarks
with special emphasis on the techniques used in the calculation 
of the features relevant for their collider implications.
\end{abstract}

\noindent 
In four space-time dimensions, Dirac fermions can be decomposed into left-
and right-handed chiralities, $\psi=\psi_L+\psi_R$, with
$\psi_{L,R}=(1\mp \gamma^5)/2 \, \psi$. Chiral fermions, for which the two
components have different charges under the electroweak $SU(2)_L\times
U(1)_Y$ gauge group, have masses proportional to the electroweak
symmetry breaking (EWSB) scale ($v\sim 174$ GeV)
and are therefore expected to be relatively
light, $m_{\mathrm{chiral}}= \lambda v \lesssim v$. They can only be
made heavy at the expense of introducing a large dimensionless
coupling $\lambda \gg 1$, thereby inducing large one loop corrections to
electroweak observables (they do \textit{not decouple}). 
Thus, new chiral fermions
are quite constrained experimentally from 
electroweak precision tests (EWPT). If the fermions are vector-like
(the two chiralities have the
same quantum numbers) we can write down
a gauge invariant (Dirac) mass that is unrelated to the
EWSB scale. They can therefore be
naturally heavier than $v$ without introducing large dimensionless
couplings. As the Dirac mass becomes large, their low energy effects
become negligible (\textit{decoupling}) and EWPT impose no constraints.
Furthermore,
new vector-like fermions are a very common prediction of theories beyond
the Standard Model (SM), from Kaluza-Klein modes in models with extra
dimensions to partners of the SM 
fermions in little Higgs models, or members of extended multiplets in
GUT theories. 

For their relevance to LHC physics, we will restrict ourselves to
vector-like quarks, transforming as triplets under the $SU(3)$ gauge
group. We will also consider only the case of new quarks that can
sizably mix (through mass terms) with SM quarks, generating
flavour changing electroweak processes that can lead to interesting
signatures at the LHC. Finally, we will assume that a (relatively)
light SM-like Higgs boson (doublet under $SU(2)_L$) 
exists in the spectrum and is the main source of EWSB.

\section{HOW TO COMPUTE THE RELEVANT FEATURES: 
MASSES AND COUPLINGS} 

New vector-like quarks with sizable mixings with the SM quarks cannot
have arbitrary quantum numbers. They can only couple in a gauge
invariant way to a SM quark and the Higgs, which gives only the seven
distinct possibilities shown in Table~\ref{vectorlike_multiplets}.
\begin{table}[ht]
\caption[Vector-like quark multiplets mixing with
the SM quarks through Yukawa couplings]
{Vector-like quark multiplets $Q$ mixing with
the SM quarks through Yukawa couplings. 
The electric charge is $Q=T_3+Y$.} 
\label{vectorlike_multiplets}
\begin{center}\begin{tabular}{|c|ccccccc|}\hline
\rule[-4.2ex]{0pt}{9.5ex}
$Q$ & $U$ & $D$ & $\begin{pmatrix} U\cr D\end{pmatrix} $ &
$\begin{pmatrix}
X\cr U \end{pmatrix}$ &$\begin{pmatrix}D\cr Y\end{pmatrix}$ & 
$\begin{pmatrix} X\cr U\cr D\end{pmatrix}$
& $\begin{pmatrix} U\cr D\cr Y\end{pmatrix} $ \\ 
\hline 
\rule[-1.3ex]{0pt}{4ex}
isospin &
0&0&${1}/{2}$&${1}/{2}$&${1}/{2}$&1&1 \\ 
\rule[-1.3ex]{0pt}{4ex}
hypercharge &
${2}/{3}$&$-{1}/{3}$&${1}/{6}$& ${7}/{6}$&
$-{5}/{6}$&${2}/{3}$&$-{1}/{3}$\\\hline
\end{tabular}%
\end{center}
\end{table}
Given these possibilities, the question is, how do we get the relevant
features of the model in order to study it at colliders. Or said
otherwise, what are the masses and couplings of these new quarks? 
The aim of this short review is to present the general algorithm to
compute these relevant features and to exemplify it in a
particular case that will be further explored below.
The general Lagrangian can be written, in the current eigenstate basis
(\textit{i.e.} in terms of fermions with well defined $SU(2)_L\times
U(1)_Y$ \textit{quantum numbers}), as
\begin{equation}
\mathcal{L}_{l}+\mathcal{L}_h+\mathcal{L}_{lh},
\end{equation}
where $\mathcal{L}_{l}$ is the SM Lagrangian, $\mathcal{L}_h$, 
contains the kinetic, Dirac mass terms (that can always be taken
diagonal) and Yukawa couplings involving only the (heavy) vector-like 
quarks, and finally $\mathcal{L}_{lh}$ contains the (linear) mass (and
Yukawa) mixing between SM and 
vector-like quarks. The general expressions for these Lagrangians can
be found in Ref.~\cite{delAguila:2000rc}.
In order to obtain the properties of the physical particles
(\textit{i.e.} those with a well defined \textit{mass})
we have to diagonalize the corresponding mass matrices that
are contained in $\mathcal{L}_h+\mathcal{L}_{lh}$. Due to
the unbroken electromagnetic $U(1)_Q$ gauge invariance, the
mass matrix will be block diagonal according to the charges of the
different fields,
\begin{equation}
\mathcal{L}_\mathrm{mass} = \sum_Q \bar{\psi}^{(0)Q}_L \mathcal{M}_Q
\psi^{(0)Q}_R 
+ \mathrm{h.c.},
\end{equation}
where $\psi^{(0)Q}_{L,R}$ is a vector that contains the $n_Q$ quarks with
charge $Q$ and $\mathcal{M}^Q$ is a $n_Q \times n_Q$ matrix. The
overscript $(0)$ denotes current eigenstates.
We can then diagonalize each block with two unitary
matrices $U^Q_{L,R}$,
\begin{equation}
(U^Q_L)^\dagger \mathcal{M}_Q U^Q_R = \mathcal{D}_Q,
\end{equation}
with $\mathcal{D}_Q$ a diagonal matrix containing the masses of the
physical particles. $U^Q_{L,R}$ can be computed as
the unitary matrices that diagonalize the mass matrix squared,
$(U_L^Q)^\dagger \mathcal{M}_Q
\mathcal{M}_Q^\dagger U_L^Q = 
(U^Q_R)^\dagger \mathcal{M}_Q^\dagger \mathcal{M}_Q U^Q_R = \mathcal{D}_Q^2$.
Once we have computed the rotation matrices $U^Q_{L,R}$, we only have
to replace in the gauge and Yukawa couplings the mass eigenstates in
terms of the physical quarks,
\begin{equation}
\psi^{(0)Q}_{L,R}=U_{L,R}^Q \psi^Q_{L,R},
\end{equation}
where $\psi^Q_{L,R}$ are now the physical states. This final step will
have two main implications. First, it will modify the SM quark couplings
to the electroweak gauge bosons and the Higgs, and second it will
introduce off-diagonal (electroweak) gauge and Yukawa couplings
between heavy and SM quarks. In general, the first effect strongly 
contraints large mixing of vector-like quarks with first and second
generation quarks (as their couplings have been measured to agree very
well with the SM prediction).~\footnote{One exception is the
  possibility of cancellations 
  between the contributions of several vector-like quarks, something
  that can happen quite naturally in some models with extra
  dimensions.} 
Only mixing with the top is at present poorly constrained and could be
large. If the new vector-like quarks play a role in
the resolution of the hierarchy problem, it is indeed natural
for them to be relatively light and to mix sizably with the
top. Note that in this case they might induce sizable corrections at
loop level to precision observables, most notably the $T$
parameter and the $Z \bar{b}b$ coupling, that should be checked in
particular models~\cite{DelAguila:2001pu,Carena:2006bn,Carena:2007ua}.

Let us make this procedure more explicit with one relevant example whose
phenomenology will be further studied in this report. 
In particular, we
will consider an extension of the SM with two vector-like quark
doublets, $q_{L,R}=(q^u,q^d)^\mathrm{T}_{L,R}$ 
and $\chi_{L,R}=(\chi^u,\chi^d)^\mathrm{T}_{L,R}$,  
with hypercharges $Y_\chi=7/6$ and $Y_q=1/6$, respectively. The
electric charge of the different components are given by $Q=T_3+Y$
($Q_{\chi^u}=5/3$, $Q_{q^u}=Q_{\chi^d}=2/3$ and $Q_{q^d}=-1/3$). Note in
particular the exotic charge of $\chi^u$. For simplicity, we will also
assume that these two new quark doublets can only mix with the top but not
with the bottom or any other light quark. 
This example is
motivated by recent ideas using a subgroup of the custodial symmetry
to protect large corrections to the $Z\bar{b}_L b_L$
coupling~\cite{Agashe:2006at} for which these simplifying assumptions
are naturally realized.
This mechanism has been successfully implemented
in composite
Higgs~\cite{Carena:2006bn,Contino:2006qr,Carena:2007ua,Medina:2007hz,Carena:2007tn}  
and Higgsless models~\cite{Cacciapaglia:2006gp} in warped extra dimensions. 

The mass
Lagrangian for the charge $2/3$ quarks (the other ones do not have any
mass mixings and are therefore mass eigenstates) can be written as,
\begin{equation}
\mathcal{L}=
\begin{pmatrix}
\bar{t}_L^{(0)} & \bar{q}_L^{u(0)} & \bar{\chi}_L^{d(0)}
\end{pmatrix}
\begin{pmatrix}
m_t & 0 & 0 \\
m_{q,t} & M_q & 0 \\
m_{\chi,t} & 0 & M_\chi
\end{pmatrix}
\begin{pmatrix}
t_R^{(0)} \\
q_R^{u(0)} \\
\chi_R^{d(0)} 
\end{pmatrix} 
+\mathrm{h.c.}\, .
\end{equation}
EWSB masses, which are proportional to the Higgs vev,
are denoted with a lower case $m$ whereas Dirac masses, that can be
arbitrarily larger than 
$v$ are denoted by a capital $M$. The hierarchy between these two types of
masses, allows us to diagonalize the mass matrix perturbatively, in a
power expansion of $m/M\ll 1$ in order to obtain simple
analytic expressions for the masses and couplings of the physical
quarks. For instance,  to leading order in $m/M$, we have 
\begin{equation}
t^{(0)}_R \approx t_R +
\frac{m_{q,t}}{M_q} q^u_R +\frac{m_{\chi,t}}{M_\chi} \chi^d_R,
\quad
q^{u(0)}_R \approx q^u_R - \frac{m_{q,t}}{M_q} t_R,
\quad
\chi^{d(0)}_R \approx \chi^d_R - \frac{m_{\chi,t}}{M_\chi} t_R,
\end{equation}
where fields without the $(0)$ superscript are physical (mass
eigenstate) fields. All other fields are not modified (\textit{i.e.}
they are already mass eigenstates) at this order.
Now we only have to
introduce these rotations in the gauge and Yukawa couplings to obtain
the couplings among the physical fields. The couplings to the $Z$ are,
in the current and mass eigenstate basis, respectively
\begin{eqnarray}
\mathcal{L}^Z 
&=& -\frac{g}{2c_W} Z_\mu 
[\bar{q}_R^{u(0)} \gamma^\mu q_R^{u(0)} 
-\bar{\chi}_R^{d(0)} \gamma^\mu \chi_R^{d(0)} 
-2 s_W^2 J^{R\mu}_{2/3}] 
\nonumber \\
&=&\mathcal{L}^Z_0-\frac{g}{2c_W} Z_\mu 
\left[
\frac{m_{\chi,t}}{M_\chi} \bar{t}_R \gamma^\mu \chi^d_R
-\frac{m_{q,t}}{M_q} \bar{t}_R \gamma^\mu q^u_R + \mathrm{h.c.}
\right] + \mathcal{O}\left(\frac{m^2}{M^2}\right),
\end{eqnarray}
where $J^{R\mu}_{2/3}
=(2/3)(\bar{t}_R^{(0)}  \gamma^\mu t^{(0)}_R
+\bar{q}_R^{u(0)} \gamma^\mu q_R^{u(0)} 
+\bar{\chi}_R^{d(0)} \gamma^\mu \chi_R^{d(0)}) 
=(2/3)(\bar{t}_R  \gamma^\mu t_R
+\bar{q}_R^{u} \gamma^\mu q_R^{u} 
+\bar{\chi}_R^{d} \gamma^\mu \chi_R^{d}) $ is the electromagnetic
current for the RH charge $2/3$ quarks and $\mathcal{L}^Z_0$ is equal
to the original Lagrangian with the replacement $\psi^{(0)}\to \psi$.
Similarly we obtain, for the charge currents
\begin{equation}
\Delta \mathcal{L}^W= \mathcal{L}^W-\mathcal{L}^W_0=\frac{g}{\sqrt{2}} W^+_\mu
\left[\frac{m_{q,t}}{M_q}\bar{t}_R \gamma^\mu q^d_R
+\frac{m_{\chi,t}}{M_\chi}\bar{\chi}^u_R \gamma^\mu t_R
\right]+\mathrm{h.c.}+ \mathcal{O}\left(\frac{m^2}{M^2}\right),
\end{equation}
and Yukawa couplings
\begin{equation}
\Delta \mathcal{L}^H=\mathcal{L}^H-\mathcal{L}^H_0
= \frac{H}{v} [ 
m_{q,t}\bar{q}^u_L t_R
+m_{\chi,t}\bar{\chi}^d_L t_R + \mathrm{h.c.}]+ \mathcal{O}\left(\frac{m^2}{M^2}\right).
\end{equation}
Diagonal couplings and masses are only modified at
$\mathcal{O}(m^2/M^2)$~\cite{delAguila:1982fs} which, for the purpose
of their collider implications can usually be considered a negligible
correction.
(Of course if the vector-like quark masses are close to the
electroweak scale, $M\sim v$, 
this expansion breaks down and a full numerical diagonalization would
be required). 
Using these couplings we can compute the decay width of the different
heavy quarks. The result is extremely simple in the large $M$ limit,
in which we obtain
\begin{eqnarray}
&&\Gamma(\chi^u\to W t)\approx 
2\Gamma(q^u\to Z t) \approx 2\Gamma(q^u\to H t) \approx
\frac{g^2}{64 \pi} \frac{m_{q,t}^2
  M_q}{M_W^2}, 
\\
&&
\Gamma(q^d\to W t) \approx 
2\Gamma(\chi^d\to Z t) \approx 2\Gamma(\chi^d\to H t) \approx
\frac{g^2}{64 \pi} \frac{m_{\chi,t}^2
  M_\chi}{M_W^2}. 
\end{eqnarray} 
Note the $2:1:1$ pattern of decays into $W,Z,H$ as predicted by the
Equivalence Theorem. This 
means that, for heavy vector-like quarks, the decay is mainly into
longitudinal gauge bosons (or equivalently in the Goldstone bosons in
a gauge other than the unitary one).

Vector-like quarks can be pair or singly produced. Pair 
production is governed by QCD and is therefore model-independent. 
There is a very strong suppression at large masses
which makes the reach at the LHC $\sim 1-2$
TeV~\cite{AguilarSaavedra:2005pv,AguilarSaavedra:2006gv}.  
However, pair production can
give rise to spectacular signatures. For instance, pair production of
$\chi^u$ or $q^d$ results in a final state with four $W$ plus two $b$
quarks. An analysis of
the reach in that case, using like-sign leptons will be discussed below.
Single production occurs through the off-diagonal couplings between SM
and heavy quarks. Although it is model-dependent, 
the suppression at large masses is milder than in pair
production. Thus, if the new quarks mix sizably with light quarks
(including the bottom), single production can give a better reach than
pair production~\cite{Azuelos:2004dm}.

\section{CONCLUSIONS}

Vector-like quarks are a common prediction of models of new
physics. They are in general compatible with EWPT and can
give rise to spectacular signals at the LHC. We have reviewed the
tools needed to compute their collider implications (masses and couplings)
and discussed a particular example relevant for realistic composite
Higgs and Higgsless models. Other possibilities have been discussed
elsewhere, with special emphasis on the case of a vector-like singlet
(for a recent review see~\cite{Buescher:2006jm}, page 72, and
references there in).

\section*{ACKNOWLEDGEMENTS}

We thank R. Contino, B. Dobrescu, C. Grojean and especially
E. Pont\'on for helpful comments. This work is supported by 
SNSF under contract 200021-117873. 

\AddToContent{J.~Santiago}
\setcounter{figure}{0}
\setcounter{table}{0}
\setcounter{section}{0}
\setcounter{equation}{0}
\setcounter{footnote}{0}
\clearpage
%

\superpart{Nonleptonic Final States}


\part{Top-partner mass reconstruction by using jets}

{\it M.M.~Nojiri and M.~Takeuchi}


\begin{abstract}
At the LHC,
the top-partner ($T_-$) and its antiparticle is  produced in pairs
in the Littlest Higgs model with T-parity.
Each top-partner decays into top quark ($t$) and the lightest 
$T$-odd gauge partner  $A_H$, and $t$ decays into three jets.
We demonstrate the reconstruction of $t$ decaying hadronically, 
and measure the top-partner mass 
from the  $m_{T2}$ distribution. 
We also discuss the dependency on  four jet reconstruction 
algorithms (simple cone, kt, Cambridge, SISCone).
\end{abstract}

\section{TOP PARTNERS IN THE LITTLEST HIGGS MODEL WITH T PARITY}
\subsection{Productions and decays}
We consider the Littlest Higgs model with T parity (LHT).
In the following, we assume 
the top partner is the lightest in the SM fermions' partners. 
The top partner may be produced in pairs  and decays at LHC  as,
\begin{eqnarray}
pp \rightarrow T_-\overline{T}_- \rightarrow t\bar{t}A_H A_H 
\rightarrow bW^+\bar{b}W^- A_H A_H 
\rightarrow 6j + E{\!\!\!/}_T.
\end{eqnarray}
This process is similar to the  top squark ($\tilde t$) production process in the MSSM.
However, the top partner production cross section 
is larger than that of top squarks when the masses are same, because 
the top partner is a fermion. 
At the LHC, the $T_-\overline{T}_-$ production cross section is 0.171 pb
for the top partner mass $m_{T_-}$ set to 800 GeV.
Once the top partner is produced,
it decays into top and the heavy photon $A_H$ (We set to 150GeV).
The branching ratio is 100\%
when the top partner is the next lightest T-odd particle.
To identify this process, it is important to identify two tops and missing transverse momentum $E{\!\!\!/}_T$.

To simulate the top partner reconstruction we generated 8,550 $\tilde{t}\tilde{t}^\ast$ events with HERWIG6.5 (expected number of events for 50fb$^{-1}$).
We use a similar MSSM model point instead of the LHT for the event generation. 
The distribution of production and decay   is almost correct 
except for the spin correlations of $T_-$ decays, so we 
do not simulate top partner polarization effects but top polarization is taken into account.
The main source of SM background is $t\bar{t}$ events,
whose production cross section is 463 pb at tree level.  
We also generated 4,630,000 $t\bar{t}$ events (for 10fb$^{-1}$).
When we compare signal and background distributions,
we simply multiply  $t\bar{t}$ distributions by a factor of 5. We 
ignore the $t\bar{t}Z$ background  because this can be calibrated with $t\bar{t}Z(\rightarrow l^+ l^-)$
events easily.

The process was already analysed in \cite{Matsumoto:2006ws},
where AcerDET1.0 \cite{RichterWas:2002ch} was used for 
the detector simulation and jet reconstruction.
We also use AcerDET1.0 for the detector simulation. 
In addtion to that, we interfaced  the calorimeter information to FastJet2.2beta \cite{Cacciari:2006sm}  which allows us to compare different jet 
reconstruction algorithms (kt, Cambridge, SISCone).
To compare the four jet algorithms under the same conditions (in Sec \ref{toppartner_label2}),
we switch off jet energy smearing of AcerDET. 
In this section, we only show the results with the Cambridge algorithm.

\subsection{Event selection and Top mass reconstruction}
\label{toppartner_label1}
We apply successive cuts to select $T_-\overline{T}_-$ events, the summary of the cuts is shown 
in Table \ref{toppartner_label3}.
First, we imposed the standard cut to collect events related with new physics,
\begin{eqnarray}
E{\!\!\!/}_T \ge 200{\rm GeV\ and}\ E{\!\!\!/}_T \ge 0.2M_{\rm eff},\ \ 
n_{\rm lepton}=0.
\end{eqnarray}
The lepton cut is for dropping $t\bar{t}$ production events,
in which large $E\!\!\!/_T$ is
dominantly caused by neutrinos from the leptonic decay of top.
We do  not assume any b tagging, because for the given kinematics 
jets are very collinear with each other, and the $b$ tagging efficiency 
has not been studied in that case. 

We applied the hemisphere analysis to find top candidates\cite{hemisphere}. 
Each of the jets was assigned to hemispheres which were defined as follows;
\begin{eqnarray}
&&\forall i \in H_1, j \in H_2 \ \ \ \ \ \  \ \ d(p_{H_1},p_{i})\le d(p_{H_2},p_{i}) {\rm \ and\  }
d(p_{H_2},p_{j})\le d(p_{H_1},p_{j}).
\cr
&&\ \ \  \ \ d(p_{1},p_2)\equiv
\frac{(E_{1}-|\mathbf{p}_{1}|\cos\theta)E_{1}}{(E_{1} + E_{2})^2}
\ \ \  
\ (\theta
\ {\rm is\  the\ angle\ between\ }
\mathbf{p}_1\ {\rm and}\ \mathbf{p}_2).
\end{eqnarray}
Here,  $p_{H_j}\equiv\sum_{i \in H_j} p_{i}$, 
and we  required both hemispheres' transverse momenta to be larger than a threshold,
\begin{eqnarray}
p_{T,H_1},p_{T,H_2} > 200 {\rm GeV}.
\end{eqnarray}

Top quarks from $T_-\overline{T}_-$ production
are highly boosted. 
Therefore decay products from two top quarks  are correctly
grouped into two hemispheres with  high probability.
We analysed  the events in which each hemisphere has up to 3 jets 
to drop contribution from other QCD jets. 

\begin{figure}[!ht]
\begin{tabular}[t]{lr}
\begin{minipage}[t]{10cm}
\centering
\includegraphics[scale=0.23]{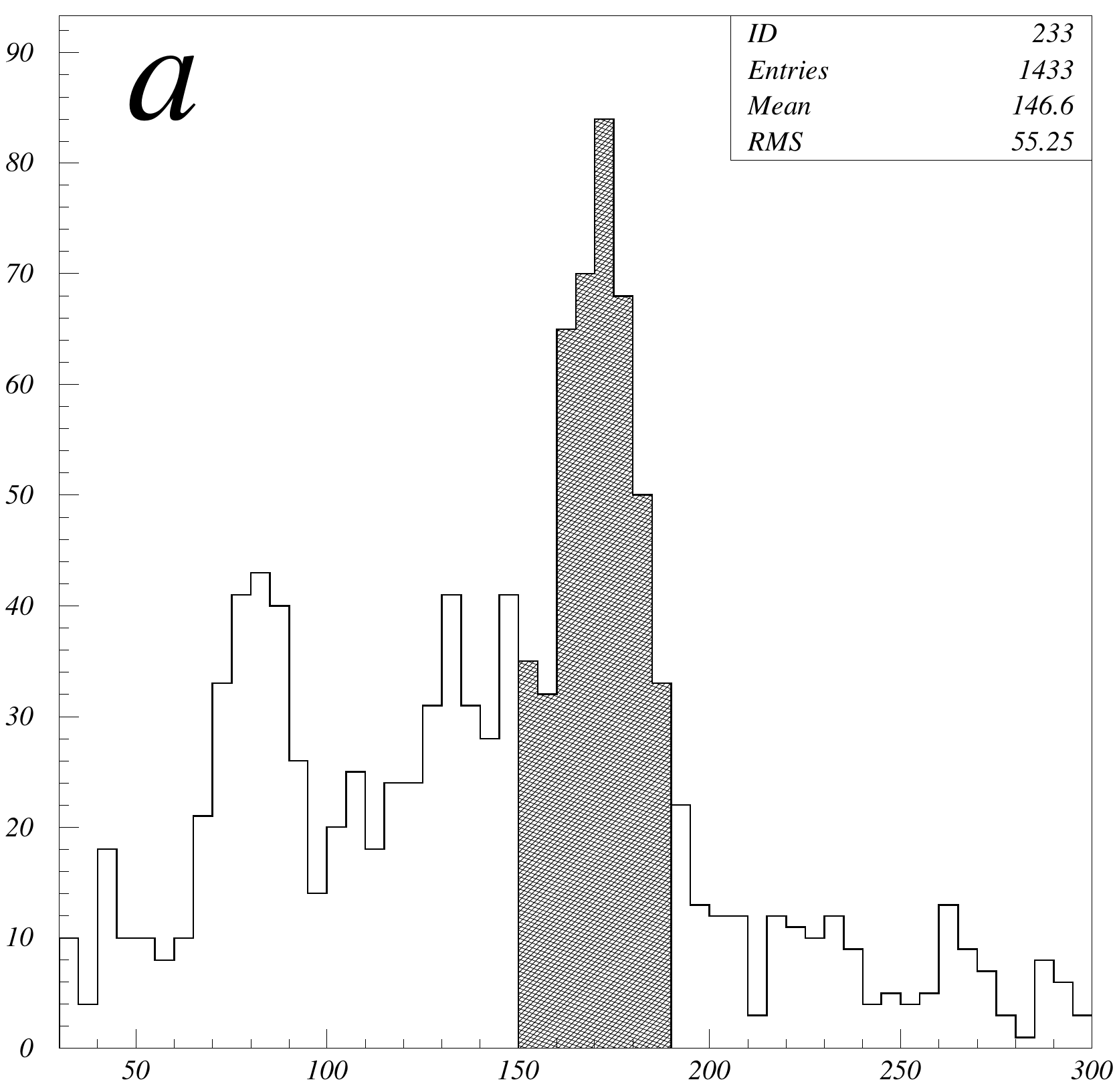}
\includegraphics[scale=0.23]{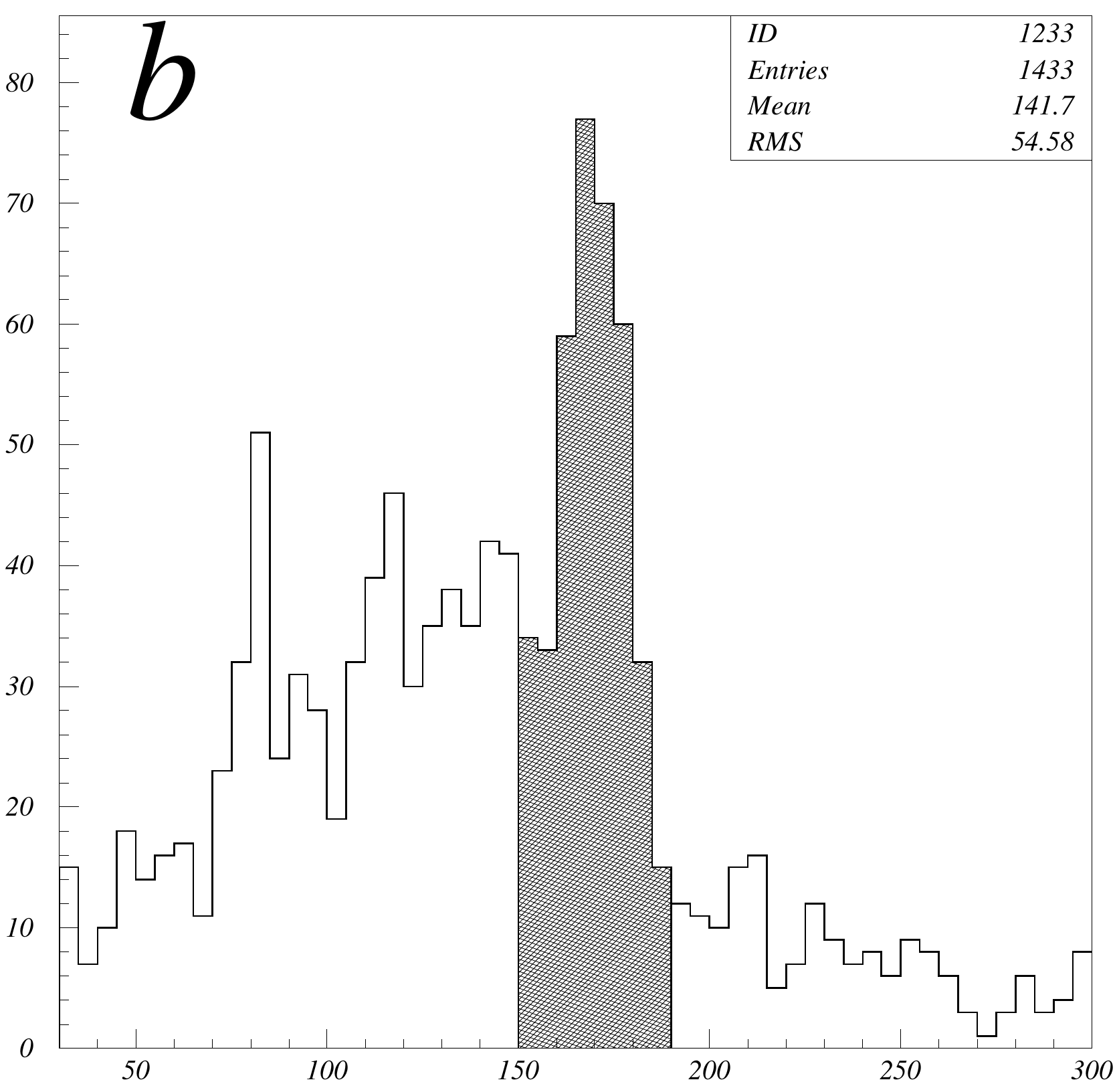}
\\
\includegraphics[scale=0.23]{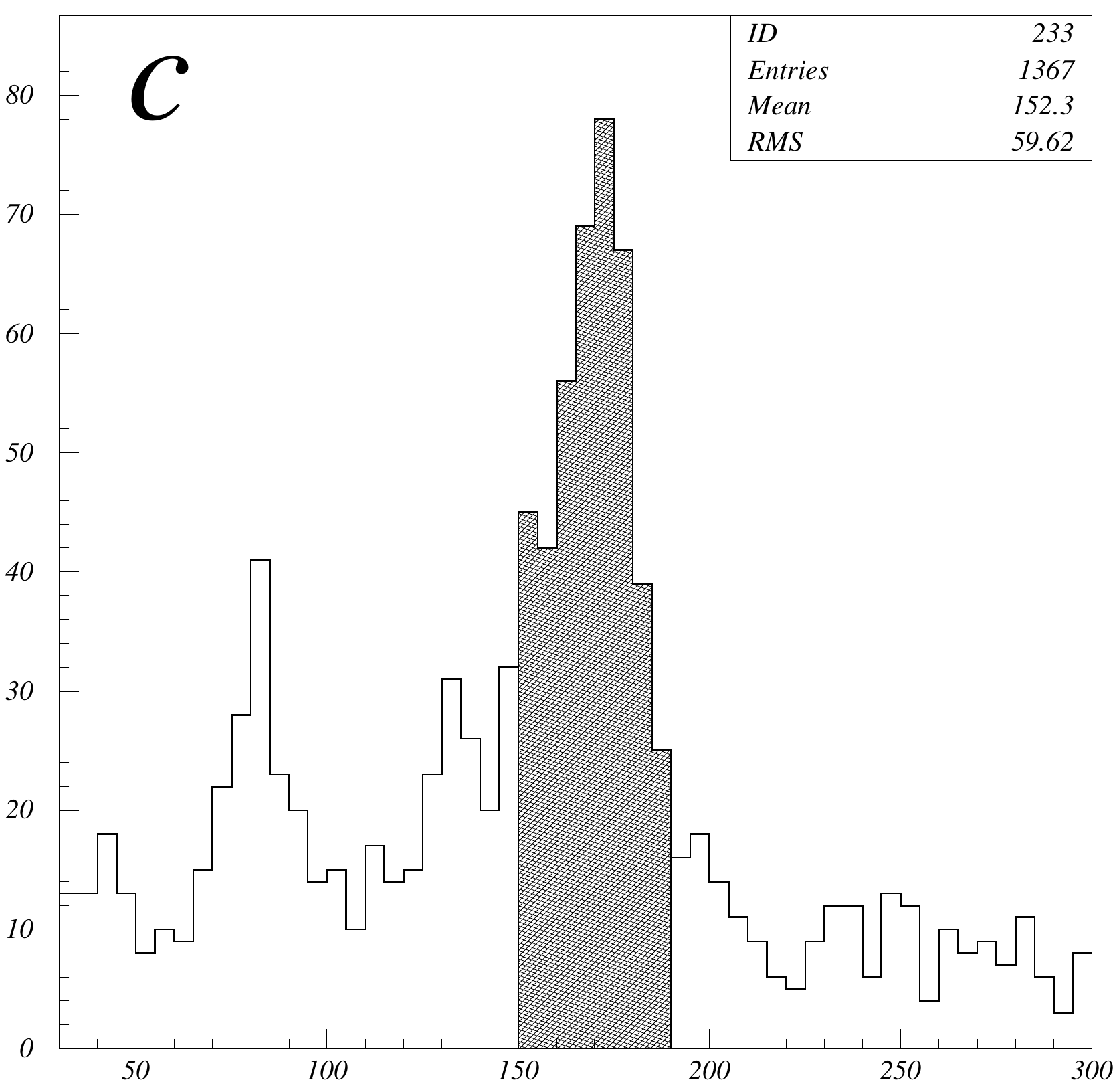}
\includegraphics[scale=0.23]{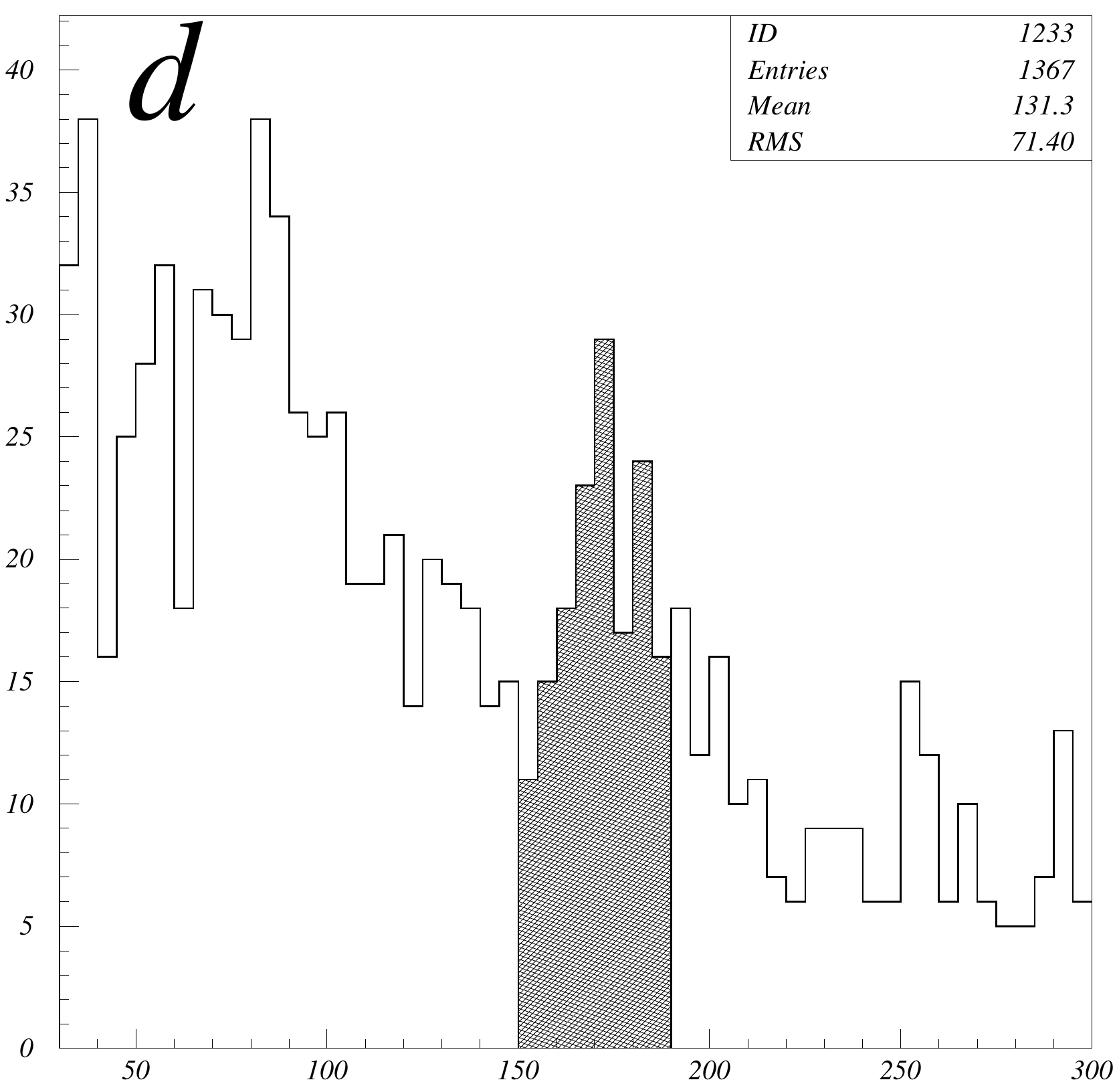}
\caption{The distributions of invariant masses of 
a) $H_1$ for the $T_-\overline{T}_-$ events,
b) $H_2$ for the $T_-\overline{T}_-$ events,
c) $H_1$ for the $t\bar{t}$ events,
d) $H_2$ for the $t\bar{t}$ events.}
\label{toppartner_label5}
\end{minipage}
&
\begin{minipage}[t]{5cm}
\centering
\includegraphics[scale=0.5]{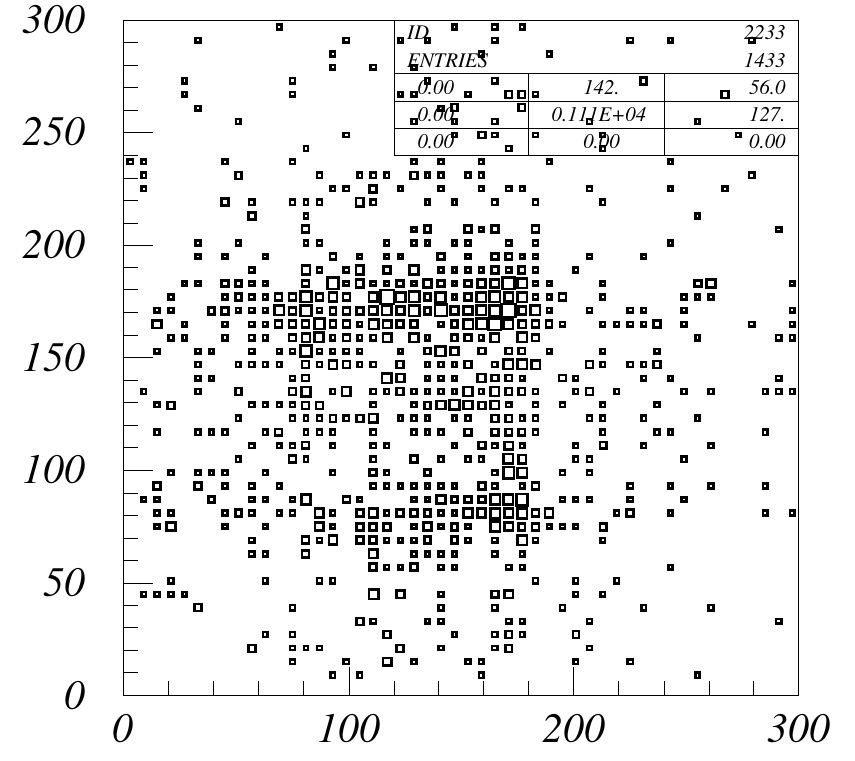}
\\
\includegraphics[scale=0.5]{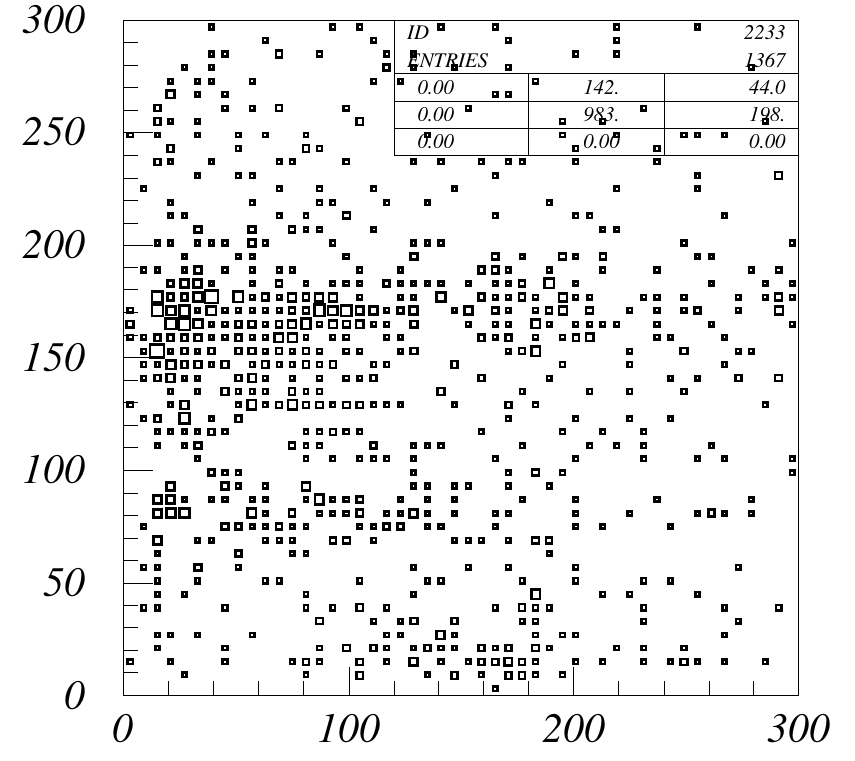}
\caption{$m_{P_{H1}}$ vs. $m_{P_{H2}}$ 
for $T_-\overline{T}_-$ (upper figure) and 
for $t\bar{t}$ (lower figure).
Cambridge algorithm is used for Jet reconstruction.}
\label{toppartner_label6}
\end{minipage}
\end{tabular}
\begin{tabular}{cc}
\begin{minipage}[c]{5cm}
\includegraphics[scale=0.25]{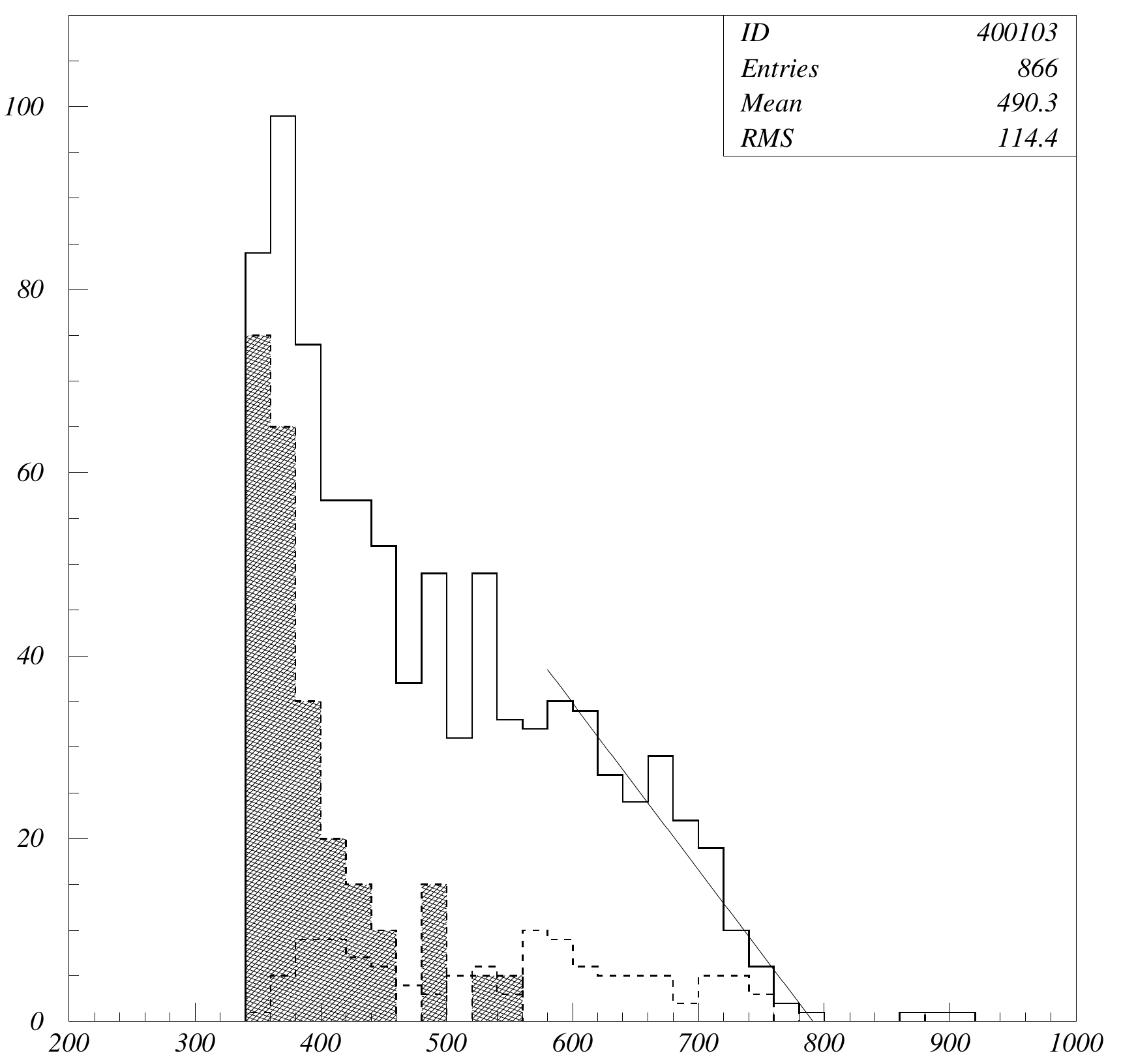}
\end{minipage}
&
\begin{minipage}[c]{10cm}
\caption{The $m_{T2}$ distribution for the Cambridge algorithms, 
where proper $m_{\tilde \chi^0_1}=150$ GeV is used.
The dashed line shows for events with $150{\rm GeV} < m_{P_{Hi}}< 190{\rm GeV}$
The solid line shows for events with $50{\rm GeV} < m_{P_{Hi}} < 190{\rm GeV}$.
The endpoints are 790.95 GeV for nominal $m_{A_H}$.
($m_{T_-}=800$ GeV).
}
\label{toppartner_label4}
\end{minipage}
\end{tabular}
\end{figure}

The distributions of invariant masses of hemispheres ($m_{P_H}\equiv \sqrt{p_H^2}$) for the $T_-\overline{T}_-$ and $t\bar{t}$ events are shown in the Fig \ref{toppartner_label5}.
We can see the top mass peak both for the  $T_-\overline{T}_-$ and $t\bar{t}$ events in the $H_1$.
On the other hand, the top mass is not well reconstructed in the $H_2$ for $t\bar{t}$ events.
This is because at least one of tops must decay leptonically to make large $E\!\!\!/_T$.
Two dimensional scattering plots of $m_{P_{H1}}$ vs. $m_{P_{H2}}$ are also shown in Fig \ref{toppartner_label6}.
We regard a hemisphere's momentum as the top momentum
in this paper if its mass is consistent with the top mass ($150$GeV $<m_{P_H}< 190$GeV).
The matching of a hemisphere's momentum with the parton momentum is discussed in Section \ref{toppartner_label7}. 

\begin{table}
\footnotesize
\begin{tabular}{|c||r|r|r|r|r||r|r|r|r|}
\hline 
   & generated & \scriptsize{$E{\!\!\!/}, M_{\rm eff}$ cut} &\scriptsize{ $n_{\rm lep}=0$} & 
\scriptsize{ $p_{T,H}$ cut }& $n_{{\rm jet},H}\! \le \! 3$ & $m_{P_{H1}}$ &$m_{P_{H2}}$ & \scriptsize{both} $m_{P_{H}}$& 
\scriptsize{relaxed $m_{P_{H}}$} \cr
\hline 
 $T_-\overline{T}_-$  &   8,550 &  6,590  & 4,384& 2276  & 1433 & 437 & 380 & 118 &708\cr
 $t\bar{t}$  & 23,150,000  &  199,640 & 88,540& 9475 & 6835 & 2105 & 765 & 235 & 1835\cr
\hline
 S/$\sqrt{N}$ &1.777  &  14.75  & 14.73 &  23.38 & 17.33 & 9.525 & 13.73 & 7.70 & 16.53  \cr
\hline
\end{tabular}
\caption{Summary of the cuts.
The number of events are for  $\int\!\! dt {\cal L} = 50\,$ fb$^{-1}$.
$p_{T,H}$ denotes the cuts that  requires both hemispheres' momenta $p_{T,H_1},p_{T,H_2}>200$GeV.
$m_{P_{Hi}}$ denotes the cuts that 
the hemisphere $i$'s momentum is consistent
 with top mass ($150$GeV $<m_{P_H}< 190$GeV).
The relaxed $m_{P_H}$ cut denotes that both hemispheres' momenta
satisfy $50$GeV $<m_{P_H}< 190$GeV.
}
\label{toppartner_label3}
\end{table}

\subsection{Measurement of the end point of $m_{T2}$ 
distribution.}
To extract the mass of $T_-$, we used the Cambridge $m_{T2}$ variable 
\cite{Barr:2003rg}.
This variable is considered in events like 
$ \zeta \zeta^\prime \to (a\alpha) (b\beta)$,
where we assume $\zeta$ and $\zeta^\prime$ have the same mass $m_\zeta$,
$a$ and $b$ are visible objects, 
and $\alpha$ and $\beta$ are invisible  particles 
and have the same mass $m_{\tilde \chi^0_1}$.
In the events, the $m_{T2}$ variable is defined as follows:
\begin{eqnarray}
\!\!\!\!
m_{T2}^2(\mathbf{p}_T^{a},\mathbf{p}_T^{b},\mathbf{p} {\!\!\!/}_T;m_{\tilde \chi^0_1}) \equiv 
\min_{\mathbf{p} {\!\!\!/}_T^{\alpha} + \mathbf{p} {\!\!\!/}_T^{\beta} = \mathbf{p} {\!\!\!/}_T}
\left[
\max \left\{
m_T^2(\mathbf{p}_T^a,\mathbf{p} {\!\!\!/}_T^{\alpha};m_{\tilde \chi^0_1}),
m_T^2(\mathbf{p}_T^b,\mathbf{p} {\!\!\!/}_T^{\beta};m_{\tilde \chi^0_1})
\right\}
\right].
\end{eqnarray}
Here, the transverse mass $m_T$ is defined as 
$m_T^2(\mathbf{p}_T^a,\mathbf{p} {\!\!\!/}_T^{\alpha};m_{\tilde \chi^0_1})\equiv m_a^2 + m_{\tilde \chi^0_1}^2 + 2 
\left[ E_T^a E {\!\!\!/}_T^\alpha -  \mathbf{p}_T ^{a}\mathbf{p} {\!\!\!/}_T^{\alpha}\right].
$
By the definition, the $m_{T2}$ variable is a function of $m_{\tilde \chi^0_1}$, and satisfies the condition
$m_{T2}(m_{\tilde \chi^0_1}) \le m_\zeta$.
Then we can measure $m_{\zeta}$ by measuring the upper endpoint of the $m_{T2}$ distribution.

For our purpose, visible particles are two top quarks  and invisible particles are two $A_H$.
We may regard hemispheres' momenta as top momenta for the hemisphere whose masses are 
$150{\rm GeV} < m_{P_{Hi}} < 190{\rm GeV}$.
However, there are not enough events left under this cut.
Therefore  we use a relaxed criterion  $50{\rm GeV} < m_{P_{Hi}} < 190{\rm GeV}$.
Indeed, when $m_{P_{H}}$ is less than $m_t$, all of the top decay products do not contribute to the hemisphere reconstruction due to the minimum jet energy cuts.
The endpoint of $m_{T2}$ does not change because $m_T$ is always underestimated compared to the true $m_T$. 
This is easy to understand if you consider the system of 
a neutralino and other sources of missing momenta 
( for example neutrino ) as one invisible particle.
The invisible particle mass  ($m_{\rm invisible}$) is always larger than the neutralino mass
which results in $m_{T2}(m_{\tilde \chi^0_1}) \le m_{T2}(m_{\rm invisible}) \le m_\zeta$.

The $m_{T2}$ distribution is shown in Fig \ref{toppartner_label4},
where the nominal  value $m_{\tilde \chi^0_1}=150$ GeV is used  for 
the calculation of $m_{T2}$.
We show only the region of $m_{T2} > 350 {\rm GeV}$.
The $t\bar{t}$ events have low $m_{T2}$ values, 
so we can neglect the BG to fit the endpoint.
The dashed lines show the $m_{T2}$ distributions for events with $150{\rm GeV} < m_{P_{Hi}} < 190{\rm GeV}$
and have the same endpoint as those with the relaxed criterion of $50{\rm GeV} < m_{P_{Hi}} < 190{\rm GeV}$ (solid line).
This fact supports the validity to relax the criterion to determine the endpoint.
We fitted the distribution near the endpoint by a linear function and 
obtained  $m_{T2}^{max}=790.95$ GeV.
This value is consistent with the proper value $m_{T_-}=800$ GeV.

It is not possible to determine the top partner mass itself unless the $A_H$ mass 
is determined. If  we assume the $A_H$ thermal relic density 
is consistent with the dark matter density in our universe, $m_{A_H}$  is 
related to the Higgs mass ($m_H$) so that it is determined with two fold ambiguity \cite{Matsumoto:2006ws}.

\section{COMPARISON AMONG JET RECONSTRUCTION ALGORITHMS}
\label{toppartner_label2}
We now make a comment on the dependence on 
the jet reconstructing algorithm.
Four algorithms for jet reconstruction 
(Simple cone, kt,
Cambridge, SISCone) 
are used in the following analysis.
For this study, we interfaced 
FastJet2.2beta \cite{Cacciari:2006sm} to AcerDET1.0.  The  energy deposits 
in calorimeter cells are regarded as "particle momentum". 
Cone sizes (or the counterparts for clustering algorithms) $R$ are chosen as 0.4 and overlap parameter $f$ as 0.5 for SISCone.
Simple cone denotes the algorithm used in AcerDET1.0. 

\subsection{Invariant mass distributions}
\label{toppartner_label7}
The results are summarized in Table \ref{toppartner_label8}.
Acerdet has an option to calibrate jet momenta, therefore 
both calibrated and non-calibrated numbers  are given for the 
simple cone algorithm in the table. The calibrated result is shown 
in  the figures.

\begin{table}[b]
\centering
\footnotesize
\begin{tabular}{|c||r|r||r|r|r|r||r|}
\hline 
For $ T_-\overline{T}_-$ events    & $E{\!\!\!/}, M_{\rm eff}$ cut &
 $n_{{\rm jet},H} \le 3$ & $m_{P_{H1}}$ &$m_{P_{H2}}$ & both $m_{P_H}$& relaxed $m_{P_H}$ &
\scriptsize{ $m_{T2}$ endpoint} \cr
\hline 
Simple Cone \scriptsize{ (calibrated)}&6673&  945 & 306 & 326 & 110 & 439 & 795.69 GeV\cr
$\ \ \ \ \ \ \ \ \ \ \ \ \ $\scriptsize{ (not calibrated)}&
\scriptsize{6673}&\scriptsize{ 1213} &\scriptsize{  283} & \scriptsize{ 323 }&\scriptsize{   82 }
& \scriptsize{ 545} & \scriptsize{ 779.37 GeV}\cr
 kt                    &6673& 1436 & 444 & 384 & 105 & 604 & 780.62 GeV\cr
 Cambridge             &6673& 1433 & 437 & 380 & 118 & 621 & 790.95 GeV\cr
 SISCone               &6673& 1656 & 512 & 437 & 150 & 608 & 803.67 GeV\cr
\hline
\end{tabular}
\caption{Summary of the cuts for various jet algorithms.
cut is the same as Table \ref{toppartner_label3}.}
\label{toppartner_label8}
\end{table}

\begin{figure}[htbp]
\begin{tabular}{ccc}
\begin{minipage}[t]{7.8cm}
\includegraphics[scale=0.45]{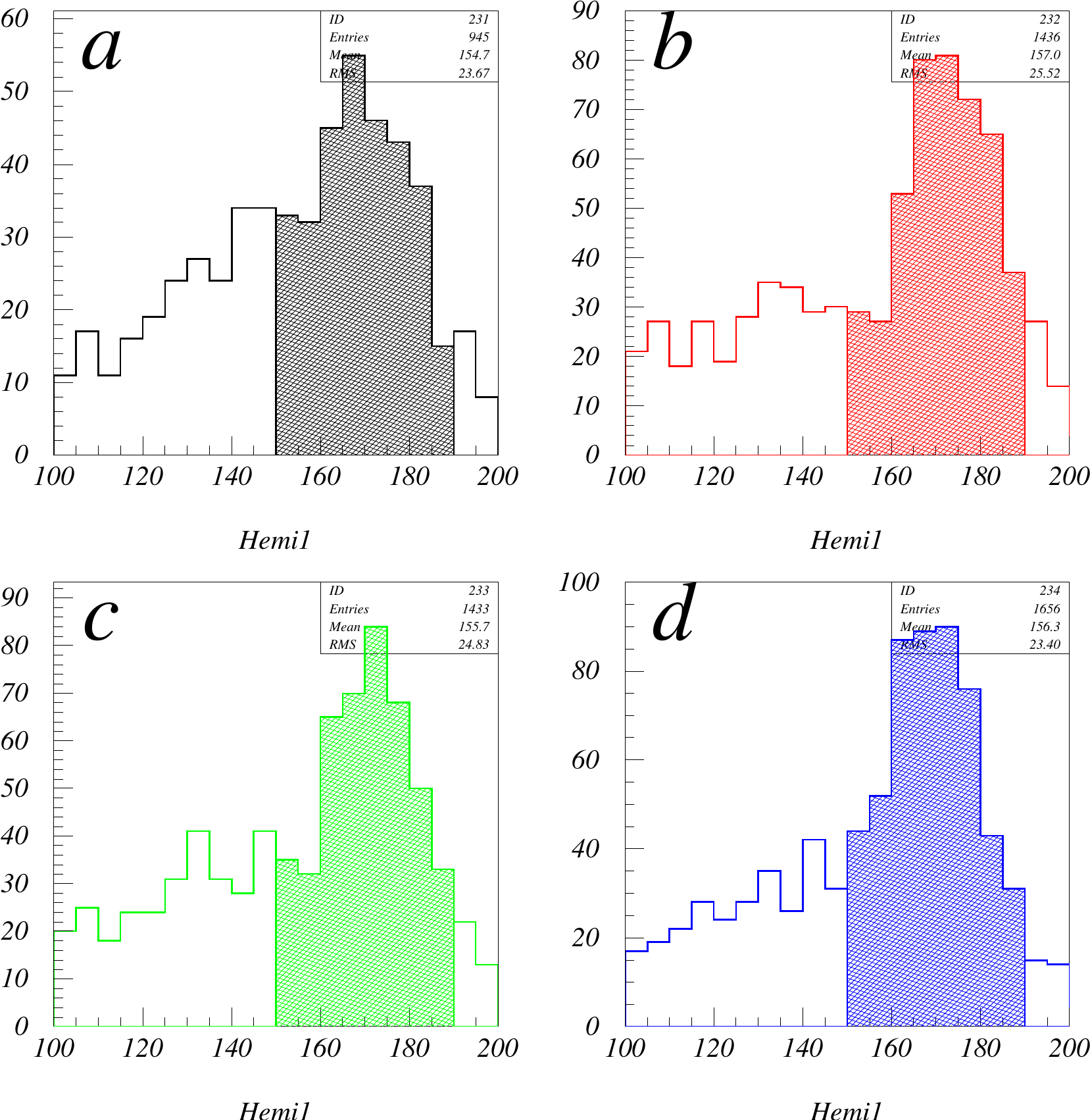}
\caption{The distributions of  $m_{P_{H_1}}$ for the $T_-\overline{T}_-$ events
for a) Simple Cone (Acerdet), b) kt, c) Cambridge and d) SISCone.
Acerdet has dull peak and broad tail 
and the position located below $m_t$}
\label{toppartner_label9}
\end{minipage}
&
\begin{minipage}[t]{4cm}
\includegraphics[scale=0.45]{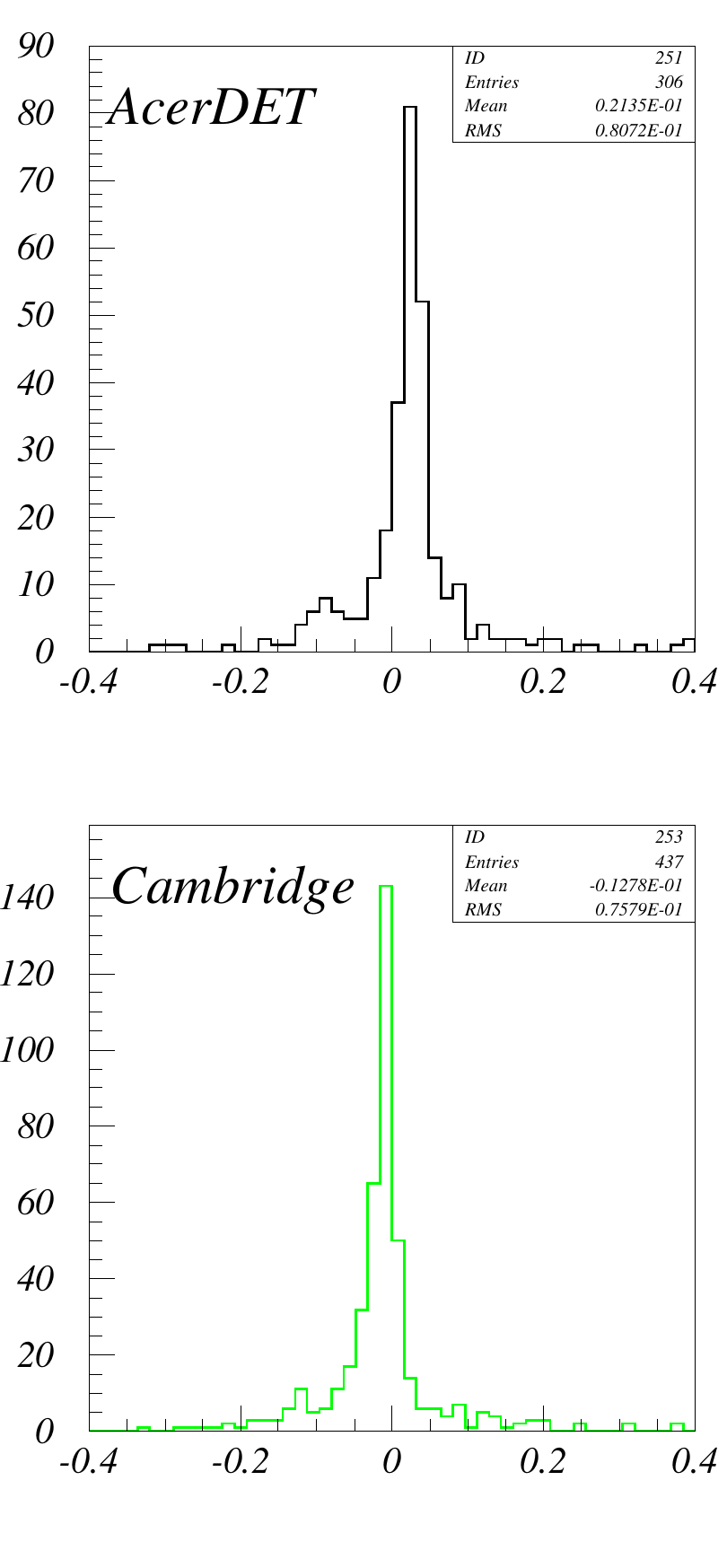}
\caption{$(\Delta p_T)/p_T$ distributions of the hemisphere $H_1$ 
selected by 150 GeV$< m_{P_{H_1}} <$ 190 GeV for AcerDET with jet calibration  (upper) and Cambridge (lower).
}
\label{toppartner_label10}
\end{minipage}
&
\begin{minipage}[t]{4cm}
\includegraphics[scale=0.45]{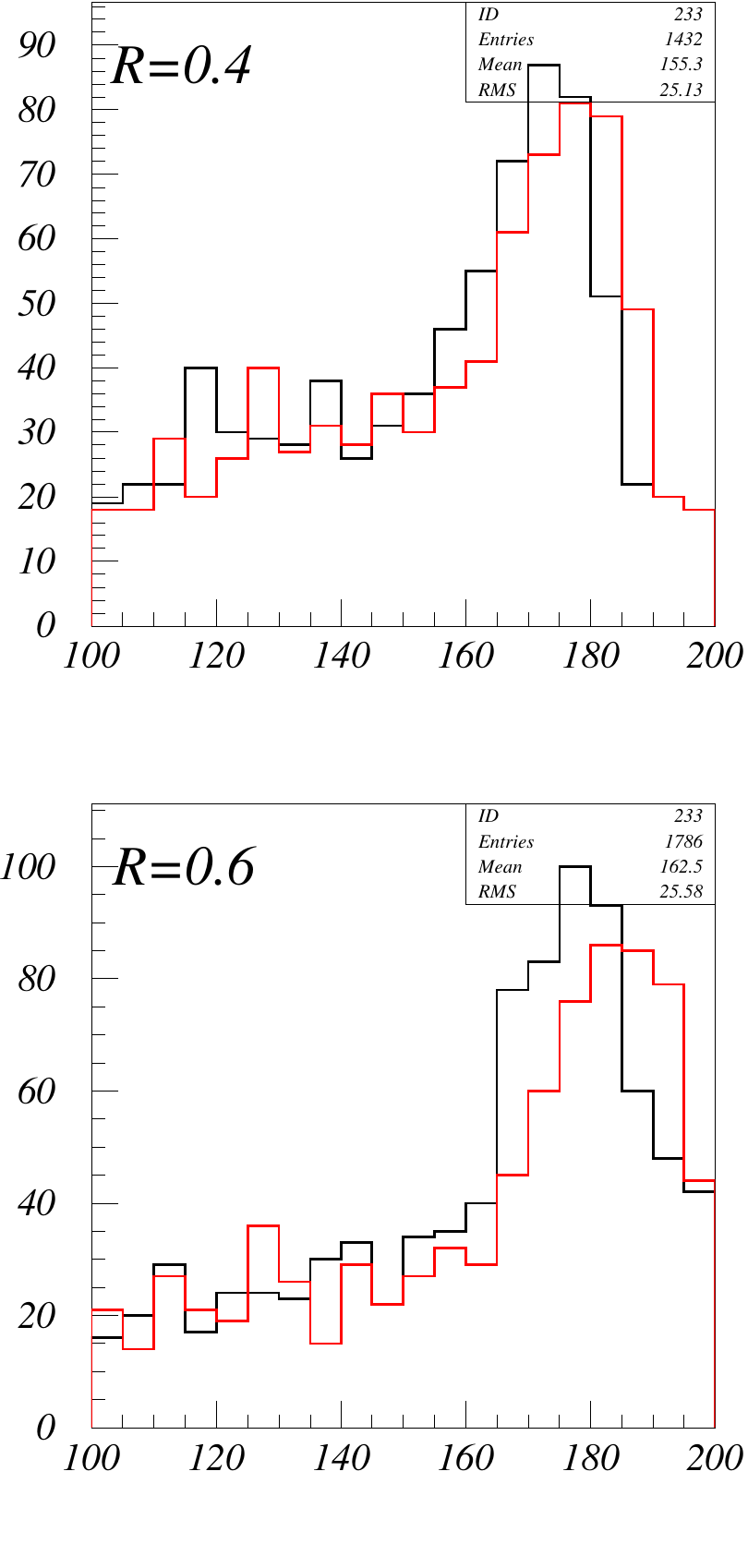}
\caption{The comparison between  Cambridge (Solid) and kt (dashed) algorithms with underlying events.
Cone sizes of jets are $\Delta R=0.4$ (upper) and $0.6$ (lower).
}
\label{toppartner_label11}
\end{minipage}
\end{tabular}
\end{figure}

The distributions of $m_{P_{H1}}$ 
are shown in Fig \ref{toppartner_label9}.
The shaded regions denote that the hemisphere's invariant mass is
consistent with $m_t$ ($150{\rm GeV} < m_{P_H} < 190{\rm GeV}$).
The peak for the Simple cone algorithm is dull and has a broad tail
and the position of the peak is located below $m_t$. 
This is because  AcerDET takes massless jets. 
Therefore the probability to find the jets consistent with the 
top mass 
is relatively low among the four algorithms. 
The acceptance becomes worse as the lower bound of an invariant mass cut for $m_{P_H}$ is increased.  
If the other jet definitions are feasible 
in the LHC environment, the reconstruction efficiency can be increased significantly. 
On the other hand, the endpoints of the $m_{T2}$ distributions are
795.69 (779.37) GeV (Simple cone),
780.62 GeV (kt),
790.95 GeV (Cambridge),
803.67 GeV (Siscone) respectively.
The dependence on jet algorithms is not severe for this analysis.

Fig \ref{toppartner_label10}
shows the  deviation of the  hemisphere momentum 
from the true top quark momentum  $(\Delta p_T)/p_T$. 
We selected signal events with 150 GeV$< m_{P_{H1}}<$ 190 GeV
for the Simple cone and Cambridge algorithms.
They are mostly distributed in a $\pm 5$\% region.
We can see the $(\Delta p_T)/p_T$ are  larger than 0  by about 2\% 
for the Simple cone. 
The positive contribution is due to  jet  $p_T$ calibration of AcerDET. 

\subsection{Effect of Underlying Events}
So far we have ignored the effect of underlying events 
when generating events. We now investigate the  reconstruction with 
underlying events by comparing the kt algorithm and Cambridge algorithm.
We generated the underlying events using 
JIMMY with HERWIG6.5.
Fig \ref{toppartner_label11} shows the distributions of invariant masses of $P_{H_1}$
for the kt and Cambridge algorithms with underlying events.
The event selection cuts are the same as in
section \ref{toppartner_label1}.
We can see that the position of the peak for the kt algorithm is larger than 
that for the Cambridge algorithm and the distribution is smeared  for increasing 
$R$. 
This is because the
kt algorithm over-collects contributions which are far from the jet 
direction (large $R_{ij}$)
due to the factor $\min(k^2_{ti},k^2_{tj})$ in the definition of distance
(splash-in effects) .
The Cambridge algorithm does not have the factor $\min(k^2_{ti},k^2_{tj})$ 
therefore it is  not too sensitive to the existence of underlying events.

\section*{CONCLUSIONS}
We have reconstructed top quarks  from $T_- \overline{T}_-$ 
production and the decay into top and stable  gauge partner 
in the LHT   for the top quark decaying hadronically. 
The main background from SM processes is $t\bar{t}$ production events.
They can be reduced by applying the hemisphere analysis and imposing a cut for 
hemisphere momenta.
The top partner mass can be  measured by using $m_{T2}$ variable.

We have also investigated the dependence on the jet reconstructing algorithms.
AcerDET takes  massless jets, so invariant masses are significantly 
underestimated for the boosted top quark. We therefore present our result 
for the Cambridge jet reconstruction algorithm. 
We find that the kt algorithm over collects far activities and the invariant masses are overestimated (Splash-in effect) compared with the Cambridge algorithm.

\AddToContent{M.M.~Nojiri and M.~Takeuchi}
\setcounter{figure}{0}
\setcounter{table}{0}
\setcounter{section}{0}
\setcounter{equation}{0}
\setcounter{footnote}{0}
\clearpage


\part[Searches for a paraphoton in $e^+e^-$collisions]{Searches for a paraphoton in associated production with $t \bar{t}$ in $e^+e^-$collisions}

{\it E.~Boos, V.~Bunichev and H.J.~Schreiber}


\begin{abstract}
We discuss prospects to search for a new massless neutral gauge boson, 
the paraphoton, in $e^+e^-$ collisions at center-of-mass energies of 0.5 
and 1 TeV. 
The paraphoton naturally appearing in models with abelian kinetic 
mixing has interactions with the Standard Model fermion fields 
being proportional to the fermion mass and growing with energy.
At the ILC, potentially the best process to search for the
paraphoton is its radiation off top quarks.
The event topology of interest is a pair of acoplanar top quark 
decaying to jets and missing energy. 
Applying a multivariate method for signal selection
expected limits for the top-paraphoton coupling are derived. 
Arguments in favor
of the missing energy as the paraphoton with spin 1 are shortly discussed.
\end{abstract}

\section{INTRODUCTION}
Although the Standard Model does not require any additional
gauge bosons it is possible to introduce gauge invariant
operators in the Lagrangian which involve new gauge
fields. An example is given in \cite{Holdom:1985ag} 
(see in addition \cite{Goldberg:1986nk,Dobroliubov:1989mr,Mohapatra:1990vq}) 
by the abelian kinetic mixing
of the SM $U_Y(1)$ field with a new $U_P(1)$ field in a gauge invariant manner. 
The mixing term of the two $U(1)$ fields 
can be diagonalized and canonically normalized by an $SL(2,R)$
transformation in a way that one linear combination
of the fields corresponds to the ordinary photon which
couples in the usual manner to all electrically charged particles within the SM.
The other linear combination appears as a massless spin-1 neutral particle,
referred to as the "paraphoton" in
\cite{Okun:1982xi,Okun:1983vw,Ignatiev:1978xj}
and denoted by $\gamma^{\prime}$ in this note.
The paraphoton couples only indirectly to the SM fields
via higher dimensional operators as was worked out in 
\cite{Dobrescu:2004wz}.
The effective interactions of the 
paraphoton with the SM fermions following from the higher 
dimensional operators have the chirality flip structure, proportional to the SM 
fermion masses and inversely proportional
to new physics scale $M$ squared: 
\begin{eqnarray}
\frac{1}{M^2} F^{\gamma^{\prime}}_{\mu\nu} \left(\bar{q}_L \sigma^{\mu\nu} C_u 
\tilde{H} u_R
+ \bar{q}_L \sigma^{\mu\nu} C_d H d_R + \bar{l}_L \sigma^{\mu\nu} C_e H e_R + h.c\right),
\end{eqnarray}
where $F^{\gamma^{\prime}}_{\mu\nu}$ is the paraphoton field strenth, $q_L,l_L$ 
are the quark and lepton doublets, $u_R,d_R$ the up and
down-type $SU(2)$ singlet quarks, $e_R$ the electrically-charged
$SU(2)$-singlet leptons, and $H$ is the Higgs doublet.
The dimensionless coupling parameters $C_f=C_u,C_d,C_e$
after the Higgs field gets its vev $v_h$ are re-expressed in the form
$C_f=c_f*m_f/(v_h/\sqrt{2})$. The coefficients $c_f$ are unknown,
but various phenomenological constraints exist. 
Discussions on possible lower limits on $\gamma^{\prime}$ interactions
with fermions can be found in ref.\cite{Dobrescu:2004wz}.

As follows from the Lagrangian (1) and existing bounds on the 
couplings $c_f$ for light fermions the paraphoton couples most 
strongly to the top quark and very weakly to light fermions.
Therefore, one expects the most interesting process to search
for the paraphoton is $\gamma^{\prime}$ radiation
off the top. Since so far no constraint on $c_t$ exists,
access to $M/\sqrt{c_t}$ seems possible or corresponding
limits might be set for the first time.
It seems a priori very difficult to perform $\gamma^{\prime}$ searches 
at hadron colliders because of very large $ t\bar{t}$ + multi-jet 
background.
The next generation $e^+e^-$ linear collider (ILC) provides 
potentially a possibility to search for the paraphoton via the channel
\begin{eqnarray}
e^+e^- \to t~\bar{t}~\gamma^{\prime}~.
\end{eqnarray}

\section{THE SIGNAL REACTION}

The parameter scan was done for various couplings and detailed
simulations of $t\bar{t}\gamma^{\prime}$ signal events
were performed for a 'reasonable' value of the effective
coupling parameter $M/\sqrt{c_t}$ ($M/\sqrt{c_t}$ = 0.2 TeV) 
when the signal is large enough to be clearly distinguishable from the SM 
background. All the computations and simulations were performed 
for the ILC collision energies $\sqrt{s} = 0.5 $ and 1.0 TeV 
and an integrated luminosity of 0.5, respectively, 1 ab$^{-1}$.

\begin{figure}[ht]
\centering
\includegraphics[width=60mm,height=40mm]{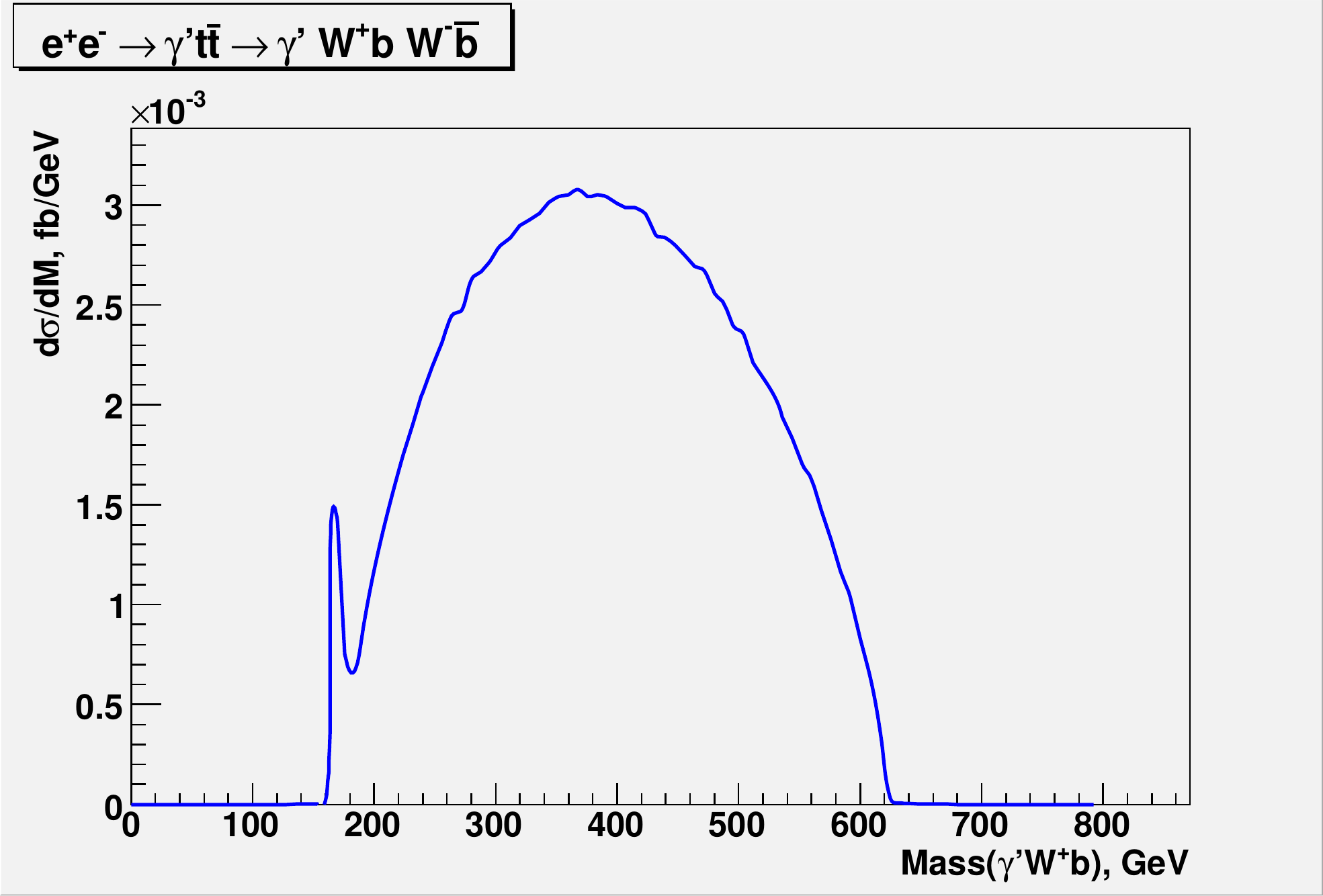}
\includegraphics[width=60mm,height=40mm]{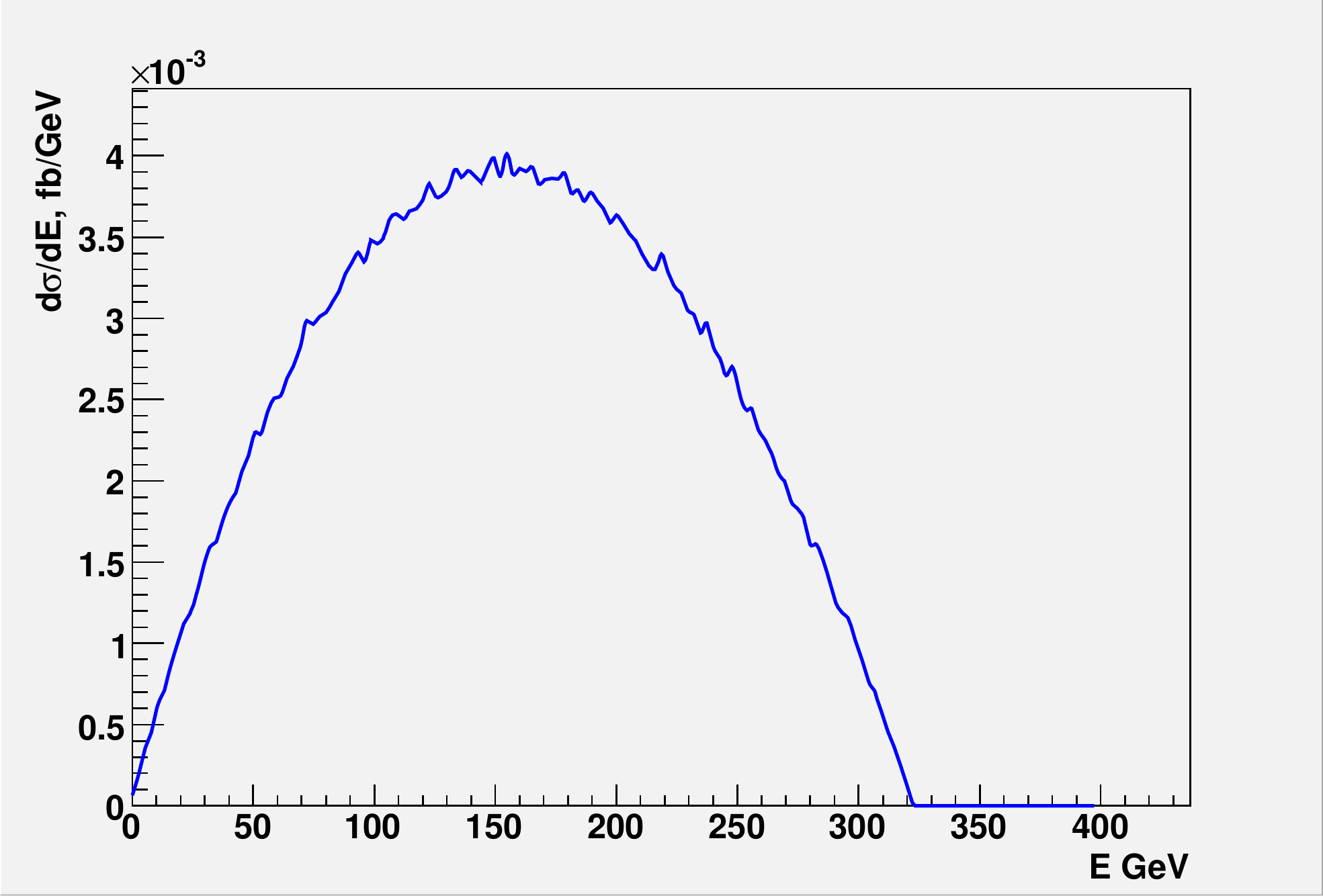}
\caption{Left: Invariant mass of the $\gamma^{\prime} W b$ system.
         Right: $\gamma^{\prime}$ energy distributions at $\sqrt{s} = 1$ TeV.}
\label{paraphoton_inv_mass}
\end{figure}

The characteristics of the signal reaction were computed
and partonic events were generated by means of
the program package CompHEP \cite{Pukhov:1999gg, Boos:2004kh}.
The Feynman rules for the fermion-fermion-$\gamma^{\prime}$  vertices
following from the effective Lagrangian (1)
\begin{eqnarray}
\frac{c_f}{M^2}\cdot m_f \cdot p_{\nu}^{\gamma^{\prime}}
\big(\gamma^\nu \gamma^\mu -\gamma^\mu \gamma^\nu \big) 
\end{eqnarray}
have been implemented into CompHEP.
An interface with PYTHIA 6.202 \cite{Sjostrand:2003wg} allows one
to simulate initial and final state radiation
and jet hadronization. Also,
beamstrahlung effects \cite{AguilarSaavedra:2001rg} are taken into account.
The signal event rate is given in the Table 1.
\begin{table}
\begin{center}
\begin{tabular}{|c|c|c|} \hline
$M/\sqrt{c_t}~~[TeV]$ & $\sqrt{s}=0.5~TeV$ & $\sqrt{s}=1~TeV$\\
\hline
$0.2$ & $5 700$ & $42 500$ \\
\hline
$0.3$ & $1 100$ & $8 500$ \\
\hline
$0.5$ & $40$ & $1 100$ \\
\hline
$1$ & $10$ & $70$ \\
\hline
\end{tabular}
\end{center}
\caption{ $t\bar{t}\gamma^{\prime}$ event rates for several values
  of $M/\sqrt{c_t}$
  at $\sqrt{s} = 0.5$ and 1 TeV and
  an integrated luminosity of 0.5 respectively 1 ab$^{-1}$.}
\end{table}
In order to establish a search strategy for the paraphoton in $t\bar{t}$ events
it is advantageous to know whether an off-shell or on-shell top quark radiates
the $\gamma^{\prime}$. Fig.~\ref{paraphoton_inv_mass} (left)
shows the invariant mass of the $\gamma^{\prime} W b$ system of that top
which radiates the paraphoton. 
Clearly, in most cases the paraphoton is radiated off a top being off-shell,
 and $\gamma^{\prime}$ search strategies should be based on 
on-shell top with $t \to W b$ decays in association with the $\gamma^{\prime}$.
The energy of the $\gamma^{\prime}$ shown in
Fig.~\ref{paraphoton_inv_mass} (right) reflects that the paraphoton-top coupling
is proportional to the paraphoton momentum,  
so that large missing energy, ~\paraphotonme, will be a tag of signal events.

\section{SIGNAL EVENT SELECTION}

After event generation using CompHEP, PYTHIA
and the CompHEP-PYTHIA interface packages with the Les Houches Accord 
implemented\cite{Boos:2001cv} 
an approximate response of an ILC detector was simulated
by means of SIMDET$_{-}$v4 \cite{Pohl:2002vk}.

The most important background consists of ~$t\bar{t}+(\gamma)$ events,
where photons from initial state radiation (ISR) are not 
detected.
The number of events expected for both energies are given in Table 2.   
They exceed substantially
the number of signal events (see Table 1) for the chosen value for the 
parameter $M/\sqrt{c_t}$.
The next significant background to consider is
$e^+e^- \to t\bar{t}+\nu\bar{\nu}$,
with the same signature as for the signal.
The corresponding event numbers also given in Table 2 are
comparable to the signal event rates
for not too small $M/\sqrt{c_t}$ values.
An invariant mass cut of e.g. $M_{\nu\bar{\nu}} <$ 80 GeV,
i.e. a cut on the event missing mass,
removes most of these events.

\begin{table}
\begin{center}
\begin{tabular}{|c|c|c|} \hline
$background$ & $\sqrt{s}=0.5~TeV$ & $\sqrt{s}=1~TeV$\\ 
\hline
$t\bar{t}(\gamma)$ & $276 675$ & $200 310$ \\
\hline
$t\bar{t}\nu\bar{\nu}$ & $75$  & $930$ \\
\hline
\end{tabular}
\end{center}
\caption{Background events at $\sqrt{s} = 0.5$ and 1 TeV for an
integrated luminosity of 0.5, respectively, 1 ab$^{-1}$.}
\end{table}
To discriminate signal and a very large background we use a multivariate 
technique based on likelihood method. 18 kinematics variables, such as missing
energy, invariant masses of various jet combinations etc.  
were combined into a global discriminant variable~$P_P$, designed to give
a measure of the 'Paraphoton-likeness' of any particular event.
As it should be the background events are preferentially distributed at 
low $P_P$ values while for signal events $P_P$ is concentrated close to 
unity. By choosing optimal values of cuts for the discriminant 
$P_P$ one gets the signal selection efficiency of 49\% (76\%) 
at $\sqrt{s}$ = 0.5 (1) TeV, while only 9\% of background events survive.
At $\sqrt{s}$ = 0.5 TeV,
S/$\sqrt{B}$ = 11.96 for $M/\sqrt{c_t} = 0.2$ TeV,
while S/$\sqrt{B}$ = 162.6 at 1 TeV, i.e. the probability of measuring 
the total event rates
as a result of a background fluctuation is $0.5\cdot10^{-12}$ 
and $< 10^{-15}$ at 0.5, respectively, 1 TeV, using Gaussian sampling of 
uncertainties.
In this way, an almost background-free signal event sample can be extracted
for further measurements at 1 TeV. The situation is much less
convenient at 0.5 TeV.

\section{DISCUSSION OF THE RESULTS}

If an excess of signal events over the SM background is established,
limits on the inverse coupling parameter $M/\sqrt{c_t}$
accessible for a significance of  $S/\sqrt{B} = 5$ can be derived
being sufficient for the paraphoton discovery.
The numbers of surviving
$\gamma^{\prime}$ events for 5$\sigma$ discovery
at 0.5 and 1 TeV energies and an integrated luminosity of 0.5 and 1.0 
ab$^{-1}$ can be converted into limits for $M/\sqrt{c_t}$. These limits 
are found to be of 0.33 and 0.61 TeV
for 0.5 and 1 TeV cases respectively. The value $M/\sqrt{c_t}=0.61$ TeV, 
is expected to be the most stringent limit accessible at the ILC.

The signal-to-background ratio, $S/B$, is about 1.79 at 1 TeV, 
sufficiently large to understand a spin assignment of the radiated 
massless particle. In order to demonstrate the spin-1 nature of the 
$\gamma^{\prime}$, we follow studies
performed to establish the vector nature of the gluon in 3-jet $e^+e^-$ 
annihilation events
at PETRA \cite{Brandelik:1980vs,Behrend:1981ng,Berger:1980fa,Burger:1982} and LEP 
\cite{Alexander:1991ak,Adeva:1991dx,Abreu:1991rc} energies,
based on predictions that a spin-$\frac{1}{2}$ quark radiates the spin-1 
gluon.
The analysis of the Ellis-Karliner angle \cite{Ellis:1978wp} distribution and 
the polar angle distribution of the normal to the reaction plane
at 1 TeV clearly shows that spin-1 assignment for the paraphoton 
is highly favored over spin 0.

\section{CONCLUSIONS}

Some realistic extensions of the Standard Model
suggest the existence of a new  massless neutral gauge boson, denoted as
the paraphoton $\gamma^{\prime}$ in this study. This particle is similar to the
ordinary photon, but the interactions of the $\gamma^{\prime}$ are very 
distinct:
couplings to SM fermions are proportional to fermion masses and 
therefore strongest to the top quark, and grow with the $\gamma^{\prime}$ 
momentum. Hence, the paraphoton radiation off
the top at the ILC is studied as the most promising process for the paraphoton 
observation.
Only the all-hadronic top decay mode was selected to ensure a high signal-to-background
ratio and to avoid complications due to final state neutrinos in leptonic W decays.
A multivariate search strategy was used to better separate the signal from 
backgrounds. 
Allowing for a 5$\sigma$
paraphoton discovery significance,
$e^+e^-$ collisions at 1 TeV
allow to bound the $\gamma^{\prime}$-top quark inverse coupling to 
$M/\sqrt{c_t} \paraphotonlsim 0.61$ TeV,
which is expected to be the most stringent limit accessible at the next 
generation colliders.
For the sake of demonstration two angular variables,
the Ellis-Karliner angle and the polar angle of the normal 
to the $t~\bar{t}~\gamma^{\prime}$ plane as a function of a thrust cut-off,
were studied to establish the vector nature of the $\gamma^{\prime}$.
Both angular distributions are in accord with the spin-1 assignment
of the paraphoton and inconsistent with e.g. a scalar hypothesis.

\section*{ACKNOWLEDGEMENTS}

The work of E.B. and V.B. is partly supported by the grant
NS.1685.2003.2 of the Russian Ministry of Education and Science.
V.B. acknowledges  support of the grant of "Dynasty" Foundation.
E.B. and V.B. are grateful to DESY and Fermilab for the kind hospitality.
We thank the organizers of the Les Houches Workshop for a productive scientific atmosphere.
Especially we would like to thank Bogdan Dobrescu for many valuable discussions.

\AddToContent{E.~Boos, V.~Bunichev and H.J.~Schreiber}
\setcounter{figure}{0}
\setcounter{table}{0}
\setcounter{section}{0}
\setcounter{equation}{0}
\clearpage

\part{High $p^T$ Hadronic Top Quark Identification}

{\it G.~Brooijmans}


\begin{abstract}
At the LHC objects with masses at the electroweak scale will for the first 
time be produced with very large transverse momenta.  In many cases, these 
objects decay hadronically, producing a set of collimated jets.
This interesting new experimental phenomenology requires the development 
and tuning of new tools, since the usual reconstruction methods would 
simply reconstruct a single jet.  This study describes the application 
of the YSplitter algorithm in conjunction with the jet mass to identify
high transverse momentum top quarks decaying hadronically.
\end{abstract}

\section{INTRODUCTION}

\begin{figure}[htbp]
\begin{center}
\subfigure[]{
\label{hpttop_fig:dr1}
\includegraphics[width= 
0.48\textwidth]{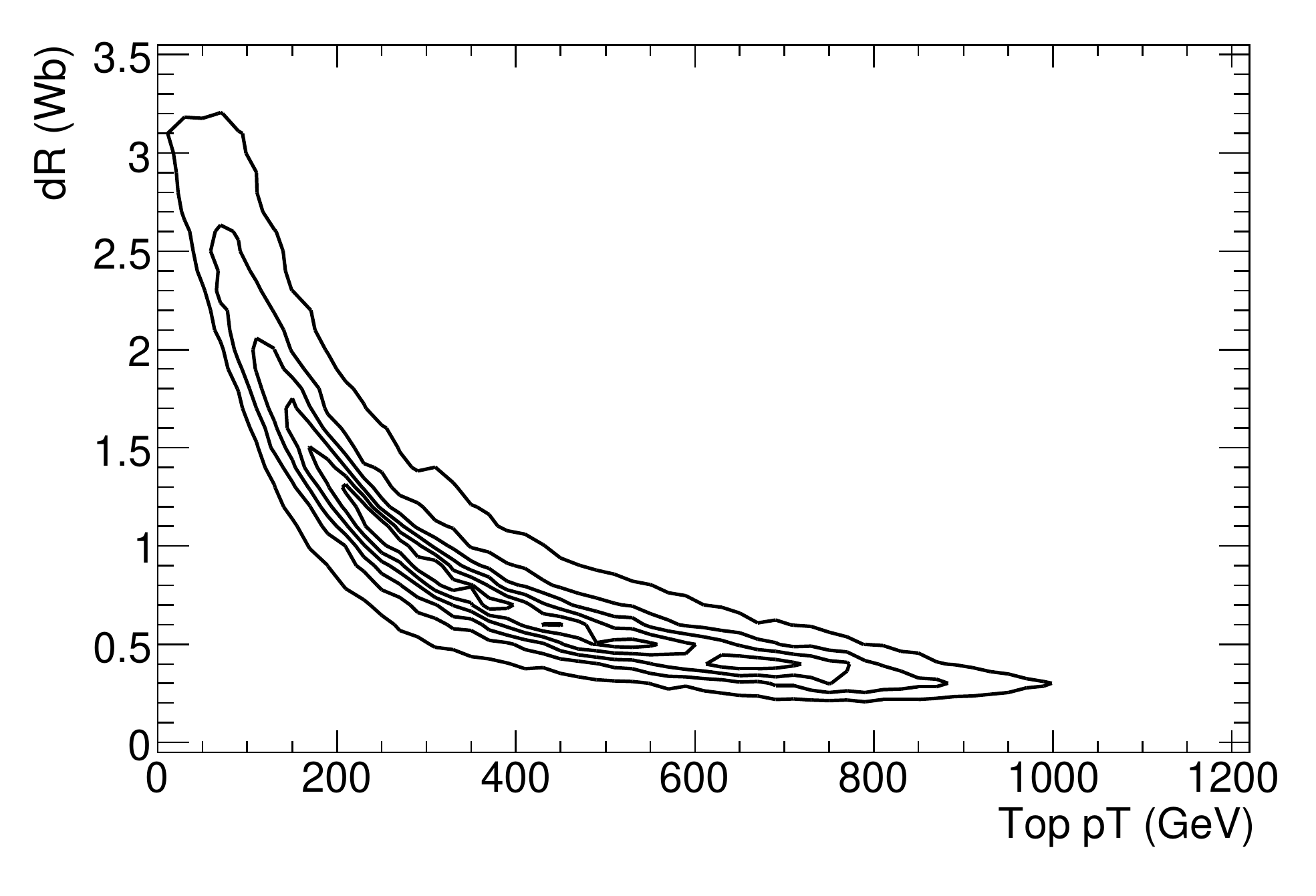}
}
\subfigure[]{
\label{hpttop_fig:dr2}
\includegraphics[width= 
0.48\textwidth]{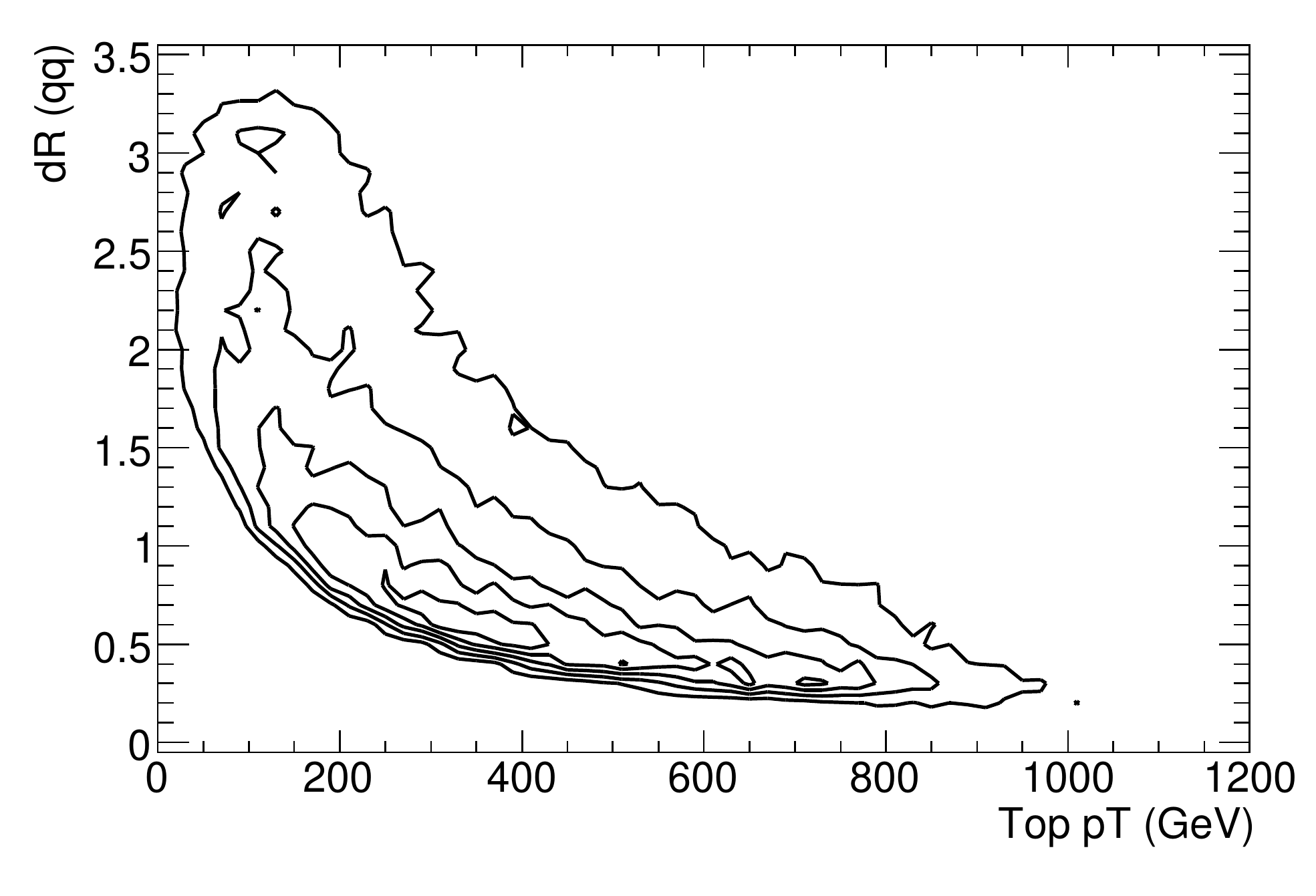}
}
\caption{Angular distances between decay products in top quark decays as a function of
top quark transverse momentum: (a)
between the b quark and $W$ boson, and (b) between quarks from $W$ boson decays.
\label{hpttop_fig:drbW}}
\end{center}
\end{figure}

At the LHC, top quarks, $W$ and $Z$ bosons are {\it relatively} light and can 
be produced with very high transverse momenta with respect to their masses.  In 
the case of hadronic decays, the quarks can be so close together in the detector that
they are in principle reconstructed as a single jet.  This is illustrated in 
Figure~\ref{hpttop_fig:drbW}, which shows generator-level distributions
of angular distance between the b quark and $W$ boson, and quarks from $W$ boson 
decays in $t\bar{t}$ events.  Here $dR = \sqrt{(\Delta \phi)^2 + (\Delta\eta)^2}$
with $\phi$ the azimuthal angle and $\eta$ the rapidity.  For top 
quark transverse momenta larger than about 200 GeV, the distance between the 
decay products is often smaller than twice the typical jet radius.  These events
were generated using PYTHIA~\cite{Sjostrand:2000wi} and no top polarization effects are included.  Such effects
are very model dependent and therefore beyond the scope of this generic study.

Identifying top quarks at high transverse momentum with high efficiency is of 
particular interest in searches for new physics.  In addition to a number of recent 
theoretical models specifically proposing the existence of high mass resonances decaying 
dominantly to top quarks (see for 
example~\cite{Agashe:2006hk,Matsumoto:2006ws,Fitzpatrick:2007qr,Djouadi:2007eg,Contino:2008hi}), 
the large top quark mass suggests it might be 
closely linked to forms of new physics that would manifest themselves at very high 
energies.  It is therefore quite probable that new heavy objects decay to top 
quarks at least some fraction of the time, if not exclusively.

\section{DATASETS AND TOOLS}

The datasets used in this study are a) sequential standard model $Z'$-bosons of 
mass 2 and 3 TeV decaying to top-antitop pairs for signal, and b) multijet events with transverse momenta 
ranging from 300 to 2200 GeV for the background.  For the signal events, one of the 
top quarks is forced to decay hadronically, and the other semileptonically.  
The events were all generated using 
PYTHIA, passed through the full ATLAS detector simulation 
and reconstructed using the ATLAS reconstruction program.

\begin{figure}[htbp]
\begin{center}
\subfigure[]{
\includegraphics[width= 
0.30\textwidth]{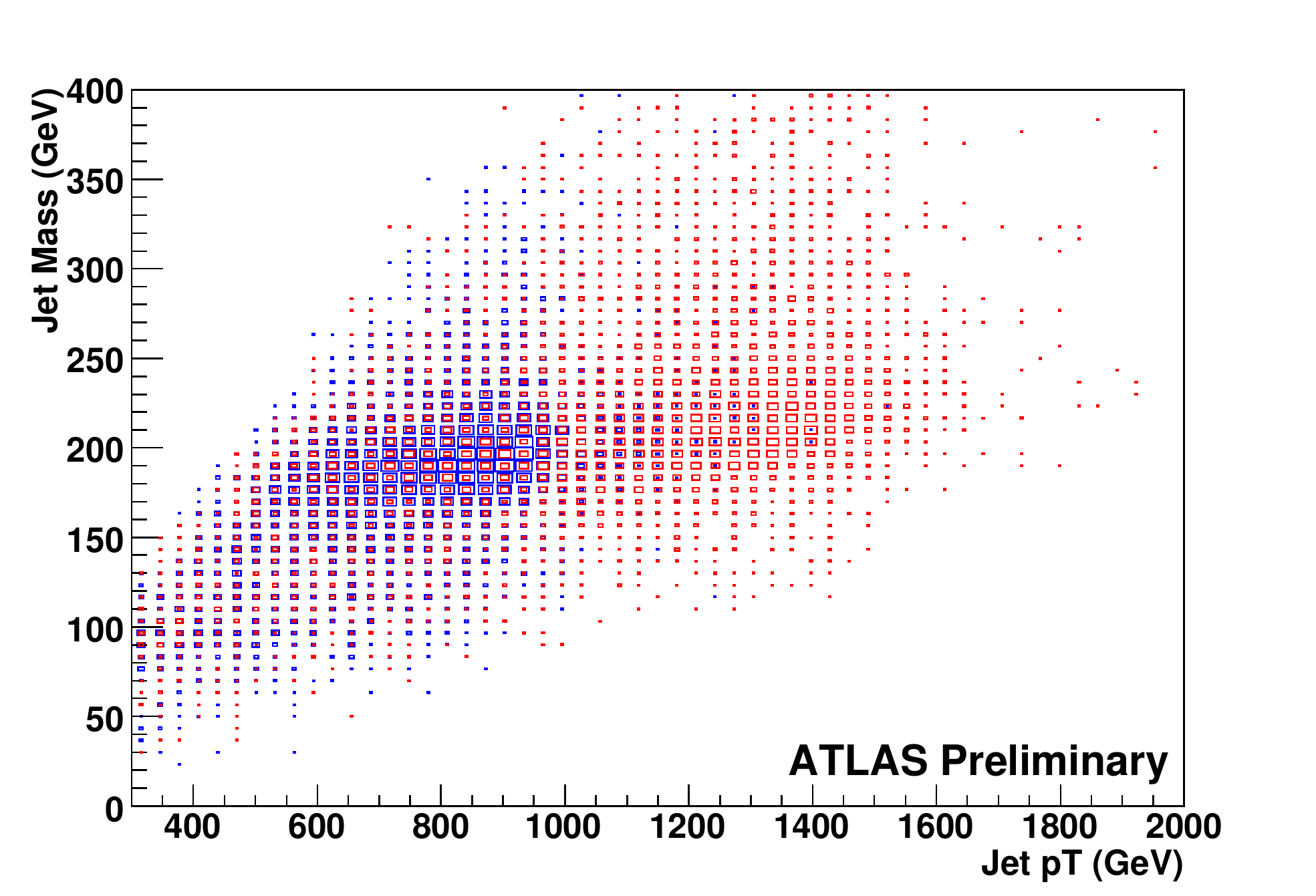}
}
\subfigure[]{
\includegraphics[width= 
0.30\textwidth]{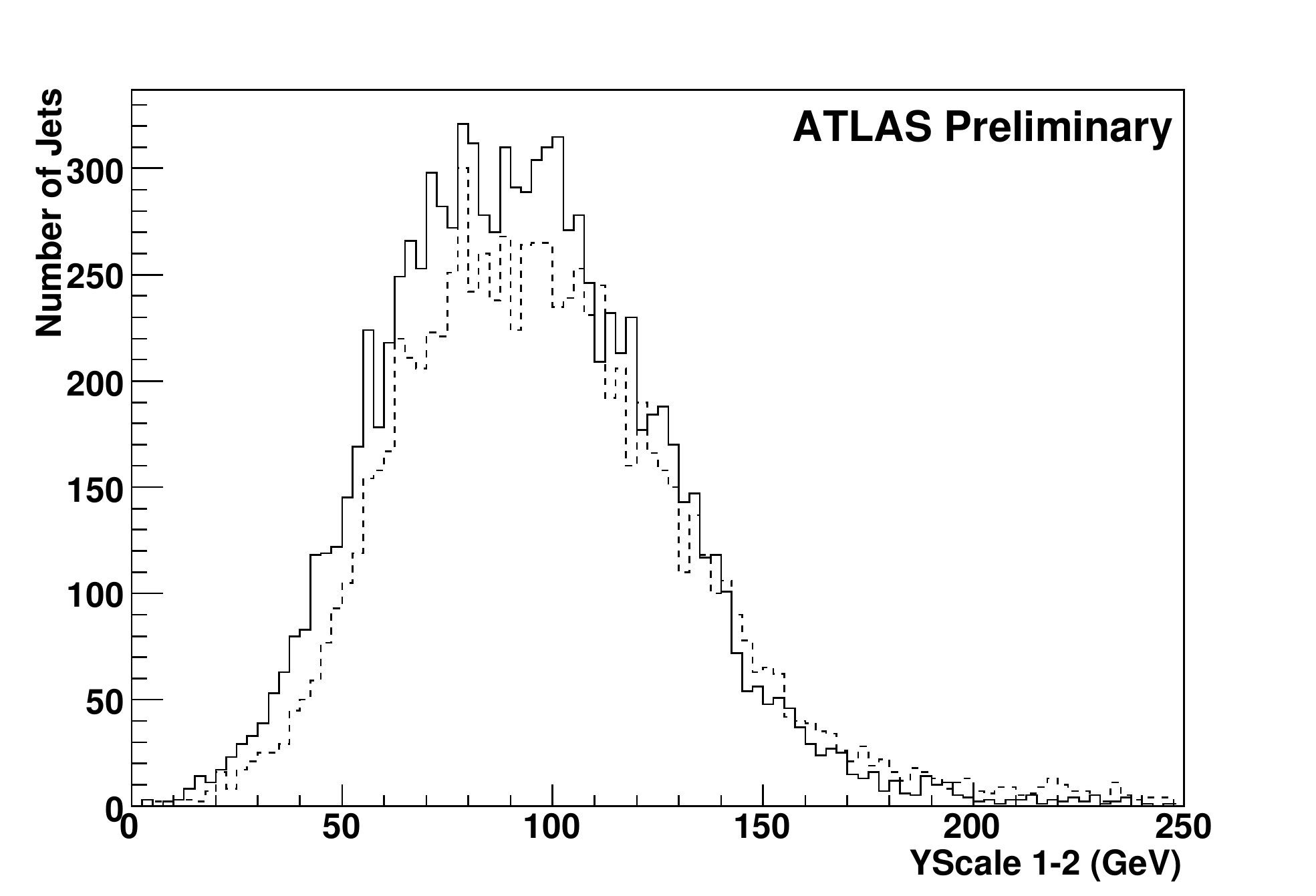}
}
\subfigure[]{
\includegraphics[width= 
0.30\textwidth]{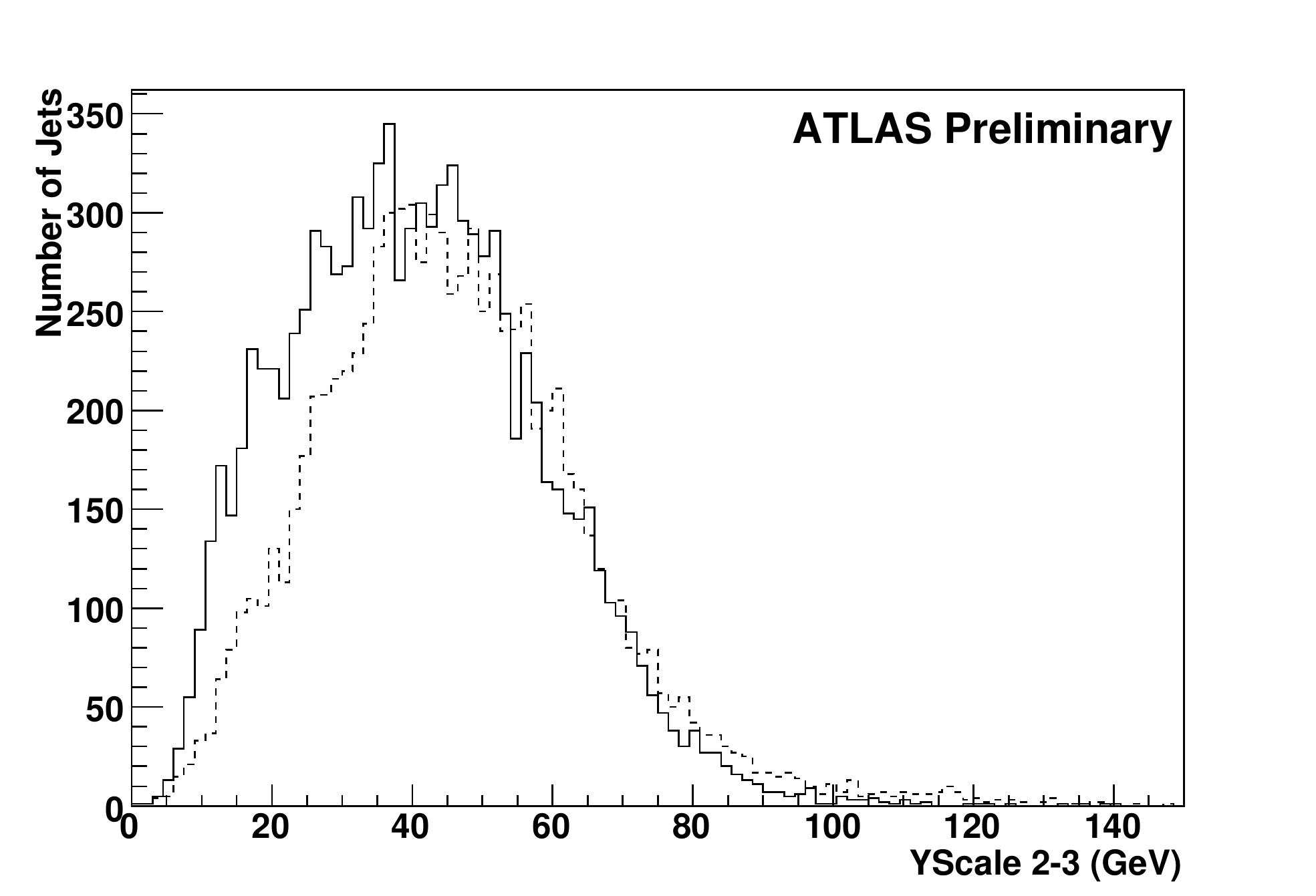}
}
\caption{(a) Jet mass as a function of transverse momentum for 
jets passing the top monojet selection: jets from the M = 2 (3) TeV $Z'$ sample in blue (red).
(b) and (c) Solid (dashed): scales at which the top monojet splits into two and three jets respectively for 
events in the $M_{Z'} = 2 (3)$ TeV samples.}
\label{hpttop_fig:signal}
\end{center}
\end{figure}

\begin{figure}[htbp]
\begin{center}
\subfigure[]{
\includegraphics[width= 
0.30\textwidth]{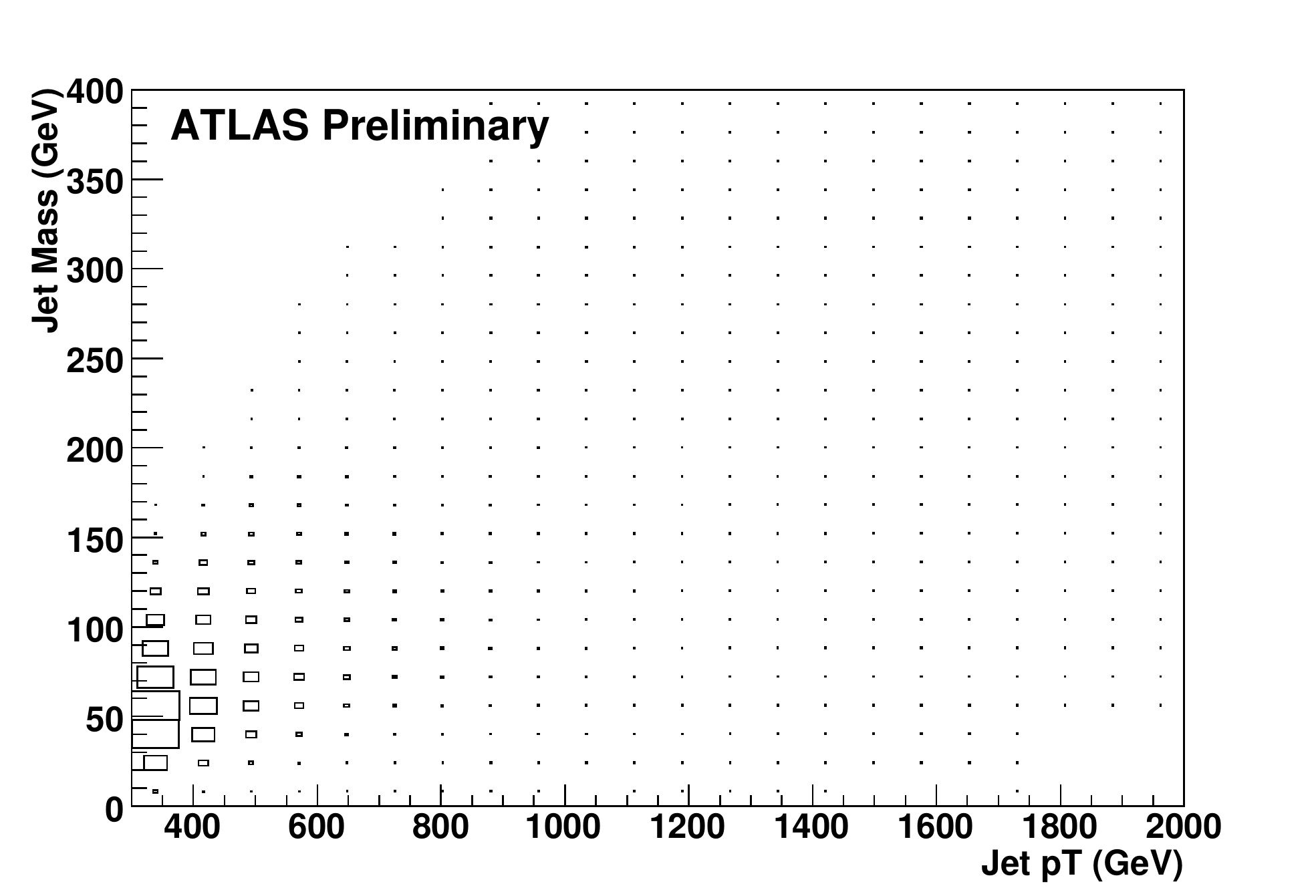}
}
\subfigure[]{
\includegraphics[width= 
0.30\textwidth]{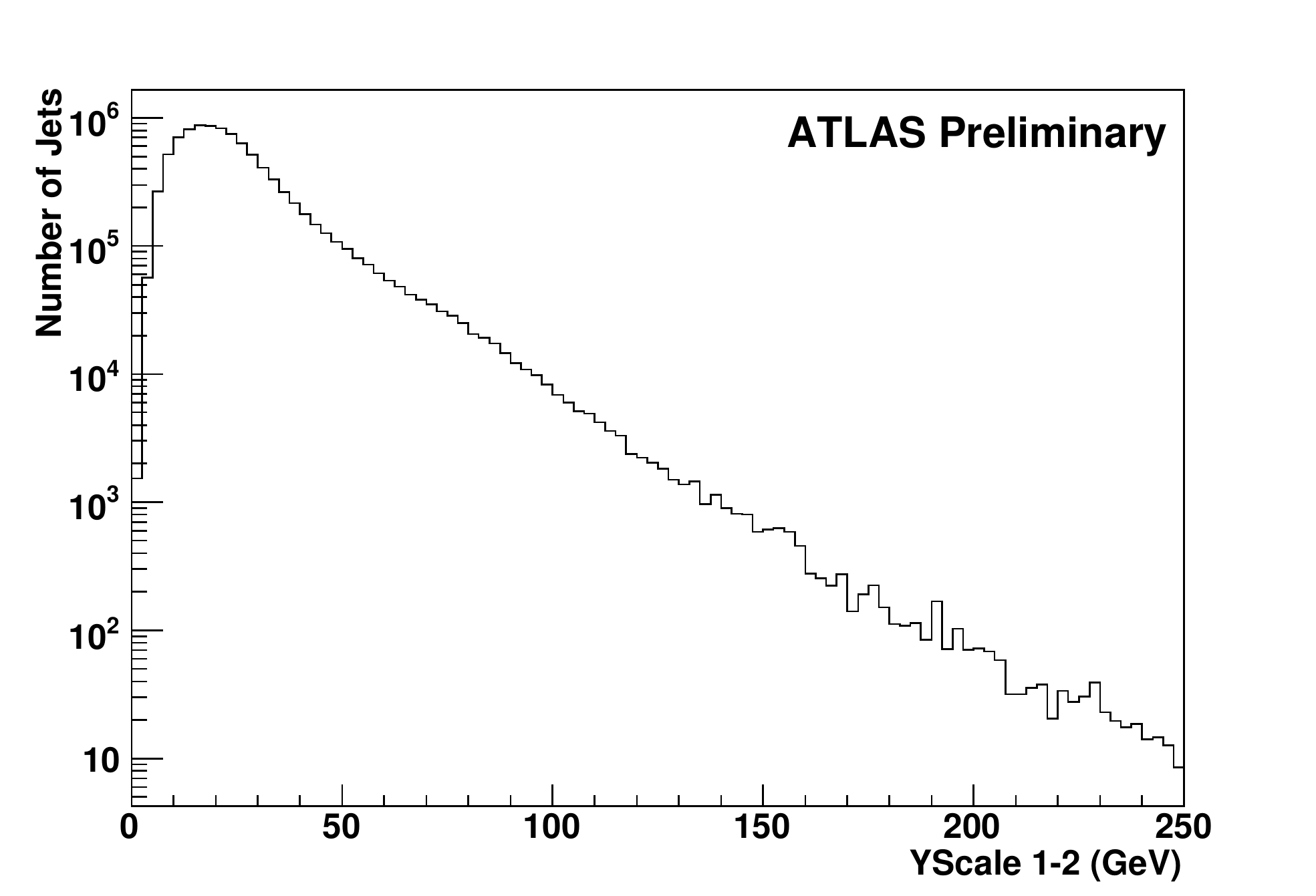}
}
\subfigure[]{
\includegraphics[width= 
0.30\textwidth]{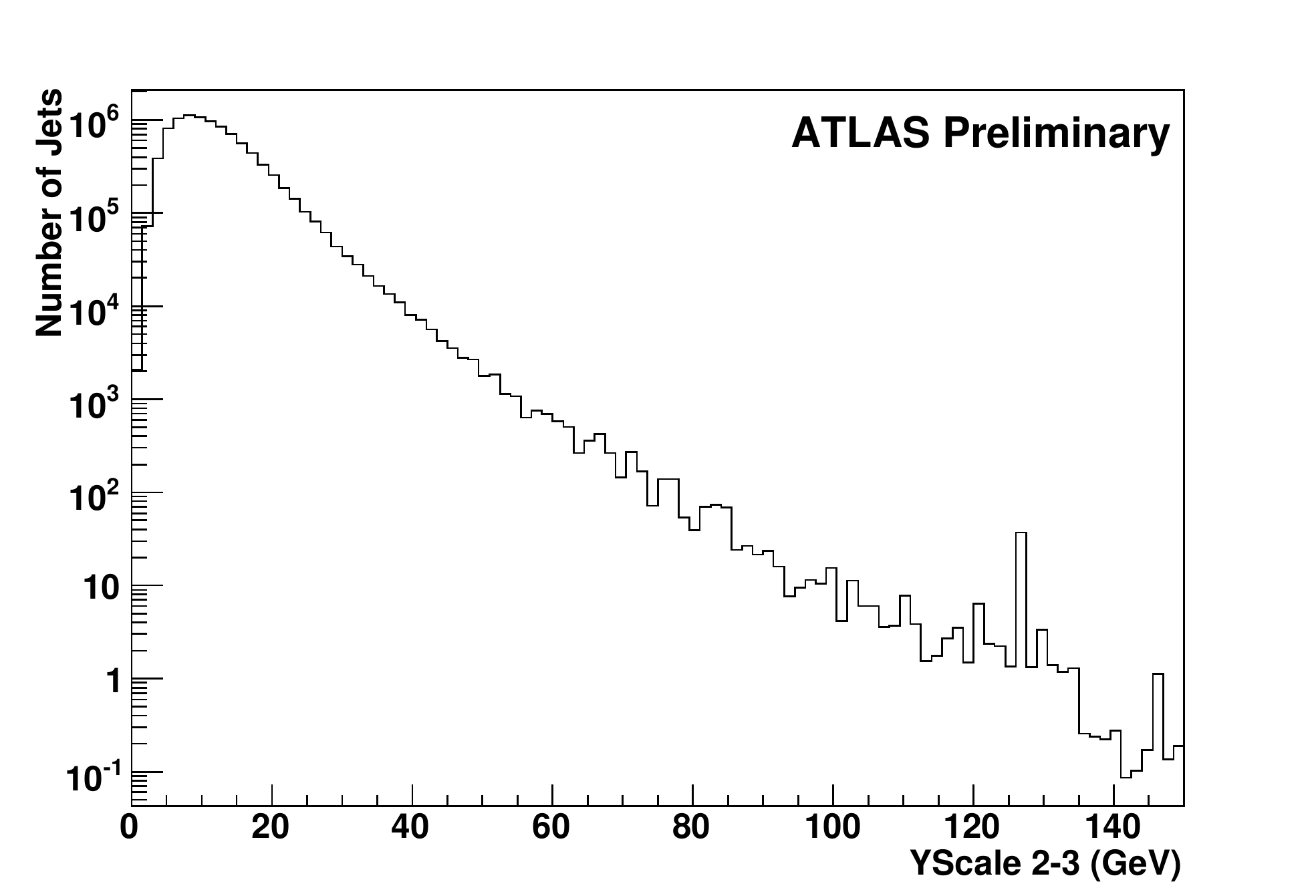}
}
\caption{(a) Jet mass as a function of transverse momentum, (b) and (c) scales 
at which the jet splits into two and three jets respectively for 
events in the background samples.}
\label{hpttop_fig:bckground}
\end{center}
\end{figure}

The jet mass, which is the 
invariant mass of all the jet's constituents (typically calorimeter cells or towers), and 
``YSplitter''~\cite{Butterworth:2002tt}, which determines the scales at which jets can be resolved 
into two or more subjets are used as discriminating variables.  
Top ``monojets'' are selected in the signal samples by selecting events in which only one 
jet with $p_T > 20$ GeV
has $dR < 1.0$ from the closest top quark and $dR < 2.0$ from the hadronically decaying 
$W$ boson.  For these jets, Figure~\ref{hpttop_fig:signal} shows the distribution of jet mass as
a function of jet transverse momentum, and the scales at which the jet splits up into two and 
three jets.
The distributions have only limited dependence on the jet transverse momentum distribution and are 
therefore well suited to the identification of top monojets over a wide spectrum.  The splitting scales
into two and three jets cluster around half the top quark mass and half the $W$-boson mass respectively as
expected.  The same 
distributions are shown for the background samples in Figure~\ref{hpttop_fig:bckground}.

\section{QUANTITATIVE ANALYSIS}

In principle, with this number of variables a multivariate tool like an artificial neural network 
might yield optimal results in terms of high/low efficiency for signal/background.  For the sake of 
clarity however this study is based on simple two-dimensional cuts, keeping in mind that the use
of multivariate tools typically leads to a factor of approximately 1.5 improvement in the signal 
over background ratio.

\begin{figure}[htbp]
\begin{center}
\subfigure[]{
\includegraphics[width= 
0.30\textwidth]{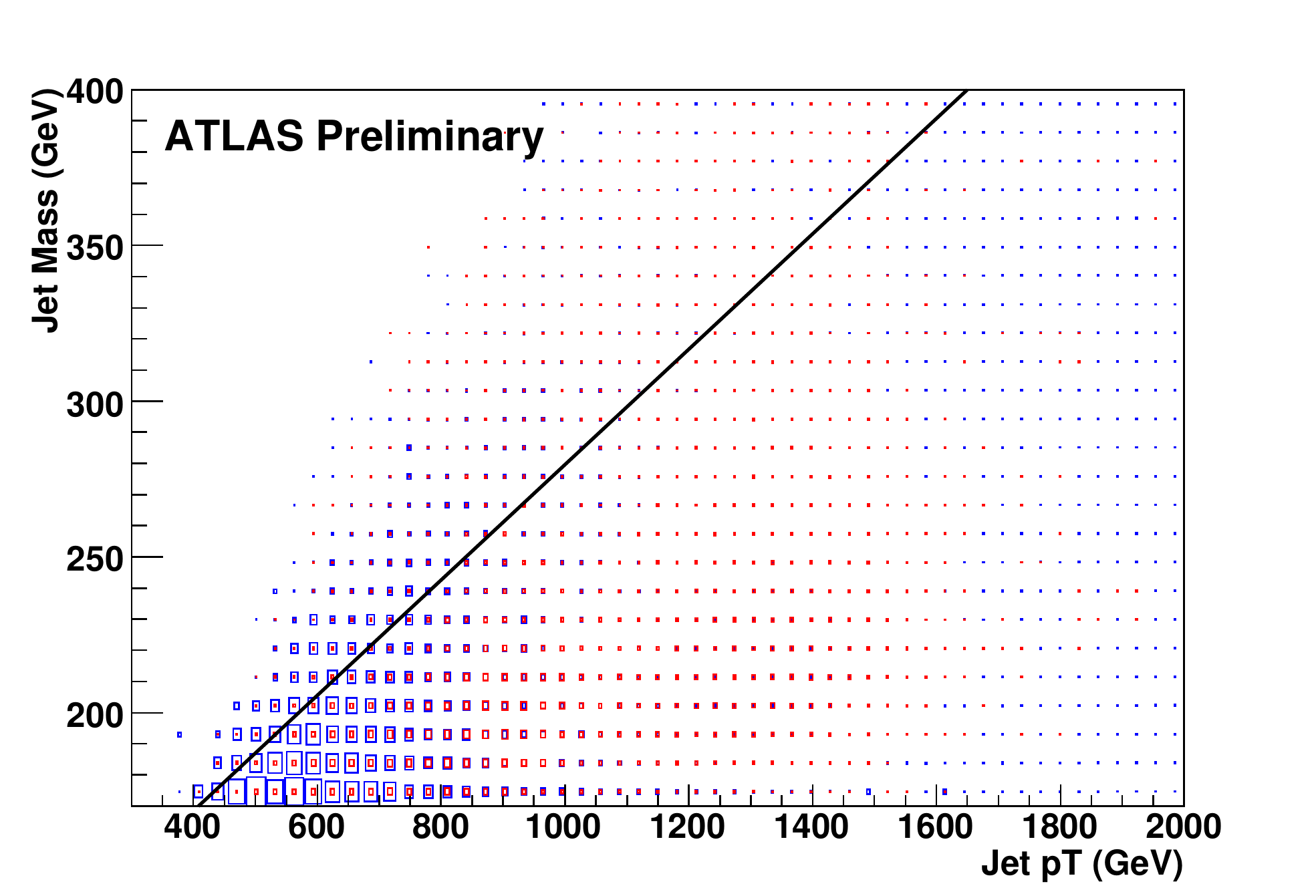}
}
\subfigure[]{
\includegraphics[width= 
0.30\textwidth]{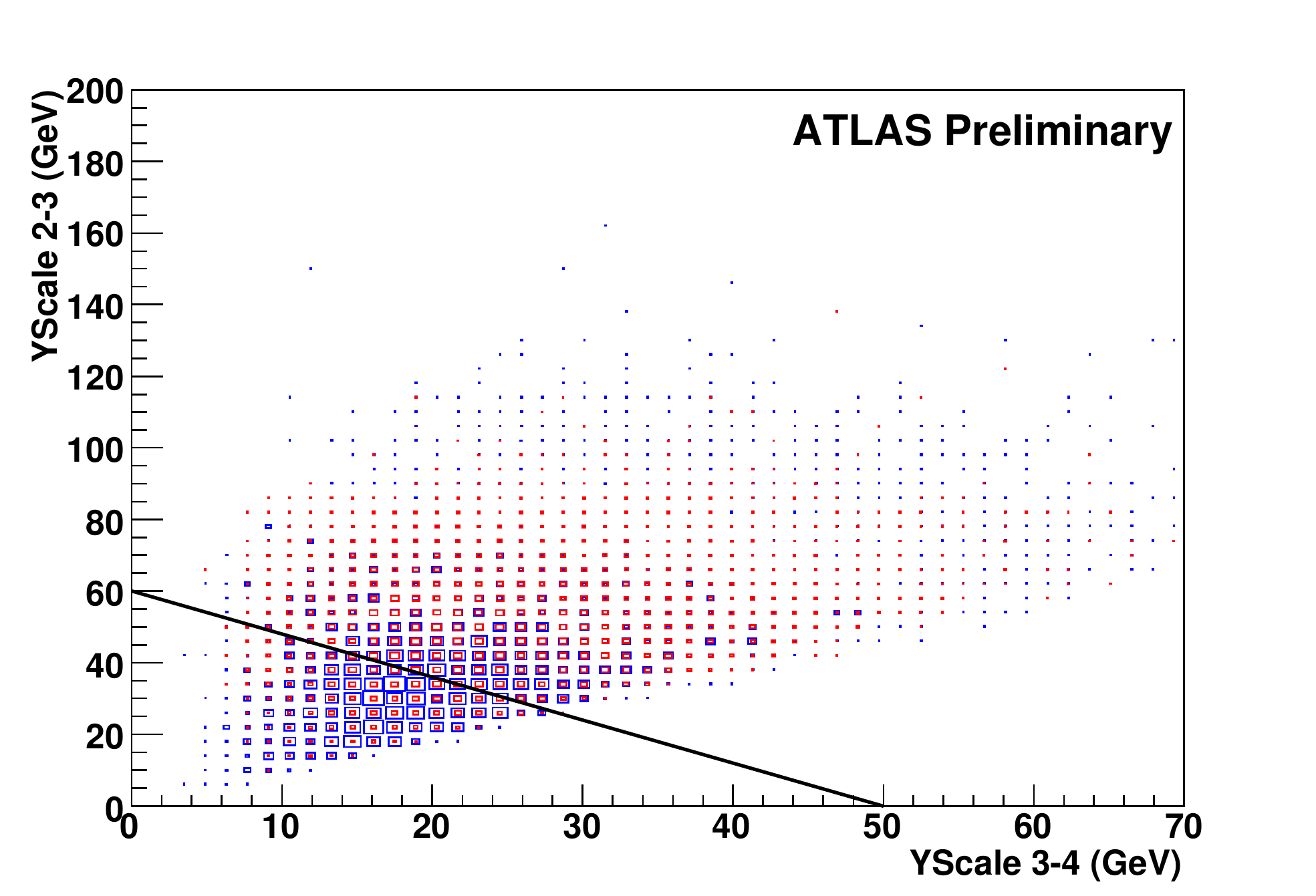}
}
\subfigure[]{
\includegraphics[width= 
0.30\textwidth]{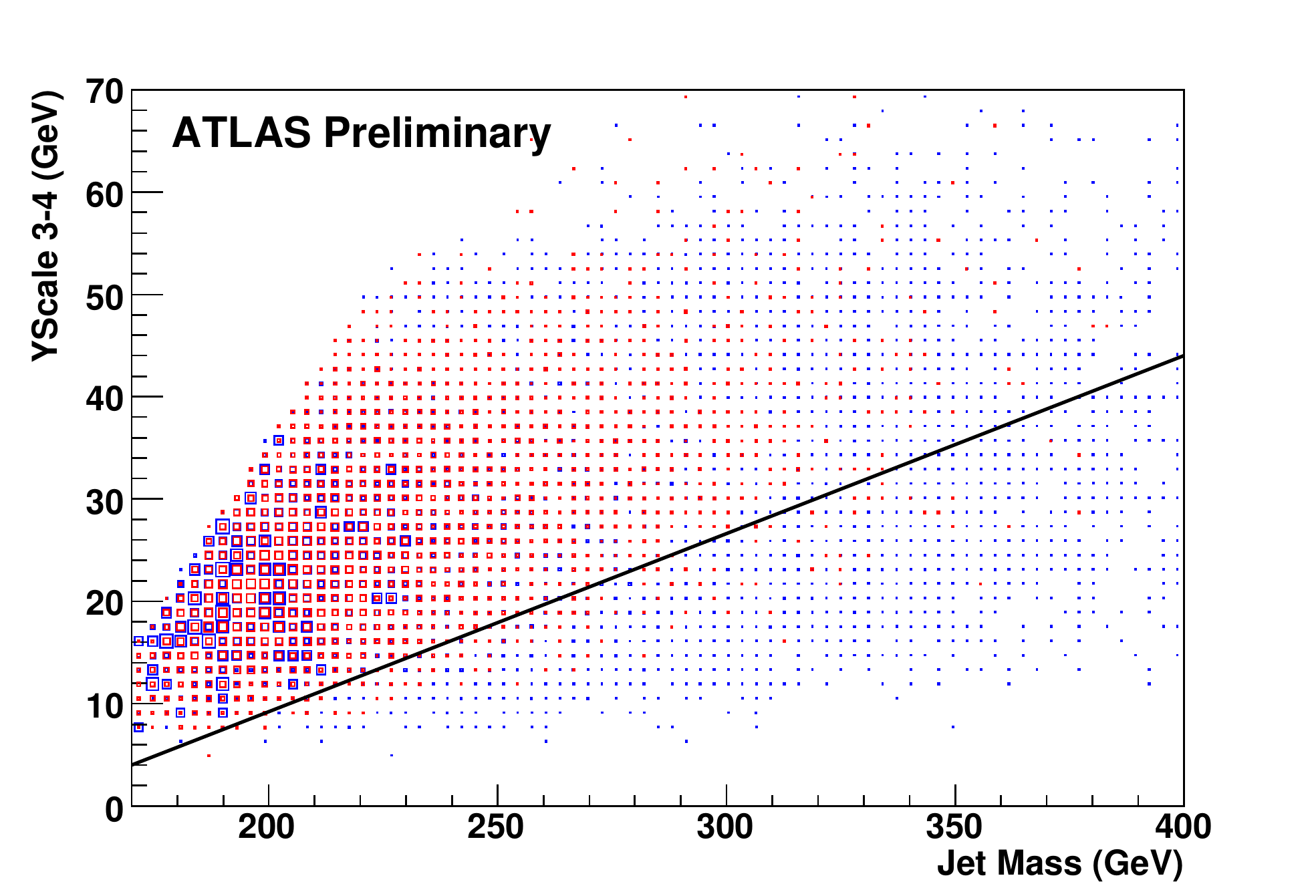}
}
\caption{Cuts applied on (a) jet mass as a function of jet transverse momentum (events are required
to lie below the line), (b) $YScale_{23}$ as a function of $YScale_{34}$ (events are required to lie 
above the line), and (c) $YScale_{34}$ as a function of jet mass (events are required to lie 
above the line).  In each plot, the background is in blue and the signal in red.}
\label{hpttop_fig:cuts}
\end{center}
\end{figure}

\begin{figure}[htbp]
\begin{center}
\includegraphics[width= 
0.60\textwidth]{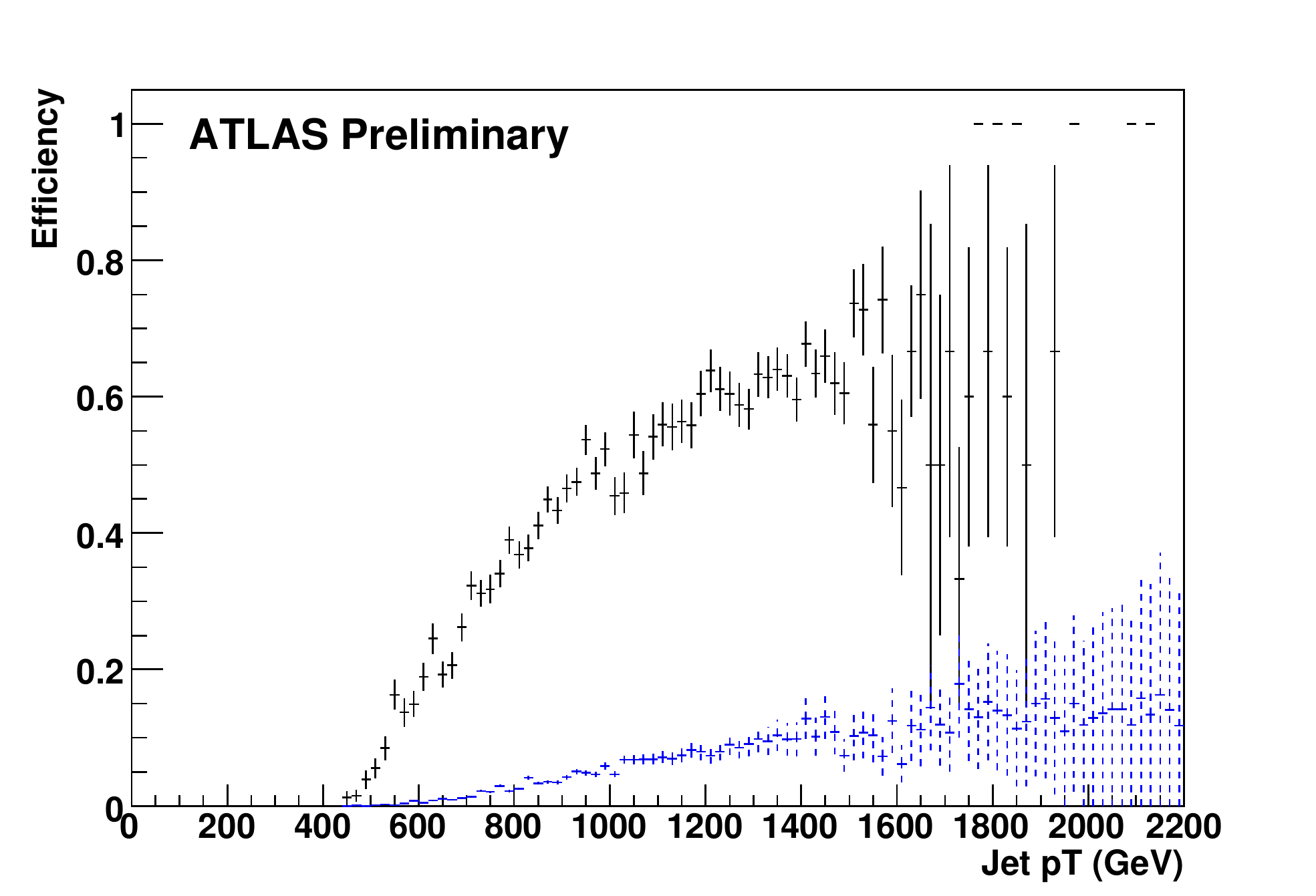}
\caption{Selection efficiency as a function of transverse momentum for top monojets (solid, black)
and jets in the background samples (blue, dashed).}
\label{hpttop_fig:results}
\end{center}
\end{figure}

\noindent Cuts are applied on 
\begin{enumerate}
\item Jet mass ($> 170$ GeV),
\item jet mass as a function of jet transverse momentum,
\item YScale 1-2 (split from one to two jets) as a function of YScale 2-3 (split from two to three jets),
\item YScale 2-3 as a function of YScale 3-4,
\item YScale 1-2 as a function of jet mass,
\item YScale 2-3 as a function of jet mass, and
\item YScale 3-4 as a function of jet mass (two cuts).
\end{enumerate}
As an illustration, three of these (2, 4 and 7) are shown in Figure~\ref{hpttop_fig:cuts}.
The resulting efficiencies for signal and background events are shown in Figure~\ref{hpttop_fig:results}.
For the background the efficiency plateaus at about 10\% starting at jet transverse momenta of 1300 GeV, 
where the signal efficiency reaches 60\%.  The ratio of signal over background efficiency increases for
smaller jet transverse momenta.
A more detailed description of this analysis is available in~\cite{hpttop_atlas}.

\section{CONCLUSIONS}

In this analysis, the measured jet mass and jet splitting scales have been used to 
distinguish high transverse momentum ``top monojets'' from jets originating from 
light quarks.  The combination of algorithms allows for good separation of signal 
and background, with the ratio of selection efficiencies for signal and background
evolving from approximately 30 for jets with $p_T =$ 600 GeV to 10 for 1000 GeV 
and 7 for 1500 GeV.  Further work using subjets and tracking information is underway.

\section*{ACKNOWLEDGEMENTS}

This work has been performed within the ATLAS Collaboration, and we thank collaboration members 
for helpful discussions. We have also made use of the physics analysis framework and tools which are 
the result of collaboration-wide efforts. 



\AddToContent{G.~Brooijmans}
\setcounter{figure}{0}
\setcounter{table}{0}
\setcounter{section}{0}
\setcounter{equation}{0}
\setcounter{footnote}{0}
\clearpage

\superpart{A Les Houches Interface}


\part{A Les Houches Interface for BSM Generators}

{\it J.~Alwall, E.~Boos, L.~Dudko, M.~Gigg, M.~Herquet,
A.~Pukhov, P.~Richardson, A.~Sherstnev and 
P.~Skands}

\begin{abstract}
We propose to combine and slightly extend two existing ``Les Houches
Accords'' to provide a simple generic interface between
beyond-the-standard-model parton-level and event-level generators. All
relevant information --- particle content, quantum numbers
of new states, masses, cross sections, parton-level
events, etc --- is collected in one single file, which adheres to the 
Les Houches Event File (LHEF) standard.
\end{abstract}

\section{INTRODUCTION}

The simulation of interactions at the LHC is characterized by the use of
many different programs specializing in different stages of
the calculation, such as matrix-element-level event generation, decay
of resonances, parton showering, hadronization, and underlying event
simulation. The communication of simulation parameters between those
stages can be complicated and program-specific. For supersymmetric
models, this situation has been greatly improved by the introduction
of the SUSY Les Houches Accord \cite{Skands:2003cj} (SLHA) and its upcoming
extension \cite{Allanach:2007zz}. For general models however, there is
still no corresponding standard. In this note, we suggest an addition
to the SLHA to allow for the specification of the
quantum numbers, masses, and decays of arbitrary new states, thus
generalizing the accord beyond its original supersymmetry-specific
scope. We also make a proposal for how to include these model
parameter files into Les Houches Accord event files 
\cite{Alwall:2006yp} (LHEF) in a standardized way. This 
both reduces the number of files that need to be passed
around and minimizes the possibility 
for error by keeping all relevant model information together with the
actual events. 

\section{DEFINITION OF THE INTERFACE}
The concrete proposal consists of the following three points:
\begin{enumerate}
\item Introduce new SLHA-like blocks \interfacettt{QNUMBERS} (for ``quantum
numbers'') with the format: 
\begin{verbatim}
BLOCK QNUMBERS 7654321 # balleron
    1     0  # 3 times electric charge
    2     1  # number of spin states (2S+1)
    3     1  # colour rep (1: singlet, 3: triplet, 8: octet)
    4     0  # Particle/Antiparticle distinction (0=own anti)
\end{verbatim}
where this example pertains to a fictitious neutral spin-0
color-singlet self-conjugate particle to which we assign ``PDG'' code
7654321 and the name ``balleron''. That is, the \interfacettt{BLOCK}
declaration should define a PDG code and, optionally, a human readable
name after the \interfacettt{\#} character (if no name is given, the
PDG code may be used). 
We advise to choose PDG numbers in excess of 
3 million for new states, to minimize the  possibility of
conflict with already agreed-upon numbers \cite{Yao:2006px}. 
The entries so far defined are: \interfacettt{1}: the electric
charge times 3 (so that most particles will have integer values, but
real numbers should also be accepted); \interfacettt{2}: 
the particle's number of spin states: $2S+1$; \interfacettt{3}: the colour
representation of the particle, e.g., 1 for a 
singlet, 3 (-3) for a triplet (antitriplet), 8 for an octet, etc.; 
\interfacettt{4}: particle/antiparticle distincition, should be 0 (zero) if the
particle is its own antiparticle, or 1 otherwise. 
\item Use the existing SLHA blocks
\interfacettt{MASS} and \interfacettt{DECAY}  \cite{Skands:2003cj} 
to define particle masses and decay tables. If the model in question
is a SUSY model, a full SLHA spectrum 
\cite{Skands:2003cj} can also be included. 
We propose that the reader should ``turn on'' SUSY whenever
the SLHA SUSY model definition block \interfacettt{MODSEL} is present. 
\item Include the information from points 1 and 2 enclosed within the subtags
\interfacettt{<slha> </slha>} in the \interfacettt{<header>} part of Les Houches event files 
\cite{Alwall:2006yp}. 
\end{enumerate}

\section{IMPLEMENTATIONS}
For the purpose of this contribution, the above proposal was tested
explicitly by interfacing \interfacetsc{MadGraph/MadEvent} with
\interfacetsc{Pythia}. Below we summarize the main aspects of these
implementations. 

\subsection{MadGraph/MadEvent implementation}
Starting from version 4 \cite{Alwall:2007st}, the multi-purpose
\interfacetsc{MadGraph/MadEvent} parton-level event generator by
default includes a detailed summary of all simulation parameters in
the output LHEF \cite{Alwall:2006yp} parton-level event file. From
version 4.1.47, this information is stored in the XML
\texttt{<header>} section. For the interface
considered here, the relevant part of this section is a copy of the
so-called \interfacettt{param\_card.dat} \interfacetsc{MG/ME} input file.

The \interfacetsc{MG/ME} \interfacettt{param\_card.dat} uses an extension of the SUSY
Les Houches Accord \cite{Skands:2003cj,Allanach:2007zz} for model
parameters in all implemented models. In particular, it always
includes the \interfacettt{SMINPUTS}, \interfacettt{MASS}, and \interfacettt{DECAY} blocks. This
file is used by \interfacetsc{MadGraph/MadEvent} as an input for cross
section computations and event generation but is not modified by the
program. The file is instead assumed to be created by an external
``Model Calculator''. Such calculators are currently available on the
web for the SM, MSSM and 2HDM models. Starting from the parameters in
the Lagrangian (primary parameters), they calculate all needed
secondary parameters (such as masses, decay widths, and auxiliary
parameters). Note that widths and branching ratios can also be
evaluated in an intermediate step by \interfacetsc{MG/ME} itself or by 
external tools like \interfacettt{DECAY} or \interfacettt{BRIDGE} \cite{Meade:2007js}.

In previous versions of \interfacetsc{MadGraph/MadEvent}, the
\interfacettt{param\_card.dat} file did not contain information regarding the
particle content of the physical model considered. This
information is stored in the \interfacettt{particle.dat} file filled by model
writers during the model creation. Starting from version 4.1.43, the
template for inclusion of user defined models (called \interfacettt{USRMOD}) in
\interfacetsc{MadGraph/MadEvent} automatically generates the \interfacettt{QNUMBERS} blocks
described above from the information contained in the
\interfacettt{particle.dat} file. These blocks are then included in the default
\interfacettt{param\_card.dat} for the new model (and from there are copied
into the LHEF output), such that no extra
intervention is required to pass them to parton shower programs after
parton-level event production. The script only outputs information for
particles which have PDG numbers not identified as standard SM or MSSM
particles, since those are assumed to be defined in the parton shower
generators.

Note that in the current version, the spin, color and
particle/antiparticle information is automatically extracted, but not
the electric charge, which is set to zero by default. This is due to
the fact that, in \interfacetsc{MadGraph/MadEvent}, the electric charge does
not appear in the list of particle properties and is only defined
through the value of the coupling to the photon. This issue will be
addressed in future versions of \interfacettt{USRMOD}, but can currently be
circumvented by fixing the electric charge information by hand at the
end of the model implementation process.

\subsection{Pythia implementation}
The following capabilities are implemented in \interfacetsc{Pythia 6.414}
\cite{Sjostrand:2006za} and subsequent versions. 

Already for some time it has been possible 
to use the \interfacettt{QNUMBERS} blocks described above to define new
particles in \interfacetsc{Pythia} via its SLHA interface \cite{Pukhov:2005je}.
What is new is that, when reading an LHEF event file,
\interfacetsc{Pythia} now automatically searches for \interfacettt{QNUMBERS} blocks in the header
part of the LHEF file, updating its internal particle data tables
accordingly. It then proceeds to search for \interfacettt{MASS} and \interfacettt{DECAY}
tables, and finally looks for other SLHA blocks contained in the header. 
If the SUSY model definition block \interfacettt{MODSEL} is found, SUSY is
automatically switched on and the remaining SLHA blocks are read, without the
user having to intervene. The read-in of LHEF files containing general
BSM states, masses, and decay tables, should therefore now be relatively
`` plug-and-play''.  

A note on decay tables: only 2- and 3-body decays can currently be handled
consistently. They are then generated with flat phase space, according
to the branching ratios input via the \interfacettt{DECAY} tables. The colour
flow algorithms have been substantially generalized, but if too many
coloured particles are involved (e.g., an octet decaying to three
octets) \interfacetsc{Pythia} will still not be able to guess which colour flow to
use, leading to errors. Please also read the warnings in the section
on decay tables in the SLHA report \cite{Skands:2003cj} 
concerning the dangers of
double counting partial widths and obliterating resonance shapes. 
To get around the
restriction to flat phase space, either 1) use \interfacetsc{Pythia}'s 
internal resonance decays whenever possible 
(e.g., do not read in decay tables for particles for which
\interfacetsc{Pythia}'s internal treatment is not desired modified), 2) perform
the decays externally, before the event is handed to 
\interfacetsc{Pythia} (e.g., with \interfacetsc{MadGraph/Bridge}
\cite{Meade:2007js} or \interfacetsc{CalcHEP} \cite{calc-lhs-website}),  
or 3) do a post facto re-weighting of the
generated events, based on the kinematics of the particle decays stored
in the event record. 

The interfaces can of course still also be used stand-alone,
independently of LHEF. The user must then manually open a 
spectrum file containing \interfacettt{QNUMBERS}
and \interfacettt{MASS} information and give \interfacetsc{Pythia} the logical
unit number in \interfacettt{IMSS(21)}. 
New states can then be read in via either of the calls\\
\interfacecode{CALL PYSLHA(0,KF,IFAIL)~~~~! look for QNUMBERS for PDG = KF}
\interfacecode{CALL PYSLHA(0,0,IFAIL)~~~~~! read in all QNUMBERS}
and \interfacettt{MASS} information can be read by\\
\interfacecode{CALL PYSLHA(5,KF,IFAIL)~~~~! look for MASS entry for PDG = KF}
\interfacecode{CALL PYSLHA(5,0,IFAIL)~~~~~! read in all MASS entries}
where \interfacettt{IFAIL} is a standard return code, which is zero if
everything went fine. (For read-in of a complete SLHA SUSY spectrum file,
these direct calls should not be used, instead  set \interfacettt{IMSS(1)=11}
before the call to \interfacetsc{PYINIT}.) For stand-alone decay table read-in, 
the unit number of the SLHA decay table file 
should be given in \interfacettt{IMSS(22)}, and the corresponding
read-in calls are\\
\interfacecode{CALL PYSLHA(2,KF,IFAIL)~~~~! look for DECAY table for PDG = KF}
\interfacecode{CALL PYSLHA(2,0,IFAIL)~~~~~! read in all DECAY tables}

\section{CONCLUSIONS AND OUTLOOK}
We have proposed a simple file-based interface between parton- and
event-level generators focusing on the particular problems encountered
in the simulation of beyond-the-standard-model collider physics.  To
deal with general BSM models, we add a new block \interfacettt{QNUMBERS} to the
SLHA structure, which defines the SM quantum numbers of new states for
use in subsequent resonance decay, parton showering, and hadronization
programs. We also integrate the SLHA file into the existing LHEF
format to minimize the number of separate files needed.  The proposal
has been tested explicitly by implementations in the
\interfacetsc{MadGraph/MadEvent} and \interfacetsc{Pythia6} Monte Carlo event generators.

In the near future, also the \interfacetsc{Herwig++} \cite{Bahr:2007ni} and
\interfacetsc{Pythia8} \cite{Sjostrand:2007gs} generators will be extended to
automatically read in SLHA spectra from LHEF headers. Likewise,
forthcoming versions of the \interfacetsc{CalcHEP} \cite{Pukhov:2004ca} and
\interfacetsc{CompHEP} \cite{Boos:2004kh} parton-level generators will include
write-out of this information in their LHEF output, including also the
\interfacettt{QNUMBERS} extension.

In the longer term, with the XML format emerging as the de
facto standard for file-based interfaces, we note that it could be worth
investigating the merits of formulating an XML-SLHA scheme, that is,
transforming the current ASCII SLHA format conventions into a native XML
form that could be parsed with standard XML packages. A concrete 
first realization of such a strategy is HepML~\cite{Belov:2007qg} which
aims to unify the description of generator information in the form of
standard XML schemes, in which an XML-SLHA scheme would form a natural part.
The first release of the public HepML library 
has been implemented into CompHEP version 4.5, including 
also HepML headers in the LHEF output.

\section*{ACKNOWLEDGEMENTS}
The proposal and implementations contained herein 
originated at the workshop ``Physics at TeV Colliders'', Les Houches,
France, 2007.  This work has been partially supported by Fermi
Research Alliance, LLC, under Contract No.\ DE-AC02-07CH11359 with the
United States Department of Energy. The work of MH was
supported by the Institut Interuniversitaire des Sciences Nucl\'eaires
and by the Belgian Federal Office for Scientific, Technical and
Cultural Affairs through the Interuniversity Attraction Pole
P6/11. The HepML project is supported by the RFBR-07-07-00365 grant.
The work of JA was was supported by the Swedish Research
Council.

\AddToContent{J.~Alwall et al.}
\setcounter{figure}{0}
\setcounter{table}{0}
\setcounter{section}{0}
\setcounter{equation}{0}
\setcounter{footnote}{0}
\clearpage
%


\providecommand{\href}[2]{#2}

\end{document}